\documentclass{article}

\usepackage{format_arxiv/arxiv}
\usepackage[english]{babel}

\usepackage{natbib}
\usepackage{amsmath}
\usepackage{bm}
\usepackage{amssymb}
\usepackage{amsthm}

\usepackage{graphicx}
\usepackage{float}
\usepackage{tabularx}
\usepackage{multirow}
\usepackage{booktabs}
\usepackage{makecell}
\usepackage{soul}
\usepackage[colorlinks=true, allcolors=blue]{hyperref}

\usepackage{tikz}
\usepackage{comment}
\usetikzlibrary{shapes.geometric, shapes.misc, arrows.meta, positioning}
\usepackage{amsmath}
\usepackage{bm}
\usepackage{placeins}
\usepackage{subcaption}

\definecolor{color1}{HTML}{0060AD} 
\definecolor{color2}{HTML}{FF4500} 
\definecolor{color3}{HTML}{FFA500} 
\definecolor{color4}{HTML}{006400} 
\definecolor{color5}{HTML}{9400D3} 
\definecolor{color6}{HTML}{800000} 
\definecolor{color7}{HTML}{000000} 
\definecolor{color8}{HTML}{0000FF} 
\definecolor{color9}{HTML}{FF0000} 
\definecolor{mycolor_blue}{RGB}{66,124,161}
\definecolor{mycolor_grey}{RGB}{198,198,198} 

\tikzstyle{line1} = [color=color7, thick]
\tikzstyle{line2} = [color=color1,densely dotted,thick]
\tikzstyle{line3} = [color=color9,dash dot,thick]
\tikzstyle{line4} = [color=color5,dash dot,thick]
\tikzstyle{line5} = [color=color4,dash dot dot,thick]
\tikzstyle{line6} = [color=color6,thick]

\tikzstyle{mark1} = [color=color4, mark=x,mark size=2pt,mark options=solid, thick] 
\tikzstyle{mark2} = [color=color7, mark=o,mark size=2pt,mark options=solid, thick]
\tikzstyle{mark3} = [color=color9, mark=triangle,mark size=2pt,mark options=solid,thick]
\tikzstyle{mark4} = [color=color8,mark=square,mark size=2pt,mark options=solid,thick]
\tikzstyle{mark5} = [color=color3,mark=diamond, mark size=2pt,mark options=solid,thick]
\tikzstyle{mark6} = [color=color5, mark=star,mark size=2pt,mark options=solid,thick]
\tikzstyle{mark7} = [color=color7,mark=*,mark size=2pt,mark options=solid,thick]
\tikzstyle{mark8} = [color=color7,mark=triangle,mark size=2pt,mark options=solid,thick]

\newcommand{\vect}[1]{\bm{#1}}
\newcommand{\tensor}[1]{\bm{\mathsf{#1}}}

\title{A Modified Moving Reference Frame Method for Propeller Resolution}

\author{
    \href{https://orcid.org/0009-0009-6909-1709}{\includegraphics[scale=0.06]{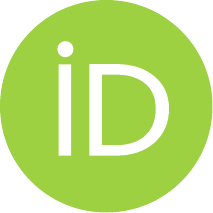}\hspace{1mm}Denis Andreev}\thanks{Corresponding author.} \\
    Institute for Fluid Dynamics and Ship Theory (M-8) \\
    Hamburg University of Technology \\
    Am Schwarzenberg-Campus 4, Hamburg, D-21073, Germany \\
    \texttt{denis.andreev@tuhh.de} \\
\And
    \href{https://orcid.org/0000-0001-8033-9861}{\includegraphics[scale=0.06]{pictures/orcid.pdf}\hspace{1mm}Georgios Bletsos} \\
    Institute for Fluid Dynamics and Ship Theory (M-8) \\
    Hamburg University of Technology \\
    Am Schwarzenberg-Campus 4, Hamburg, D-21073, Germany \\
    \texttt{george.bletsos@tuhh.de} \\
\And
    \href{https://orcid.org/0009-0009-3119-0218}{\includegraphics[scale=0.06]{pictures/orcid.pdf}\hspace{1mm}Antonios Kritikos} \\
    Institute for Fluid Dynamics and Ship Theory (M-8) \\
    Hamburg University of Technology \\
    Am Schwarzenberg-Campus 4, Hamburg, D-21073, Germany \\
    \texttt{antonios.kritikos@tuhh.de} \\
\And
    \href{https://orcid.org/0000-0002-4229-1358}{\includegraphics[scale=0.06]{pictures/orcid.pdf}\hspace{1mm}Niklas K\"uhl} \\
    Hamburg Ship Model Basin \\
    Bramfelder Strasse 164, Hamburg, D-22305, Germany \\
    \texttt{kuehl@hsva.de} \\
\And
    \href{https://orcid.org/0000-0002-3454-1804}{\includegraphics[scale=0.06]{pictures/orcid.pdf}\hspace{1mm}Thomas Rung} \\
    Institute for Fluid Dynamics and Ship Theory (M-8) \\
    Hamburg University of Technology \\
    Am Schwarzenberg-Campus 4, Hamburg, D-21073, Germany \\
    \texttt{thomas.rung@tuhh.de} \\
}

\renewcommand{\shorttitle}{A Modified MRF Method for Propeller Resolution}

\hypersetup{
    pdftitle={A Modified Moving Reference Frame Method for Propeller Resolution},
    pdfauthor={Denis Andreev, Georgios Bletsos, Antonios Kritikos, Niklas Kuehl, Thomas Rung},
    pdfkeywords={Propeller-hull interaction, Moving reference frame, Self-propulsion simulation, Computational fluid dynamics, RANS, Sliding interface},
}

\begin{document}
\maketitle

\begin{abstract}
    Accurate resolution of propeller-hull interaction is essential for predicting the self-propulsion point in ship CFD, yet motion-resolving methods such as sliding interfaces (SI) are computationally expensive, while the classical Moving Reference Frame (MRF) approach cannot capture unsteady interaction effects. Partially rotating grid methods bridge this gap by splitting the propeller rotation into a grid-resolved and an MRF component, but the abrupt transition between the rotating and stationary domains introduces discontinuities in the velocity field. This work presents a modified MRF (mMRF) formulation in which the reference-frame rotation rate is scaled by a spatially varying function that decays smoothly from unity near the propeller to zero at the domain interface, restoring velocity and pressure continuity across the boundary. The governing equations are derived and implemented in the RANS solver FreSCo$^+$, verified against the analytical Taylor--Couette solution, and applied to open-water propeller and Japan Bulk Carrier self-propulsion simulations at model scale. Both MRF and mMRF reproduce the principal integral propulsion quantities ($n$, $K_{\mathrm{T}}$, $K_{\mathrm{Q}}$, $1-t$, $1-w_{\mathrm{T}}$, $\eta_{\mathrm{R}}$) accurately, but the mMRF markedly reduces interface discontinuities and non-physical artifacts in the local flow field, particularly at large MRF fractions, at essentially the same computational cost.
\end{abstract}

\keywords{Propeller-hull interaction \and Moving reference frame \and Self-propulsion simulation \and Computational fluid dynamics \and RANS \and Sliding interface}

\FloatBarrier
\section{Introduction}
%
%

The primary mode of transportation for global trade is maritime transport, and ship efficiency has a direct effect on both its environmental footprint and cost. Fuel consumption is a major part of operating expenses, while engine emissions contribute to atmospheric pollutants.
As a result, ecological and economic concerns are increasingly important in ship design and operation.
Global regulatory organizations are more and more focusing on these issues. The International Maritime Organization (IMO) introduced the Energy Efficiency Design Index (EEDI, \citeyear{imo-2014}) for new ships and later the Energy Efficiency Existing Ship Index (EEXI, \citeyear{imo-2021}) for existing vessels. These regulations increase pressure to reduce emissions and encourage the use of high-fidelity methods for the simulation of ship propulsion.
In this context, computational fluid dynamics (CFD) is an important tool in the process of designing the shape of a hull, a propeller or an energy saving device (ESD) as well as optimizing 
their interaction.
This directly links CFD to environmental and economic goals.

%
%


%
%

Accurate modeling of the flow around the ship's stern and propeller is required for a reliable prediction of the self-propulsion point. Propellers generate complex three-dimensional and inherently unsteady flow structures that interact strongly with the hull, the rudder and the ESD, directly affecting thrust, torque and hull resistance. Consequently, the method used to approximate propeller flow has a decisive influence on the determination of the self-propulsion point.

Simulations that resolve the propeller motion, such as sliding interface (SI, \cite{blades-2007})  and overset grid,  cf. \cite{benek-1983}, provide high-fidelity resolution of the rotating blades and their interaction with the surrounding flow. These methods are capable of capturing unsteady hydrodynamic effects but are associated with high computational cost. In contrast, the Multiple Reference Frame (MRF) approach, cf. \cite{luo-1994},  reduces computational effort by assuming a stationary, fixed representation of the propeller's position in a coordinate system --typically that of the cylindrical propeller domain-- which rotates relative to the ship's hull.  However, this assumption compromises an accurate resolution of unsteady propeller-hull-rudder/ESD interactions. 
Simpler approaches, such as the body-force method and the actuator-disc method, cf., for example, \cite{kawamura-1997} or \cite{krasilnikov-2013}, do not take the propeller geometry into account but instead replace it with propeller-induced forces on the fluid, which are usually averaged over the circumference and further reduce the computational effort.
Such methods require prior knowledge of the open-water characteristics of the propeller. Although they do not capture local flow features, 
they remain useful for preliminary design.
%
%
Furthermore, marine CFD simulations  present a multiscale challenge arising from the different time scales associated with propeller- and hull-related flow phenomena.
Simulations that resolve the propeller motion require very small time steps to resolve blade motion and the associated unsteady phenomena, typically limiting the rotation to only a few degrees per time step. In contrast, hull-related phenomena, including, for example, the establishment of flotation in self-propulsion conditions, evolve over significantly larger time scales and allow the use of substantially larger time steps, which are often several orders of magnitude larger than those in motion-resolving simulations, cf. \cite{bakica-2019}.

To address these conflicting  requirements, partially rotating grid  approaches have been proposed by \citeauthor{durasevic-2022} (\citeyear{durasevic-2022, durasevic-2023}). In such methods, the rotational speed of the resolved propeller motion is artificially reduced and the resulting  deficit is compensated by using an MRF formulation. 
This strategy enables the use of larger time steps while preserving sufficient resolution of the near-propeller flow and thereby offer an effective trade-off between computational cost and temporal resolution.
However, their application raises additional issues at the interface between the rotating and stationary domains. These issues limit their accuracy and warrant further methodological research, which is discussed in detail in this paper. 
%
%
The present work introduces a modification to the partially rotating grid method to mitigate the issues at the boundary between rotating and non-rotating domains.
The modification ensures a seamless and consistent transition of the momentum equations between rotating and non-rotating regions, thereby reducing the discrepancies associated with the conventional (hybrid) MRF formulation. In particular, 
the proposed method enforces a continuous velocity field across the interface. 
Numerical studies are conducted to evaluate the improvements in prediction accuracy for propeller-hull interaction.
%
%
%
%
%
The application examples given refer to the Japan Bulk Carrier (JBC) and its model-scale propeller, for which a large amount of experimental and numerical data is available,  see \cite{wang-2019, hino-2020}. 
%

The remainder of the paper is organized as follows. Section~\ref{sec:methodology} presents the methodology including the governing equations and the proposed modified MRF approach. The subsequent third section is devoted to a numerical verification using a simple, generic Taylor--Couette flow example. 
Section~\ref{sec:application} is concerned with applications to the JBC vessel at model scale. The section discusses results obtained from open-water, resistance, and self-propulsion simulations. The paper closes with Sec.~\ref{sec:conclusions} where conclusions are drawn and future research direction is discussed.

\FloatBarrier
\section{Methodology}
\label{sec:methodology}

This section presents the governing equations and the numerical method employed in the study. 
The focus is on reviewing various formulations for fixed grids, moving grids, and rotating coordinate systems based on incompressible single- and two-phase flows in Sec. \ref{sec:goveqns}.
A  compatibility condition at the interface between rotating and stationary regions is derived, highlighting the challenges that arise when using an MRF approach. Based on this, the formulation and derivation of the modified partially moving grid method (mMRF) equations are presented in Sec. \ref{sec:secmmrf}.
Finally, a brief overview of the discretization methods used in the simulations is provided in Sec.~\ref{sec:nummeth}.

\subsection{Governing Equations}
\label{sec:goveqns}
The governing equations are the conservation of mass, momentum, turbulence parameters and, if necessary, a balance equation for determining the volume concentration of the fluid phases under consideration. In the current project, only single- and two-phase flows of air and water are considered. The latter are modeled as a mixture of two immiscible, incompressible  fluid phases, which can be determined using a single (common) velocity field via the  volume-of-fluid approach (VOF), see 
\cite{hirt-1981}.
The conservation of mass for an incompressible fluid is expressed by the continuity equation, viz.
\begin{equation}
    \label{eq:continuity}
    \nabla \cdot \vect{u} = 0
    \: \text{.}
\end{equation}
Here $\vect{u}$ represents the velocity vector field and $\nabla$ is the spatial gradient vector in a spatially fixed Eulerian reference frame. 
The momentum conservation is governed by the Navier-Stokes equations 
\begin{equation}
\label{eq:ns}
\begin{split}
    \rho
    \frac{d\vect{u}}{dt}
    =
    \nabla \cdot \tensor{\sigma}
    + \vect{g}\: \text{.}
\end{split}
\end{equation}
In this expression, $t$ is the time, $\rho$ denotes the fluid density, $\vect{g}$ is a volume-specific force vector and $\tensor{\sigma}$ is the Cauchy stress tensor, i.e.
\begin{equation}
    \label{eq:stress_tesor}
    \tensor{\sigma}
    =
    -  p \tensor{I}
    +
    \mu
    \left({
        \nabla \vect{u}
        +
        \left({
            \nabla \vect{u}
        }\right)^{T}
    }\right)
    \: \text{.}
\end{equation}
In Eq. (\ref{eq:stress_tesor}) $p$ is the static pressure field, $\tensor{I}$ is the identity tensor, and $\mu$ represents the dynamic viscosity.
An alternative formulation substitutes the material derivative by a local temporal derivative and convective term, i.e.

\begin{equation}
\label{eq:ALE}
    \frac{d\vect{u}}{dt}
    =
    \frac{\partial \vect{u}}{\partial t}
    +
    \vect{u} \cdot (\nabla \vect{u}) \:.
\end{equation}
where the last term is often converted into $\nabla \cdot (\vect{u} \vect{u})$ based on (\ref{eq:continuity}).
In simulations of incompressible two-phase flows, the fluid properties are modified to account for the distribution of the two fluid phases:
\begin{equation}
\label{eq:rho_eff}
    \rho = c \rho_1 + (1-c)\rho_2
 \, , \qquad    \mu = c \mu_1 + (1-c)\mu_2
\end{equation}
Here  $c$ denotes the concentration of the first fluid phase, e.g. the air phase, 
and the $\rho_{1[2]}$
[$\mu_{1[2]}$] refer to the constant bulk densities [viscosities] of the two fluid phases. 
Restricting ourselves to immiscible fluid phases, the volume concentration of the first phase follows from the Eulerian representation of the immiscibility condition  $Dc/Dt=0$, which, using the continuity equation (\ref{eq:continuity}), is given by 
\begin{align}
        \cfrac{\partial {c}}{\partial t}
        + 
        \nabla
        \cdot
        \left(
            \vect{u} \: c
        \right) = 0\, .
\end{align}
Furthermore, in turbulent flows, we consider the Reynolds Averaged Navier-Stokes (RANS) equations, and the viscosity $\mu$ is supplemented by a turbulent viscosity $\mu_t = \rho k/\omega$ based on the eddy viscosity approach which models the additional contributions to the tensor $\sigma$ in (\ref{eq:stress_tesor}), i.e.,  
\begin{align}
    \tensor{\sigma} \to  
    \tensor{\sigma} 
    -\rho \overline{\vect{u}' \vect{u}'}, \qquad \textrm{with} \quad 
    \rho \overline{\vect{u}' \vect{u}'}= \frac{2 \rho k}{3} \tensor{I} -\mu_t 
    \left({
        \nabla \vect{u}
        +
        \left({
            \nabla \vect{u}
        }\right)^{T}
    }\right) \, . 
\end{align}
The turbulence parameters 
are derived from two separate transport equations, such as the widely used SST $k-\omega$ turbulence model from \cite{menter-2003}, which we employ in our studies. 

\subsubsection{Moving Grid Formulation}
To simulate the rotating motion of the propeller-fixed cylindrical domain, a sliding interface (SI) method is frequently applied together with an arbitrary Lagrangian-Eulerian (ALE) formulation, see for example \cite{blades-2007}. 
Within the ALE framework, the material derivative of a fluid property $\phi$ on a moving grid is defined as
\begin{equation}
\label{eq:material_deriv}
    \frac{d \phi}{dt}
    =
    \left[\frac{\partial \phi}{\partial t}\right]^{\mathrm{GR}}
    +
    \left(
        \vect{u}
        -
        \vect{u}^{\mathrm{GR}}
    \right)
    \cdot
    \nabla \phi
    \: ,
\end{equation}
where 
$[\cdot]^{\mathrm{GR}}$ denotes the time derivative evaluated on moving grids.
In addition, $\vect{u}$ is the fluid velocity and $\vect{u}^{\mathrm{GR}}$ denotes the grid velocity. 
The reason for this is the determination of the local time derivative in (\ref{eq:ALE}). When evaluating the local time derivatives on a moving mesh, the solution from the previous time level must first be interpolated onto the updated mesh positions. Consequently, the local time derivative computed on a moving grid differs from the corresponding derivative on a stationary grid. For the specific case of rigid-body rotation, the grid velocity is defined as 
$\vect{u}^{\mathrm{GR}} = \vect{\Omega}^{\mathrm{GR}} \times \left(\vect{r} - \vect{r}^{0}\right)$
, where $\vect{r}^{0}$ is the origin of grid rotation, $\vect{r}$ is the position vector, and $\vect{\Omega}^{\mathrm{GR}}$ is the vector of the moving grid rotational velocity. The relationship between the stationary and moving local time derivatives of the velocity field $\vect{u}$ can thus be expressed as
\begin{equation}
\label{eq:loc_accel_moving_grid}
        \frac{\partial \vect{u}}{\partial t}
    =
    \left[
        \frac{\partial \vect{u}}{\partial t}
    \right]^{\mathrm{GR}}
    -
    \left( \vect{\Omega}^{\mathrm{GR}}
        \times
        \left(
            \vect{r}
            -
            \vect{r}^{0}
        \right)
    \right)
    \cdot
    \nabla \vect{u}.
\end{equation} 
Note that the rate of volume change of a grid element and the grid velocities in general -i.e., for non-rigid-body motions- are linked by the space conservation law (SCL) to avoid continuity errors; see \cite{demirdzic-1988}.

By substituting (\ref{eq:loc_accel_moving_grid}) into 
(\ref{eq:ns},\ref{eq:ALE}), the incompressible momentum conservation  takes the following form
\begin{equation}
\label{eq:ns_moving_grid}
\begin{split}
    \rho
        \left[
            \cfrac{\partial \vect{u}}{\partial t}
        \right]^{\mathrm{GR}}
        +
        \rho \nabla
        \cdot
        \bigg[
                 \bigg({ \vect{u}
                - \underbrace{
               \vect{\Omega}^{\mathrm{GR}}
                \times
                \left({
                    \vect{r} - \vect{r}^{0}
                }\right)}_{\vect{u}^{\mathrm{GR}}}\bigg)
            \vect{u}
        }\bigg]
    =
    \nabla \cdot \tensor{\sigma} + \vect{g}
    \: .
\end{split}
\end{equation}
To reduce the complexity of the subsequent derivations, the moving grid modification is  not considered in the discussion of MRF formulations in Secs. \ref{sec:mrf} \& \ref{sec:secmmrf} but can always be introduced through the local time derivative relation  (\ref{eq:loc_accel_moving_grid}).
\subsubsection{Moving Reference Frame Formulation}
\label{sec:mrf}
\begin{figure}[htbp]
    \centering
    \begin{minipage}[c]{0.9\textwidth}
    \begin{subfigure}[t]{0.33\linewidth}
        \includegraphics[width=1.0\linewidth]{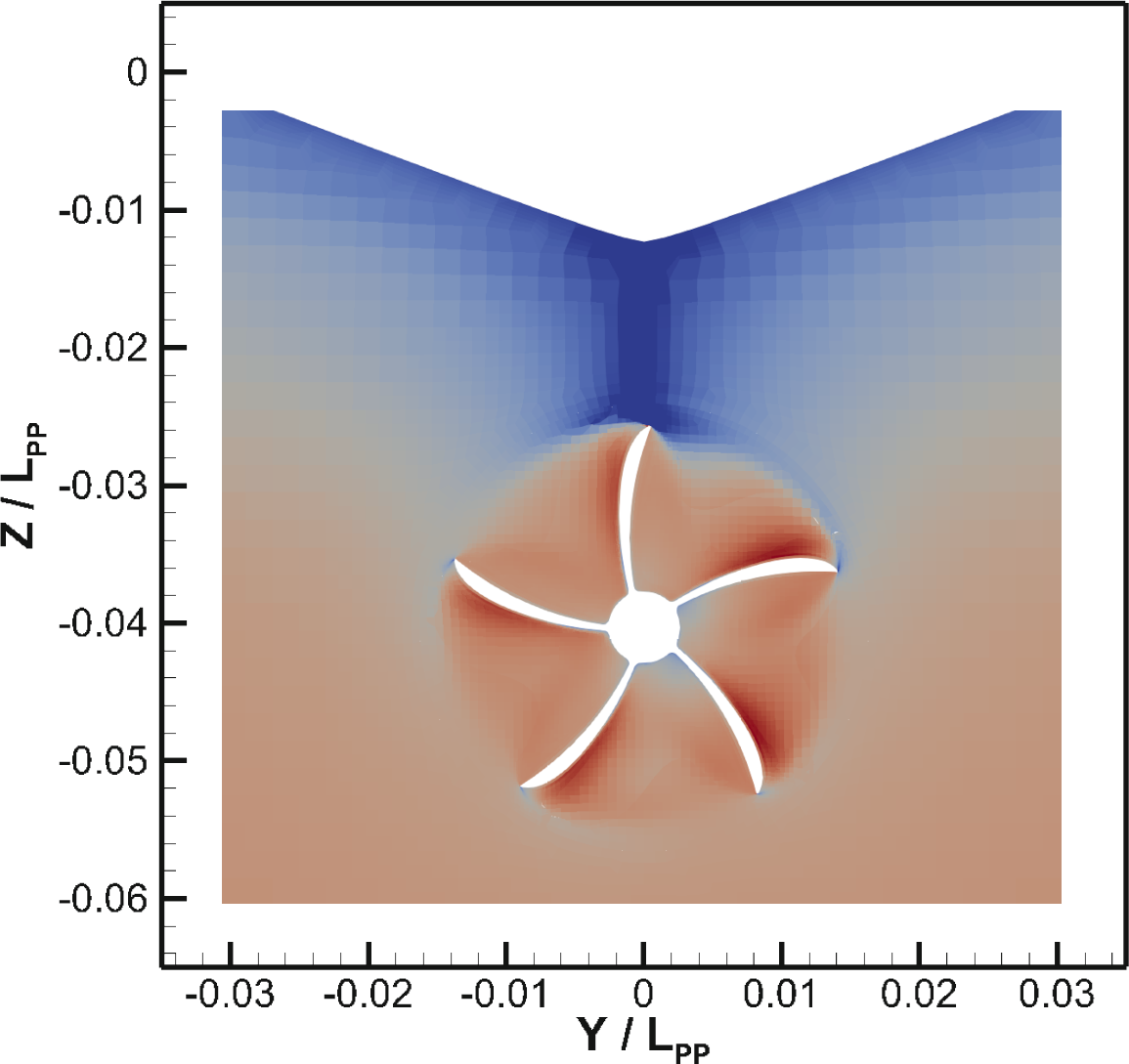}
        \captionsetup{width=0.9\linewidth}
        \caption{SI}
        \label{fig:vx_prop_plane_si}
    \end{subfigure}
    \begin{subfigure}[t]{0.33\linewidth}
        \includegraphics[width=1.0\linewidth]{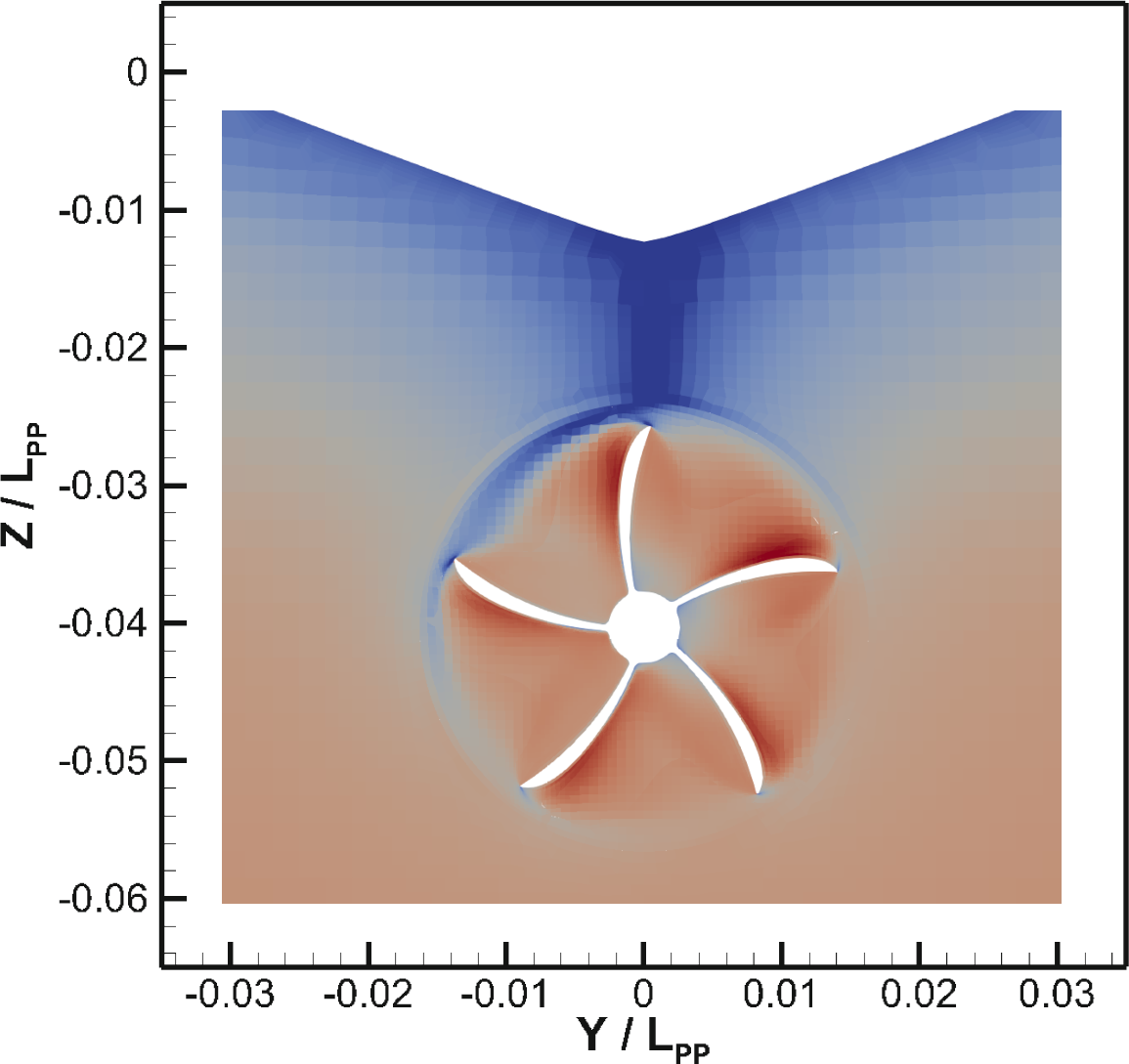}
        \captionsetup{width=0.9\linewidth}
        \caption{MRF, $n_{\mathrm{MRF}} / n = 0.5$}
        \label{fig:vx_prop_plane_mrf}
    \end{subfigure}
    \begin{subfigure}[t]{0.33\linewidth}
        \includegraphics[width=1.0\linewidth]{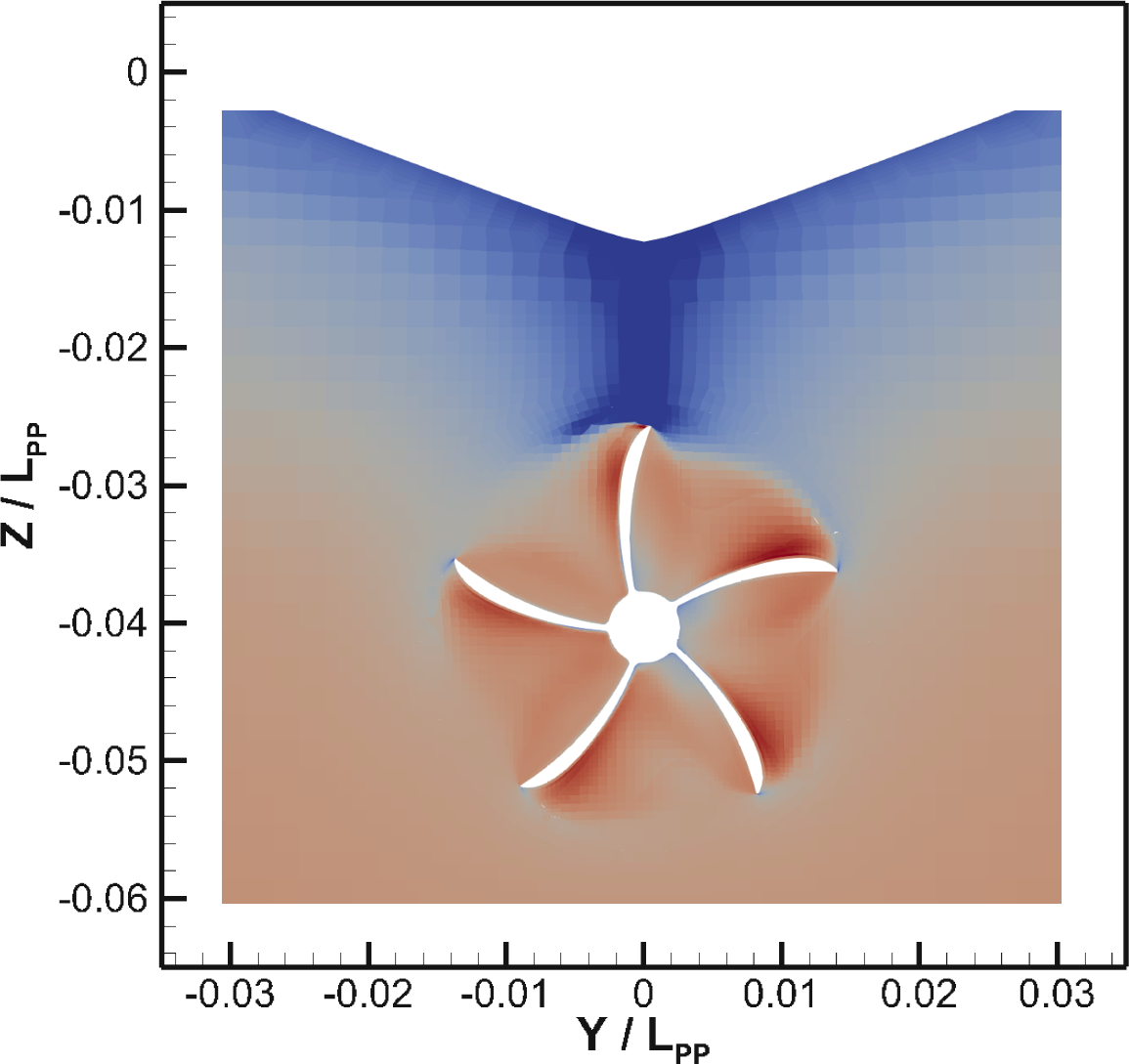}
        \captionsetup{width=0.9\linewidth}
        \caption{mMRF, $n_{\mathrm{MRF}} / n = 0.5$}
        \label{fig:vx_prop_plane_mmrf}
    \end{subfigure}
    \end{minipage}
    \begin{minipage}[c]{0.09\textwidth}
    \begin{subfigure}[t]{0.35\linewidth}
        \includegraphics[width=1.0\linewidth]{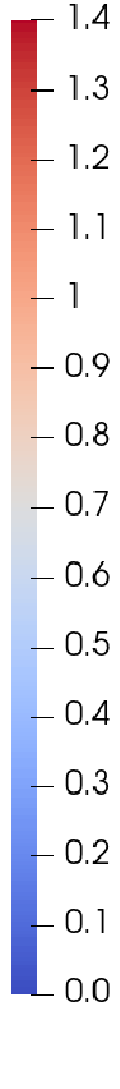}
        \vspace*{8pt}
    \end{subfigure}
    \end{minipage}
    \caption{Comparison of  axial velocity snapshots obtained from a simulation with a sliding interface (left), an unmodified simulation with a partially rotating grid (center), in which 50\% of the rotation is resolved, and a corresponding modified simulation with a partially rotating grid (right). All figures are extracted in the propeller plane for a self-propulsion simulation, cf. Sec.~\ref{ssec:self-prop}.
    }
    \label{fig:vx_prop_plane}
\end{figure}

The MRF (Multiple Reference Frame) approach reduces
the computational effort for a moving propeller mesh by assuming a stationary representation of the propeller within a cylindrical propeller domain, which in turn rotates relative to the ship's domain grid. 
To account for the rotation of the reference frame within the propeller region, the momentum conservation  of an MRF  approach is typically supplemented by an additional Coriolis-like source term, i.e., $ \vect{\Omega}^{\mathrm{MRF}} \times \vect{u}$, to Eq. (\ref{eq:ns}).
The incompressible momentum equations for the absolute velocity $\vect{u}$ inside a propeller domain that performs a solid-body rotation with the MRF velocity 
\begin{equation}
\label{eq:v_mrf_2}
    \vect{u}^{\mathrm{MRF}}
    = 
    \vect{\Omega}^{\mathrm{MRF}}
    \times
    \left({
        \vect{r} - \vect{r}^{0}
    }\right),  
\end{equation}
reads
\begin{equation}
\label{eq:ns_mrf_abs}
\begin{split}
    \rho
        \cfrac{\partial \vect{u}}{\partial t}
        +
        \rho \nabla
        \cdot
        \bigg[{
            \bigg({
                \vect{u}
                -
               \underbrace{\vect{\Omega}^{\mathrm{MRF}}
                \times
                \left({
                    \vect{r} - \vect{r}^{0}
                }\right)}_{\vect{u}^{\mathrm{MRF}}}
            }\bigg)
            \vect{u}
        }\bigg]
        +
        \rho \vect{\Omega}^{\mathrm{MRF}} \times \vect{u}
    =
    \nabla \cdot \tensor{\sigma} + \vect{g}
    \: \text{,}
\end{split}
\end{equation}
where $ \vect{r}^{0} $ is the position vector of the center of rotation, $ \vect{r} $ is the position  vector, and $ \vect{\Omega}^{\mathrm{MRF}} $ is the vector of the reference frame rotational velocity.
Subtracting the momentum equations  (\ref{eq:ns}) from  (\ref{eq:ns_mrf_abs}) yields a compatibility condition at the interface between the rotating and stationary grid domains, viz. 
\begin{equation}
    \label{eq:err}
    \vect{\Omega}^{\mathrm{MRF}}
    \times
    \vect{u}
    = 
    \left({
        \vect{\Omega}^{\mathrm{MRF}}
        \times
        \left({
            \vect{r} - \vect{r}^{0}
        }\right)
    }\right)
    \cdot
    \nabla
    \vect{u}
    .
\end{equation}
This compatibility condition is not automatically satisfied by an arbitrary velocity field at the interface between the two domains. While Eq. (\ref{eq:err}) holds true, for example, for a uniform velocity field aligned with the axis of rotation, it is generally not satisfied in the complex, non-uniform wake behind a ship. Furthermore, as the rotational speed of the propeller domain's reference frame increases,  the discrepancy at the interface becomes increasingly larger. The phenomenon associated with this is shown in Figure~\ref{fig:vx_prop_plane}\subref{fig:vx_prop_plane_mrf}, which presents the axial velocity in the propeller plane during the self-propulsion simulation outlined in Sec.~\ref{sec:application}.
\subsection{Modified Partially Moving Grid Formulation}
\label{sec:secmmrf}

The partially rotating grid approach 
of \cite{durasevic-2022,durasevic-2023} combines a classical sliding interface and a multiple reference frame approach by decomposing the propeller rotation into a grid-resolved component $\vect{\Omega}^{\mathrm{GR}}$ and an MRF component $\vect{\Omega}^{\mathrm{MRF}}$. This decomposition allows for the use of larger time steps and enables users to run simulations at any intermediate state between a motion-resolving simulation and an MRF simulation. 
To address the shortcomings illustrated in Fig.~\ref{fig:vx_prop_plane}, we propose a modification of the model which aims to eliminate discontinuities at the interface between rotating and non-rotating regions. 
The primary goal is to maintain the solid-body motion near the rotating propeller surface, while the behavior gradually converges with that in the stationary reference frame as one approaches the outer boundary of the MRF region.
This modification, labeled mMRF, ensures the continuity of both the velocity and pressure fields across the interface.

We suggest to spatially vary the vector of the rotational speed $\vect{\Omega}^{\mathrm{MRF}}$, i.e., 
$\vect{\Omega}^{\mathrm{MRF}} \to f \vect{\Omega}^{\mathrm{MRF}}$, 
allowing different points to rotate around the center at different speeds. Here $f$ is an additional scaling function that ranges from zero at the interface between the rotating and non-rotating domain to one inside the rotating domain and in the vicinity of the propeller. 
For the derivation of the modified MRF method, two sets of orthonormal basis vectors are defined: $\vect{e}_i$ for the stationary frame and $\vect{e}^R_i$ for the rotating frame. The position vector $\vect{r}$ in the stationary frame is expressed in terms of the rotating frame position $\vect{r}^R$, similar to the standard MRF formulation and consistent with the co-moving tensorial framework of \cite{huckins-1974}, i.e. 
\begin{equation}
    \label{eq:pos1}
    \vect{r}
    =
    \tensor{\Theta}
    \cdot
    \vect{r}^{\mathrm{R}}
    +
    \vect{r}^{0}
    \: \text{,}
\end{equation}
where $\tensor{\Theta}$ represents the rotation-deformation tensor.
Under the assumption of zero rotation angle, the rotation-deformation reduces to the identity matrix, which allows the time derivative of the rotating basis vectors to be expressed using a vector-product form
\begin{equation}
    \label{eq:mMRF-basis}
    \cfrac{ d \vect{e}^{\mathrm{R}}_{i} }{d t}
    =
    f^{\mathrm{R}} \left({ \vect{r}^{\mathrm{R}} }\right)
    \vect{\Omega}^{\mathrm{MRF}}
    \times
    \vect{e}^{\mathrm{R}}_{i}
    =
    f \left({ \vect{r} }\right)
    \vect{\Omega}^{\mathrm{MRF}}
    \times
    \vect{e}^{\mathrm{R}}_{i}
    \: \text{.}
\end{equation}
Equation (\ref{eq:mMRF-basis}) is needed to convert velocities and accelerations between the rotating and non-rotating reference frames, 
similar to the derivation of the classical MRF method. 
Since $f$ depends solely on the time-invariant position in the (partially) rotating reference frame, the field is defined such that $f(\vect{r}) = f^{\mathrm{R}}(\vect{r}^R)$. 
Introducing $f$ into the velocity transformation between stationary and partially moving reference frames allows to account for a spatially varying rotation, i.e., 
\begin{equation}
\label{eq:vel_relation}
\begin{split}
    \vect{u}
    & =
    \cfrac{d \vect{r} }{ d t }
    = 
    \cfrac{d \vect{r}^{\mathrm{R}}}{ d t }
    +
    \cfrac{d \vect{r}^{0}}{ d t }
    =
    \sum_{i} 
    \left(
        \cfrac{d {r}^{\mathrm{R}}_{i}}{ d t } \, \vect{e}^{\mathrm{R}}_{i}
        +
        {r}^{\mathrm{R}}_{i} \, \cfrac{d \vect{e}^{\mathrm{R}}_{i}}{ d t } 
    \right)
    =
    \left[
        \cfrac{d \vect{r}^{\mathrm{R}}}{ d t }
    \right]^{\mathrm{R}}
    +
    \sum_{i} 
    {r}^{\mathrm{R}}_{i} \, \cfrac{d \vect{e}^{\mathrm{R}}_{i}}{ d t } 
    =
    \vect{u}^{\mathrm{R}}
    +
    \underbrace{f^{\mathrm{R}} \left( \vect{r}^{\mathrm{R}} \right) \, \vect{\Omega}^{\mathrm{MRF}} \times \vect{r}^{\mathrm{R}}}_{\vect{u}^{\mathrm{mMRF}}} \, . 
\end{split}
\end{equation}
In this expression, $[\cdot]^R$ denotes differentiation within the (partially) rotating frame. The velocity of the modified partially moving reference frame is therefore written as 
\begin{equation}
    \label{eq:v_mrf_1}
    \vect{u}^{\mathrm{mMRF}}
    = 
    f^{\mathrm{R}} \left( \vect{r}^{\mathrm{R}} \right)
    \vect{\Omega}^{\mathrm{MRF}}
    \times 
    \vect{r}^{\mathrm{R}}
    = 
    f \left( \vect{r} \right)
    \vect{\Omega}^{\mathrm{MRF}}
    \times 
    \left({ \vect{r} - \vect{r}^{0} }\right)
    \: \text{.}
\end{equation}
In the following, the terms $f^R =  f^{\mathrm{R}} \left( \vect{r}^{\mathrm{R}} \right)$ and $f= f \left( \vect{r} \right)$ are used as abbreviations. To simplify the derivations, an unmodified MRF velocity (\ref{eq:v_mrf_2}) is also used.
Following the derivation detailed in Appendix~\ref{app:mMRF_derivation}, 
the acceleration in the modified frame is expanded and substituted into 
the incompressible momentum equations~(\ref{eq:ns}), yielding the 
governing equations for velocity in the relative frame, viz.
\begin{equation}
\label{eq:ns_mMRF_rel}
\begin{split}
    &
    \rho
    \left[{
        \cfrac{\partial \vect{u}^{\mathrm{R}}}{ \partial t }
        + 
        \nabla
        \cdot
        \left({
            \vect{u}^R
             \vect{u}^R
        }\right)
        +
        2 f^{\mathrm{R}} \,
        \vect{\Omega}^{\mathrm{MRF}}
        \times
        \vect{u}^{\mathrm{R}}
        + 
        (f^{\mathrm{R}})^{2} \,
        \vect{\Omega}^{\mathrm{MRF}}
        \times
        \left({
            \vect{\Omega}^{\mathrm{MRF}}
            \times
            \vect{r}^{\mathrm{R}}
        }\right)
    }\right.
    \\
    &  
    \left.{
        \hspace{5cm}
        + f^{\mathrm{R}} \,
        \cfrac{ \partial \vect{u}^{\mathrm{MRF}} }{ \partial t}
        +
        \nabla f^{\mathrm{R}}
        \cdot
        \left({
            \vect{u}^{\mathrm{R}}
            \vect{u}^{\mathrm{MRF}}
            +
            \vect{u}^{\mathrm{MRF}}
            \vect{u}^{\mathrm{R}}
        }\right)
    }\right]
    =
    \nabla \cdot \tensor{\sigma} + \vect{g}
    \: \text{.}
\end{split}
\end{equation}
While Eq.~(\ref{eq:ns_mMRF_rel}) is solved in the relative frame, 
it is convenient to reformulate the momentum equation in terms of 
the absolute velocity $\vect{u}$. Using the velocity relation~(\ref{eq:vel_relation}), and following the steps outlined in 
Appendix~\ref{app:mMRF_derivation}, the absolute-velocity formulation 
of the momentum equations reads
\begin{equation}
\label{eq:ns_mMRF_abs_03}
\begin{split}
    \rho
    \left[{
        \cfrac{\partial \vect{u} }{ \partial t }
        +
        \nabla
        \cdot
        \bigg({
            \left({
                \vect{u} - \vect{u}^{\mathrm{mMRF}}
            }\right)
            \vect{u}
        }\bigg)
        +
        \bigg({
            \nabla f 
            \cdot
            \vect{u}^{\mathrm{MRF}}
        }\bigg)
        \vect{u}
        +
        f \, 
        \vect{\Omega}^{\mathrm{MRF}} 
        \times 
        \vect{u}
    }\right]
    =
    \nabla \cdot \tensor{\sigma} + \vect{g}
    \: \text{.}
\end{split}
\end{equation}
The spatial variation of the rotation rate is enforced through the damping function $f$, which scales the local rotational velocity. It is noted that for the limiting case $f = 1$, the governing equations recover the standard MRF formulation. 
Apart from the (implicit) appearance of the scaling function $f$ within the source term and the convection term, the only new term in Eq.~(\ref{eq:ns_mMRF_abs_03}) compared to the unmodified Eq. (\ref{eq:ns_mrf_abs}) is the one proportional to the gradient of $f$.
This result is not surprising, since the material derivative of arbitrary property $\phi$ of an incompressible fluid can be translated from a  Lagrangian  to a moving Eulerian  system as follows:
$d\phi /dt 
= \partial \phi /\partial t
 + \nabla \cdot (( \vect{u}-\vect{u}^{\mathrm{mMRF}}) \phi) - (\nabla \cdot \vect{u}^{\mathrm{mMRF}}) \phi$. 
The compatibility condition at the interface between the rotating and stationary subdomains takes the following form, 
\begin{equation}
\label{eq:mMRF-non-mMRF}
    f \, 
    \vect{\Omega}^{\mathrm{MRF}}
    \times
    \vect{u} 
    = 
    f \, 
    \vect{u}^{\mathrm{MRF}}
    \cdot
    \nabla
    \vect{u}, 
\end{equation}
which resembles Eq. (\ref{eq:err}).
If the scaling function vanishes at the outer  boundary of the rotating propeller domain, Eq. (\ref{eq:mMRF-non-mMRF}) is satisfied automatically. On the other hand, the scaling function should satisfy $f = 1$ in the vicinity of the moving propeller surfaces.

\subsubsection{Scaling Function}
\begin{figure}[htbp]
    \centering
    %
    %
    \begin{subfigure}[t]{0.29\linewidth}
        \includegraphics[width=1.0\linewidth]{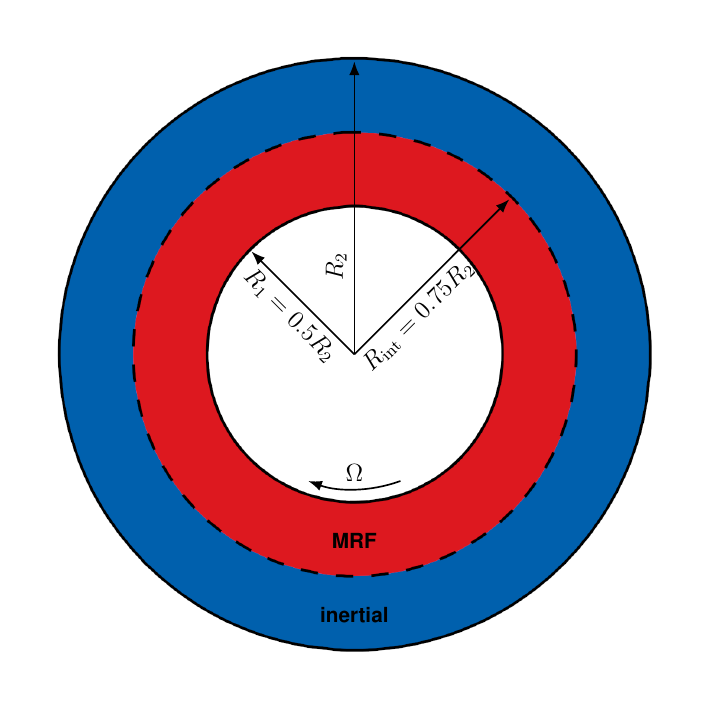}
        \captionsetup{width=1.0\linewidth}
        \caption{Taylor–Couette flow configuration: MRF subdomain and the inertial subdomain separated by the grid interface (dashed line).}
        \label{fig:couette_setup}
    \end{subfigure}
    \begin{subfigure}[t]{0.70\linewidth}
        \includegraphics[width=1.0\linewidth]{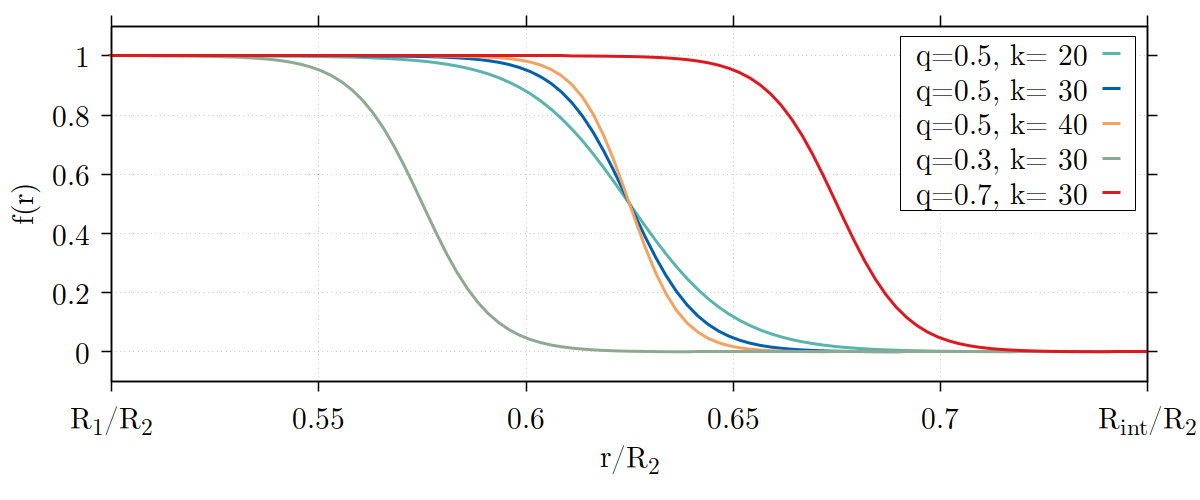}
        \captionsetup{width=0.9\linewidth}
        \caption{Evolution of the scaling function $f(r)$ (\ref{eq:sigmoid}) for different parameter values $q$ and $k$, evaluated between the inner cylinder radius $R_1$ and the grid interface radius $R_{\mathrm{int}}$.
        }
        \label{fig:couette_ramping}
    \end{subfigure}
    %
    \caption{Computational setup for the Taylor–Couette flow verification case and evolution of the scaling function $f(r)$.}
    \label{fig:ramping}
\end{figure}

The functional form of $f(\vect{r})$ is not unique and may be chosen freely. 
The function transitions from 0 at the domain interface to 1 at the rotating surface. The proposed ramping function depends on both the wall distance -- already employed in the turbulent viscosity computation -- and the distance to the grid interface, computed analogously. The center of this transition is controlled by a predefined parameter $q$. The sigmoid-based function employed here is selected for its convenience and its ability to ensure a controllable and smooth spatial variation of the reference frame rotation,
\begin{equation}
    \label{eq:sigmoid}
    f \left({ \vect{r} }\right) = 
    \cfrac{
        1
    }{
        1 +
        e^{
            k
            \left[{ 1 - a \left({ \vect{r} }\right) - q }\right] 
        }
    }.
\end{equation}
Here the parameter $k$ defines the steepness of the transition.
The function $[1-a(\vect{r})-q]$ is used to position the transition from one to zero, where the distance function $a$ depends on both the distance from the wall $d_{\mathrm{wall}}$ and the distance to the grid interface $d_{\mathrm{int}}$, viz. 
\begin{equation}
    \label{eq:sigmoid-arg}
    a\left({ \vect{r} }\right) = 
    \cfrac{
        d_{\mathrm{int}} \left({ \vect{r} }\right)
    }{
        d_{\mathrm{int}} \left({ \vect{r} }\right)
        +
        d_{\mathrm{wall}} \left({ \vect{r} }\right)
    }
\end{equation}
and the parameter $q$ is used to control the relative position of the transition center. Larger $k$ values yield steeper ramping function, while larger $q$ values shift the transition center away from the rotating surface toward the domain interface. The effect of the parameters $k$ and $q$ on the shape of the scaling function is illustrated in Fig.~\ref{fig:couette_ramping} using the {example of the flow between two coaxial circular cylinders; see Section \ref{ssec:couette}.}
\subsection{Numerical Method}
\label{sec:nummeth}

The computational framework uses the Finite Volume (FV) software FreSCo$^+$
(\cite{rung-2009}).
The segregated algorithm utilizes a cell-centered, collocated storage arrangement in combination with arbitrary polyhedral unstructured grids. Spatial integration employs a 2nd-order accurate mid-point rule. The approximation of diffusive fluxes employs 2nd-order central differencing while higher-order upwind-biased methods are used to approximate convective fluxes. Time integration follows a 1st-order accurate, implicit Euler approach. The solution is iterated to convergence using a pressure-correction scheme.  Parallelization refers to a domain decomposition approach and inter-processor communication uses the MPI protocol.
\section{Verification}
\label{sec:verification}
\subsection{Taylor–Couette Flow }
\label{ssec:couette}
A Taylor–Couette flow configuration, depicted in  Fig.~\ref{fig:couette_setup}, serves as a verification case. The setup consists of two concentric cylindrical walls, where the inner wall rotates at a constant angular velocity ${\Omega}$ and the outer wall remains stationary. The computational domain is divided into an inner region associated with the rotating wall and an outer stationary region. The flow is assumed to be incompressible and laminar.
The geometric and physical parameters are
\begin{equation}
    R_{1} = 0.5~\mathrm{m}, \quad
    R_{2} = 1.0~\mathrm{m}, \quad
    R_{\mathrm{int}} = 0.75~\mathrm{m}, \quad
    \Omega = 1.745~\mathrm{rad/s}, \quad
    \rho = 1000~\mathrm{kg/m^3}, \quad
    \mu = 1.0~\mathrm{Pa \cdot s}.
\end{equation}
Here, $R_{1}$ is the radius of the inner rotating wall, $R_{2}$ is the radius of the stationary outer wall, and $R_{\mathrm{int}}$ is the radius of the interface between the rotating and stationary computational domains. The parameter $\Omega$ denotes the angular velocity of the inner wall.
For the steady flow between two concentric cylinders with a rotating inner wall, the analytical solution for the azimuthal velocity component is obtained from \cite{taylor-1923}, viz.
\begin{equation}
    u_{\theta}\left({ r }\right) = A r + \cfrac{B}{r}, \quad
    p\left({ r }\right) 
    =
    \rho
    A^{2}
    \left({
        \cfrac{{r}^{2}}{2}
        -
        2 {R}_{2}^{2} \ln r
        -
        \cfrac{ {R}_{2}^{4} }{ 2 {r}^{2} }
    }\right)
    + C .
\end{equation}
The constants $A$, $B$, and $C$ are determined from the boundary conditions, i.e. , 
\begin{equation}
    u_{\theta}\left({R_{1}}\right) = \Omega \: R_{1} , \qquad
    u_{\theta}\left({R_{2}}\right) = 0 , \qquad
    p\left({R_{1}}\right) = 0 ,
\end{equation}
which yields 
\begin{equation}
    A = -\cfrac{\Omega R_{1}^{2}}{R_{2}^{2} - R_{1}^{2}}, \qquad
    B = \cfrac{\Omega R_{1}^{2} R_{2}^{2}}{R_{2}^{2} - R_{1}^{2}}, \quad
    C = 
    - \rho
    \cfrac{
        {\Omega}^{2} {R}_{1}^{4}
    }{
        \left({
            {R}_{2}^{2} - {R}_{1}^{2}
        }\right)^{2}
    } 
    \left({
        \cfrac{{R}_{1}^{2}}{2}
        -
        2 {R}_{2}^{2} \ln {R}_{1}
        -
        \cfrac{{R}_{2}^{4}}{2 {R}_{1}^{2}}
    }\right)
    .
\end{equation}
This results in the final form of the analytical velocity and pressure profiles 
\begin{equation}
 \label{eq:taylorcuette}
    u_{\theta} \left({r}\right)
    =
    \frac{\Omega R_{1}^2}{R_{2}^2 - R_{1}^2}
    \left(
        \frac{R_{2}^2}{r} - r
    \right)
    , \quad
    p \left({r}\right)
    =
    \rho
    \cfrac{
        {\Omega}^{2} {\mathrm{R}}_{1}^{4}
    }{
        \left({
            {\mathrm{R}}_{2}^{2} - {\mathrm{R}}_{1}^{2}
        }\right)^{2}
    }
    \left({
        \cfrac{
            {r}^{2} - {\mathrm{R}}_{1}^{2}
        }{2} 
        -
        2 {\mathrm{R}}_{2}^{2} \ln\left({
            \cfrac{r}{{\mathrm{R}}_{1}}
        }\right)
        -
        \cfrac{{\mathrm{R}}_{2}^{4}}{2}
        \left({
            \cfrac{1}{{r}^{2}} - \frac{1}{{\mathrm{R}}_{1}^{2}}
        }\right)
    }\right) .
\end{equation}
The radial and axial velocity components are zero. This analytical solution serves as the reference for the verification of the numerical implementation.
To this end, five different MRF and mMRF configurations are assessed. 
Table~\ref{tab:couette-setup} provides an overview of the investigated scenarios. 
\begin{table}[htbp]
    \centering
    \begin{tabular}{l c c c c}
    \toprule
    Case
    & Formulation
    & $\Omega$ [rad/s]
    & $\Omega^{\mathrm{MRF}}/\Omega$
    & $\Delta t$ [s] \\
    \midrule

    C1 & SI   & \multirow{5}{*}{1.745} & 0.0 & 0.02 \\[1mm]   
    C2 & MRF  &                        & 0.5 & 0.04 \\        
    C3 & MRF  &                        & 1.0 & ---  \\[1mm]   
    C4 & mMRF &                        & 0.5 & 0.04 \\        
    C5 & mMRF &                        & 1.0 & ---  \\        

    \bottomrule
    \end{tabular}
    \caption{Investigated  Taylor-Couette flow verification cases.}
    \label{tab:couette-setup}
\end{table}

Similar to the derivation given above, the total angular velocity of the system, ${\Omega}$, is decomposed into two components in these cases 
\begin{equation}
\label{eq:couette_rot}
\Omega = \Omega^{\mathrm{GR}} + \Omega^{\mathrm{MRF}},
\end{equation}
where $\Omega^{\mathrm{GR}}$ denotes the grid-resolved (physical) rotation and $\Omega^{\mathrm{MRF}}$ represents the rotation modeled through the moving reference frame formulation. 
As shown in Table \ref{tab:couette-setup}, we also provide the solution of an unsteady sliding interface method in C1, which represents the gold standard for propeller flow simulations. The time-step size 
corresponds to a $2^{\circ}$ rotation of the inner wall per time step, which ensures both an integer number of full revolutions and sufficient physical time for the solution to reach a steady state. 
Two different scenarios involving moving reference frames are focused on below.
In the first scenario, the entire rotation for cases C3 and C5 is modeled using the moving reference frame approach, wherein neither of the two grid domains moves and the simulations are conducted in a steady state. Investigations are carried out for both the unmodified MRF formulation (C3) and the suggested modified mMRF formulation (C5).
In the second scenario, half of the rotational motion is attributed to the moving reference frame approach, while the remaining half is resolved by the rotation of interior domain grid.  Investigations are again carried out for both the unmodified MRF formulation (C2) and the modified mMRF formulation (C4). Consequently, the time step is doubled in these cases compared to the SI approach.

\begin{figure}[htbp]
    \centering
    \begin{subfigure}[t]{0.9\linewidth}
        \includegraphics[width=1.0\linewidth]{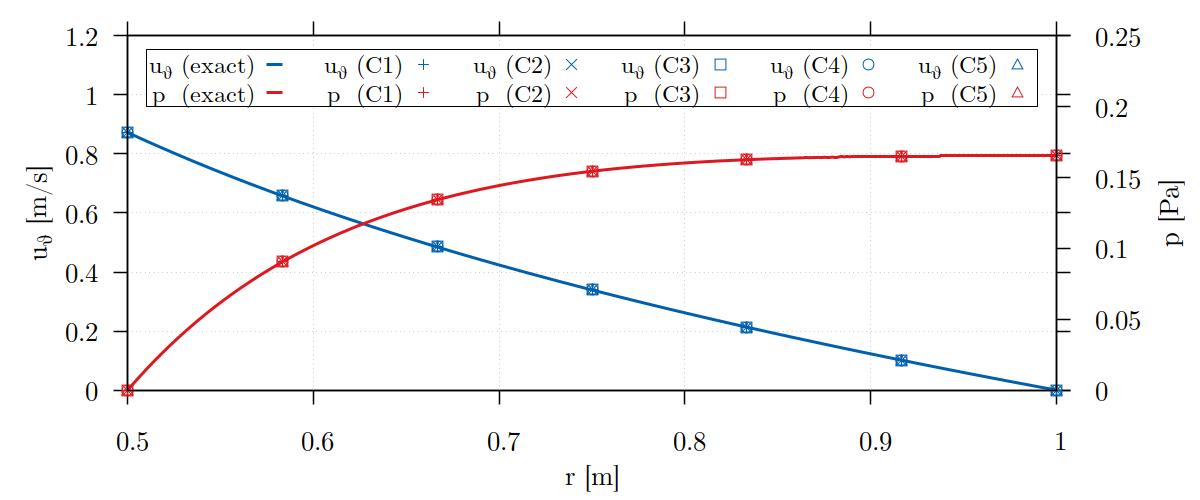}
        \captionsetup{width=0.9\linewidth}
    \end{subfigure}
    \caption{Comparison of computed azimuthal velocities and pressure values 
    with analytical results (\ref{eq:taylorcuette}) for the axisymmetric Taylor-Couette verification case.}
    \label{fig:couette-vel-p}
\end{figure}

Due to the axial symmetry of the flow field (\ref{eq:taylorcuette}),  the compatibility relation (\ref{eq:err}) is maintained and $\vect{u}^{\mathrm{MRF}} \cdot \nabla f = 0$. 
Therefore, the sliding interface (SI), moving reference frame (MRF), modified moving reference frame (mMRF) simulations should always produce identical velocity fields. 
As outlined by Fig.~\ref{fig:couette-vel-p}, 
numerical results --represented by symbols and obtained 
with varying ratios of grid-resolved rotation and MRF rotation-- demonstrate strong agreement with the analytical solution, represented by lines.
Minor discrepancies are observed and are attributed to the spatial discretization of the computational domain. These differences are, however,  negligible and yield virtually  indistinguishable velocity profiles. 

\FloatBarrier
\section{Application}
\label{sec:application}
\begin{figure}[htbp]
    \centering
        \includegraphics[width=0.99\linewidth]{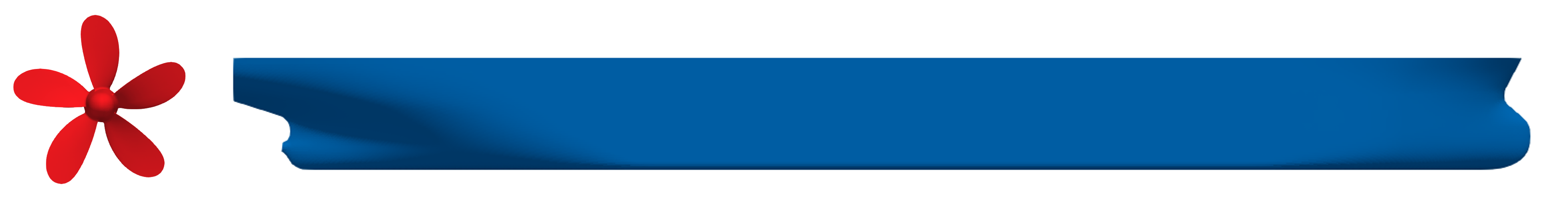}
    \caption{Illustration of the investigated JBC hull and propeller geometry.}
    \label{fig:jbc-geom}
\end{figure}

\FloatBarrier
The present study is applied to the Japan Bulk Carrier (JBC) equipped with a propeller. Figure \ref{fig:jbc-geom} displays the geometries of the investigated hull and the propeller.  The specific configuration is chosen because detailed experimental data of both, local flow features and integral performance are publicly available.   
The corresponding towing tank experiments, which included measurements for resistance, mean flow fields, free-surface elevation and the self-propulsion scenario, were conducted  at the National Maritime Research Institute (NMRI), cf. \cite{hino-2020}. 
 Table~\ref{tab:jbc-data} summarizes the principal dimensions of the hull and propeller. All numerical simulations refer to the geometry at a model scale of $\lambda = 40$. These simulations consider the ship without the rudder. 
 The global coordinate system origin is assigned to the aft perpendicular on the undisturbed waterline. The origin for the five-bladed propeller is located at $ \left({ 0.09375, \: 0.0, \: -0.283}\right) $ in model scale. Both the hull and the propeller surfaces are modeled as hydraulically smooth. 

\begin{table}[htbp]
\centering
\begin{tabular}{llcc}
    \toprule
    \textbf{Hull Particulars} & \textbf{Symbol, Unit} & \textbf{Full-scale} & \textbf{Model-scale} \\ 
    \midrule
    Length between the perpendiculars & $L_{\mathrm{PP}}$, m & 280.0 & 7.0000 \\
    Length of waterline & $L_{\mathrm{WL}}$, m & 285.0 & 7.1250 \\
    Maximum beam of waterline & $B_{\mathrm{WL}}$, m & 45.0 & 1.1250 \\
    Depth & $D$, m & 25.0 & 0.6250 \\
    Draft & $T$, m & 16.5 & 0.4125 \\
    Wetted surface area & $S_{\mathrm{W}}$, m$^2$ & 19556.1 & 12.2230 \\
    Displacement volume & $\nabla$, m$^3$ & 178369.9 & 2.7870 \\
    Design speed & $V$, m/s (knots) & 7.4594 (14.5) & 1.1790 \\
    Froude number & $Fr$ & 0.1420 & 0.1420 \\
    Block coefficient & $C_{\mathrm{B}}$ & 0.8580 & 0.8580 \\
    \midrule
    \textbf{Propeller Particulars} & & & \\
    Propeller diameter & $D_{\mathrm{P}}$, m & 8.120 & 0.2030 \\
    Number of blades & $Z$ & 5 & 5 \\
    Hub ratio & $d_{\mathrm{H}}/D_{\mathrm{P}}$ & 0.180 & 0.180 \\
    Pitch ratio ($0.7 R$) & $P/D_{\mathrm{P}}$ & 0.750 & 0.750 \\
    Expanded area ratio & $A_{\mathrm{E}}/A_\mathrm{{0}}$ & 0.500 & 0.500 \\
    \bottomrule
\end{tabular}
\caption{Principal particulars of the investigated JBC configuration.}
\label{tab:jbc-data}
\end{table}

\FloatBarrier

\subsection{Propeller open-water test}

The initial open-water studies address the simulation of the flow around a deeply submerged propeller of diameter $D_p$,  subjected to  homogeneous inflow $V_{\mathrm{A}}$ in axial direction.
To determine the propeller open-water characteristics and to validate the suggested mMRF method, simulations are once again conducted using the transient sliding-grid approach (OW1) and compared with two different simulations based on rotating reference frames. Specifically, both the classical MRF (OW2, OW3) and the modified mMRF (OW4, OW5) formulations are considered, each applied in a time-dependent partially rotating configuration (OW2, OW4) and a steady-state configuration in which the entire rotation is modeled by the reference frame (OW3, OW5), as summarized in Table~\ref{tab:ow-setup}. The parameters of the scaling function (\ref{eq:sigmoid})  are assigned to $q=0.8$ and $k=30$. For the open-water case, only one set of the parameters $q$ and $k$ is used, as the result is not highly sensitive to their selection. Instead, the parameters are discussed in detail for the propeller in behind conditions, cf.  Sec.~\ref{sec:qkbehav}.

\begin{table}[htbp]
    \centering
    \begin{tabular}{l c c c c c c}
    \toprule
    Case
    & Formulation
    & $q$
    & $k$
    & $n$ [rev/s]
    & $n_{\mathrm{MRF}}/n$
    & $\Delta t$ [s] \\
    \midrule

    OW1 & SI   & --  & --  & \multirow{5}{*}{8} & 0.0  & $5.0 \times 10^{-4}$ \\[1mm]   
    OW2 & MRF  & --  & --  &                    & 0.75 & $2.0 \times 10^{-3}$ \\        
    OW3 & MRF  & --  & --  &                    & 1.0  & ---                  \\[1mm]   
    OW4 & mMRF & 0.8 & 30  &                    & 0.75 & $2.0 \times 10^{-3}$ \\        
    OW5 & mMRF & 0.8 & 30  &                    & 1.0  & ---                  \\        

    \bottomrule
    \end{tabular}
    \caption{Rotation decomposition and time steps used for the five open-water propeller investigations.}
    \label{tab:ow-setup}
\end{table}

The total propeller rotation rate $n$ is decomposed into a grid-resolved component $n_{\mathrm{GR}}$ and a moving reference frame component, i.e., $n = n_{\mathrm{GR}} + n_{\mathrm{MRF}}$, where 
 $n^{\mathrm{GR}}$ represents the physically resolved grid rotation, while $n^{\mathrm{MRF}}$ denotes the rotation introduced through the moving reference frame. To this end, the total rotational speed of the propeller remains constant across all cases. 
For simulations involving grid-resolved rotation, the time step is chosen to limit the propeller rotation to less than $2^\circ$ per step. Accordingly, $N_{\mathrm{ts}} = 250$ time steps per full revolution are used, corresponding to the time step of 
 $   \Delta t = 1/(N_{\mathrm{ts}} \: n_{\mathrm{GR}}).$
\begin{figure}[htbp]
    \centering
    \includegraphics[width=0.8\linewidth]{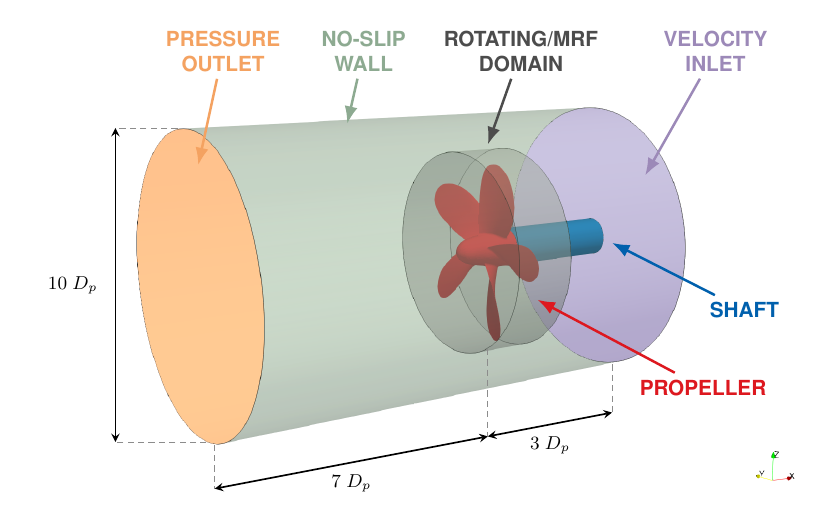}
    \caption[Open-water propeller domain]{Computational domain and boundary
    conditions for the open-water propeller simulation. The propeller is
    enclosed in a cylindrical rotating/MRF subdomain within a stationary outer
    domain. All distances are normalized by the propeller diameter $D_p$.
    The figure is not in scale.}
    \label{fig:ow-domain}
\end{figure}
The computational domain is presented in Fig.~\ref{fig:ow-domain}. The  dimensions of the considered domain follow the ITTC recommendations (\cite{ittc-perf-pred}), with the inlet located at a distance of $3D_{\mathrm{P}}$ upstream of the propeller plane, the outlet is positioned at $7D_{\mathrm{P}}$ downstream the propeller plane and the outer boundary set to $5D_{\mathrm{P}}$. 
Computations are carried out for five different advance coefficients $J = 0.2,\; 0.4,\; 0.6,\; 0.7,$ and $0.8$ defined as 
\begin{equation}
\label{eq:advance_coeff}
    J = \frac{V_{\mathrm{A}}}{n D_{\mathrm{P}}}. 
\end{equation}
These simulations are conducted by varying the advance/inlet velocity $V_A$ while maintaining a constant propeller rotation rate $n=8 \, \mathrm{Hz}$. 
To ensure the reliability of the numerical results, a verification study of temporal and spatial discretization uncertainties is carried out for the open-water case at $J=0.6$, following the ITTC recommended procedures \cite{ittc-uncertainty-2, ittc-uncertainty}. 
Further details regarding the spatial and temporal discretization of the open-water simulations are provided in Appendix~\ref{app:validation}.

\begin{figure}[htbp]
    \centering
    \begin{subfigure}[t]{1.0\linewidth}
        \includegraphics[width=1.0\linewidth]{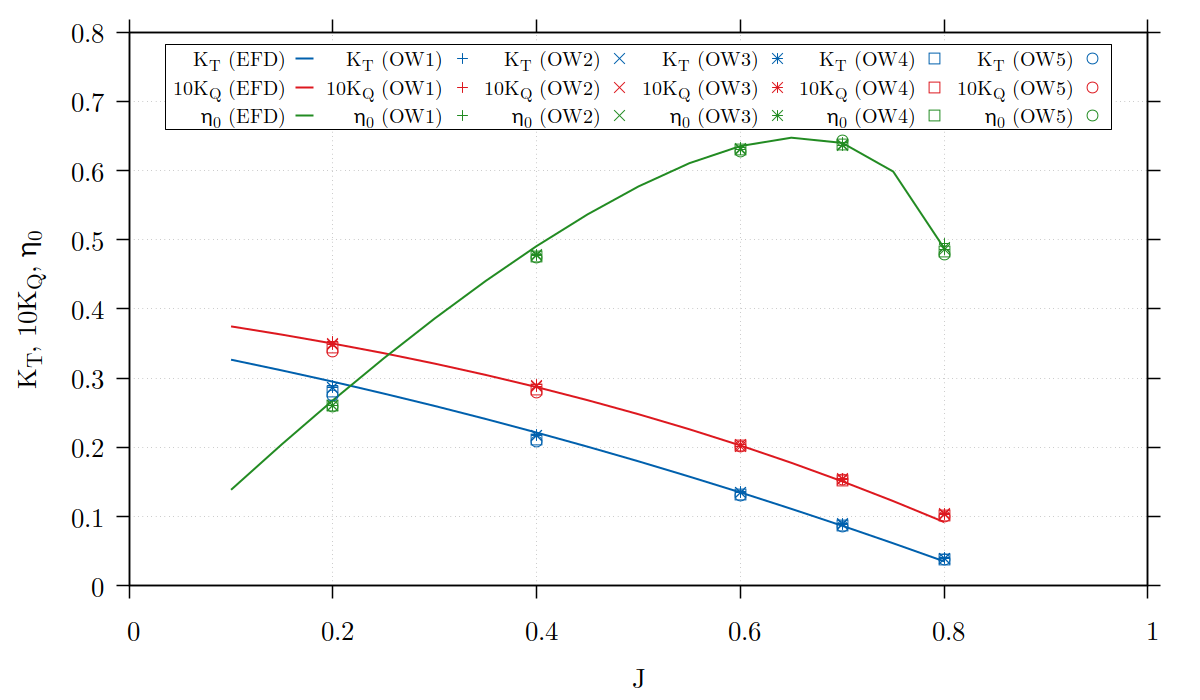}
        \captionsetup{width=0.9\linewidth}
    \end{subfigure}
    \caption{Open water propeller curves for the non-dimensional thrust $K_{\mathrm{T}}$, torque $K_{\mathrm{Q}}$ and efficiency $\eta_0$ (see Table \ref{tab:ow-setup} for the notation of legends).}
    \label{fig:ow-curves}
\end{figure}

\begin{figure}[htbp]
    \centering
    \begin{subfigure}[t]{1.0\linewidth}
        \includegraphics[width=1.0\linewidth]{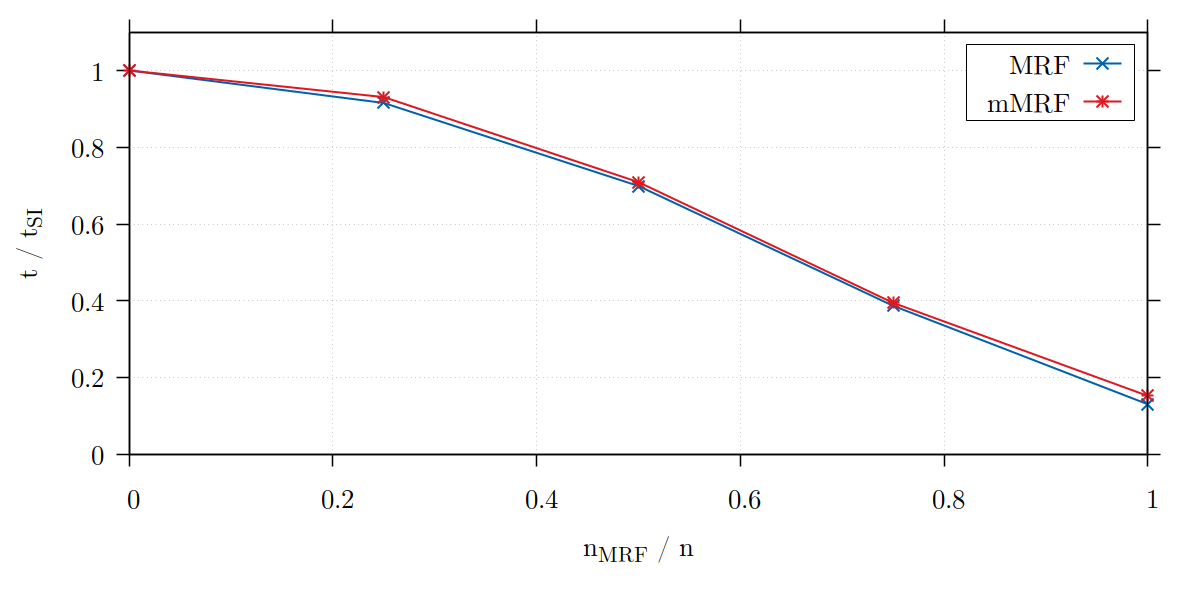}
        \captionsetup{width=0.9\linewidth}
    \end{subfigure}
    \caption{Comparison of non-dimensional computational efforts associated with the MRF and the mMRF approaches for the open-water case.}
    \label{fig:ow-time}
\end{figure}

The results for the calculated values of thrust, torque, and efficiency under open-water conditions are presented in Fig. \ref{fig:ow-curves} and compared with measurements published by  \cite{hino-2020}. 
The thrust and torque coefficients and the efficiency are defined as
\begin{equation}
\label{eq:kt}
    K_{\mathrm{T}} = \frac{T}{\rho n^2 D_{\mathrm{P}}^4},
 \qquad 
    K_{\mathrm{Q}} = \frac{Q}{\rho n^2 D_{\mathrm{P}}^5}, \qquad 
    \eta_0 = \frac{J}{2\pi} \, \frac{K_{\mathrm{T}}}{K_{\mathrm{Q}}} \, ,
\end{equation}
where $T$ is the thrust and $Q$ is the torque.
As expected for the uniform inflow of the open-water setup, both the MRF and the mMRF methods generally reproduce the SI solution and the experimental data across the full range of advance coefficients $J$. However, as the advance coefficient approaches zero (bollard pull condition), the thrust and torque values  predicted by the mMRF fall slightly below the MRF results, particularly when the entire rotation is modeled by MRF. Since under uniform inflow all non-uniformities of the velocity field originate from the propeller itself, the frame transition of the standard MRF, located further away from the propeller, introduces a smaller error than that of the mMRF, which lies closer to the propeller. 
This larger discrepancy with the EFD and SI solution is likely related to the choice of the mMRF ramping function parameters $q$ and $k$, which 
were not specifically investigated for the open-water case, since the standard MRF formulation already achieves good agreement with the reference solution at low $J$. For the self-propulsion simulations, however, a dedicated investigation of $q$ and $k$ is carried out (Sec.~\ref{sec:qkbehav}), and the resulting values ($q=0.8$, $k=30$) are adopted here for consistency.
Overall, the results demonstrate that the rotation decomposition preserves the main integral propeller characteristics.
The computational efficiency of the mMRF method is assessed by comparing total simulation times for different fractions of the moving frame rotation ($n_{\mathrm{MRF}}/n$), as reported in Fig.~\ref{fig:ow-time}. All simulations were run in parallel on two nodes (192 cores) of a distributed memory cluster with 2x Intel Xeon "Cascade Lake" Platinum 9242 processors. The results show that the proposed mMRF approach 
 achieves similar accuracy without adding extra computational cost.
\FloatBarrier

\subsection{Bare hull wave resistance estimation}
\label{sec:wavedrag}
\begin{figure}[htbp]
    \centering
    \begin{subfigure}[t]{0.9\linewidth}
        \includegraphics[width=1.0\linewidth]{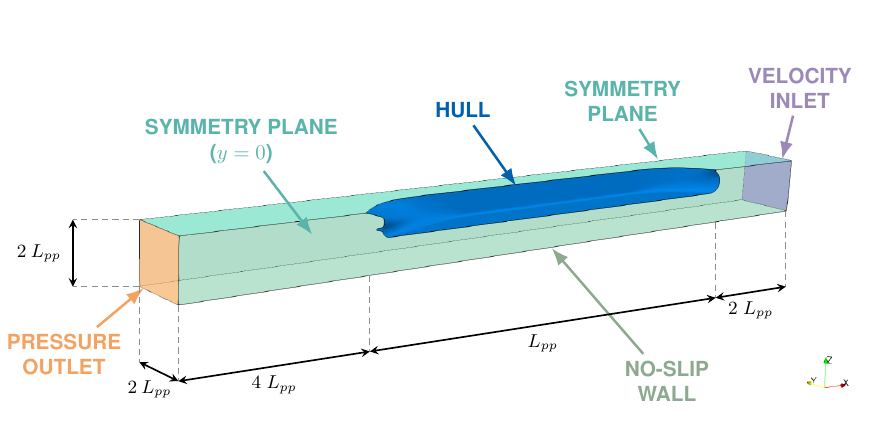}
        \captionsetup{width=0.9\linewidth}
        \caption{Illustration of the single-phase configuration for the bare-hull analysis, in which the free surface is replaced by a symmetry plane at the undisturbed waterline.}
    \end{subfigure}
    \begin{subfigure}[t]{0.9\linewidth}
        \includegraphics[width=1.0\linewidth]{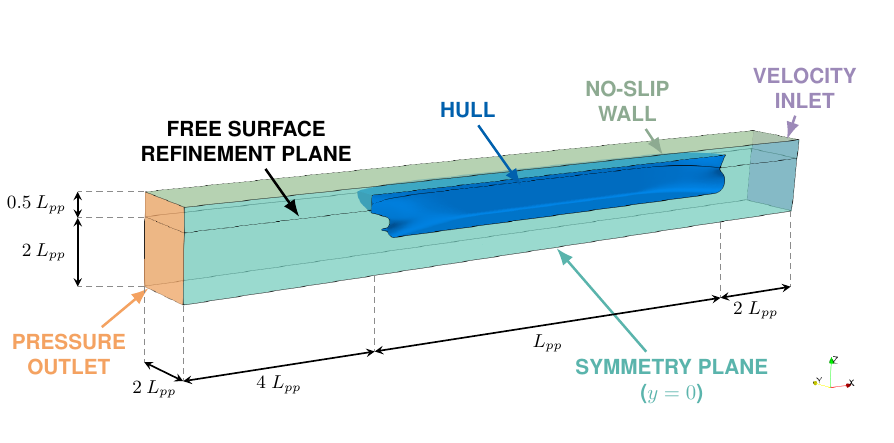}
        \captionsetup{width=0.9\linewidth}
        \caption{Illustration of the two-phase configuration for the bare-hull analysis, in which the free surface is resolved using a refinement zone around the undisturbed waterline.}
    \end{subfigure}
    \caption[Computational domain for the JBC bare-hull configurations]{Computational
    domain and boundary conditions for the JBC bare-hull configurations, exploiting a port/starboard symmetry ($y=0$). All distances are normalised by the length between perpendiculars $L_{pp}$. Figures are not in scale.}
    \label{fig:jbc-resistance-domains}
\end{figure}

\begin{figure}[htbp]
    \centering
    \begin{subfigure}[t]{0.9\linewidth}
        \includegraphics[width=1.0\linewidth]{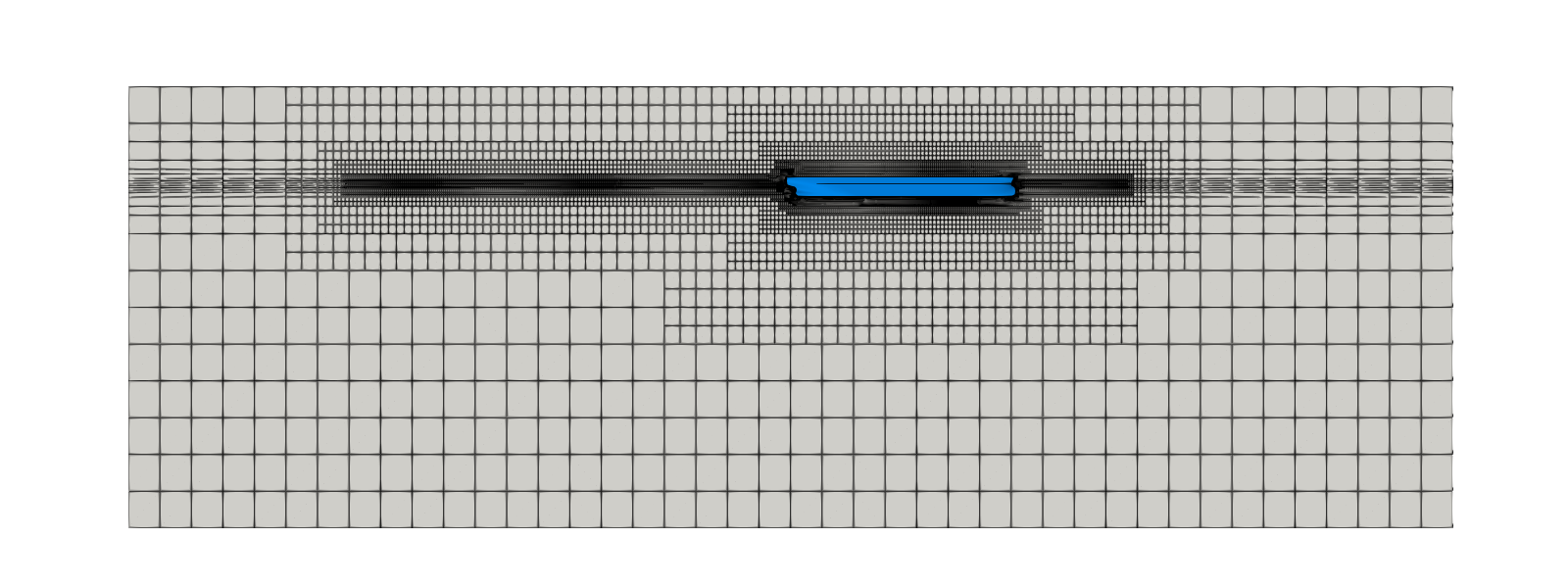}
        \captionsetup{width=0.9\linewidth}
        \caption{Longitudinal slice located in the symmetry plane ($y=0$) showing the free-surface refinement and the hull/wake refinement.}
    \end{subfigure}
    \begin{subfigure}[t]{0.9\linewidth}
        \includegraphics[width=1.0\linewidth]{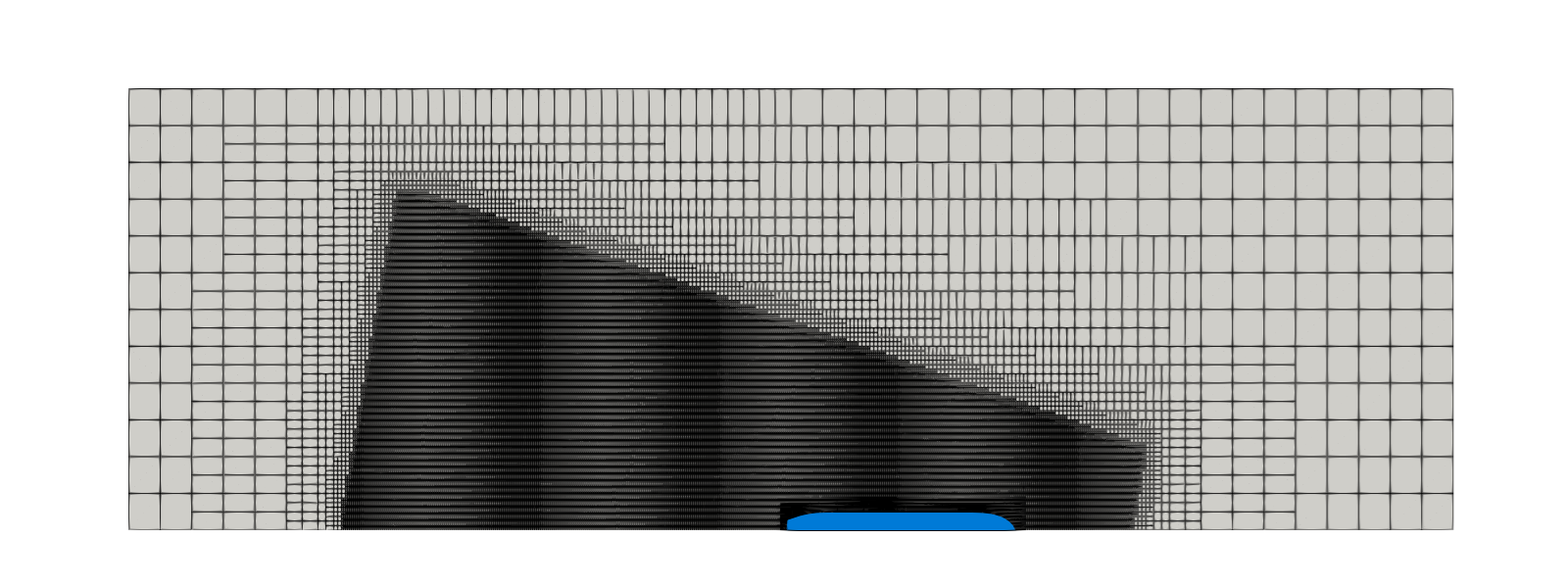}
        \captionsetup{width=0.9\linewidth}
        \caption{Horizontal slice located in the undisturbed waterline showing the refinement.}
    \end{subfigure}
    \caption{Computational grid for the two-phase flow resistance case. The mesh is refined in nested zones around the hull and in a horizontal regime  embedding the free surface.}
    \label{fig:jbc-resistance-grid}
\end{figure}

To reduce the computational effort involved in comparing the proposed mMRF model with the classical MRF approach and the sliding-grid method, single-phase simulations under self-propulsion conditions are performed in Sec. \ref{ssec:self-prop} for a double-body configuration. 
 Since these simulations do not include a wave resistance contribution, this is determined by comparing a two-phase flow and a single-phase flow around the bare hull in a preliminary study. It is assumed that the wave field is not significantly affected by the propeller 
and the wave-drag contribution is therefore assumed independent of the propeller rotation rate and is
  added to the single-phase resistance in Sec.~\ref{ssec:self-prop}.
To determine the wave resistance component, two   simulations are carried out.
First, a two-phase flow model with a resolved free surface is simulated. Second, a single-phase simulation is performed using a symmetry-plane boundary condition to replace the free surface. In both simulations the hull is held fixed at the sinkage and trim measured in the experiment~\cite{hino-2020}, i.e., no dynamic flotation (free sinkage and trim) is computed.
To reduce the computational cost, port/starboard symmetry is exploited and only half of the hull is computed. The domain extends $2\,L_{\mathrm{PP}}$ upstream of the hull to the velocity inlet and $4\,L_{\mathrm{PP}}$ downstream to the pressure outlet; its total breadth (modeled half-domain)  is $4\,L_{\mathrm{PP}}$ and its total depth is $2\,L_{\mathrm{PP}}$. For the two-phase simulation the depth is increased by  $0.5\,L_{\mathrm{PP}}$ above the undisturbed waterline to resolve the free surface. Both domains and the free-surface mesh refinements for the double phase case are illustrated in Fig.~\ref{fig:jbc-resistance-domains} and Fig.~\ref{fig:jbc-resistance-grid}.
The grids consist of approximately $1.03$ million
and $2.25$ million
control volumes for the single-phase and two-phase simulations, respectively. A high Reynolds wall-function approach is used for both simulations and the non-dimensional near wall distance approximately reads $Y^+ = 30$ for both meshes.
\begin{figure}[htbp]
    \centering
    \begin{subfigure}[t]{0.9\linewidth}
        \includegraphics[width=1.0\linewidth]{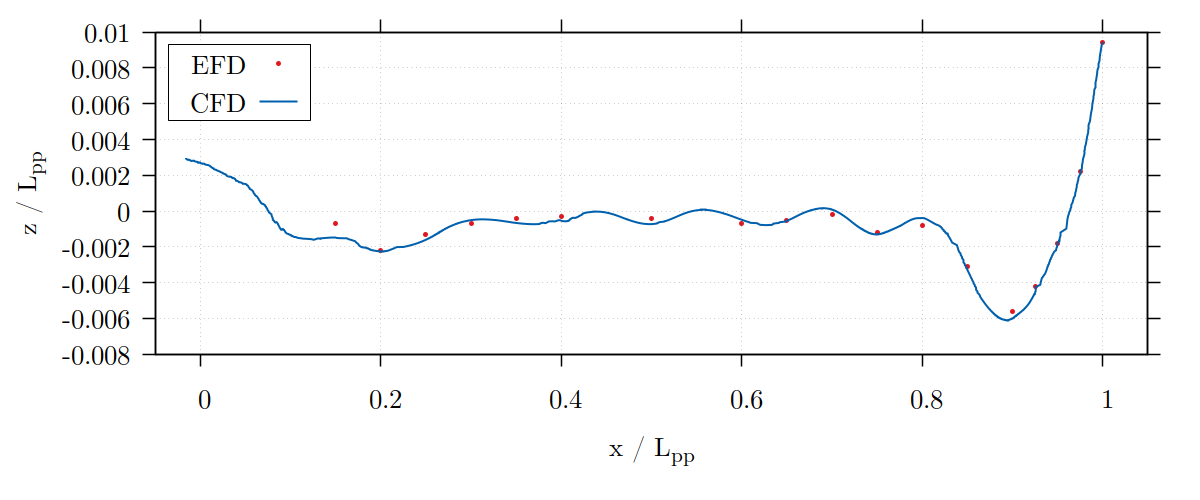}
        \captionsetup{width=0.9\linewidth}
        \caption{Hull surface}
    \end{subfigure}
    \begin{subfigure}[t]{0.9\linewidth}
        \includegraphics[width=1.0\linewidth]{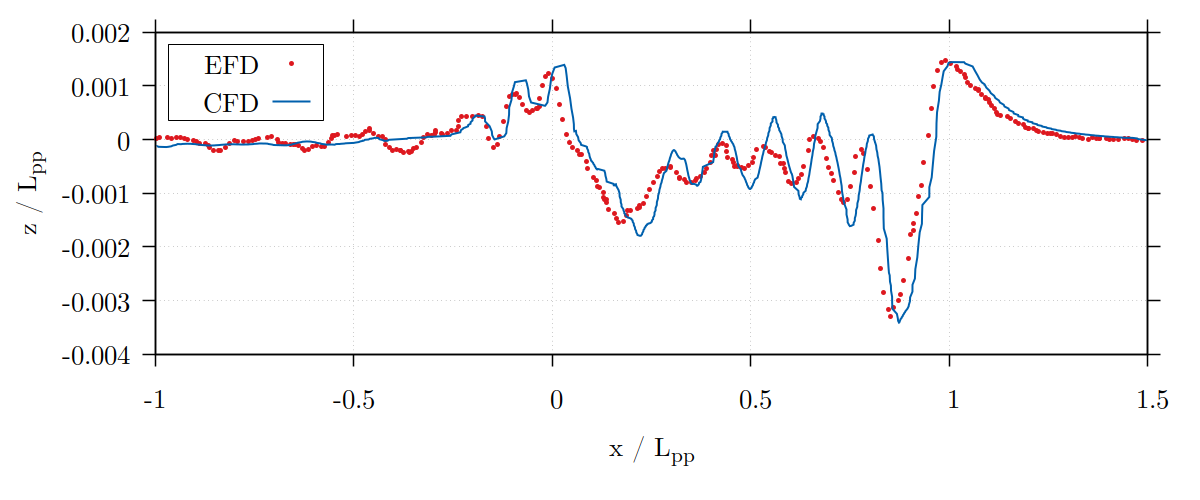}
        \captionsetup{width=0.9\linewidth}
        \caption{$y/L_{pp} = 0.1043$}
    \end{subfigure}
    \begin{subfigure}[t]{0.9\linewidth}
        \includegraphics[width=1.0\linewidth]{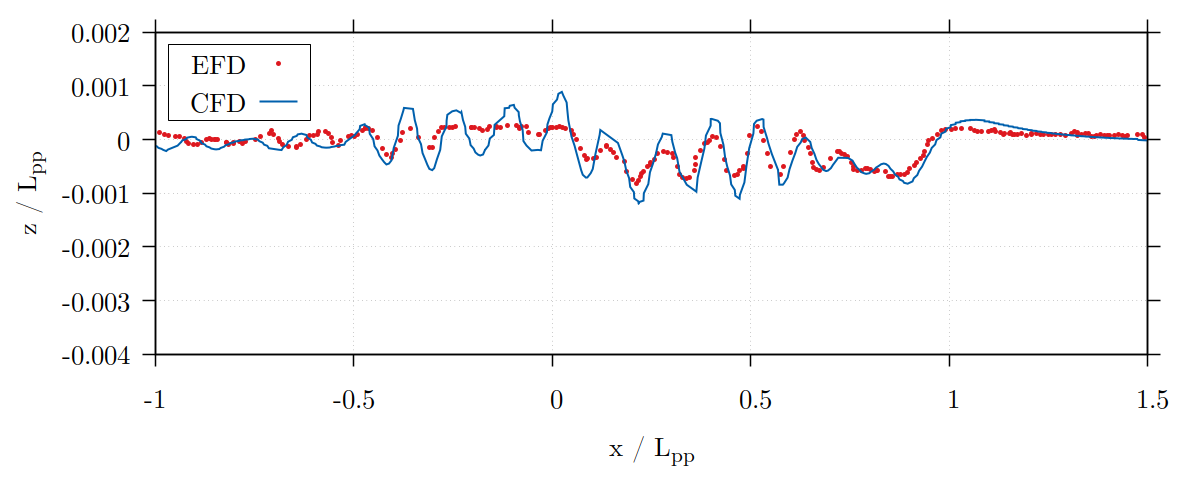}
        \captionsetup{width=0.9\linewidth}
        \caption{$y/L_{pp} = 0.1900$}
    \end{subfigure}
    \caption{Comparison of predicted and measured JBC bare hull towing test wave elevations.}
    \label{fig:jbc-waves}
\end{figure}
\begin{table}[htbp]
    \centering
    \begin{tabular}{l c c}
        \toprule
        Method & $R_{\mathrm{T}}$ [N] & $R_{\mathrm{T}}/R_{\mathrm{T}_{(\mathrm{EFD})}} - 1 $\\
        \midrule

        EFD &  18.05 & -- \\
        \midrule

        Two-phase CFD & 18.21 & +0.89 \% \\

        Single-phase CFD & 15.99 & -11.4 \%  \\
\midrule

        Wave-induced Drag $R_W$ & 2.22 & \\

        \bottomrule
    \end{tabular}
    \caption{Bare hull resistance values.}
    \label{tab:jbc_bare_resistance}
\end{table}

The resulting resistance values for the two simulations are summarized in Table~\ref{tab:jbc_bare_resistance}, where the wave-induced resistance is estimated by subtracting the total resistance of the single-phase simulation from the total resistance of the two-phase flow simulation. The difference between these two values represents the wave-induced resistance contribution, i.e., $R_{\mathrm{W}}=2.22\mathrm{N}$, which is applied in the single-phase self-propulsion calculations.
As indicated by Table \ref{tab:jbc_bare_resistance}, the two-phase simulation reproduces the experimental resistance value with good accuracy. 
In addition to the integral resistance, the free-surface flow was examined through wave elevation distributions. Figure~\ref{fig:jbc-waves} presents the wave heights along the hull and at the transverse planes $y/L_{\mathrm{PP}} = 0.1043$ and $y/L_{\mathrm{PP}} = 0.19$. The experimental data shown in the figure were digitized and adapted from \cite{hino-2020}. The computed wave profiles 
indicate 
that the numerical setup can capture the spatial development of the wave system from the near-hull region towards the far field. 

\subsection{Parameterization of the mMRF scaling function}
\label{sec:qkbehav}
\FloatBarrier
\begin{figure}[htbp]
    \begin{minipage}[c]{0.9\textwidth}
        \centering
        \begin{subfigure}[t]{0.32\linewidth}
            \includegraphics[width=1.0\linewidth]{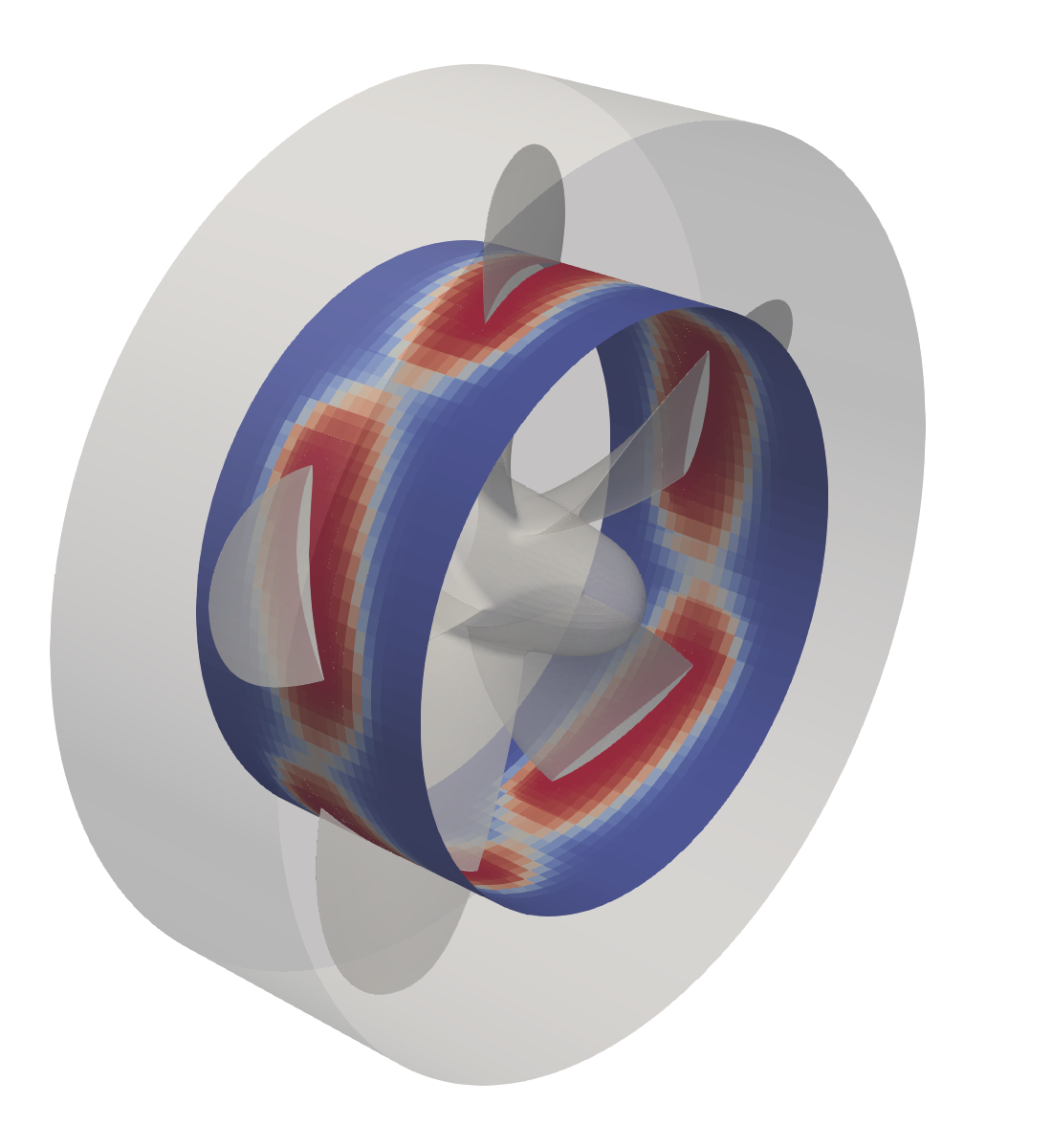}
            \captionsetup{width=0.9\linewidth}
            \caption{F03 ($q = 0.4, k = 20$)}
        \end{subfigure}
        \begin{subfigure}[t]{0.32\linewidth}
            \includegraphics[width=1.0\linewidth]{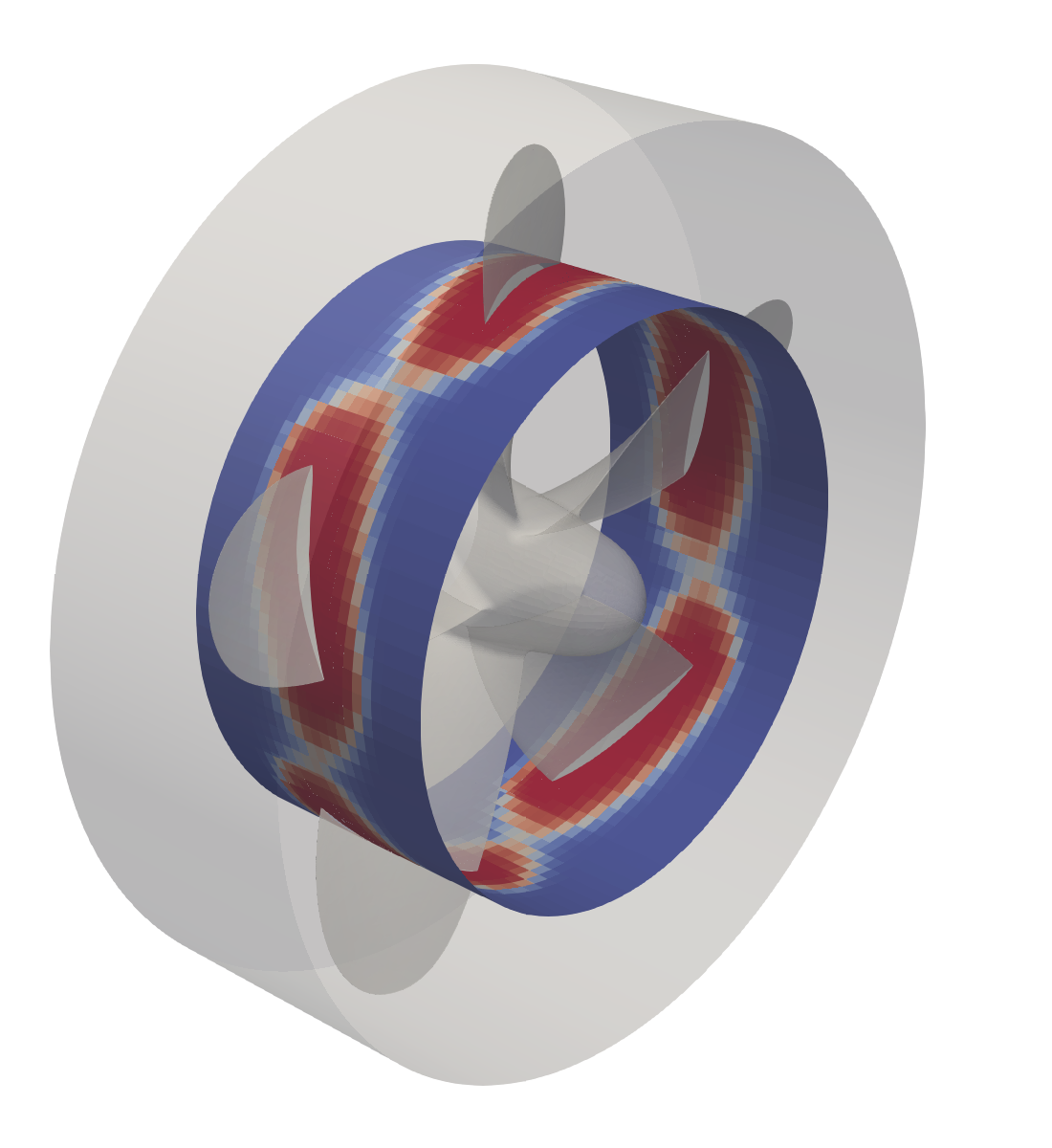}
            \captionsetup{width=0.9\linewidth}
            \caption{F04 ($q = 0.4, k = 30$)}
        \end{subfigure}
        \begin{subfigure}[t]{0.32\linewidth}
            \includegraphics[width=1.0\linewidth]{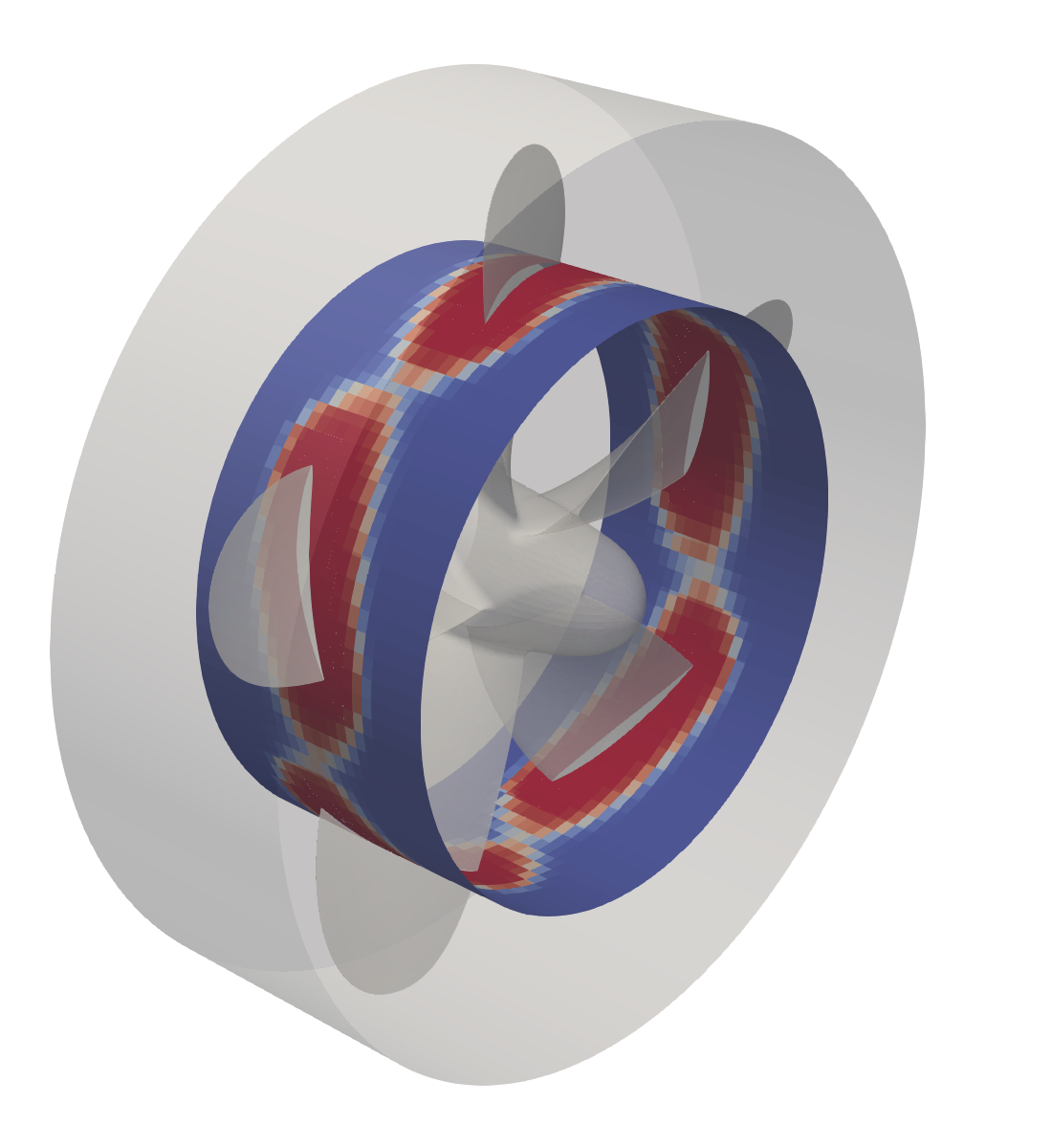}
            \captionsetup{width=0.9\linewidth}
            \caption{F05 ($q = 0.4, k = 40$)}
        \end{subfigure}
        \begin{subfigure}[t]{0.32\linewidth}
            \includegraphics[width=1.0\linewidth]{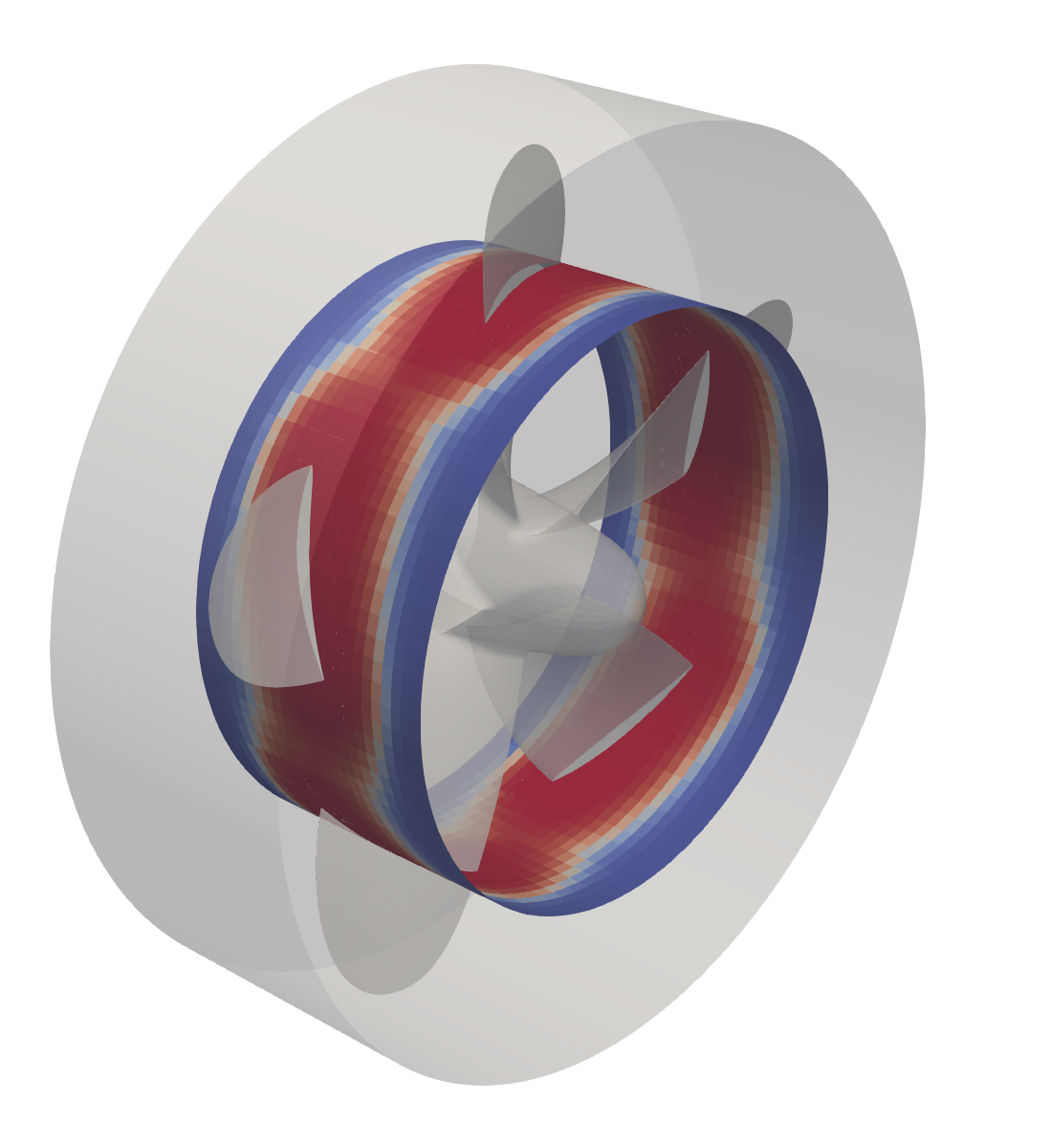}
            \captionsetup{width=0.9\linewidth}
            \caption{F06 ($q = 0.6, k = 20$)}
        \end{subfigure}
        \begin{subfigure}[t]{0.32\linewidth}
            \includegraphics[width=1.0\linewidth]{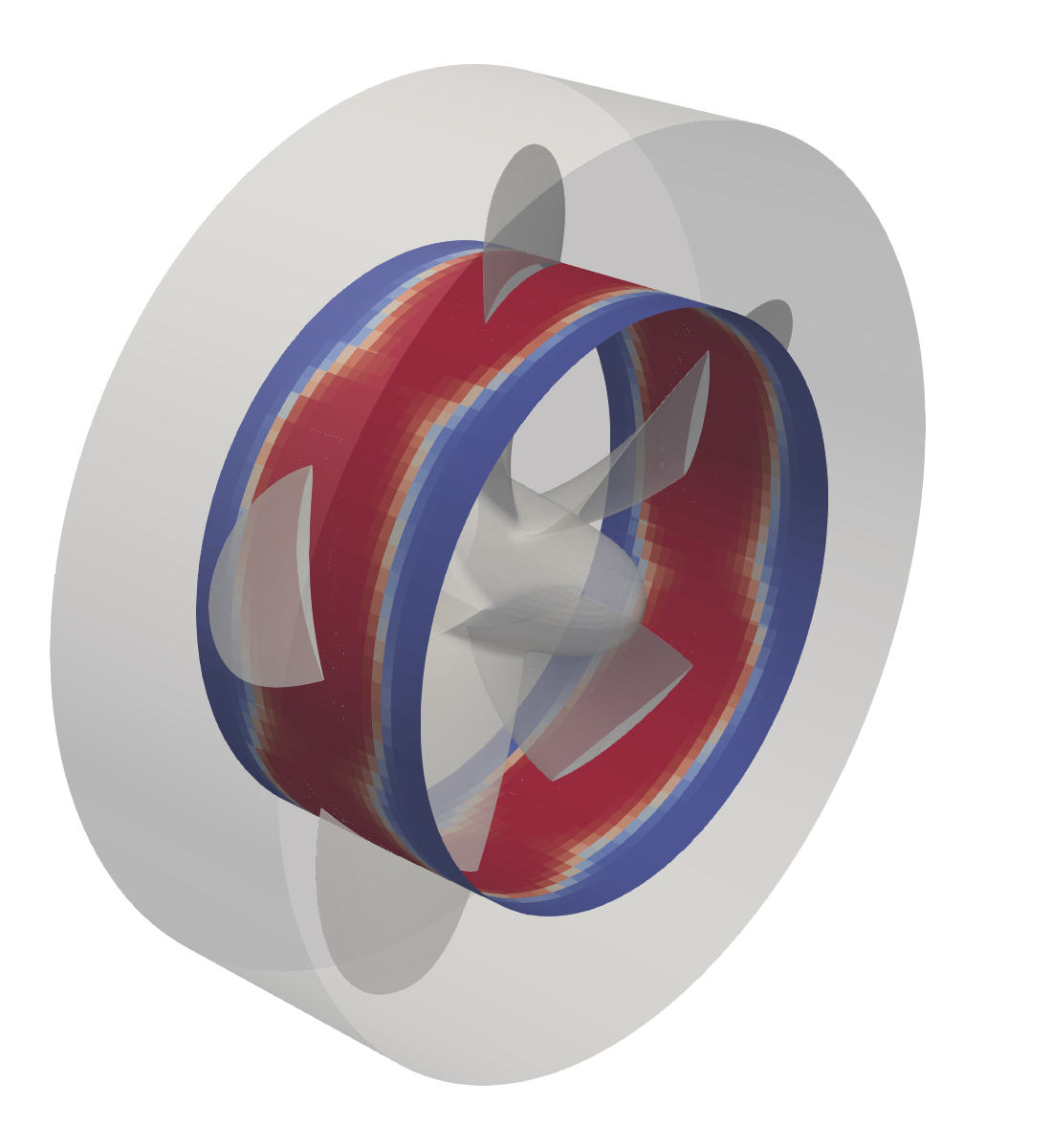}
            \captionsetup{width=0.9\linewidth}
            \caption{F07 ($q = 0.6, k = 30$)}
        \end{subfigure}
        \begin{subfigure}[t]{0.32\linewidth}
            \includegraphics[width=1.0\linewidth]{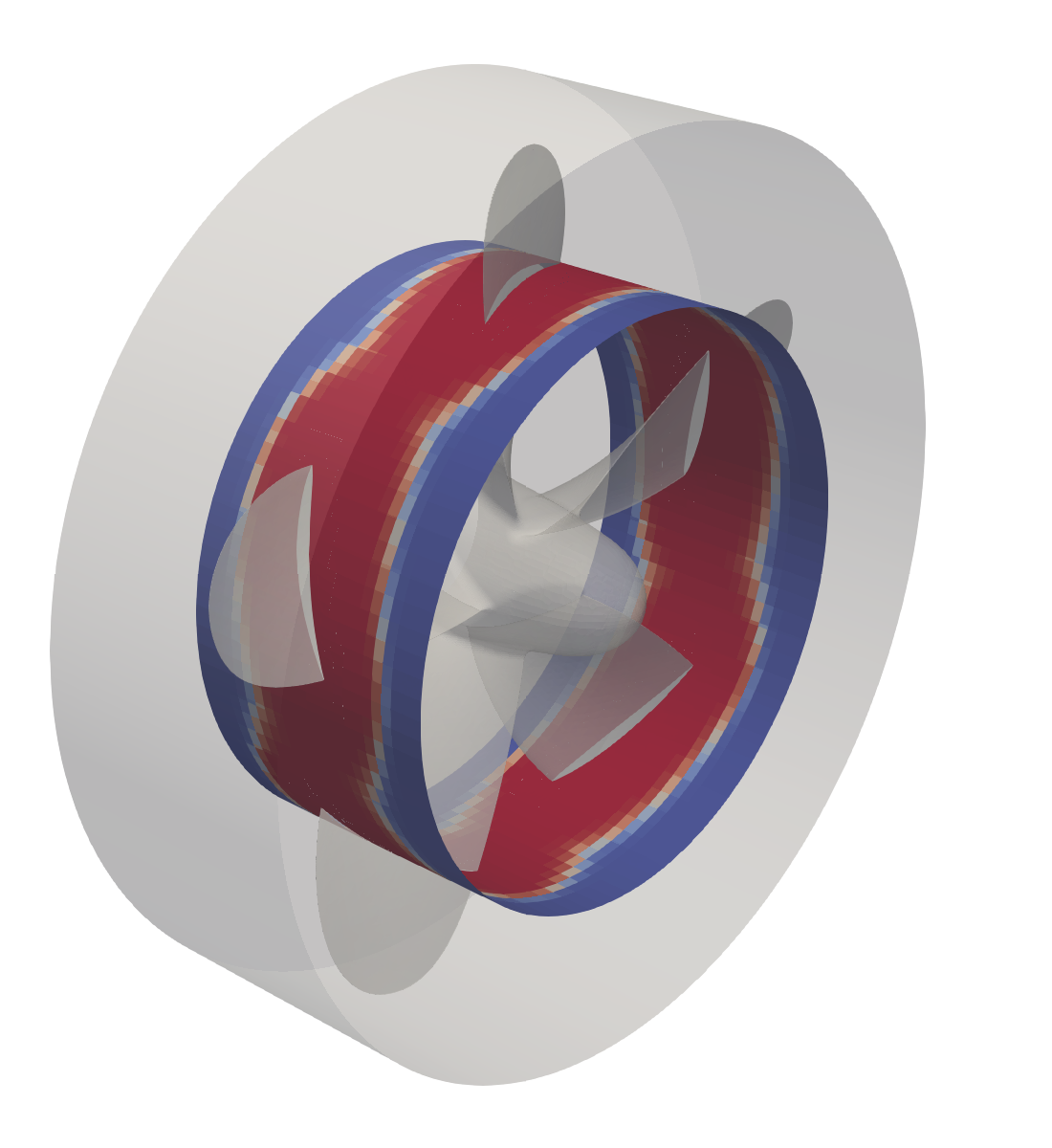}
            \captionsetup{width=0.9\linewidth}
            \caption{F08 ($q = 0.6, k = 40$)}
        \end{subfigure}
        \begin{subfigure}[t]{0.32\linewidth}
            \includegraphics[width=1.0\linewidth]{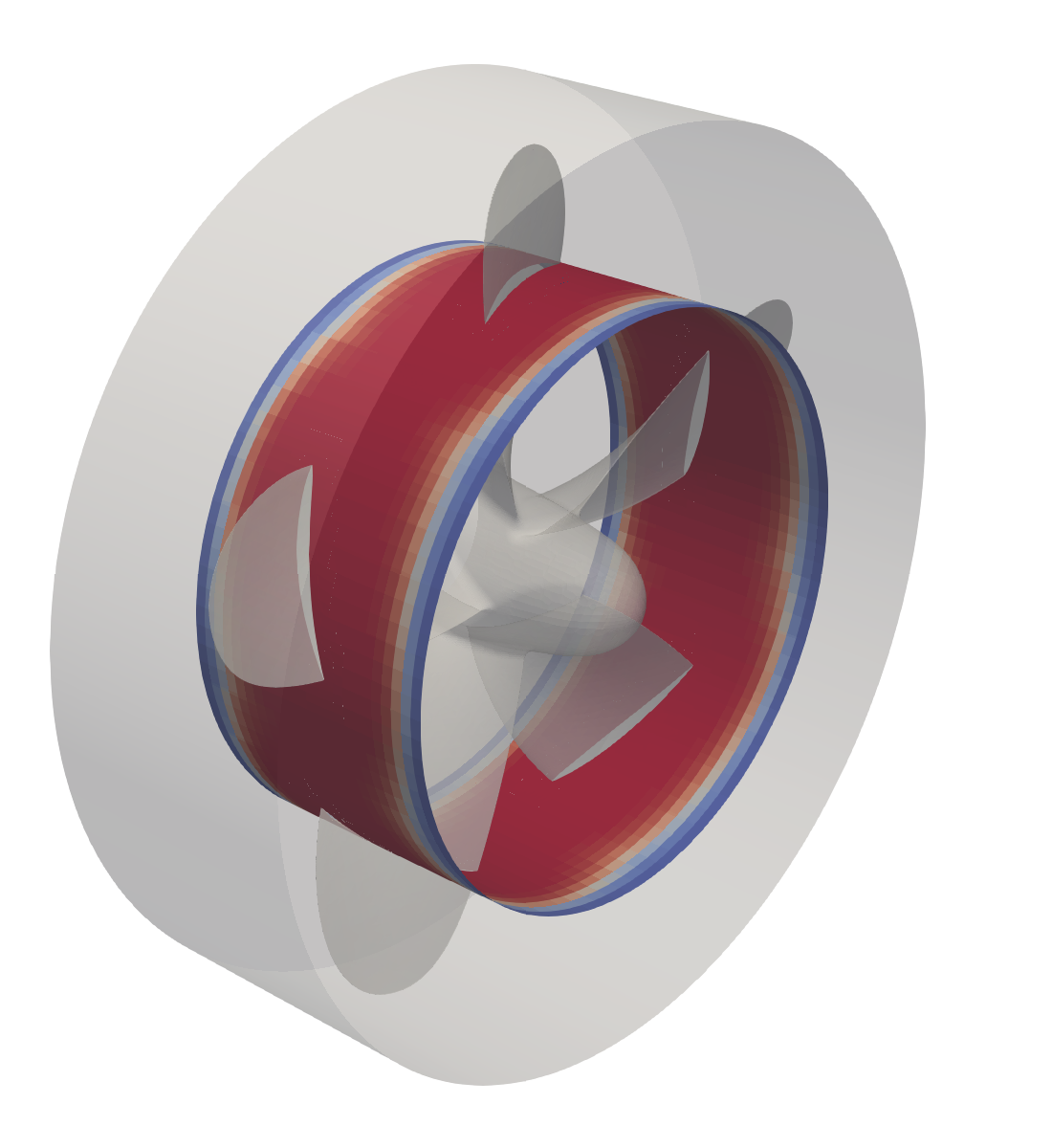}
            \captionsetup{width=0.9\linewidth}
            \caption{F09 ($q = 0.8, k = 20$)}
        \end{subfigure}
        \begin{subfigure}[t]{0.32\linewidth}
            \includegraphics[width=1.0\linewidth]{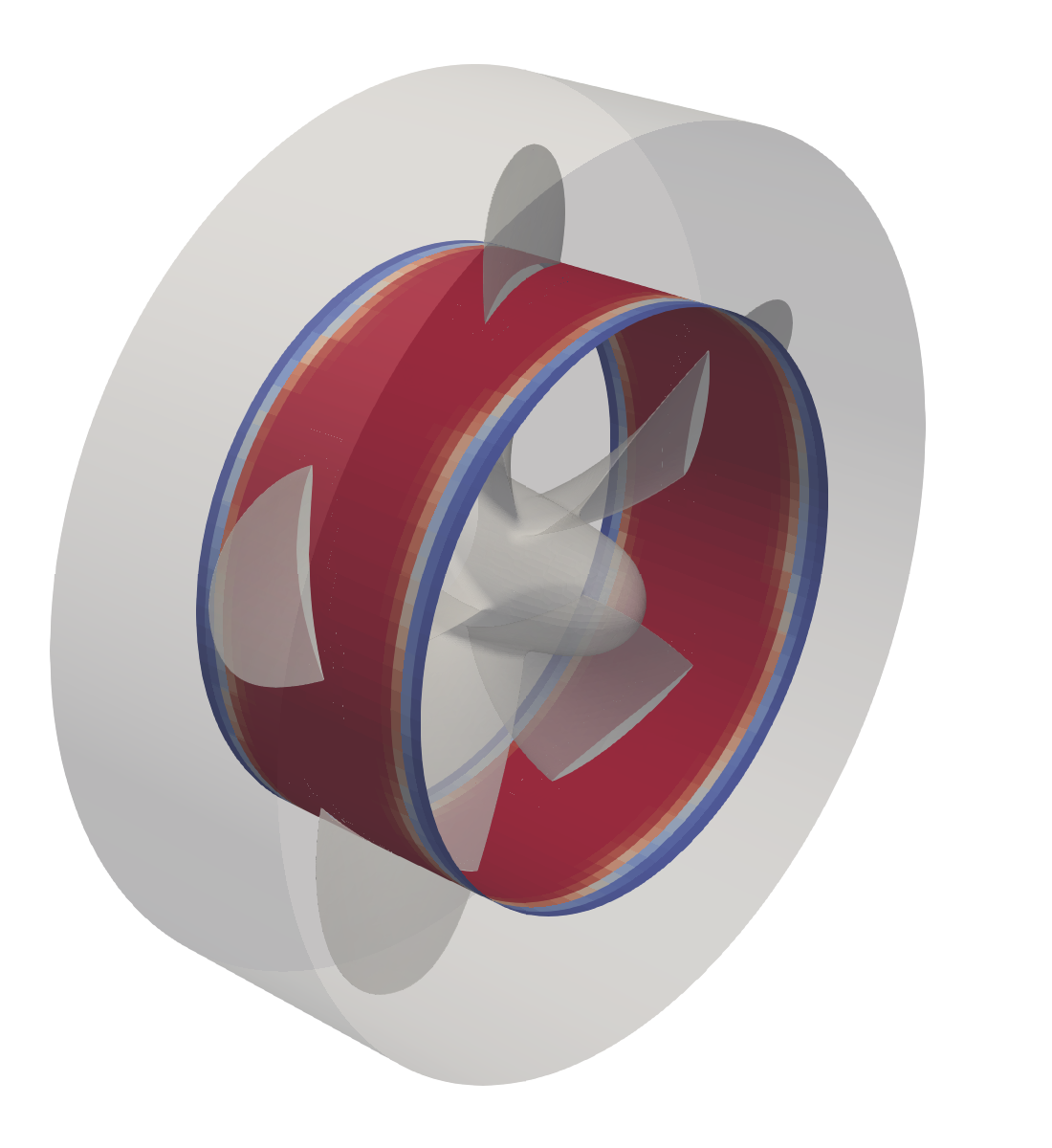}
            \captionsetup{width=0.9\linewidth}
            \caption{F10 ($q = 0.8, k = 30$)}
        \end{subfigure}
        \begin{subfigure}[t]{0.32\linewidth}
            \includegraphics[width=1.0\linewidth]{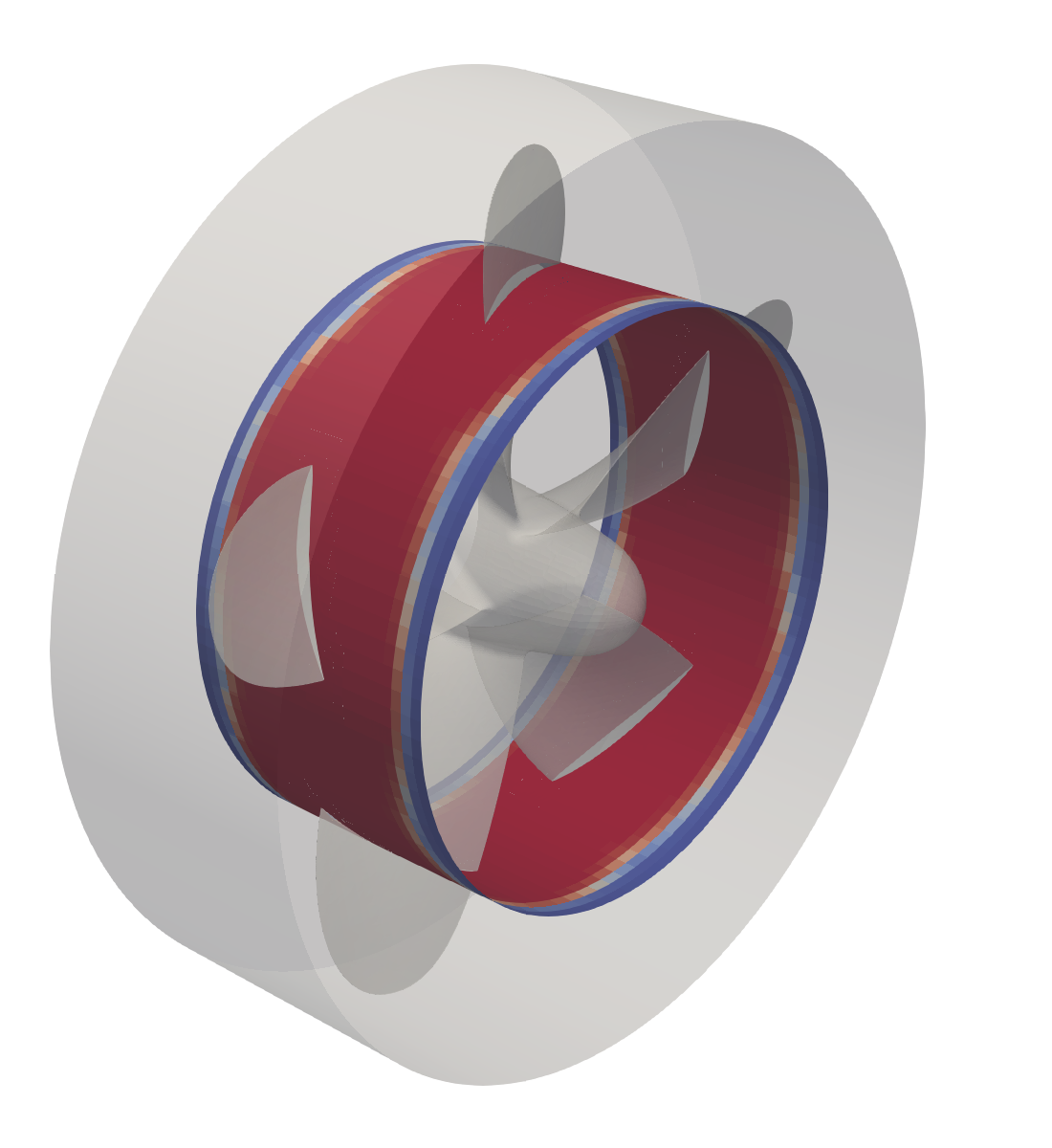}
            \captionsetup{width=0.9\linewidth}
            \caption{F11 ($q = 0.8, k = 40$)}
        \end{subfigure}
    \end{minipage}
    \begin{minipage}[c]{0.09\textwidth}
        \begin{subfigure}[t]{1.0\linewidth}
            \includegraphics[width=1.0\linewidth]{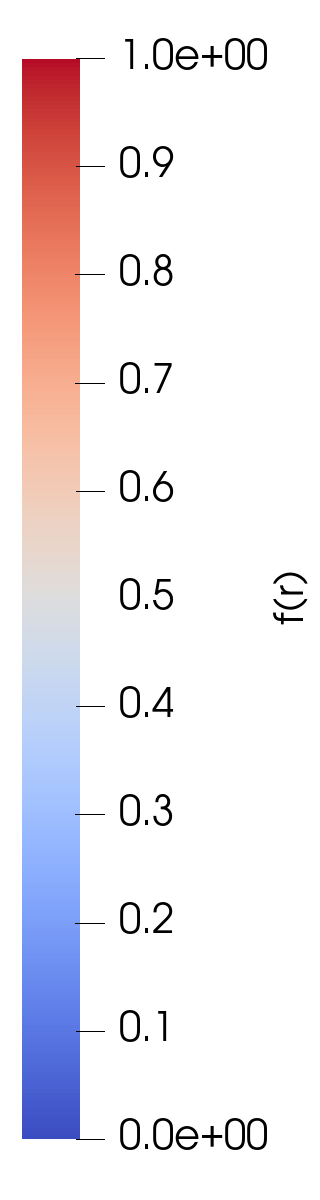}
            \captionsetup{width=0.9\linewidth}
        \end{subfigure}
    \end{minipage}
    \caption{Contour plots of the scaling function value at an exemplary cylindrical cross-section at $r/D_\mathrm{P} = 0.3$.}
    \label{fig:ramping_functions}
\end{figure}
\begin{table}[htbp]
    \centering

    \setcellgapes{4pt}
    \makegapedcells
    \begin{tabular}{l c c c c c c c c c}
        \toprule
        Case & Formulation & $n$ [rev/s] & $n_{\mathrm{MRF}}/n$ & $q$ & $k$ & $K_\mathrm{T}$ & $\Delta K_\mathrm{T}, \%$ & $K_\mathrm{Q}$ & $\Delta K_\mathrm{Q}, \%$ \\
        \midrule

        F01
        & SI
        & 8
        & 0.0
        & -
        & -
        & 0.20758
        &
        & 0.02812
        & \\

        \midrule

        F02
        & MRF
        & 8
        & 0.50
        & -
        & -
        & 0.20954 & 0.944\%
        & 0.02832 & 0.706\% \\

        \midrule

        F03
        & \multirow{9}{*}{mMRF}
        & \multirow{9}{*}{8}
        & \multirow{9}{*}{0.50}
        & 0.4
        & 20
        & 0.20209 & -2.646\%
        & 0.02736 & -2.693\% \\

        F04
        &
        &
        &
        & 0.4
        & 30
        & 0.20257 & -2.414\%
        & 0.02740 & -2.553\% \\

        F05
        &
        &
        &
        & 0.4
        & 40
        & 0.20284 & -2.286\%
        & 0.02742 & -2.471\% \\

        F06
        &
        &
        &
        & 0.6
        & 20
        & 0.20696 & -0.298\%
        & 0.02777 & -1.246\% \\

        F07
        &
        &
        &
        & 0.6
        & 30
        & 0.20698 & -0.289\%
        & 0.02777 & -1.246\% \\

        F08
        &
        &
        &
        & 0.6
        & 40
        & 0.20699 & -0.284\%
        & 0.02777 & -1.240\% \\

        F09
        &
        &
        &
        & 0.8
        & 20
        & 0.20804 & 0.232\%
        & 0.02809 & -0.113\% \\

        F10
        &
        &
        &
        & 0.8
        & 30
        & 0.20793 & 0.177\%
        & 0.02809 & -0.107\% \\

        F11
        &
        &
        &
        & 0.8
        & 40
        & 0.20797 & 0.196\%
        & 0.02809 & -0.107\% \\

        \bottomrule
    \end{tabular}
    \caption{Propeller performance results in the ship wake field obtained from mMRF simulations (F03-F11) for different scaling-function parameters $q$ and $k$ at a fixed propeller speed. Thrust ($K_\mathrm{T}$) and torque ($K_\mathrm{Q}$) coefficients are compared against the SI reference solution (F01) and the classical MRF approach (F02); relative errors $\Delta K_\mathrm{T}$ and $\Delta K_\mathrm{Q}$ are given with respect to F01.}
\label{tab:ramp_comp_table}
\end{table}

\begin{figure}[htbp]
    \centering
    \begin{subfigure}[t]{1.0\linewidth}
        \includegraphics[width=1.0\linewidth]{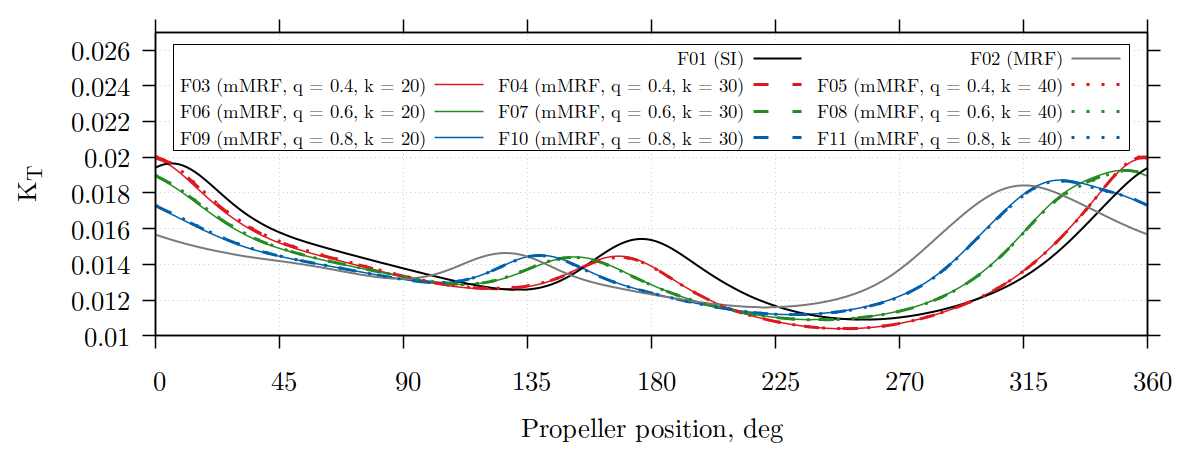}
        \captionsetup{width=0.9\linewidth}
    \end{subfigure}
    %
    %

    \caption{Blade thrust coefficients over propeller position for different mMRF parameters.}
    \label{fig:blade_over_rot}
\end{figure}
The evaluation and subsequent selection of the mMRF parameters $k$ and $q$,
 which describe the transition between the rotating and non-rotating regions,
 cf. Eqn.~(\ref{eq:sigmoid}), are performed using the self-propulsion configuration (Fig.~\ref{fig:sp-domain}). The setup employed is identical to that used in the subsequent Sec. \ref{ssec:self-prop}, except that a fixed propeller speed of $n = 8~\text{rev/s}$ is used to ensure a consistent comparison.
Table~\ref{tab:ramp_comp_table} compares results obtained from the mMRF method for different $k, q$ parameter combinations with sliding grid results (SI; F1) and classical MRF results (MRF; F2). The comparison is confined to the thrust and torque coefficients and the related deviations from the sliding grid result. 
It comprises 9 mMRF results (F03-F11) which involve three different transition locations, i.e., $q=0.4, 0.6, 0.8$, in combination with three different steepness parameters, i.e., $k=20, 30, 40$.  
Figure \ref{fig:ramping_functions} depicts contour plots of the scaling function values in an exemplary cylindrical plane for these 9 cases. 
Table~\ref{tab:ramp_comp_table} shows that, among the tested parameter combinations, the combination $q = 0.8$ and $k = 30$ (F10) exhibits the best agreement with the SI solution and is therefore selected for further use. It is also observed that increasing the extent of the rotating region by shifting the transition outward (i.e., larger $q$ values) leads to a noticeable phase shift in the blade thrust, cf. Fig.~\ref{fig:blade_over_rot}. This effect is attributed to the enlarged MRF region altering the local temporal response of the flow field. In contrast, variations in the sharpness of the transition (controlled by $k$) display a minor influence on the results within the investigated range.
\FloatBarrier
\subsection{Self-propulsion simulations -- Global flow features}
\label{ssec:self-prop}
\FloatBarrier
\begin{figure}[htbp]
    \centering
    \includegraphics[width=1.0\linewidth]{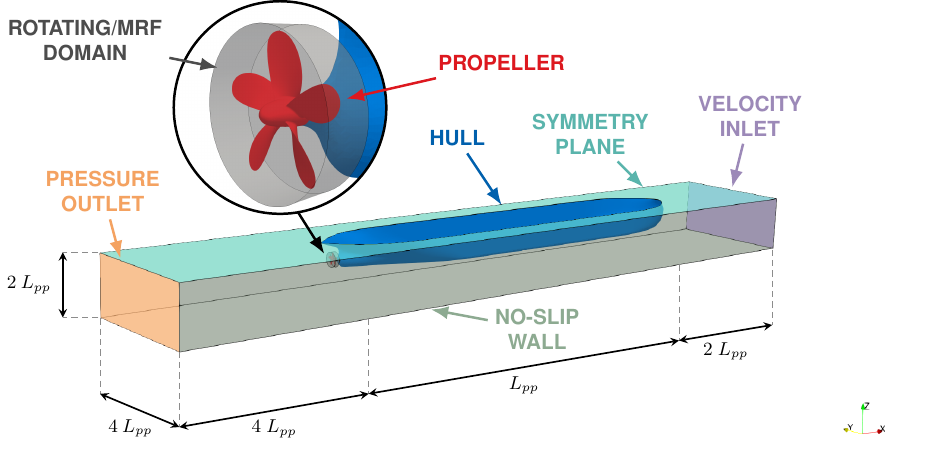}
    \caption[Self-propulsion domain]{Computational domain and boundary conditions for the JBC self-propulsion simulation. The full breadth is modeled; the propeller is enclosed in a rotating/MRF subdomain, shown enlarged in the figure. All distances are normalized by the length between perpendiculars $L_{pp}$. The figure is not in scale.}
    \label{fig:sp-domain}
\end{figure}
%
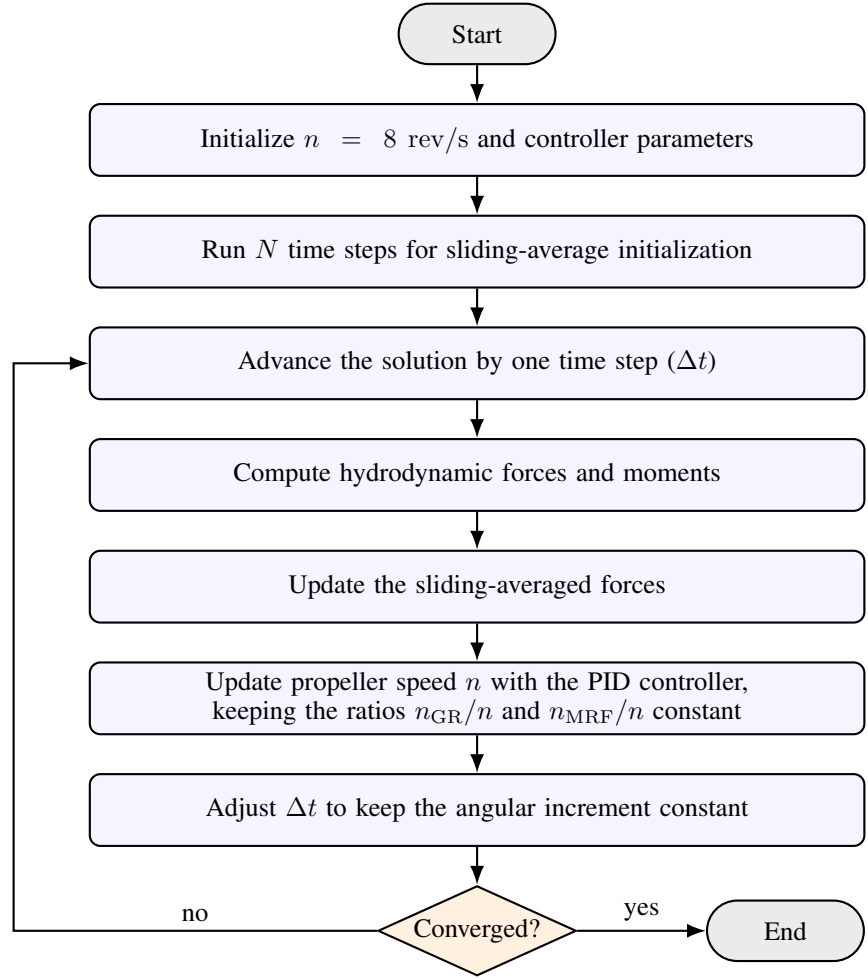
\begin{figure}[htbp]
\centering
\begin{tikzpicture}[
  node distance=0.5cm and 1.7cm,
  process/.style={rectangle, rounded corners, draw=black, thick,
                  text width=10cm, align=center, minimum height=0.95cm, fill=blue!4},
  terminal/.style={rounded rectangle, draw=black, thick,
                   text width=1.8cm, align=center, minimum height=0.8cm, fill=black!8},
  decision/.style={diamond, aspect=2.2, draw=black, thick,
                   inner sep=1pt, text width=1.7cm, align=center, fill=orange!12},
  arrow/.style={-{Latex[length=2.5mm]}, thick}
]
\node (start)   [terminal] {Start};
\node (param)   [process, below=of start]    {Initialize $n=8  \: \mathrm{rev/s}$ and controller parameters};
\node (warmup)  [process, below=of param]    {Run $N$ time steps for sliding-average initialization};
\node (advance) [process, below=of warmup]   {Advance the solution by one time step ($\Delta t$)};
\node (forces)  [process, below=of advance]  {Compute hydrodynamic forces and moments};
\node (avg)     [process, below=of forces]   {Update the sliding-averaged forces};
\node (pid)     [process, below=of avg]      {Update propeller speed $n$ with the PID controller, keeping the ratios $n_{\mathrm{GR}}/n$ and $n_{\mathrm{MRF}}/n$ constant};
\node (dt)      [process, below=of pid]      {Adjust $\Delta t$ to keep the angular increment constant};
\node (decision)[decision, below=of dt]      {Converged?};
\node (end)     [terminal, right=of decision] {End};
\draw[arrow] (start)--(param);
\draw[arrow] (param)--(warmup);
\draw[arrow] (warmup)--(advance);
\draw[arrow] (advance)--(forces);
\draw[arrow] (forces)--(avg);
\draw[arrow] (avg)--(pid);
\draw[arrow] (pid)--(dt);
\draw[arrow] (dt)--(decision);
\draw[arrow] (decision)-- node[above]{yes} (end);
\draw[arrow] (decision.west) -- node[above]{no} ++(-4.8,0) |- (advance.west);
\end{tikzpicture}
\caption{Flowchart of the propeller rotation-rate control algorithm.
}
\label{fig:pid_scheme}
\end{figure}
\begin{table}[ht!]
    \centering
    \begin{tabular}{l c c c c c}
    \toprule
    Case
    & Formulation
    & \makecell{Initial \\ $n$ [rev/s] }
    & $n_{\mathrm{MRF}}/n$
    & \makecell{Initial \\ $\Delta t$ [s] }
    & \makecell{Revolution \\ per time step [$^\circ$] } \\
    \midrule

    SP1 & SI & 8 & 0.00 & $5.0 \times 10^{-4}$ & 1.44 \\   

    \midrule

    SP2 & \multirow{4}{*}[-0.0em]{MRF}   & \multirow{4}{*}{8} & 0.25 & $6.67 \times 10^{-4}$ & 1.44 \\   
    SP3 &  & & 0.50 & $1.0  \times 10^{-3}$ & 1.44 \\                                  
    SP4 &  & & 0.75 & $2.0  \times 10^{-3}$ & 1.44 \\                                  
    SP5 &  & & 1.00 & $2.0  \times 10^{-3}$ & --   \\                                  

    \midrule

    SP6 & \multirow{4}{*}[-0.0em]{mMRF}  & \multirow{4}{*}{8} & 0.25 & $6.67 \times 10^{-4}$ & 1.44 \\   
    SP7 & & & 0.50 & $1.0  \times 10^{-3}$ & 1.44 \\                                   
    SP8 & & & 0.75 & $2.0  \times 10^{-3}$ & 1.44 \\                                   
    SP9 & & & 1.00 & $2.0  \times 10^{-3}$ & --   \\                                   

    \bottomrule
    \end{tabular}
    \caption{Rotation decomposition and time steps for self-propulsion simulation.}
    \label{tab:sp-setup}
\end{table}

\begin{table}[htbp]
    \centering
    \renewcommand{\arraystretch}{2.0}
    \setcellgapes{4pt}
    \makegapedcells
    \begin{tabular}{l c c c c c c c c c}
        \toprule
        \makecell{Test Case \\ (Formulation)} 
            & $n_{\mathrm{MRF}}/n$
            & $R_\mathrm{W}$ [N] & $F_{\mathrm{SFC}}$ [N] & $n$ [1/s] & $K_\mathrm{T}$ & $K_\mathrm{Q}$ & $1 - t$ & $1 - w_{\mathrm{T}}$ & $ \eta_{\mathrm{R}} $\\
        \midrule
        \makecell{EFD} & --- & --- & 18.2 & 7.810 & 0.2170 & 0.02790 & 0.804 & 0.552 & 1.015 \\
        \midrule
        \makecell{SP1 \\ (SI)}
                & 0.0
                & \multirow{9}{*}[-3.434em]{2.22}
                & \multirow{9}{*}[-3.434em]{18.2}
                & \makecell{7.7897 \\ (-0.26\%)} 
                & \makecell{0.2162 \\ (-0.38\%)} 
                & \makecell{0.02771 \\ (-0.68\%)}
                & \makecell{0.819 \\ (+1.87\%)} 
                & \makecell{0.562 \\ (+1.76\%)} 
                & \makecell{1.009 \\ (-0.59\%)} \\

        \makecell{SP2 \\ (MRF)} & 0.25 & &
                & \makecell{7.7917 \\ (-0.23\%)} 
                & \makecell{0.2154 \\ (-0.74\%)} 
                & \makecell{0.02762 \\ (-1.00\%)}
                & \makecell{0.819 \\ (+1.87\%)} 
                & \makecell{0.559 \\ (+1.22\%)} 
                & \makecell{1.011 \\ (-0.39\%)} \\

        \makecell{SP3 \\ (MRF)} & 0.50 & &
                & \makecell{7.7641 \\ (-0.59\%)} 
                & \makecell{0.2186 \\ (+0.74\%)} 
                & \makecell{0.02794 \\ (+0.14\%)}
                & \makecell{0.814 \\ (+1.24\%)} 
                & \makecell{0.557 \\ (+0.93\%)} 
                & \makecell{1.013 \\ (-0.20\%)} \\
                
        \makecell{SP4 \\ (MRF)} & 0.75 & &
                & \makecell{7.8329 \\ (+0.29\%)} 
                & \makecell{0.2209 \\ (+1.80\%)} 
                & \makecell{0.02822 \\ (+1.15\%)}
                & \makecell{0.791 \\ (-1.62\%)} 
                & \makecell{0.560 \\ (+1.41\%)} 
                & \makecell{1.014 \\ (-0.10\%)} \\

        \makecell{SP5 \\ (MRF)} & 1.0 & &
                & \makecell{7.8531 \\ (+0.55\%)} 
                & \makecell{0.2245 \\ (+3.46\%)} 
                & \makecell{0.02862 \\ (+2.58\%)}
                & \makecell{0.774 \\ (-3.73\%)} 
                & \makecell{0.566 \\ (+2.40\%)} 
                & \makecell{1.016 \\ (+0.10\%)} \\
        
        \makecell{SP6 \\ (mMRF)} & 0.25 & &
                & \makecell{7.7923 \\ (-0.23\%)}  
                & \makecell{0.2153 \\ (-0.78\%)} 
                & \makecell{0.02760 \\ (-1.08\%)}
                & \makecell{0.819 \\ (+1.87\%)} 
                & \makecell{0.558 \\ (+1.16\%)} 
                & \makecell{1.011 \\ (-0.39\%)} \\
        
        \makecell{SP7 \\ (mMRF)} & 0.50 & &
                & \makecell{7.7456 \\ (-0.82\%)} 
                & \makecell{0.2168 \\ (-0.09\%)} 
                & \makecell{0.02768 \\ (-0.79\%)}
                & \makecell{0.826 \\ (+2.73\%)} 
                & \makecell{0.554 \\ (+0.27\%)} 
                & \makecell{1.016 \\ (+0.10\%)} \\
        
        \makecell{SP8 \\ (mMRF)} & 0.75 & &
                & \makecell{7.7854 \\ (-0.31\%)} 
                & \makecell{0.2165 \\ (-0.23\%)} 
                & \makecell{0.02764 \\ (-0.93\%)}
                & \makecell{0.818 \\ (+1.74\%)} 
                & \makecell{0.552 \\ (+0.04\%)} 
                & \makecell{1.017 \\ (+0.20\%)} \\
        
        \makecell{SP9 \\ (mMRF)} & 1.0 & &
                & \makecell{7.8527 \\ (+0.55\%)}  
                & \makecell{0.2181 \\ (+0.51\%)} 
                & \makecell{0.02790 \\ (0.00\%)}
                & \makecell{0.799 \\ (-0.63\%)} 
                & \makecell{0.559 \\ (+1.22\%)} 
                & \makecell{1.015 \\ (0.00\%)} \\
        \bottomrule
    \end{tabular}
    \caption{Integral characteristics at self-propulsion point. Relative errors w.r.t EFD}
    \label{tab:sp_res_integral}
\end{table}
\begin{figure}[htbp]
    \centering
    %
    %
    \begin{subfigure}[t]{1.0\linewidth}
        \includegraphics[width=1.0\linewidth]{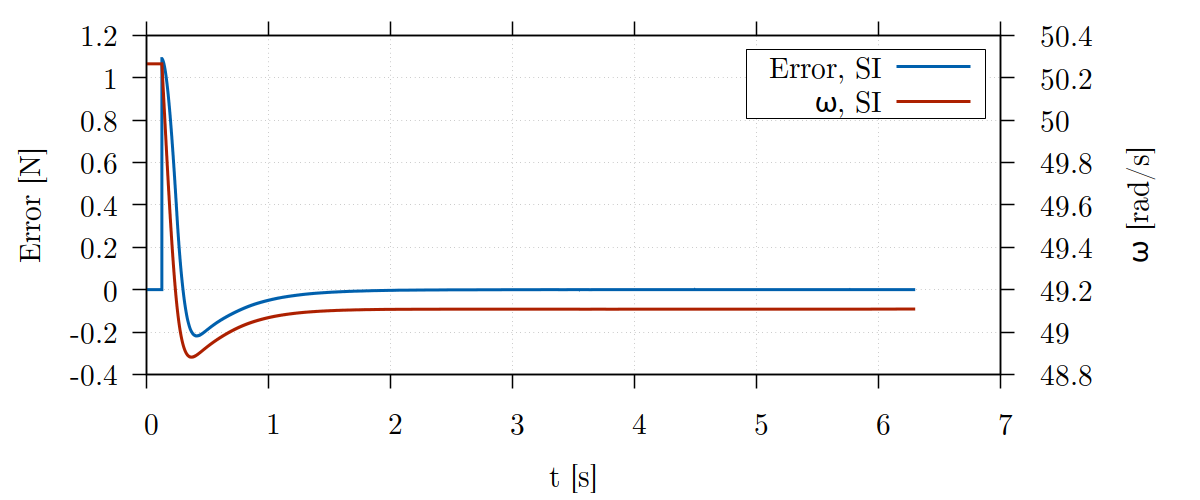}
        \captionsetup{width=0.9\linewidth}
    \end{subfigure}
    \caption[PID controller operation in the self-propulsion simulation]{Example operation of the PID controller used to reach the self-propulsion point. The thrust--drag imbalance (error, in~$N$) and the propeller rotation rate (in~$rad \cdot s^{-1}$) are shown as functions of time; the controller adjusts the rotation rate until the error converges to zero, i.e. the self-propulsion point is reached.}
    \label{fig:pid-conv}
\end{figure}
As mentioned before, self-propulsion simulations are performed using single-phase flow in combination with a double-body model, and the wave resistance contribution is supplemented by the value determined in Tab. \ref{tab:jbc_bare_resistance} of Sec. \ref{sec:wavedrag}. Table~\ref{tab:sp-setup} provides an overview of the investigated 9 cases and Fig.~\ref{fig:sp-domain} outlines the details of the employed computational domain and the boundary conditions. A symmetry-plane boundary condition is applied to represent the free surface. Model-scale computations are conducted and validated against available experimental data. For all simulations, the computational domain has a total breadth of $4\,L_{\mathrm{PP}}$, where the entire breadth of the domain is considered, and a total depth of $2\,L_{\mathrm{PP}}$.
The domain extends $2\,L_{\mathrm{PP}}$ upstream of the hull to the velocity inlet and $4\,L_{\mathrm{PP}}$ downstream to the pressure outlet.  

Simulations are performed at the ship propulsion point, incorporating a skin friction correction (SFC). The latter is used to correct the simulated (or measured) forces and 
 account for the difference between full-scale and model-scale frictional resistance, viz.  
\begin{equation}
\label{eq:sfc}
    \mathrm{SFC}
    =
    \left[{
        (1 + \beta)(C_{F0MS} - C_{F0FS}) - \Delta C_F
    }\right]
    \cdot 0.5
    \rho_w V_{MS}^2 S_W,
\end{equation}
where $(1+\beta)$ is a ship-specific form factor, $\Delta C_{\mathrm{F}}$ is the roughness allowance, $V_{\mathrm{MS}}$ denotes the ship speed at model scale, and $S_W$ is the wetted surface area. The model-scale ($C_{\mathrm{F0MS}}$) and full-scale ($C_{\mathrm{F0FS}}$) frictional coefficients are calculated using the ITTC
frictional correlation line 
\begin{equation}
\label{eq:c-frict}
    C_{F0} = \frac{0.075}{(\log_{10} Re - 2)^2}.
\end{equation}
The experimental form factor reads $(1 + \beta) = 1.314$, and the roughness allowance is $\Delta C_F = 0.12 \cdot 10^{-3}$, cf. \cite{hino-2020}. Details regarding the background and formalism of this procedure are published by the \cite{ittc-perf-pred}. 

A proportional-integral-derivative (PID) controller is used to regulate the propeller rotation rate $n$ such that a balance of forces exists between the thrust ($T$), the total resistance of the double-body configuration ($R_T$), the wave  resistance ($R_W$) outlined in Tab.\ref{tab:jbc_bare_resistance} and the SFC correction. To this end, an error $e$ is defined 
\begin{equation}
\label{eqn:pid-err}
    e = R_\mathrm{T} - T - \mathrm{SFC} + R_\mathrm{W},
\end{equation}
The PID controller employs the temporal evolution of $e$ to adjust the rotation rate $n$ according to 
\begin{equation}
\label{eq:pid}
    n
    =
    n_{\mathrm{old}}
    +
    K_\mathrm{P} e
    +
    K_\mathrm{I} \int_{t_{\mathrm{start}}}^{t_{\mathrm{end}}} e \, dt
    +
    K_\mathrm{D} \cfrac{e - e_{\mathrm{old}}}{\Delta t},
\end{equation}
where  $n_{\mathrm{old}}$ and $e_{\mathrm{old}}$ denote the rotation rate and the error (\ref{eqn:pid-err}) at the previous time step, respectively. 
  The constants $K_P$, $K_I$, and $K_D$ are the proportional, integral, and differential gains, determined empirically through manual tuning to achieve stable and sufficiently fast convergence of the propeller rotation rate. The employed values for $K_P$, $K_I$, and $K_D$ read $0.5 \cdot 10^{-2}$, $0.5 \cdot 10^{-5}$, and $0.5 \cdot 10^{-4}$, respectively.  A flowchart of the algorithm for propeller motion control is shown in Fig.~\ref{fig:pid_scheme}.

Self-propulsion simulations are performed for a sequence of fixed values of 
the normalized moving reference frame rotation $n_{\mathrm{MRF}}/n$ and the corresponding normalized grid-resolved rotation $n_{\mathrm{GR}}/n = 1-n_{\mathrm{MRF}}/n$. Hence, the case-specific relative contributions of $n_{\mathrm{GR}}/n$ and $n_{\mathrm{MRF}}/n$ remain constant, while the PID controller updates $n$ 
according to the algorithm in Fig.~\ref{fig:pid_scheme}.
For cases with non-zero grid-resolved rotation $n_{\mathrm{GR}}$, the time step is adjusted to maintain a constant rotation angle per step. If only $n_{\mathrm{MRF}}$ is used, the time step remains constant throughout the simulation.

Self-propulsion parameters and propulsion factors 
are reported in Table~\ref{tab:sp_res_integral}, where the relative differences with respect to available model-scale experimental data published by \cite{hino-2020} are given in brackets.
Self-propulsion factors are determined using the thrust identity method, cf. \cite{ittc-perf-pred}, in which the thrust coefficient $K_{\mathrm{T}}$ at self-propulsion is assumed equal to that obtained from the open-water test. The propeller open-water data obtained from the numerical simulations are used in the thrust identity calculation to evaluate the corresponding self-propulsion factors ($1 - w_{\mathrm{T}}$, $\eta_{\mathrm{R}}$).

Figure~\ref{fig:pid-conv} shows the convergence of the error $e$ and the propeller rotation rate during the PID-controlled simulation for the sliding grid case SP1 (SI), serving as a representative example of the convergence behavior observed in all cases.

The integral self-propulsion results summarized in Table~\ref{tab:sp_res_integral} indicate close agreement with the available experimental data (EFD). 
Across all simulated cases, the thrust coefficient $K_{\mathrm{T}}$ and torque coefficient $K_{\mathrm{Q}}$ remain within a few percent of the reference values, while the propeller rotational velocity $n$ is predicted with deviations of less than 1\%.
The relative efficiency $\eta_{\mathrm{R}}$ is captured with very high accuracy, showing only marginal differences across all cases. These results show that both the classical MRF and the proposed mMRF approach can reproduce the main integral characteristics of the self-propulsion point. However, their accuracy differs more markedly as the moving reference frame fraction $n_\mathrm{MRF}/n$ increases. In this regard, the mMRF approach is closer to the experimental values.
Another  clear trend is observed when increasing the fraction $n_{\mathrm{MRF}}/n$: both $K_{\mathrm{T}}$ and $K_{\mathrm{Q}}$ tend to increase, while the thrust deduction factor $1-t$ decreases. This behavior is consistent for both, the classical MRF and mMRF formulations and reflects the influence of the rotation modeling on the propeller--hull interaction. At high MRF ratios (e.g., SP4--SP5), the deviations from experimental data become more pronounced, especially for $K_{\mathrm{T}}$ and $1-t$, indicating reduced accuracy when the rotation is predominantly modeled by the steady moving reference frame approach.
In comparison, the mMRF results (SP6--SP9) show improved agreement with experimental data at higher MRF ratios. In particular, for cases with $n_{\mathrm{MRF}}/n \geq 0.5$, the mMRF formulation provides more consistent predictions of $K_{\mathrm{T}}$, $K_{\mathrm{Q}}$, and $\eta_{\mathrm{R}}$ compared to the classical MRF. This suggests that the gradual transition of the rotational velocity in the mMRF approach helps to better preserve the interaction between the propeller and the surrounding flow. Overall, the results demonstrate that the proposed method maintains accuracy while offering additional robustness for cases with a large MRF contribution.
\FloatBarrier
\FloatBarrier
\subsection{Self-propulsion simulations -- Local flow features}
\label{sec:results}
\FloatBarrier
\begin{figure}[htbp]
    \centering
    \begin{minipage}[c]{0.9\textwidth}
    \centering
    \begin{subfigure}[t]{0.32\linewidth}
        \includegraphics[width=1.0\linewidth]{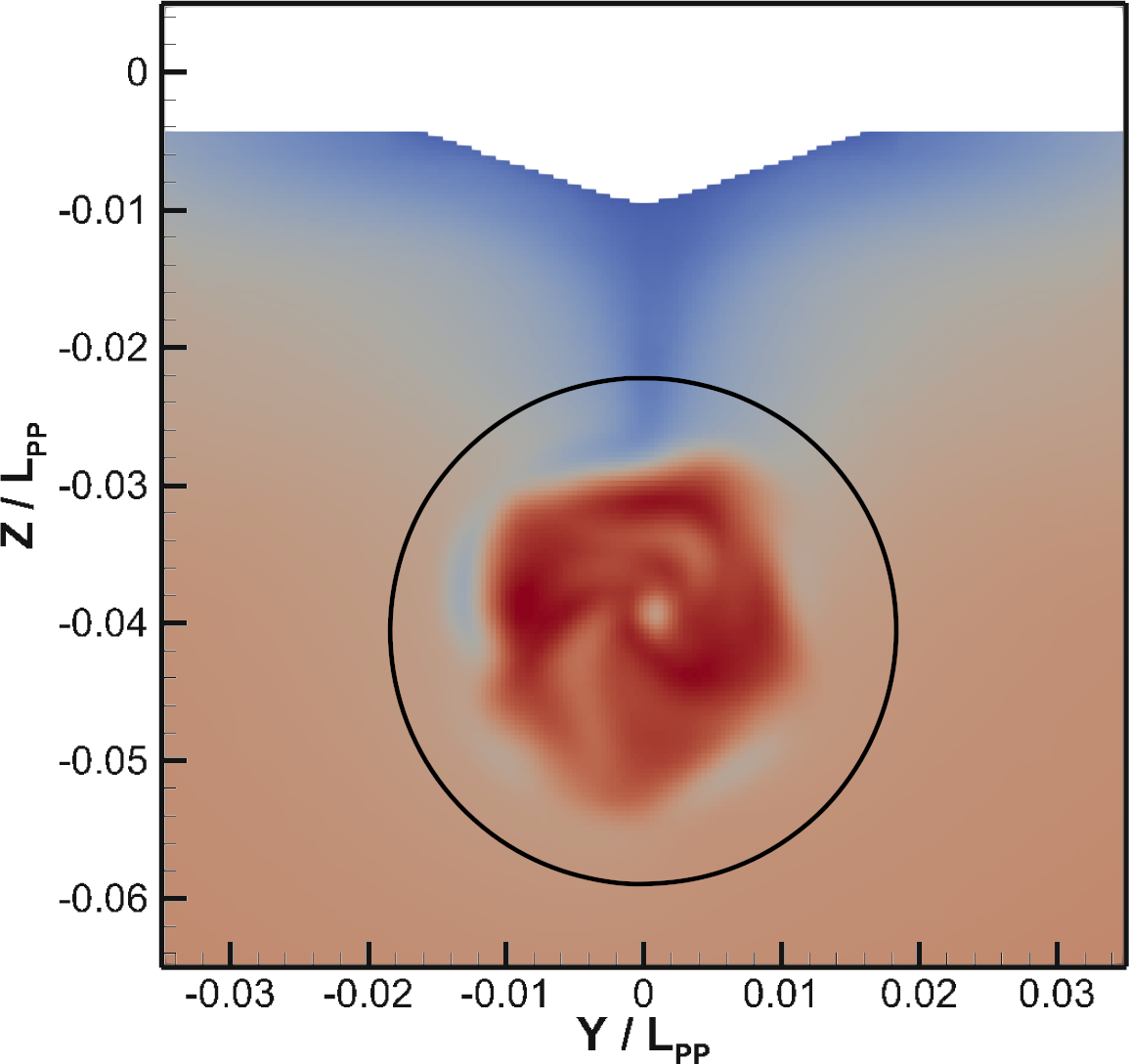}
        \captionsetup{width=0.9\linewidth}
        \caption{S1, $x / L_{\mathrm{PP}} = 0.0$}
    \end{subfigure}
    \begin{subfigure}[t]{0.32\linewidth}
        \includegraphics[width=1.0\linewidth]{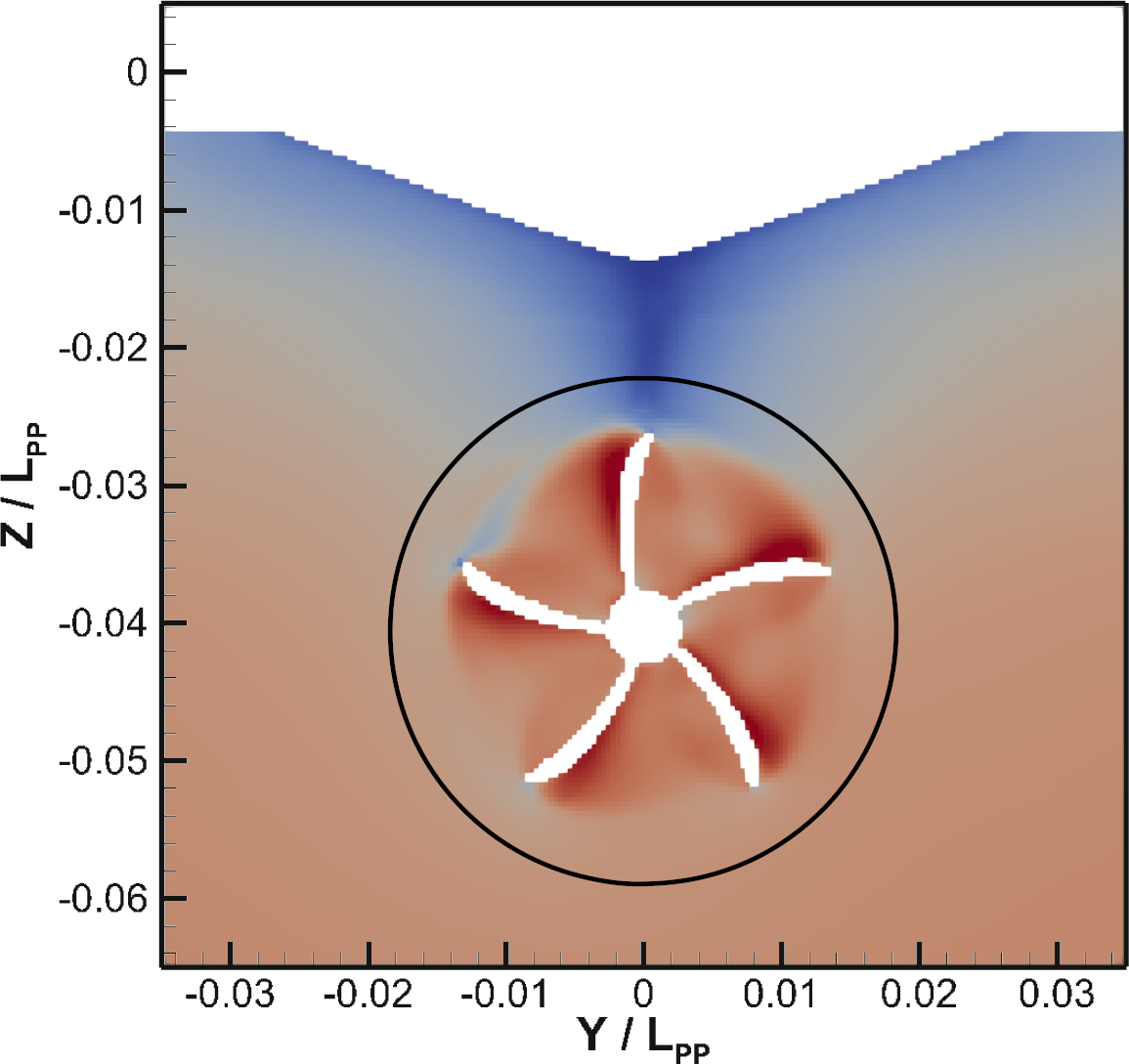}
        \captionsetup{width=0.9\linewidth}
        \caption{S1, $x / L_{\mathrm{PP}} = 0.0134$}
    \end{subfigure}
    \begin{subfigure}[t]{0.32\linewidth}
        \includegraphics[width=1.0\linewidth]{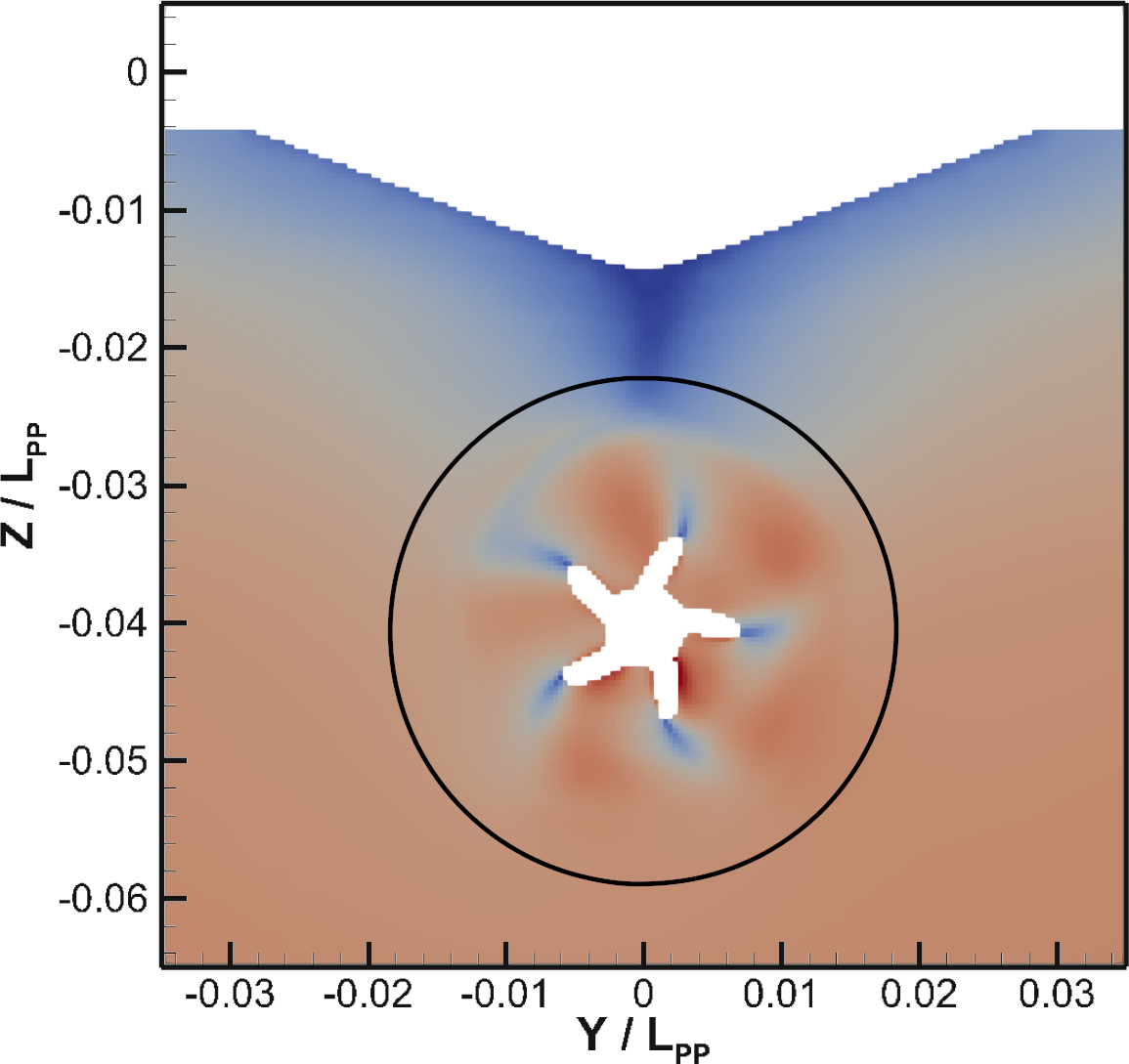}
        \captionsetup{width=0.9\linewidth}
        \caption{S1, $x / L_{\mathrm{PP}} = 0.0157$}
    \end{subfigure}
    \end{minipage}
    \begin{minipage}[c]{0.09\textwidth}
    \begin{subfigure}[t]{0.67\linewidth}
        \includegraphics[width=1.0\linewidth]{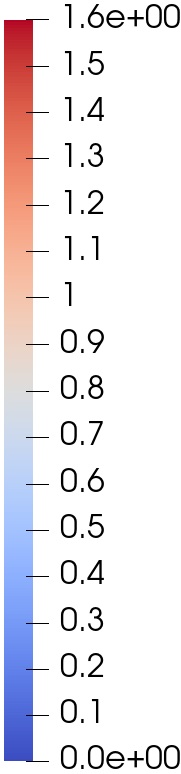}
        \vspace*{17pt}
    \end{subfigure}
    \end{minipage}
    \caption{Velocity-magnitude contour plots in the downstream ($x/L_{\mathrm{PP}}=0$), propeller ($x/L_{\mathrm{PP}}=0.0134$) and upstream ($x/L_{\mathrm{PP}}=0.0157$) planes, predicted by the sliding grid method (SP1). The circle marks the $1.5\,R_{\mathrm{P}}$ disc over which the deviation of Eq.~\eqref{eq:rms} is averaged.
    }
    \label{fig:v_magnitude}
\end{figure}
\begin{figure}[htbp]
    \centering
    \begin{minipage}[c]{0.9\textwidth}
    \centering
    \begin{subfigure}[t]{0.32\linewidth}
        \includegraphics[width=1.0\linewidth]{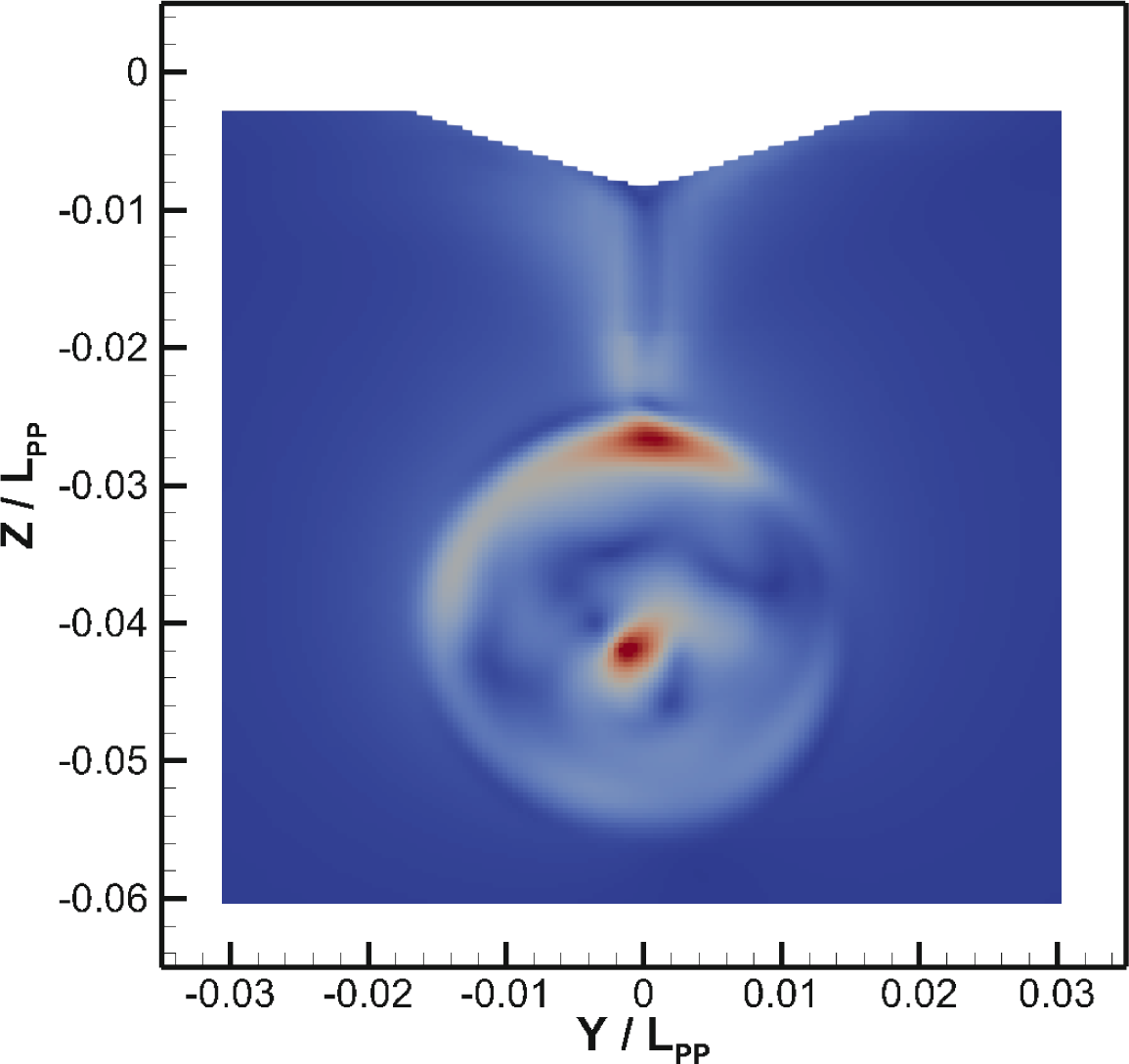}
        \captionsetup{width=0.9\linewidth}
        \caption{S3 (MRF, $n_{\mathrm{MRF}}/n = 0.50$)}
    \end{subfigure}
    \begin{subfigure}[t]{0.32\linewidth}
        \includegraphics[width=1.0\linewidth]{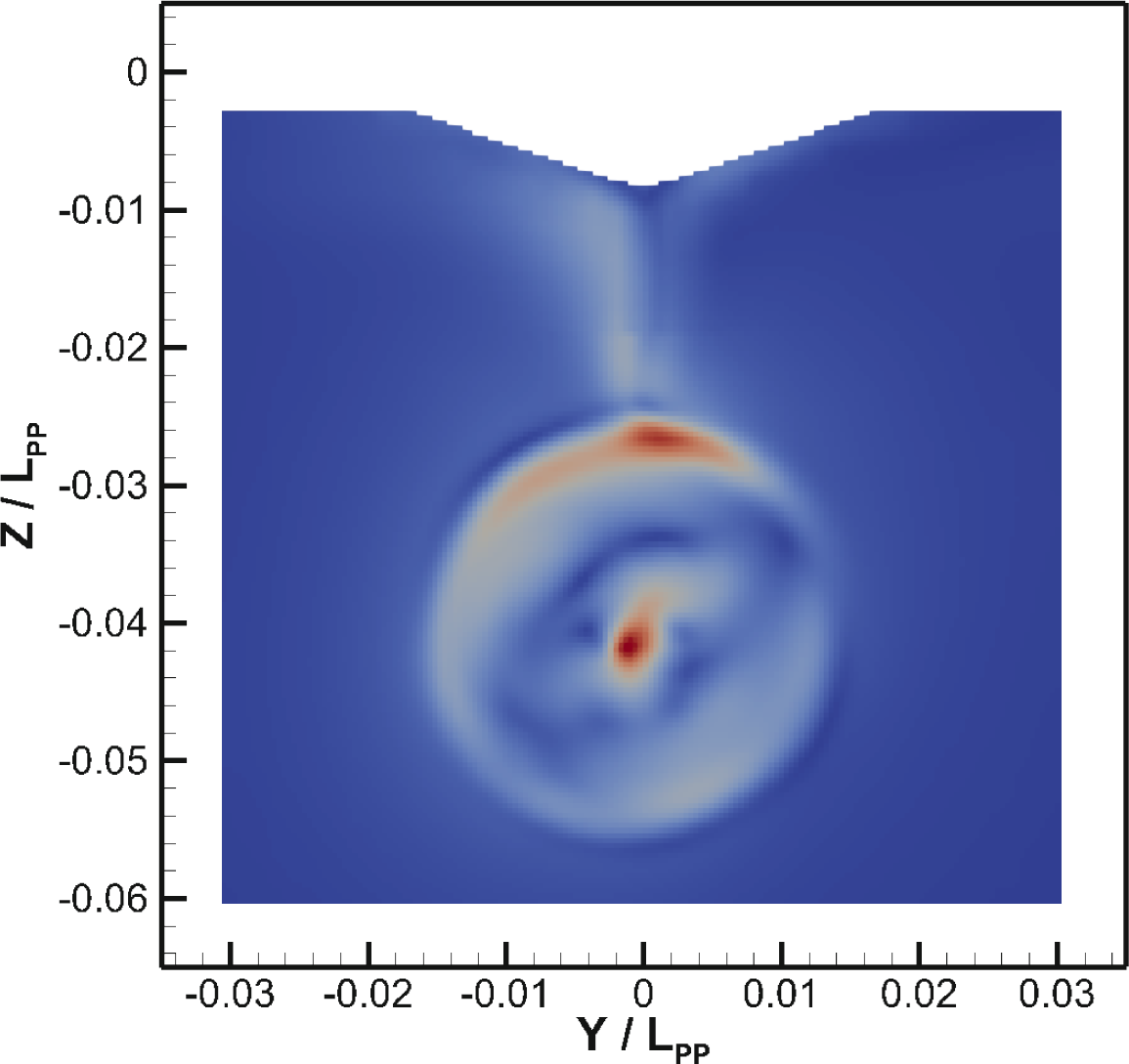}
        \captionsetup{width=0.9\linewidth}
        \caption{S4 (MRF, $n_{\mathrm{MRF}}/n = 0.75$)}
    \end{subfigure}
    \begin{subfigure}[t]{0.32\linewidth}
        \includegraphics[width=1.0\linewidth]{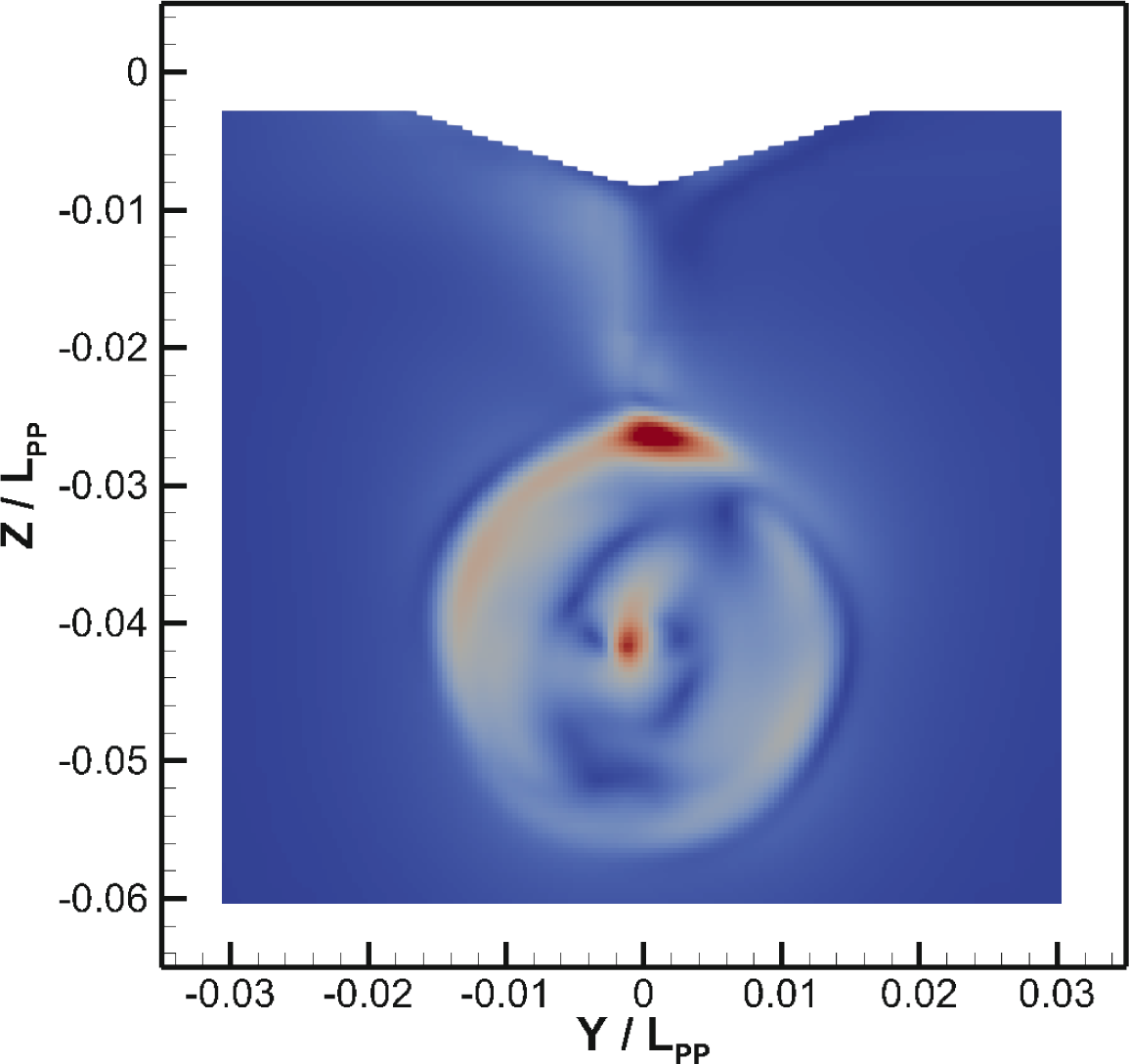}
        \captionsetup{width=0.9\linewidth}
        \caption{S5 (MRF, $n_{\mathrm{MRF}}/n = 1.0$)}
    \end{subfigure}
    \begin{subfigure}[t]{0.32\linewidth}
        \includegraphics[width=1.0\linewidth]{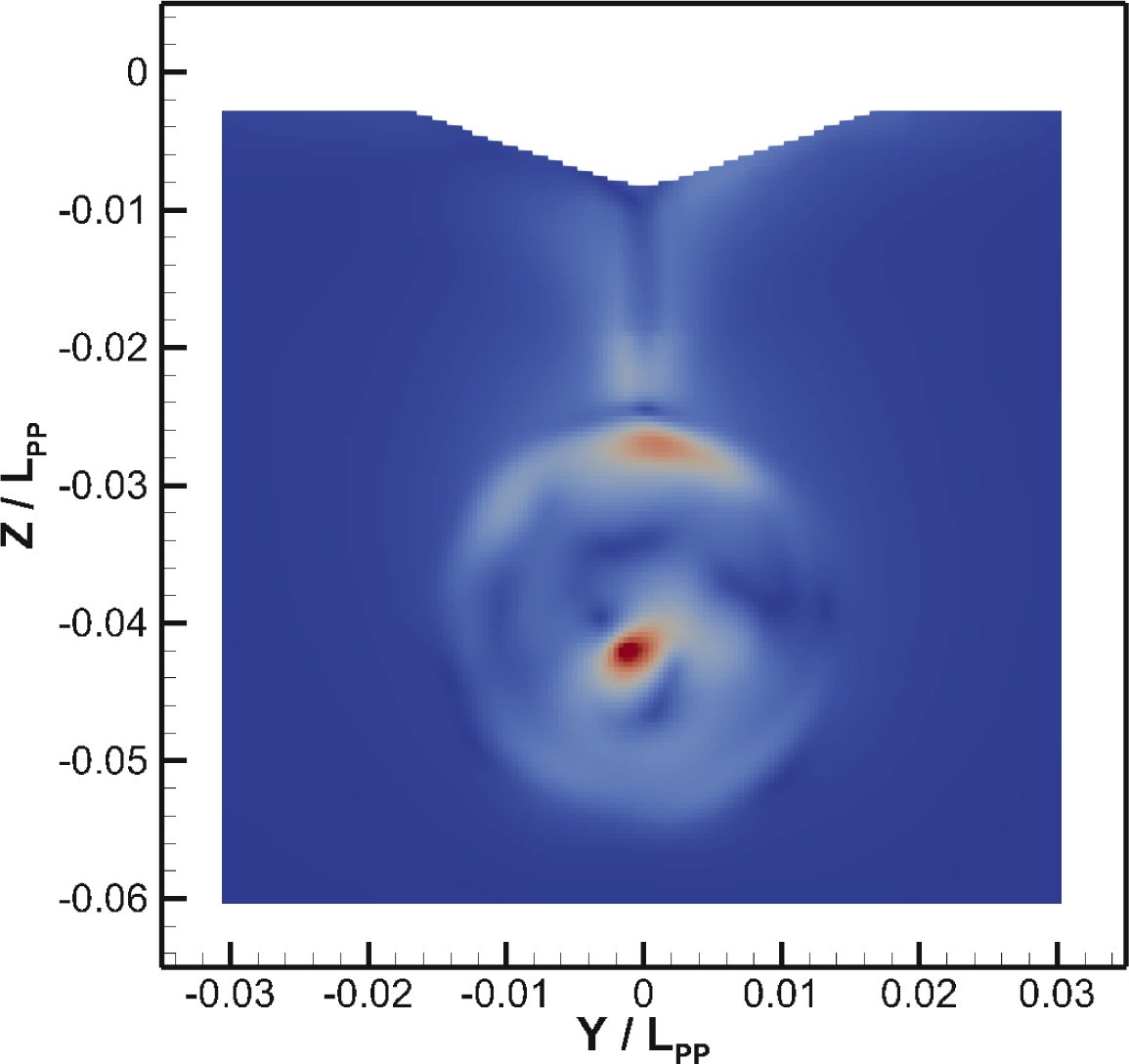}
        \captionsetup{width=0.9\linewidth}
        \caption{S7 (mMRF, $n_{\mathrm{MRF}}/n = 0.50$)}
    \end{subfigure}
    \begin{subfigure}[t]{0.32\linewidth}
        \includegraphics[width=1.0\linewidth]{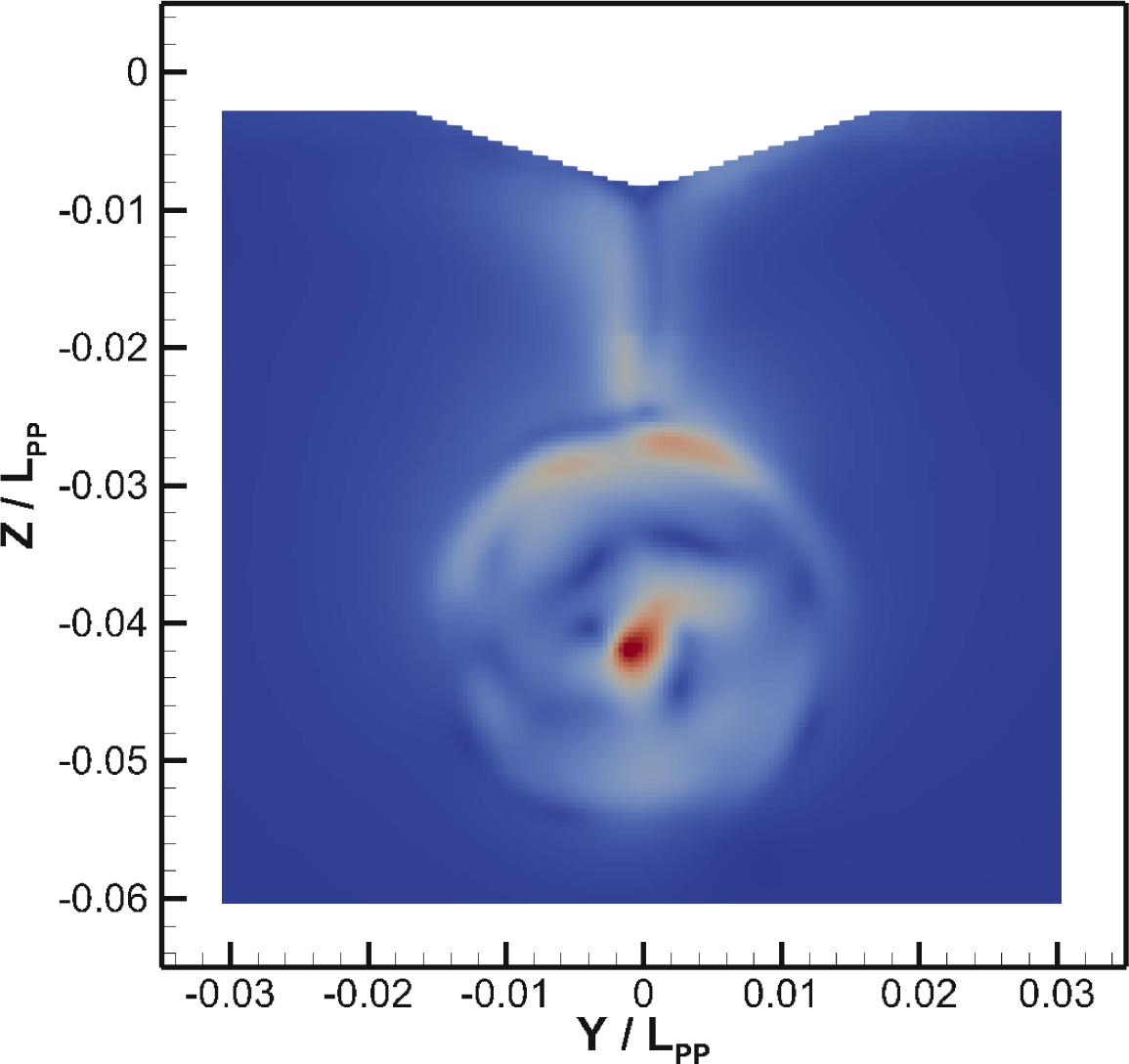}
        \captionsetup{width=0.9\linewidth}
        \caption{S8 (mMRF, $n_{\mathrm{MRF}}/n = 0.75$)}
    \end{subfigure}
    \begin{subfigure}[t]{0.32\linewidth}
        \includegraphics[width=1.0\linewidth]{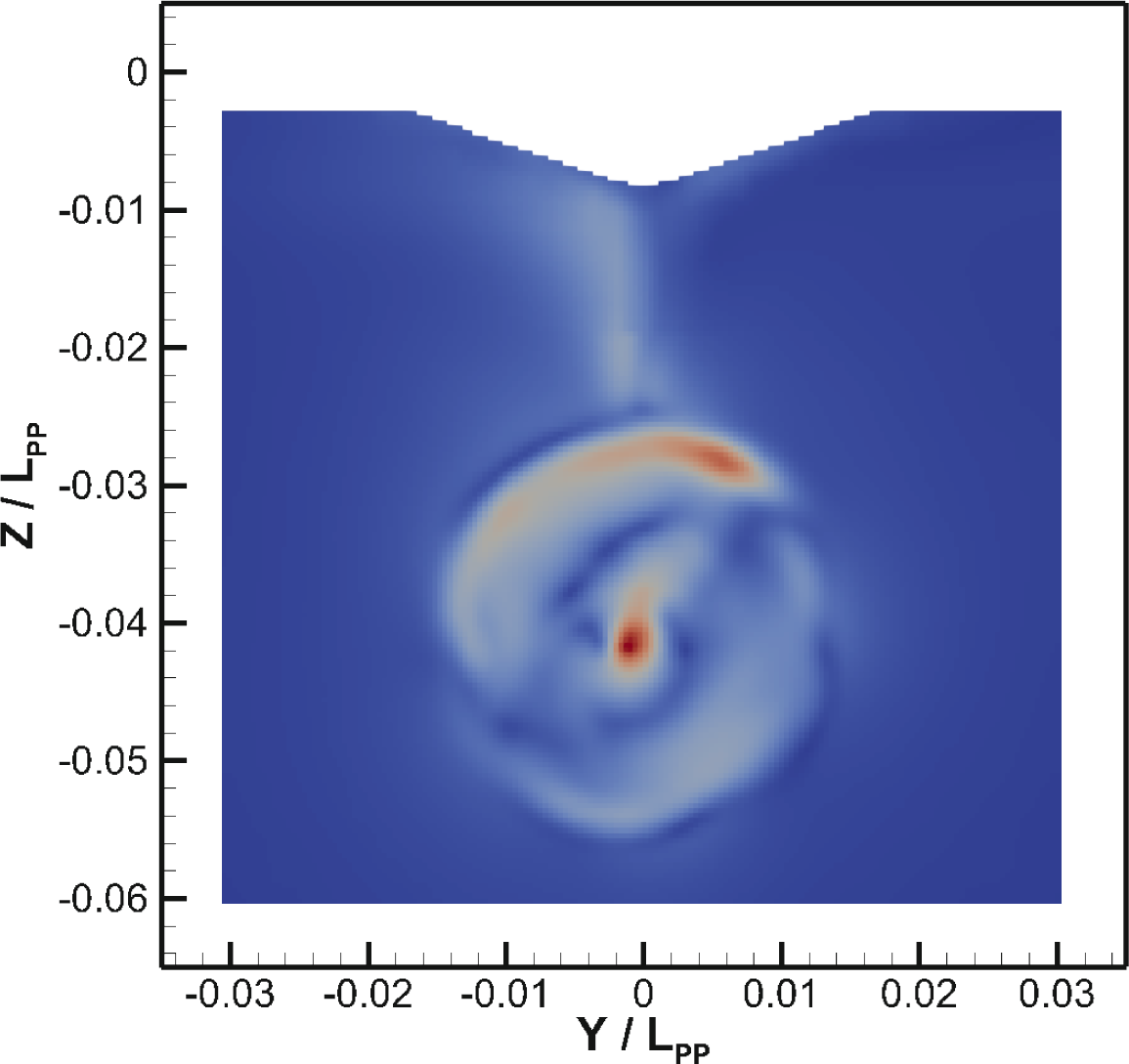}
        \captionsetup{width=0.9\linewidth}
        \caption{S9 (mMRF, $n_{\mathrm{MRF}}/n = 1.0$)}
    \end{subfigure}
    \end{minipage}
    \begin{minipage}[c]{0.09\textwidth}
    \begin{subfigure}[t]{1.0\linewidth}
        \includegraphics[width=1.0\linewidth]{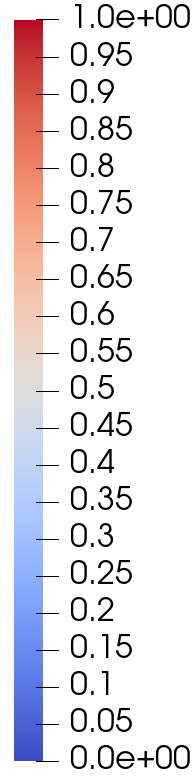}
    \end{subfigure}
    \end{minipage}
    \caption{Contour plots of the normalized velocity-magnitude deviation from the SI solution (SP1), cf. Eqn.~\eqref{eq:rms}. Results refer to the downstream plane ($x/L_{\mathrm{PP}}=0$) and compare the classical MRF (top) with the mMRF (bottom) formulation for different normalized MRF rates $n_{\mathrm{MRF}}/n=0.5$, $0.75$ and $1.0$.}
    \label{fig:rms_x0}
\end{figure}
\begin{figure}[htbp]
    \centering
    \begin{minipage}[c]{0.9\textwidth}
    \centering
    \begin{subfigure}[t]{0.32\linewidth}
        \includegraphics[width=1.0\linewidth]{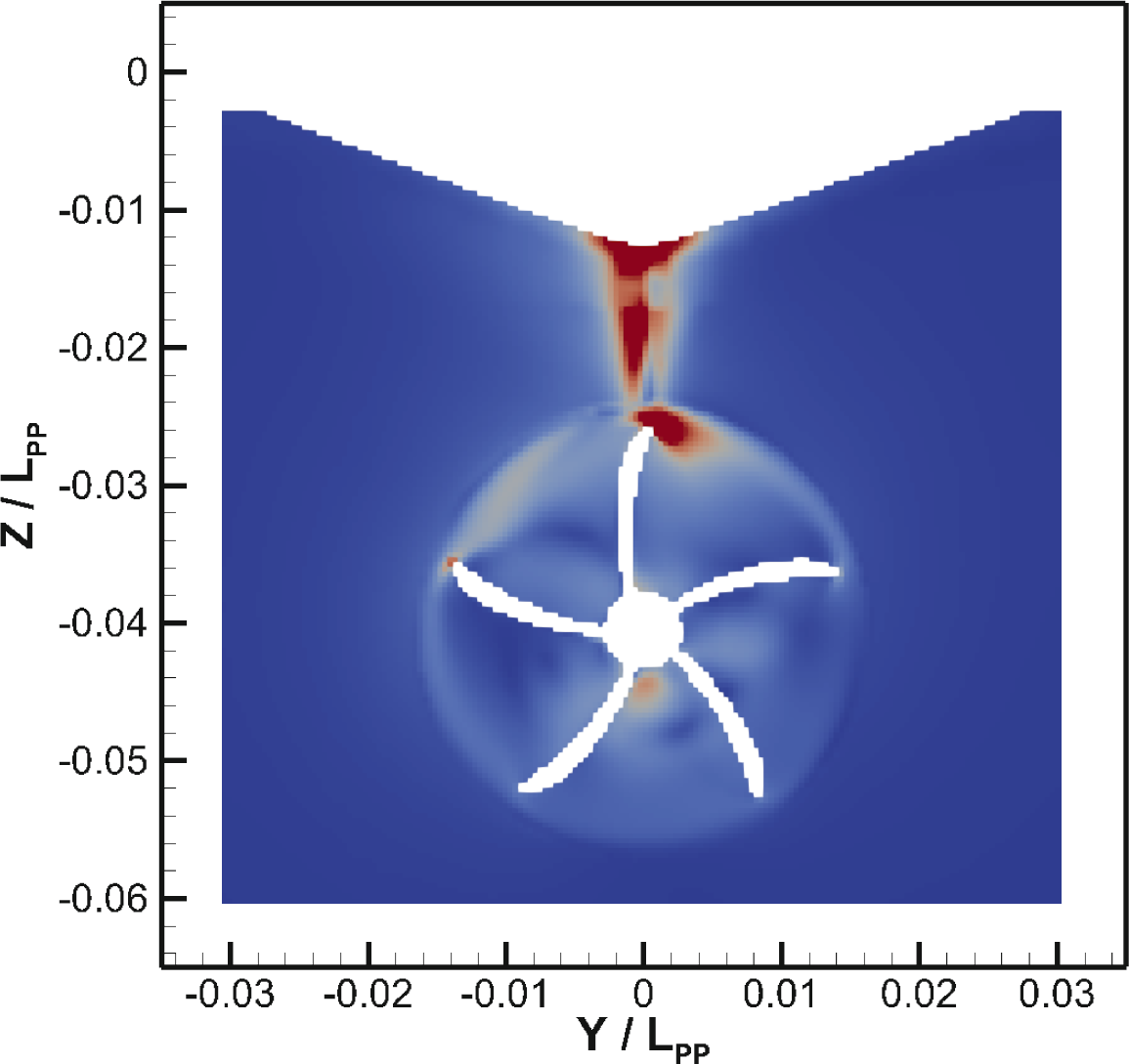}
        \captionsetup{width=0.9\linewidth}
        \caption{S3 (MRF, $n_{\mathrm{MRF}}/n = 0.50$)}
    \end{subfigure}
    \begin{subfigure}[t]{0.32\linewidth}
        \includegraphics[width=1.0\linewidth]{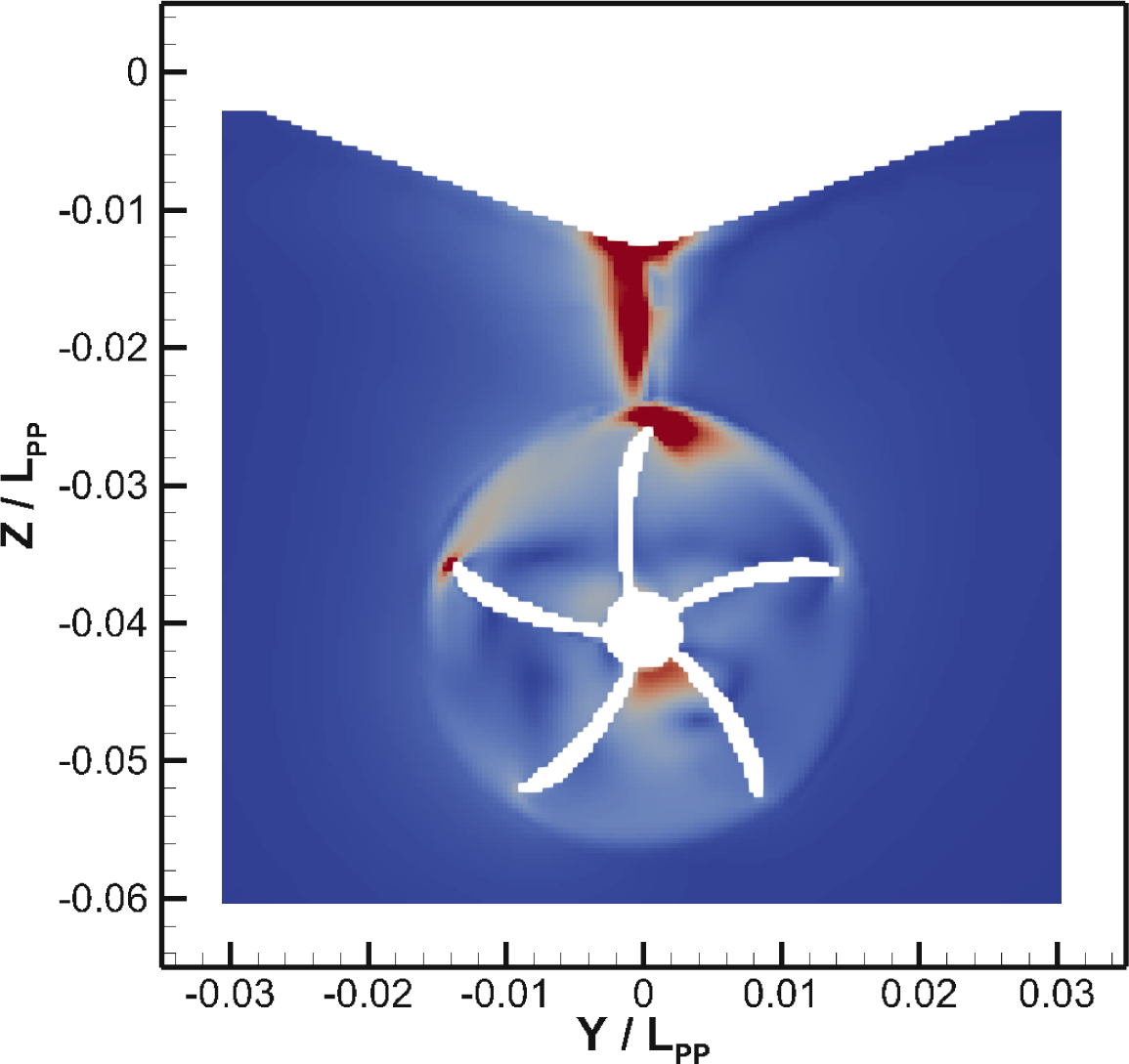}
        \captionsetup{width=0.9\linewidth}
        \caption{S4 (MRF, $n_{\mathrm{MRF}}/n = 0.75$)}
    \end{subfigure}
    \begin{subfigure}[t]{0.32\linewidth}
        \includegraphics[width=1.0\linewidth]{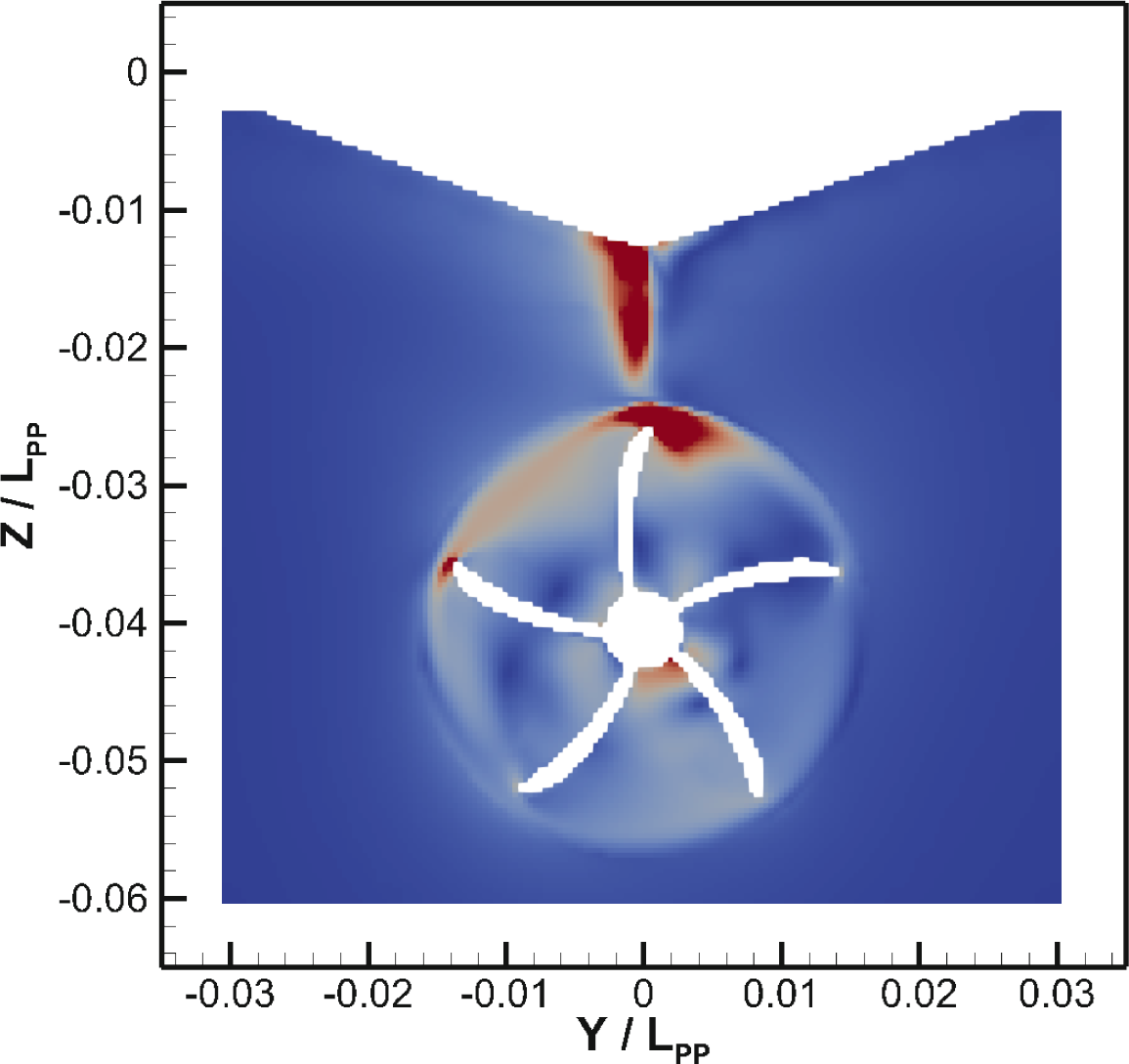}
        \captionsetup{width=0.9\linewidth}
        \caption{S5 (MRF, $n_{\mathrm{MRF}}/n = 1.0$)}
    \end{subfigure}
    \begin{subfigure}[t]{0.32\linewidth}
        \includegraphics[width=1.0\linewidth]{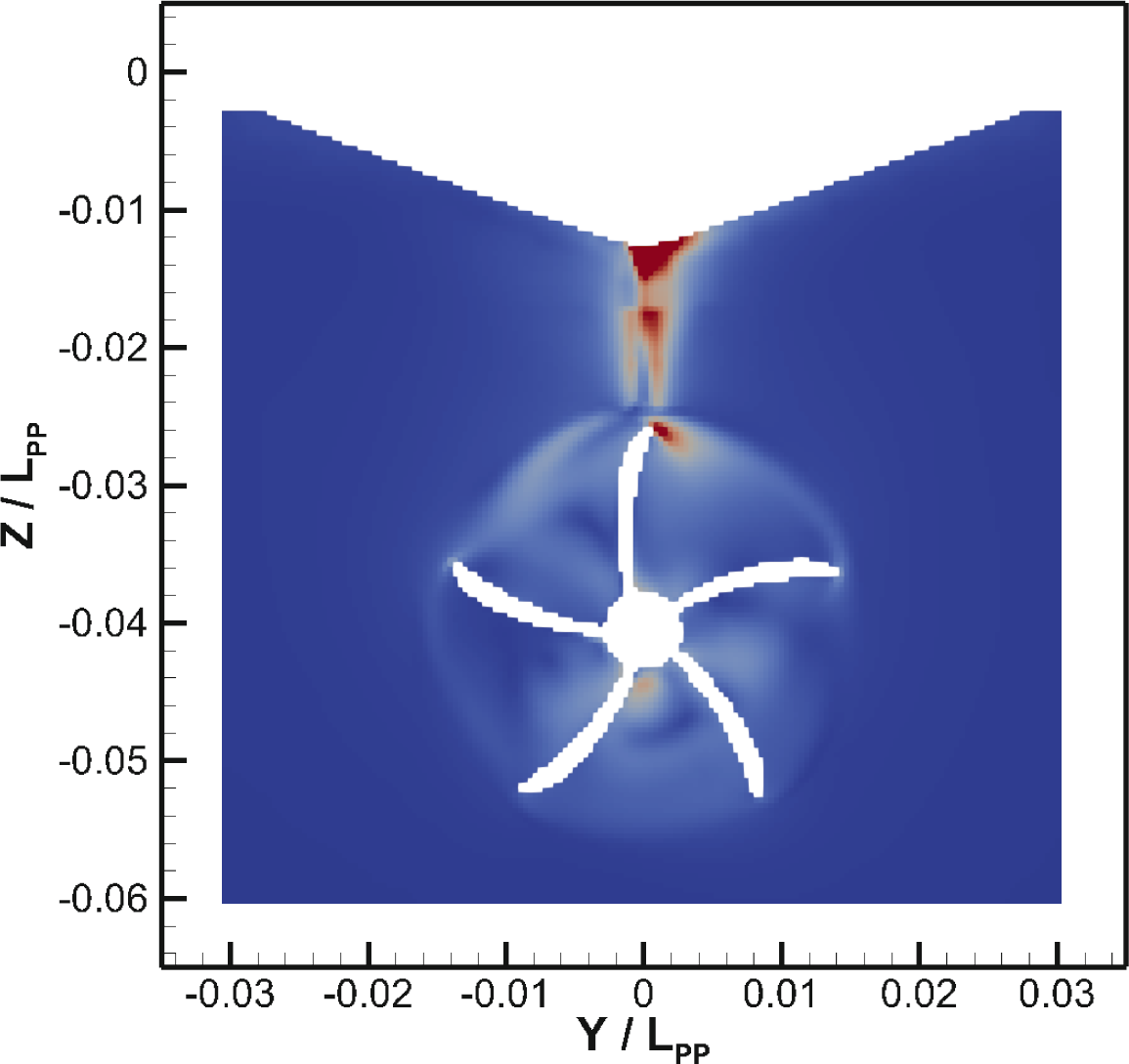}
        \captionsetup{width=0.9\linewidth}
        \caption{S7 (mMRF, $n_{\mathrm{MRF}}/n = 0.50$)}
    \end{subfigure}
    \begin{subfigure}[t]{0.32\linewidth}
        \includegraphics[width=1.0\linewidth]{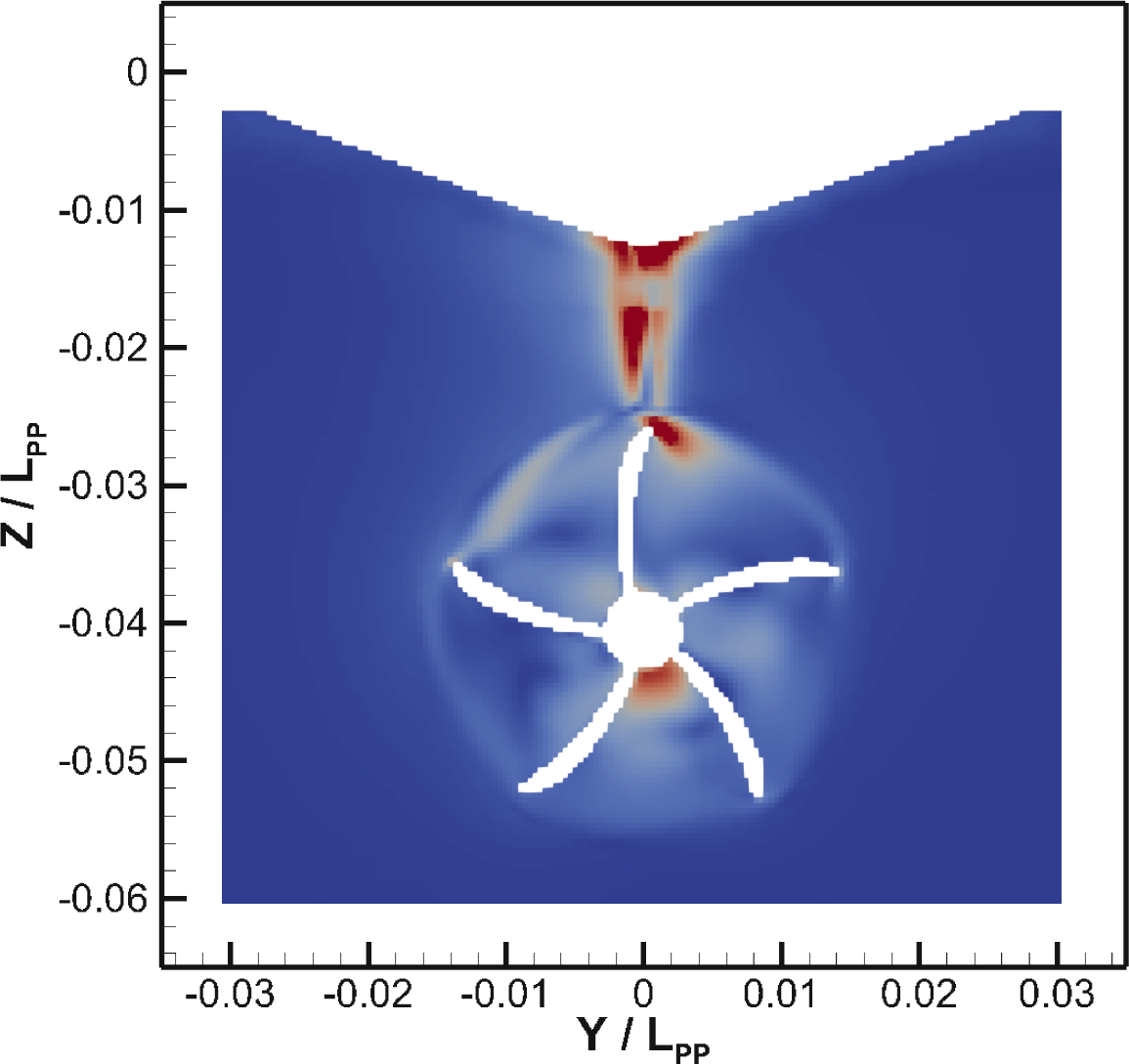}
        \captionsetup{width=0.9\linewidth}
        \caption{S8 (mMRF, $n_{\mathrm{MRF}}/n = 0.75$)}
    \end{subfigure}
    \begin{subfigure}[t]{0.32\linewidth}
        \includegraphics[width=1.0\linewidth]{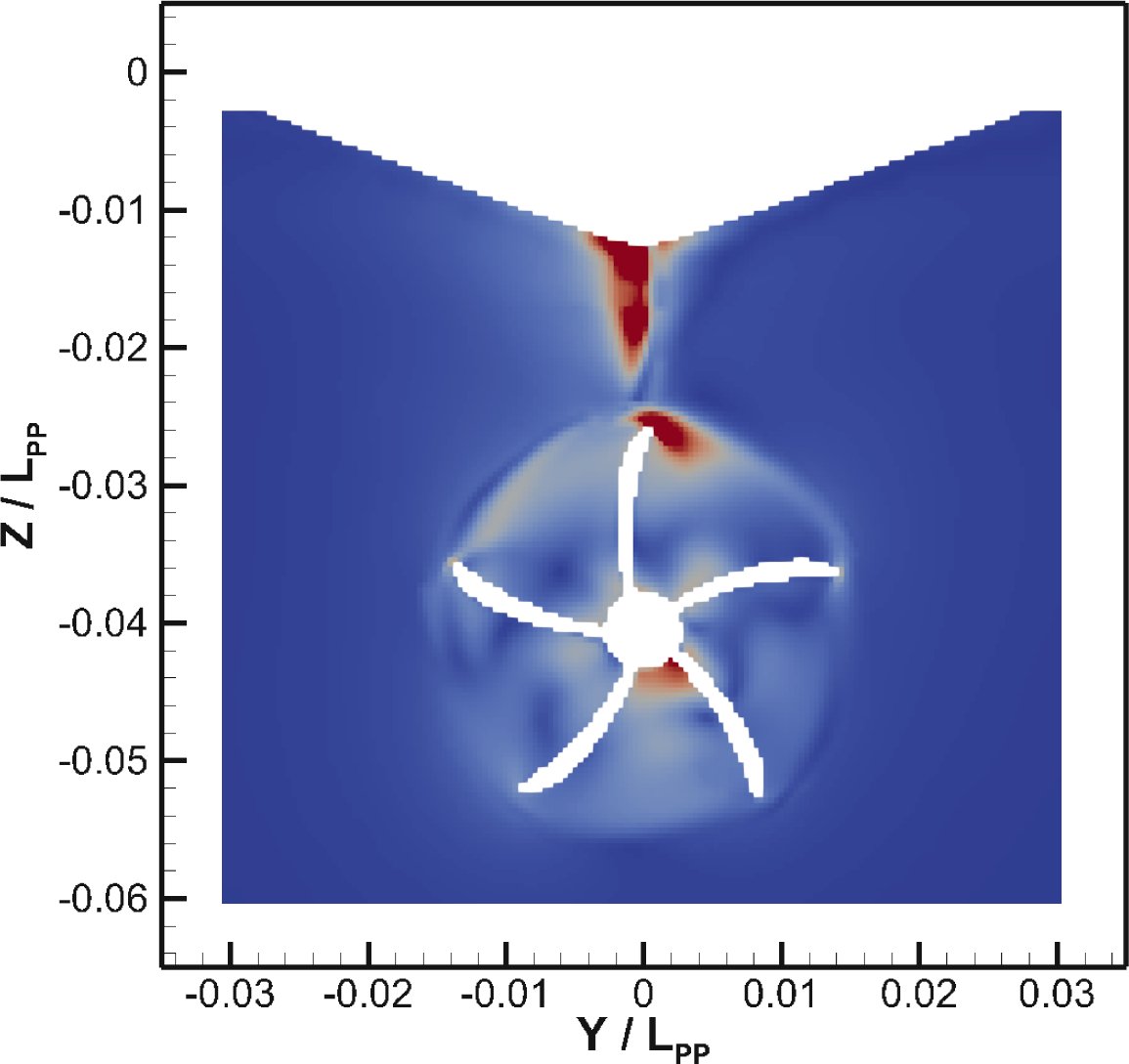}
        \captionsetup{width=0.9\linewidth}
        \caption{S9 (mMRF, $n_{\mathrm{MRF}}/n = 1.0$)}
    \end{subfigure}
    \end{minipage}
    \begin{minipage}[c]{0.09\textwidth}
    \begin{subfigure}[t]{1.0\linewidth}
        \includegraphics[width=1.0\linewidth]{pictures/rms_legend.png}
    \end{subfigure}
    \end{minipage}
    \caption{Contour plots of the normalized velocity-magnitude deviation from the SI solution (SP1), cf. Eqn.~\eqref{eq:rms}. Results refer to the propeller plane ($x/L_{\mathrm{PP}}=0.0134$) and compare the classical MRF (top) with the mMRF (bottom) formulation for different normalized MRF rates $n_{\mathrm{MRF}}/n=0.5$, $0.75$ and $1.0$.}
    \label{fig:rms_prop}
\end{figure}
\begin{figure}[htbp]
    \centering
    \begin{minipage}[c]{0.9\textwidth}
        \centering
        \begin{subfigure}[t]{0.32\linewidth}
            \includegraphics[width=1.0\linewidth]{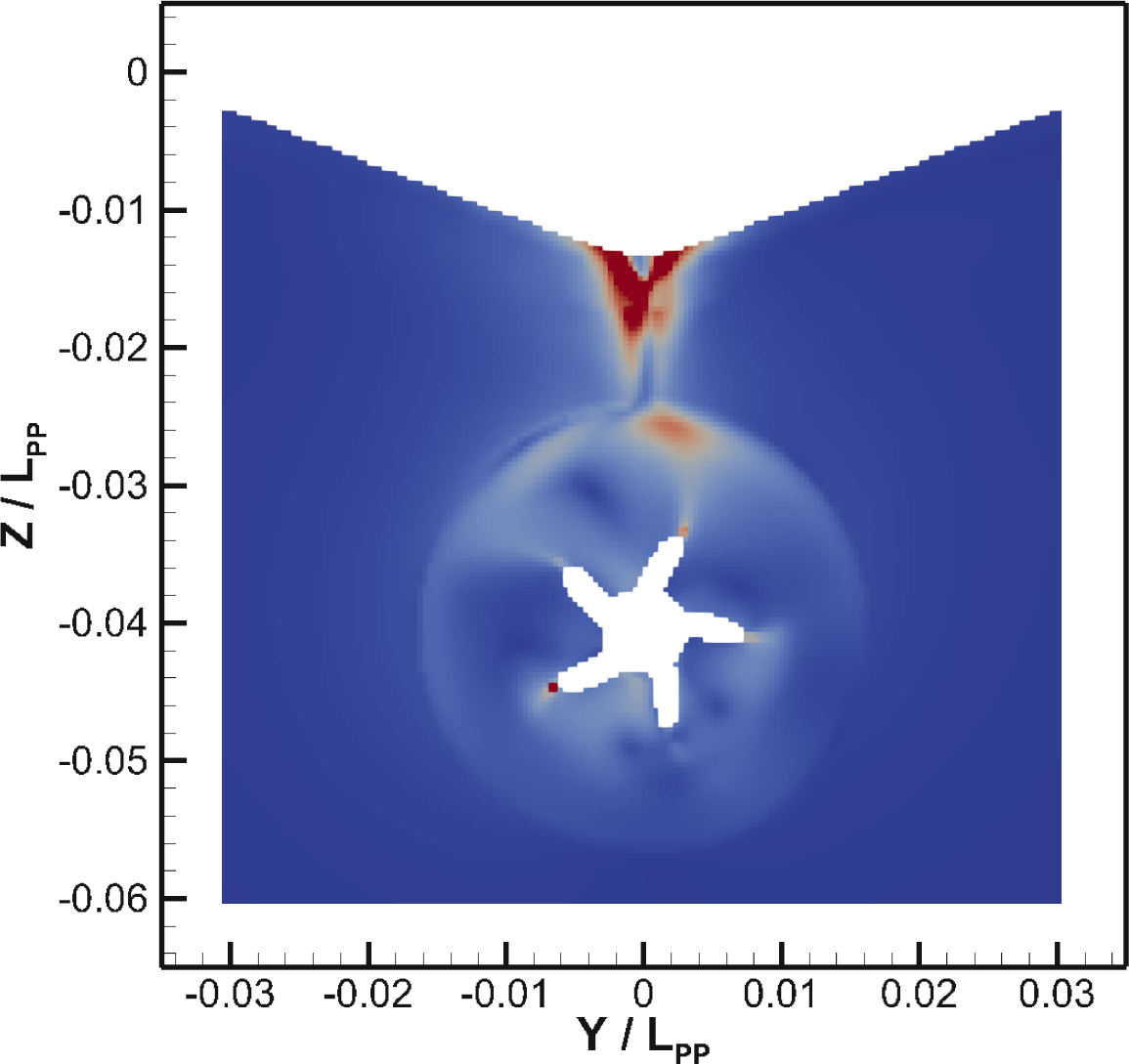}
            \captionsetup{width=0.9\linewidth}
            \caption{S3 (MRF, $n_{\mathrm{MRF}}/n = 0.50$)}
        \end{subfigure}
        \begin{subfigure}[t]{0.32\linewidth}
            \includegraphics[width=1.0\linewidth]{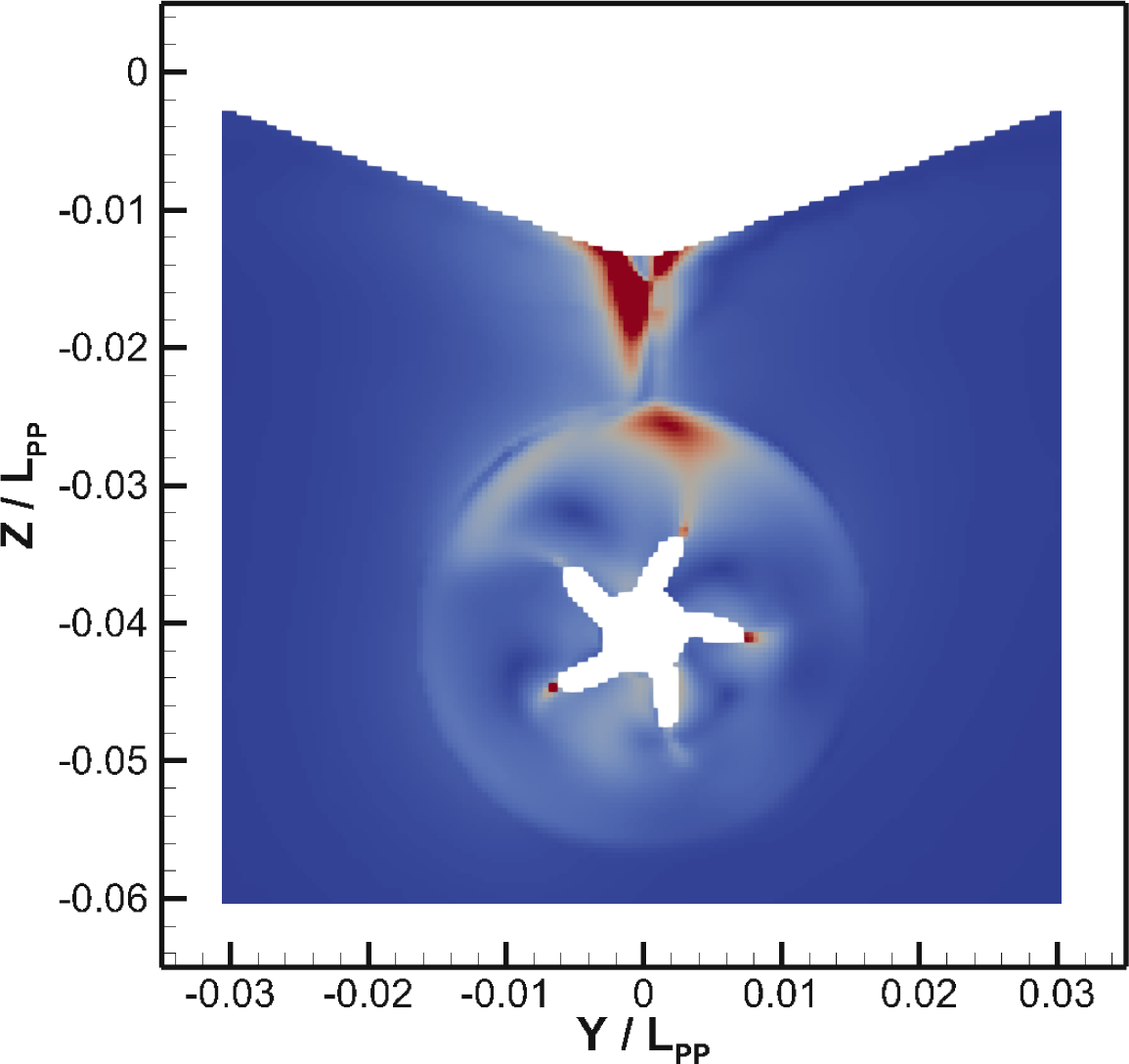}
            \captionsetup{width=0.9\linewidth}
            \caption{S4 (MRF, $n_{\mathrm{MRF}}/n = 0.75$)}
        \end{subfigure}
        \begin{subfigure}[t]{0.32\linewidth}
            \includegraphics[width=1.0\linewidth]{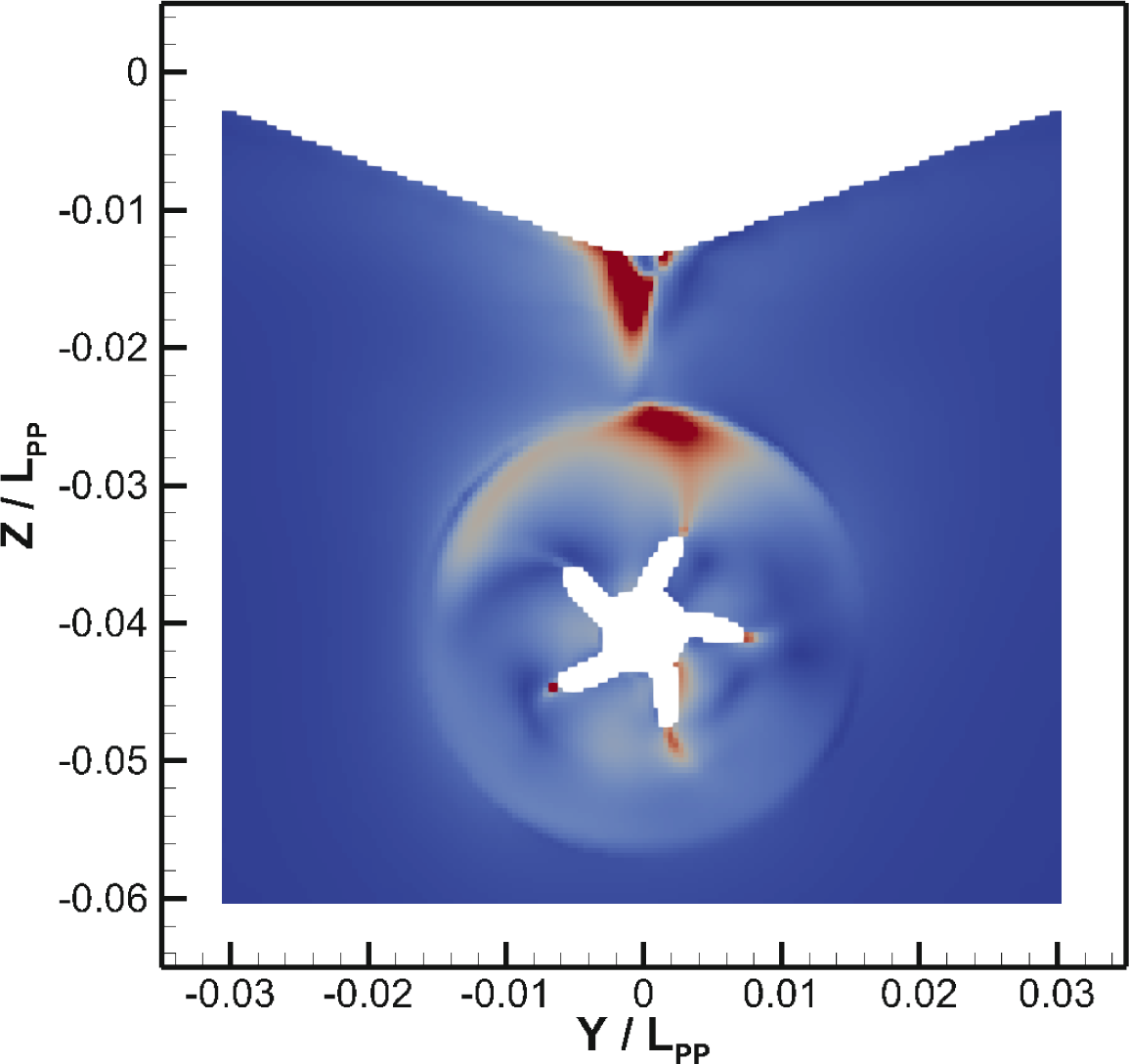}
            \captionsetup{width=0.9\linewidth}
            \caption{S5 (MRF, $n_{\mathrm{MRF}}/n = 1.0$)}
        \end{subfigure}
        \begin{subfigure}[t]{0.32\linewidth}
            \includegraphics[width=1.0\linewidth]{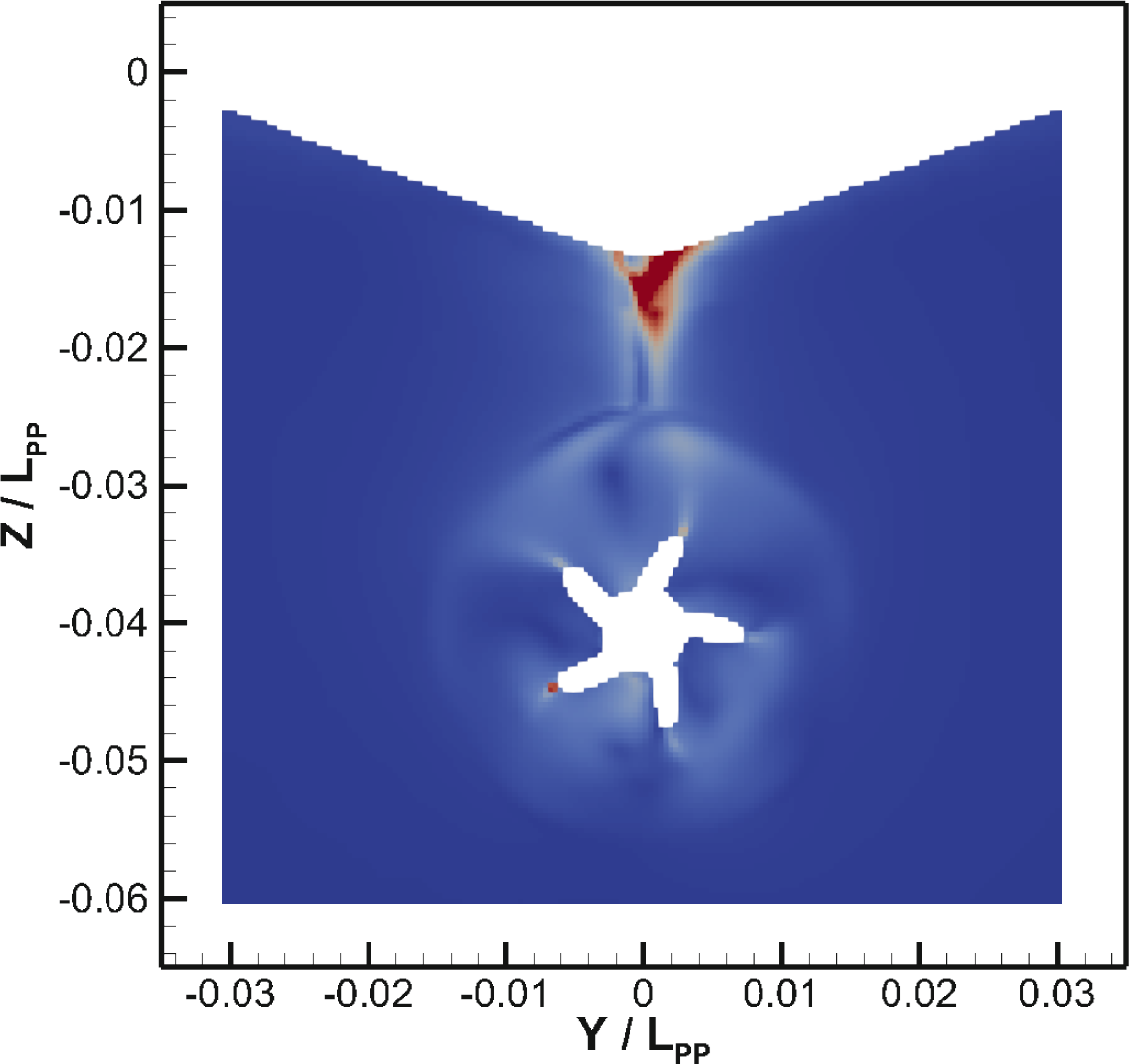}
            \captionsetup{width=0.9\linewidth}
            \caption{S7 (mMRF, $n_{\mathrm{MRF}}/n = 0.50$)}
        \end{subfigure}
        \begin{subfigure}[t]{0.32\linewidth}
            \includegraphics[width=1.0\linewidth]{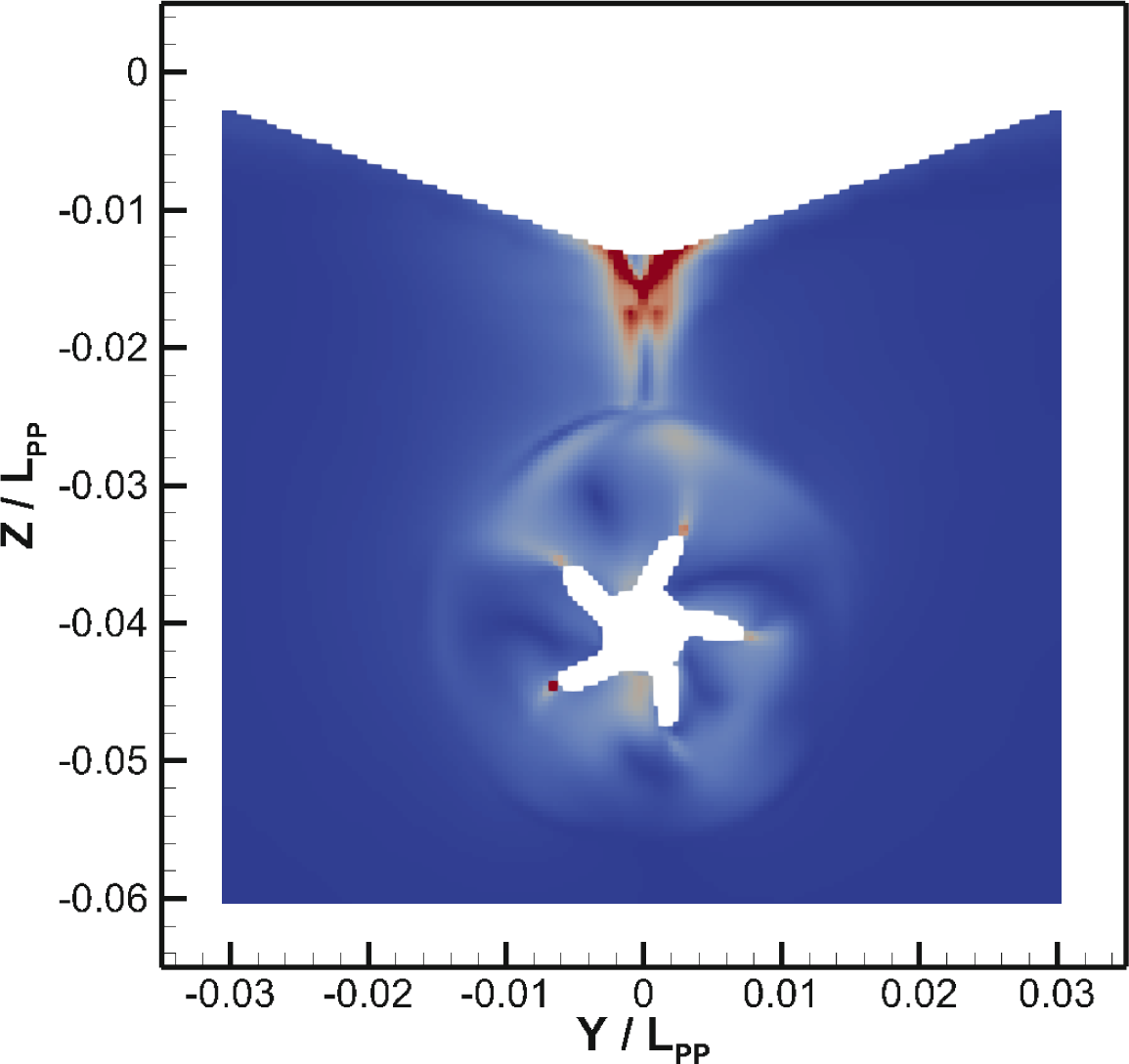}
            \captionsetup{width=0.9\linewidth}
            \caption{S8 (mMRF, $n_{\mathrm{MRF}}/n = 0.75$)}
        \end{subfigure}
        \begin{subfigure}[t]{0.32\linewidth}
            \includegraphics[width=1.0\linewidth]{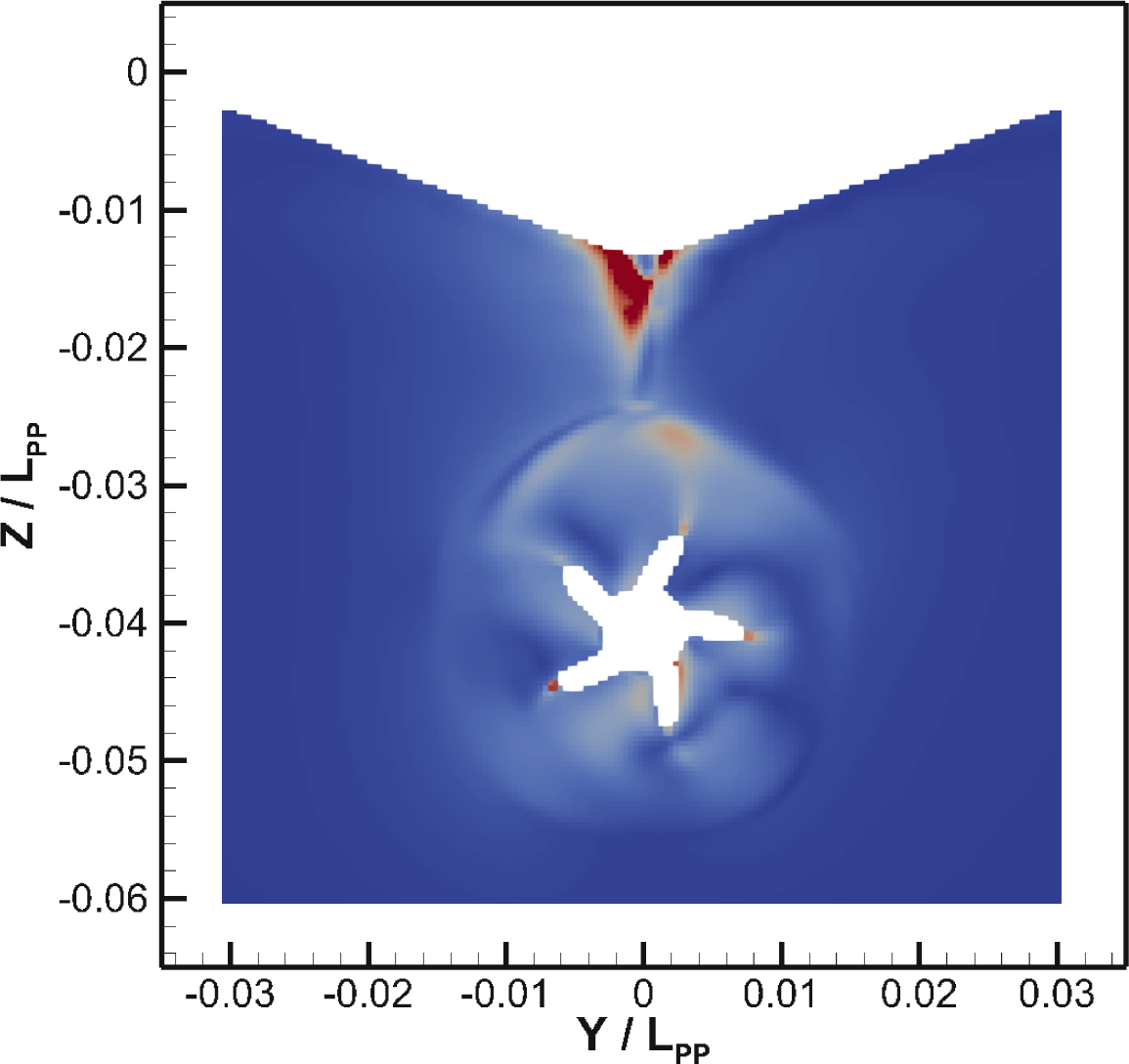}
            \captionsetup{width=0.9\linewidth}
            \caption{S9 (mMRF, $n_{\mathrm{MRF}}/n = 1.0$)}
        \end{subfigure}
    \end{minipage}
    \begin{minipage}[c]{0.09\textwidth}
        \begin{subfigure}[t]{1.0\linewidth}
            \includegraphics[width=1.0\linewidth]{pictures/rms_legend.png}
        \end{subfigure}
    \end{minipage}
    \caption{Contour plots of the normalized velocity-magnitude deviation from the SI solution (SP1), cf. Eqn.~\eqref{eq:rms}. Results refer to the upstream plane ($x/L_{\mathrm{PP}}=0.0157$) and compare the classical MRF (top) with the mMRF (bottom) formulation for different normalized MRF rates $n_{\mathrm{MRF}}/n=0.5$, $0.75$ and $1.0$.}
    \label{fig:rms_x1}
\end{figure}

The local flow field is analyzed at three cross-sectional planes, i.e.,  $x/L_{\mathrm{PP}}=0.0157$ (upstream of the propeller), $x/L_{\mathrm{PP}}=0.0134$ (in the propeller plane), and $x/L_{\mathrm{PP}}=0.0$ (downstream of the propeller).
Experimental data (EFD) are available only at $x/L_{\mathrm{PP}}=0.0$ and $x/L_{\mathrm{PP}}=0.0157$. At these two planes the axial velocity contours and the cross-plane velocity vectors collected in Appendix~\ref{app:efd-comp} (Figs.~\ref{fig:contour_x0}, ~\ref{fig:contour_x1}, ~\ref{fig:cross_x0}, ~\ref{fig:cross_x1}) are compared with both the EFD and the reference sliding-grid solution (SP1). The velocity fields shown in these four figures are time-averaged over one propeller revolution. For the numerical results (SI, MRF, mMRF), a set of instantaneous snapshots sampled at successive time steps within a single revolution is mapped onto a common reference mesh and averaged. The EFD data represent the corresponding time-averaged experimental measurements reported by \cite{hino-2020}.
For each evaluation plane the discrepancy of a rotating-frame formulation from the sliding-interface reference is quantified by the local normalized velocity-magnitude deviation $\varepsilon = |\vect{u}-\vect{u}_{\mathrm{SI}}|/|\vect{u}_{\mathrm{SI}}|$, shown in Figs.~\ref{fig:rms_x0}--\ref{fig:rms_x1}. Its area average over the propeller disc, reported in Table~\ref{tab:sp_rms}, reads 
\begin{equation}
  \bar{\varepsilon}
  = \frac{1}{A_{\mathrm{D}}}\iint_{A_{\mathrm{D}}}
      \frac{\left|\vect{u}-\vect{u}_{\mathrm{SI}}\right|}
           {\left|\vect{u}_{\mathrm{SI}}\right|}\,\mathrm{d}A ,
  \qquad
  A_{\mathrm{D}} = \pi\,(1.5\,R_{\mathrm{P}})^{2} ,
  \label{eq:rms}
\end{equation}
where $\vect{u}$ is the velocity field of the evaluated formulation (MRF or mMRF), $\vect{u}_{\mathrm{SI}}$ is the sliding-interface solution in the same location at the disc, and $A_{\mathrm{D}}$ is the area of the disc of radius $1.5\,R_{\mathrm{P}}$ centered on the propeller axis (illustrated by a circle in Fig.~\ref{fig:v_magnitude}). Unlike the phase-averaged fields used in Figs.~\ref{fig:contour_x0}--\ref{fig:cross_x1}, the deviation field $\varepsilon$ and the disc-averaged values $\bar{\varepsilon}$ in Table~\ref{tab:sp_rms} are evaluated from instantaneous velocity fields, extracted at a common propeller phase angle for all cases; no EFD reference is required for this comparison, since it is defined relative to the SI solution.
At the downstream section ($x/L_{\mathrm{PP}}=0$, Figs.~\ref{fig:contour_x0} and~\ref{fig:rms_x0}; vectors in Fig.~\ref{fig:cross_x0}), the flow is governed by the propeller slipstream and the developing wake. The SI solution agrees well with the experiment, reproducing the velocity deficit and wake contraction. As the MRF or mMRF contribution grows, the deviation from SI increases in both the axial velocity and the cross-plane components, appearing as changes in wake shape and velocity distribution; closer to the hull the mMRF results remain nearer to SI than the classical MRF. For the fully modeled cases ($n_{\mathrm{MRF}}/n=1$) additional structures and non-physical artifacts appear in the wake.
At the upstream section ($x/L_{\mathrm{PP}}=0.0157$, Figs.~\ref{fig:contour_x1} and~\ref{fig:rms_x1}; vectors in Fig.~\ref{fig:cross_x1}), the interface between the rotating and non-rotating regions is clearly visible. The classical MRF formulation produces a distinct discontinuity in the axial velocity whose magnitude grows with the modeled fraction $n_{\mathrm{MRF}}/n$; the effect is weaker in the cross-plane components. The mMRF approach yields a markedly smoother transition across the interface and substantially reduces this discontinuity, although some distortion of the velocity field develops as the mMRF contribution increases. The propeller plane ($x/L_{\mathrm{PP}}=0.0134$, Fig.~\ref{fig:rms_prop}) shows the same trend.
These observations are confirmed quantitatively by the disc-averaged deviations of Table~\ref{tab:sp_rms}. At every plane and every rotation ratio the mMRF deviation is smaller than the corresponding MRF value, with error ratios of $R_{\bar{\varepsilon}}\approx0.58$--$0.83$; the reduction is largest at the upstream plane closest to the interface. Overall, the mMRF method improves the continuity of the solution at the interface and gives more consistent local predictions than the classical MRF; the largest absolute reduction of the deviation occurs at high rotation ratios, where the interface artifacts are most pronounced.
\begin{table}[htbp]
    \centering
    \renewcommand{\arraystretch}{1.8}
    \setcellgapes{3pt}
    \makegapedcells
    \small 
    \begin{tabular}{l c c c c c c c}
        \toprule
        \multirow{2}{*}{\makecell{Test Case \\ (Formulation)}} 
            & \multirow{2}{*}{\makecell{$n_{\mathrm{MRF}}/n$}}
            & \multicolumn{2}{c}{$x/L_{\mathrm{PP}} = 0$}
            & \multicolumn{2}{c}{$x/L_{\mathrm{PP}} = 0.0134$}
            & \multicolumn{2}{c}{$x/L_{\mathrm{PP}} = 0.0157$} \\
        \cmidrule(lr){3-4} \cmidrule(lr){5-6} \cmidrule(lr){7-8}
            & & $\bar{\varepsilon}$ & $R_{\bar{\varepsilon}}$ & $\bar{\varepsilon}$ & $R_{\bar{\varepsilon}}$ & $\bar{\varepsilon}$ & $R_{\bar{\varepsilon}}$ \\
        \midrule
        SP2 (MRF)  & \multirow{2}{*}{0.25} & 0.1205974 & \multirow{2}{*}{0.699} & 0.0828058 & \multirow{2}{*}{0.647} & 0.0668114 & \multirow{2}{*}{0.575} \\
        SP6 (mMRF) &                       & 0.0842870 &                        & 0.0535918 &                        & 0.0384309 &                        \\
        \midrule
        SP3 (MRF)  & \multirow{2}{*}{0.50} & 0.1887735 & \multirow{2}{*}{0.778} & 0.1523050 & \multirow{2}{*}{0.695} & 0.1269541 & \multirow{2}{*}{0.631} \\
        SP7 (mMRF) &                       & 0.1467956 &                        & 0.1057920 &                        & 0.0800543 &                        \\
        \midrule
        SP4 (MRF)  & \multirow{2}{*}{0.75} & 0.2203891 & \multirow{2}{*}{0.833} & 0.2048130 & \multirow{2}{*}{0.739} & 0.1725515 & \multirow{2}{*}{0.676} \\
        SP8 (mMRF) &                       & 0.1835158 &                        & 0.1514470 &                        & 0.1166730 &                        \\
        \midrule
        SP5 (MRF)  & \multirow{2}{*}{1.00} & 0.2466522 & \multirow{2}{*}{0.823} & 0.2460550 & \multirow{2}{*}{0.725} & 0.2116804 & \multirow{2}{*}{0.660} \\
        SP9 (mMRF) &                       & 0.2028720 &                        & 0.1784290 &                        & 0.1397529 &                        \\
        \bottomrule
    \end{tabular}
    \caption{Disc-averaged normalized velocity deviation obtained from Eqn.~\eqref{eq:rms} and error ratio $\mathrm{R_{\bar{\varepsilon}}}=\bar{\varepsilon}_{\mathrm{mMRF}}/\bar{\varepsilon}_{\mathrm{MRF}}$ at the downstream ($x/L_{\mathrm{PP}} = 0$), propeller
    ($x/L_{\mathrm{PP}} = 0.0134$) and upstream ($x/L_{\mathrm{PP}} = 0.0157$) planes.
    }
    \label{tab:sp_rms}
\end{table}

\FloatBarrier
\FloatBarrier
\section{Conclusions}
\label{sec:conclusions}
A modified moving reference frame (mMRF) method for propeller simulation has been introduced and evaluated in the context of ship self-propulsion at model scale. The method combines a resolved rotation with a spatially varying MRF contribution, providing a smooth transition between rotating and non-rotating regions. The approach was applied to the JBC configuration using a single-phase double-body setup, with wave-making resistance accounted for through additional towing simulations. The propeller rotational speed was determined using a PID controller to satisfy the longitudinal force equilibrium. The numerical framework was based on the in-house RANS solver FreSCo$^+$, and the results were validated against available experimental data.
Both integral and local flow characteristics were analyzed to assess the performance of the proposed method. Open-water simulations confirmed that the mMRF formulation preserves the main propeller characteristics. Bare-hull resistance and wave profiles showed good agreement with experiments, supporting the reliability of the numerical setup. Self-propulsion simulations demonstrated that the integral quantities, such as $K_{\mathrm{T}}$, $K_{\mathrm{Q}}$, $1 - t$, $1 - w_{\mathrm{T}}$, $\eta_{\mathrm{R}}$, and $n$, are predicted with good accuracy across all tested cases.
A detailed comparison between classical MRF and mMRF formulations showed that the distribution of the propeller rotation has a direct impact on both global and local results. Increasing the MRF contribution generally leads to larger deviations from the reference sliding interface (SI) solution, especially with respect to local flow features. At the same time, the mMRF method provides a smoother transition at the interface between rotating and stationary regions, reducing discontinuities that are clearly visible in the classical MRF approach. This improvement is most evident in the upstream section, where the interface effects are strongest.
The analysis of local flow fields highlighted that axial velocity is more sensitive to the rotation modeling than cross-plane velocity components. While both MRF and mMRF approaches show increasing deviations with higher MRF ratios, the mMRF method maintains better consistency with the SI solution, particularly in regions close to the hull. However, for cases where the propeller rotation is modeled entirely by the MRF formulation, additional flow structures and non-physical artifacts appear, indicating reduced reliability of such configurations.
Based on the results obtained in this study, the following main conclusions can be drawn:
\begin{itemize}
    \item The proposed mMRF method provides accurate predictions of integral self-propulsion characteristics, including $K_{\mathrm{T}}$, $K_{\mathrm{Q}}$, $1 - t$, $1 - w_{\mathrm{T}}$, $\eta_{\mathrm{R}}$, and $n$, while maintaining computational efficiency comparable to the classical MRF approach.
    \item The accuracy of the solution depends on the ratio between resolved and modeled rotation. Integral quantities are less sensitive to this ratio, while local flow features are more strongly affected.
    \item The mMRF formulation improves the continuity of the solution at the interface between rotating and non-rotating regions, reducing numerical artifacts compared to the classical MRF method.
    \item For high MRF contributions, the mMRF approach shows significantly better agreement with the sliding grid  reference solution, indicating improved robustness of the rotation representation.
    \item Cases where the propeller rotation is modeled entirely by the MRF formulation lead to increased flow distortions and should be avoided when accurate local flow prediction is required.
\end{itemize}
Future work will focus on extending the proposed method to more complex applications, including configurations with energy-saving devices and pre-swirl ducts, two-phase self-propulsion simulations, full-scale computations, and cases involving body motions. Beyond these extensions, the mMRF formulation also offers a favorable trade-off for adjoint-based optimization: compared to the fully resolved, unsteady SI approach, it requires fewer time steps to capture the flow evolution, correspondingly reducing the number of flow states that must be stored for the backward integration, while retaining greater accuracy than the steady-state MRF approach.

\section*{CRediT authorship contribution statement}
\textbf{Denis Andreev:} Conceptualization, Methodology, Formal analysis, Software, Validation, Investigation, Visualization, Writing -- original draft.
\textbf{Georgios Bletsos:} Formal analysis, Software, Validation, Writing -- original draft.
\textbf{Antonios Kritikos:} Software, Validation, Writing -- review \& editing.
\textbf{Niklas K\"uhl:} Software, Resources, Funding acquisition.
\textbf{Thomas Rung:} Conceptualization, Methodology, Validation, Resources, Supervision, Project administration, Funding acquisition, Writing -- original draft.
\section*{Declaration of competing interest}
The authors declare that they have no known competing financial interests or personal relationships that could have appeared to influence the work reported in this paper.
\section*{Acknowledgments}
The authors acknowledge the support of the "Propulsion Optimization of Ships and Appendages" research project, which is funded by the German Federal Ministry for Economics and Climate Action [Grant No. 03SX599D - \textbf{DA}; \textbf{GB}; \textbf{TR} / Grant No. 03SX599C - \textbf{NK}] and the support by the Deutsche Forschungsgemeinschaft (DFG) within the Research Training Group GRK 2583 "Modeling, Simulation and Optimization of Fluid Dynamic Applications" [\textbf{DA}; \textbf{TR}]. \textbf{AK} has received funding from the European Union’s Horizon 2021 research and innovation program under the Marie Sklodowska Curie Grant Agreement No. 101072851 (MFLOPS). Views and opinions expressed are however those of the author(s) only and do not necessarily reflect those of the European Union or European Research Executive Agency (REA). Neither the European Union nor the granting authority can be held responsible for them. 

The authors gratefully acknowledge the computing time made available to them on the high-performance computer Lise at the NHR@ZIB. This center is jointly supported by the Federal Ministry of Research, Technology and Space and the state governments participating in the National High-Performance Computing (NHR) joint funding program (\url{http://www.nhr-verein.de/en/our-partners}).
\section*{Declaration of generative AI and AI-assisted technologies in the writing process}
During the preparation of this work the authors used Claude (Anthropic) in order to improve grammar and spelling and to rephrase sentences for clarity. After using this tool, the authors reviewed and edited the content as needed and take full responsibility for the content of the publication.

\FloatBarrier
\appendix
\section*{Appendices}

\section{Detailed Derivation of the Modified Partially Moving Grid model}
\label{app:mMRF_derivation}
%
Following the notation and velocity decomposition introduced in 
Sec.~\ref{sec:secmmrf}, the acceleration in the modified partially 
moving reference frame is obtained analogously to the conventional 
MRF derivation \cite{luo-1994}. Expanding the time derivative of 
Eq.~(\ref{eq:vel_relation}) gives
\begin{equation}
\label{eq_app:accel-relation}
\begin{split}
    \cfrac{d \vect{u} }{ d t }
    &  = 
    \cfrac{ d }{ d t }
    \left({
        \vect{u}^{\mathrm{R}}
        +
        f^{\mathrm{R}} \, 
        \vect{\Omega}^{\mathrm{MRF}}
        \times
        \vect{r}^{\mathrm{R}}
    }\right)
    \\
    & =
    \left[{
        \cfrac{ d \vect{u}^{\mathrm{R}}}{ d t }
        +
        f^{\mathrm{R}} \, 
        \vect{\Omega}^{\mathrm{MRF}}
        \times
        \cfrac{d \vect{r}^{\mathrm{R}}}{dt}
        +
        \cfrac{ d f^{\mathrm{R}} 
                }{ d t } \, 
        \vect{\Omega}^{\mathrm{MRF}}
        \times
        \vect{r}^{\mathrm{R}}
    }\right]^{\mathrm{R}}
    \\
    & \hspace{6cm}
    +
    f^{\mathrm{R}} \, 
    \cfrac{d \vect{\Omega}^{\mathrm{MRF}}}{d t} 
    \times 
    \vect{r}^{\mathrm{R}}
    +
    f^{\mathrm{R}} \, 
    \vect{\Omega}^{\mathrm{MRF}} 
    \times
    \underbrace{\bigg[{
        \vect{u}^{\mathrm{R}}
        +
        f^{\mathrm{R}} \, 
        \vect{\Omega}^{\mathrm{MRF}}
        \times
        \vect{r}^{\mathrm{R}}
    }\bigg]}_{d\vect{r}^R/dt \; (\ref{eq:vel_relation})}
\end{split}
\end{equation}
Owing to
$
    \left[ d \vect{u}^{\mathrm{R}}/ d t \right]^{\mathrm{R}}
    =
    \partial \vect{u}^{\mathrm{R}}/\partial t 
    +
    \vect{u}^{\mathrm{R}} \cdot \nabla \vect{u}^{\mathrm{R}}
$
and 
$
    \left[f^{\mathrm{R}} \: \vect{\Omega}^{\mathrm{MRF}} \times d \vect{r}^{\mathrm{R}}/dt \right]^{\mathrm{R}}
    =
    f^{\mathrm{R}} \: \vect{\Omega}^{\mathrm{MRF}} \times \vect{u}^{\mathrm{R}}
$
one obtains 
\begin{equation}
\label{eq_app:accel-relation-1}
\begin{split}   
   \cfrac{d \vect{u} }{ d t }
    & =
    \cfrac{ \partial \vect{u}^{\mathrm{R}}}{ \partial t }
    +
    \vect{u}^{\mathrm{R}} \cdot \nabla \vect{u}^{\mathrm{R}}
    +
    2 \, f^{\mathrm{R}} \, 
    \vect{\Omega}^{\mathrm{MRF}}
    \times
    \vect{u}^{\mathrm{R}}
    +
    (f^{\mathrm{R}})^{2} \, 
    \vect{\Omega}^{\mathrm{MRF}}
    \times
    \left({
        \vect{\Omega}^{\mathrm{MRF}}
        \times
        \vect{r}^{\mathrm{R}}
    }\right)
    \\
    & \hspace{8cm}
    +
    f^{\mathrm{R}} \cfrac{ d \vect{\Omega}^{\mathrm{MRF}} }{ d t } \times \vect{r}^{\mathrm{R}}
    +
    \left[
        \cfrac{ d (f^{\mathrm{R}} ) }{ d t }
        \vect{\Omega}^{\mathrm{MRF}} \times \vect{r}^{\mathrm{R}}
    \right]^R
    \\
    & = 
    \cfrac{ \partial \vect{u}^{\mathrm{R}}}{ \partial t }
    +
    \vect{u}^{\mathrm{R}} \cdot \nabla \vect{u}^{\mathrm{R}}
    +
    2 \, f^{\mathrm{R}} \, 
    \vect{\Omega}^{\mathrm{MRF}}
    \times
    \vect{u}^{\mathrm{R}}
    +
    (f^{\mathrm{R}})^{2} \, 
    \vect{\Omega}^{\mathrm{MRF}}
    \times
    \left({
        \vect{\Omega}^{\mathrm{MRF}}
        \times
        \vect{r}^{\mathrm{R}}
    }\right)
    \\
    & \hspace{8cm}
    +
    f^{\mathrm{R}} \cfrac{ \partial \vect{u}^{\mathrm{MRF}} }{ \partial t}
    +
    \left({
        \nabla f^{\mathrm{R}} 
        \cdot
        \vect{u}^{\mathrm{R}}
    }\right)
    \vect{u}^{\mathrm{MRF}}
    \: \text{.}
\end{split}
\end{equation}
Using the relation
\begin{equation}
\label{eq_app:convection_r}
\begin{split}
    \nabla \cdot (\vect{u}^R\vect{u}^R)
    & = 
    \vect{u}^R \cdot (\nabla \vect{u}^R)
    +
    \left(
        \nabla \cdot \left( \vect{u} - \vect{u}^{\mathrm{mMRF}} \right)
    \right)
    \vect{u}^R
    \\
    & =
    \vect{u}^R \cdot \left( \nabla \vect{u}^R \right)
    -
    (\nabla \cdot \vect{u}^{\mathrm{mMRF}}) \vect{u}^R 
    =
    \vect{u}^R \cdot \left( \nabla \vect{u}^R \right)
    -
    (\nabla f^{\mathrm{R}} \cdot \vect{u}^{\mathrm{MRF}}) \vect{u}^R 
    \: \text{,}
\end{split}
\end{equation}
the convective contribution may be rewritten prior to substituting~(\ref{eq_app:accel-relation-1}) into the incompressible momentum equation~(\ref{eq:ns}). This yields the governing equation for the velocity in the relative frame, viz.
\begin{equation}
\label{eq_app:ns_mMRF_rel}
\begin{split}
    &
    \rho
    \left[{
        \cfrac{\partial \vect{u}^{\mathrm{R}}}{ \partial t }
        + 
        \nabla
        \cdot
        \left({
            \vect{u}^R
             \vect{u}^R
        }\right)
        +
        2 f^{\mathrm{R}} \,
        \vect{\Omega}^{\mathrm{MRF}}
        \times
        \vect{u}^{\mathrm{R}}
        + 
        (f^{\mathrm{R}})^{2} \,
        \vect{\Omega}^{\mathrm{MRF}}
        \times
        \left({
            \vect{\Omega}^{\mathrm{MRF}}
            \times
             \vect{r}^R
        }\right)
    }\right.
    \\
    & \hspace{4cm} 
    \left.{
        + 
        f^{\mathrm{R}} \,
        \cfrac{ \partial \vect{u}^{\mathrm{MRF}} }{ \partial t}
        +
        \nabla f^{\mathrm{R}}
        \cdot
        \left({
            \vect{u}^R
            \vect{u}^{\mathrm{MRF}}
            +
            \vect{u}^{\mathrm{MRF}}
            \vect{u}^R
        }\right)
    }\right]
    =
    \nabla \cdot \tensor{\sigma} + \vect{g}
    \: \text{.}
\end{split}
\end{equation}
To arrive at an expression for the absolute velocity $\vect{u}$, we refer to Eq.~(\ref{eq:vel_relation}) and first rearrange the local time derivatives
$\vect{u}= \vect{u^R} + f^R \vect{u}^{\mathrm{MRF}}$
as well as the source term expression 
$
    f \vect{\Omega}^{\mathrm{MRF}} \times \vect{u}
    =
    f^{\mathrm{R}} \left[
        \vect{\Omega}^{\mathrm{MRF}}
        \times
        \left(
            \vect{u}^{\mathrm{R}} + f^{\mathrm{R}} \vect{\Omega}^{\mathrm{MRF}} \times \vect{r}^{\mathrm{R}}
        \right)
    \right]
$
of Eq.~(\ref{eq_app:ns_mMRF_rel}) to obtain 
\begin{equation}
\label{eq_app:ns_mMRF_abs_0}
\begin{split}
    \rho
    \left[{
        \cfrac{\partial \vect{u} }{ \partial t }
        +
        \nabla
        \cdot
        \left({
            \vect{u}^R
             \vect{u}^R
        }\right)
        +
        f \, 
        \vect{\Omega}^{\mathrm{MRF}} 
        \times 
        \vect{u}
        +
        f^{\mathrm{R}} \, 
        \vect{\Omega}^{\mathrm{MRF}}
        \times
        \vect{u}^{\mathrm{R}}
        +
        \nabla f^{\mathrm{R}}
        \cdot
        \left({
            \vect{u}^R
            \vect{u}^{\mathrm{MRF}}
            +
            \vect{u}^{\mathrm{MRF}}
            \vect{u}^R
        }\right)
    }\right]
    =
    \nabla \cdot \tensor{\sigma} + \vect{g}
    \: \text{.}
\end{split}
\end{equation}
Using the continuity relation
$ \nabla \cdot \vect{u}^R = - \nabla \cdot \vect{u}^{\mathrm{mMRF}} $
and the mMRF velocity divergence relation
$ \nabla \cdot \vect{u}^{\mathrm{mMRF}} = \nabla f \cdot \vect{u}^{\mathrm{MRF}} $,
the convection term can be rewritten as
\begin{equation}
\label{eq_app:conv_r_to_a}
    \nabla \cdot \left( \vect{u}^{\mathrm{R}} \vect{u}^{\mathrm{R}} \right)
    =
    \nabla \cdot \left( \vect{u}^{\mathrm{R}} \vect{u} \right)
    -
    f^{\mathrm{R}} \vect{\Omega}^{\mathrm{MRF}} \times \vect{u}^{\mathrm{R}}
    +
    \nabla f^{\mathrm{R}}
    \cdot
    \left({
        \vect{u}^{\mathrm{MRF}} \vect{u}^{\mathrm{mMRF}}
        -
        \vect{u}^{\mathrm{R}} \vect{u}^{\mathrm{MRF}}
    }\right)
    \: \text{.}
\end{equation}
Substituting into Eq.~(\ref{eq_app:ns_mMRF_abs_0}) leads to
\begin{equation}
\label{eq_app:ns_mMRF_abs_01}
\begin{split}
    \rho
    \left[{
        \cfrac{\partial \vect{u} }{ \partial t } 
        +
        \nabla
        \cdot
        \left({
            \vect{u}^R
             \vect{u}
        }\right)
        +
        \nabla f^{\mathrm{R}}
        \cdot
        \left({
            \vect{u}^{\mathrm{MRF}} \vect{u}^{\mathrm{R}}
            +
            \vect{u}^{\mathrm{MRF}} \vect{u}^{\mathrm{mMRF}}
        }\right)
    }\right]
    =
    \nabla \cdot \tensor{\sigma} + \vect{g}
    \: \text{.}
\end{split}
\end{equation}
Expressing Eq.~(\ref{eq_app:ns_mMRF_abs_01}) in terms of the absolute velocity via Eq.~(\ref{eq:vel_relation}) recovers
the absolute-velocity formulation of the momentum equation, viz. 
\begin{equation}
\label{eq_app:ns_mMRF_abs_03}
\begin{split}
    \rho
    \left[{
        \cfrac{\partial \vect{u} }{ \partial t }
        +
        \nabla
        \cdot
        \bigg({
            \left({
                \vect{u} - \vect{u}^{\mathrm{mMRF}}
            }\right)
            \vect{u}
        }\bigg)
        +
        \bigg({
            \nabla f 
            \cdot
            \vect{u}^{\mathrm{MRF}}
        }\bigg)
        \vect{u}
        +
        f \, 
        \vect{\Omega}^{\mathrm{MRF}} 
        \times 
        \vect{u}
    }\right]
    =
    \nabla \cdot \tensor{\sigma} + \vect{g}
    \: \text{.}
\end{split}
\end{equation}

\section{Temporal and spatial discretization sensitivity study}
\label{app:validation}

\FloatBarrier
\begin{figure}[htbp]
    \centering
    \begin{subfigure}[t]{0.25\linewidth}
        \includegraphics[width=1.0\linewidth]{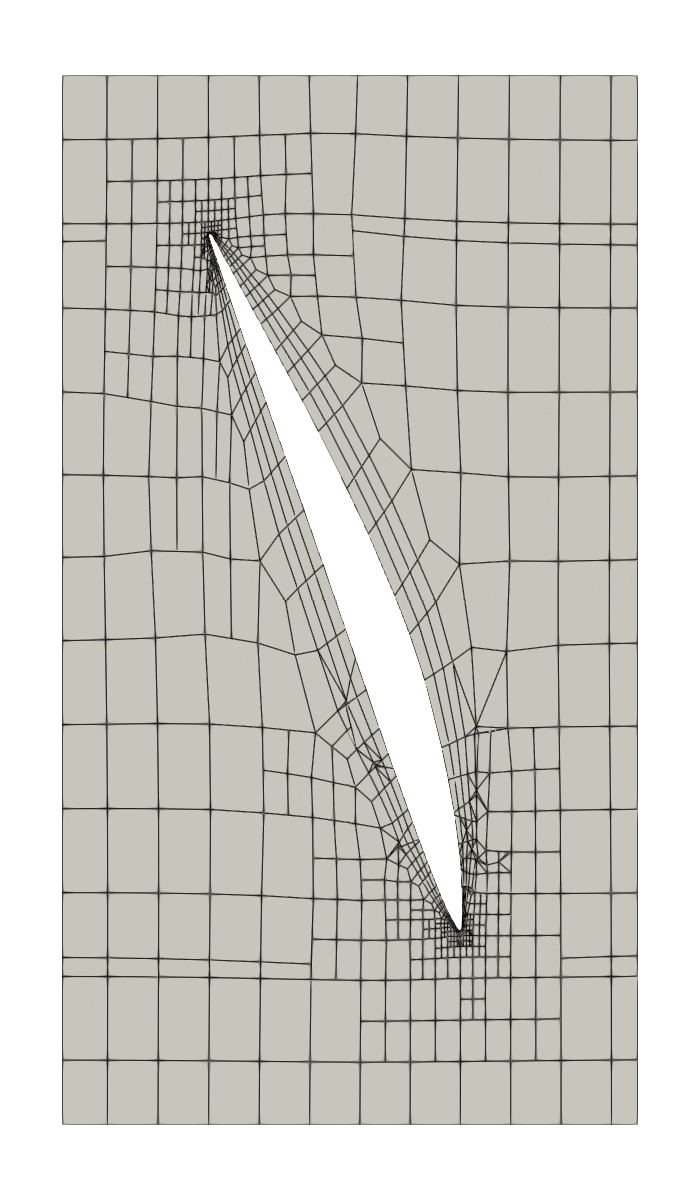}
        \captionsetup{width=0.9\linewidth}
        \caption{Coarse grid}
    \end{subfigure}
    \begin{subfigure}[t]{0.25\linewidth}
        \includegraphics[width=1.0\linewidth]{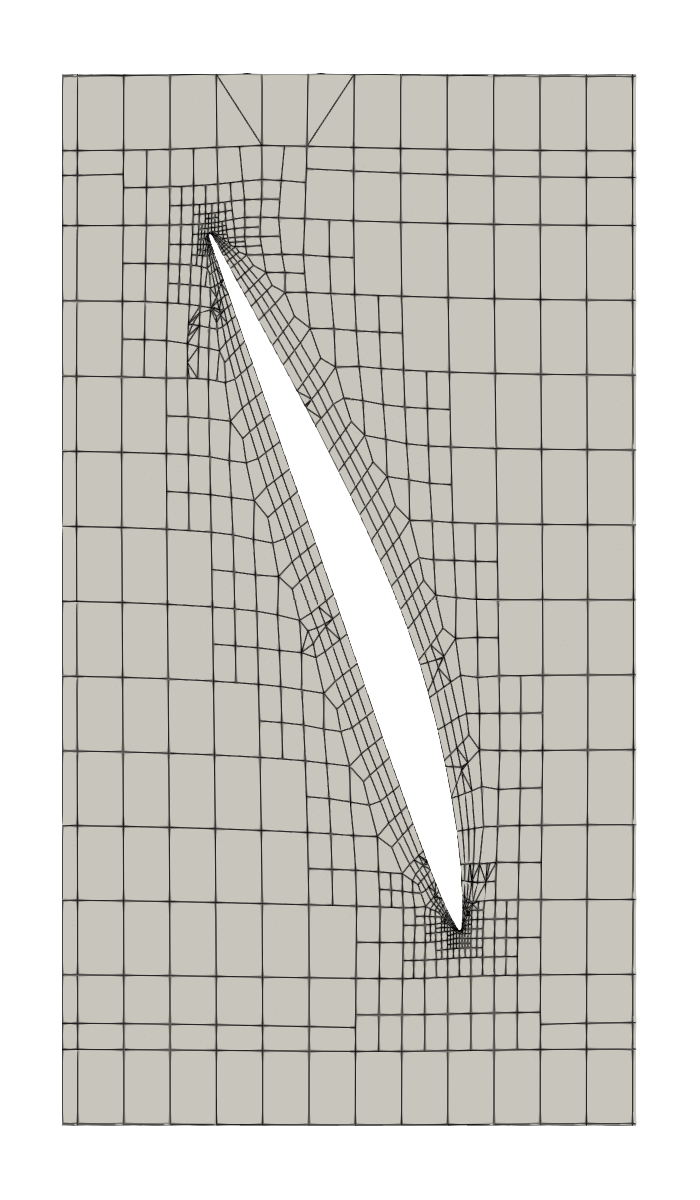}
        \captionsetup{width=0.9\linewidth}
        \caption{Medium grid}
    \end{subfigure}
    \begin{subfigure}[t]{0.25\linewidth}
        \includegraphics[width=1.0\linewidth]{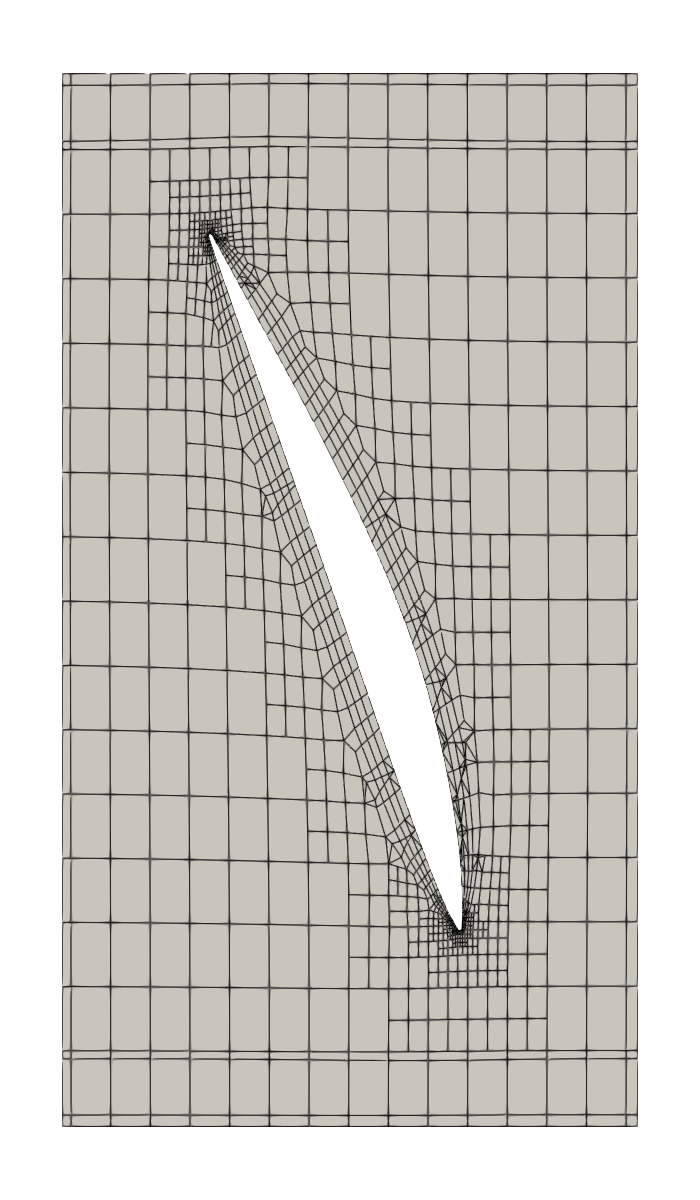}
        \captionsetup{width=0.9\linewidth}
        \caption{Fine grid}
    \end{subfigure}
    \caption{Propeller grid cross-sections at $ r/R_\mathrm{P} = 0.344 $ at three refinement levels.}
    \label{fig:valid-grids}
\end{figure}
\begin{table}[htbp]
\centering
\small
\begin{tabular}{lccc}
    \toprule
    Level & $N$ & $K_\mathrm{T}$ & $K_\mathrm{Q}$ \\
    \midrule
    1 (fine)   & 1\,345\,359 & 0.1352 & 0.02011 \\
    2 (medium) & 1\,114\,107 & 0.1358 & 0.02020 \\
    3 (coarse) & 954\,414    & 0.1361 & 0.02026 \\
    \midrule
    $\delta_{21}$ & & 0.44\,\% & 0.45\,\% \\
    $\delta_{32}$ & & 0.22\,\% & 0.30\,\% \\
    $R$ & & 0.50 & 0.67 \\
    \bottomrule
\end{tabular}
\caption{Grid sensitivity study (OW1, $J = 0.6$): cell counts, computed coefficients, relative changes between consecutive levels, and convergence ratio.}
\label{tab:spatial_uncertainty}
\end{table}
\begin{table}[htbp]
\centering
\small
\begin{tabular}{lccc}
    \toprule
    Level & $\Delta t$ [s] & $K_\mathrm{T}$ & $K_\mathrm{Q}$ \\
    \midrule
    1 (fine)   & 0.00025 & 0.1348 & 0.02009 \\
    2 (medium) & 0.00050 & 0.1358 & 0.02020 \\
    3 (coarse) & 0.00100 & 0.1362 & 0.02024 \\
    \midrule
    $\delta_{21}$ & & 0.74\,\% & 0.55\,\% \\
    $\delta_{32}$ & & 0.29\,\% & 0.20\,\% \\
    $R$ & & 0.40 & 0.36 \\
    \bottomrule
\end{tabular}
\caption{Time-step sensitivity study (OW1, $J = 0.6$): time-step sizes, computed coefficients, relative changes between consecutive levels, and convergence ratio.}
\label{tab:temporal_uncertainty}
\end{table}

The sensitivity study of the numerical results to the spatial and temporal discretization is performed for the OW1 open-water propeller case (Table~\ref{tab:ow-setup}) at $J = 0.6$, monitoring the thrust coefficient $K_{\mathrm{T}}$ and the torque coefficient $K_{\mathrm{Q}}$. Three grid levels and three time-step levels are studied, denoted by subscripts $1$ (fine), $2$ (medium), and $3$ (coarse). The corresponding propeller surface meshes for the three grid levels are shown in Figure~\ref{fig:valid-grids}, complete grid and time-step parameters are summarized in Tables~\ref{tab:spatial_uncertainty} and~\ref{tab:temporal_uncertainty}. The grid study is carried out with the medium time step and the time-step study with the medium grid, so that both studies share the level-2 solution.
 
The convergence ratio is defined as
\begin{equation}
\label{eq:conv_ratio}
    R = \frac{\phi_3 - \phi_2}{\phi_2 - \phi_1}, 
\end{equation}
where $\phi_i$ denotes the monitored quantity at the $i$-th level of refinement. Monotonic ($0 < R < 1$) convergence is obtained for both quantities in both studies (Tables~\ref{tab:spatial_uncertainty} and~\ref{tab:temporal_uncertainty}). The relative change between consecutive levels, 
$\delta_{i+1,i} = \left| \phi_{i+1} - \phi_i \right| / \phi_i \times 100\,\%$
, does not exceed $0.45\,\%$ in the grid study and $0.74\,\%$ in the time-step study, and thus remains below the experimental uncertainty of approximately $1\,\%$ indicated by the ITTC recommended procedure~\cite{ittc-uncertainty-2} for open-water tests. The level-2 results deviate from the experimental values of~\cite{hino-2020} by $0.67\,\%$ for $K_{\mathrm{T}}$ and $0.35\,\%$ for $K_{\mathrm{Q}}$. The medium resolution is therefore considered sufficient, and the time step $\Delta t = 0.0005~\mathrm{s}$, is adopted for all remaining computations.

\FloatBarrier
\section{Axial Velocity Contours and Cross-Plane Velocity Vectors}
\label{app:efd-comp}
\FloatBarrier
\begin{figure}[htbp]
    \centering
    \begin{subfigure}[t]{0.16\linewidth}
        \hfill
    \end{subfigure}
    \begin{subfigure}[t]{0.32\linewidth}
        \includegraphics[width=1.0\linewidth]{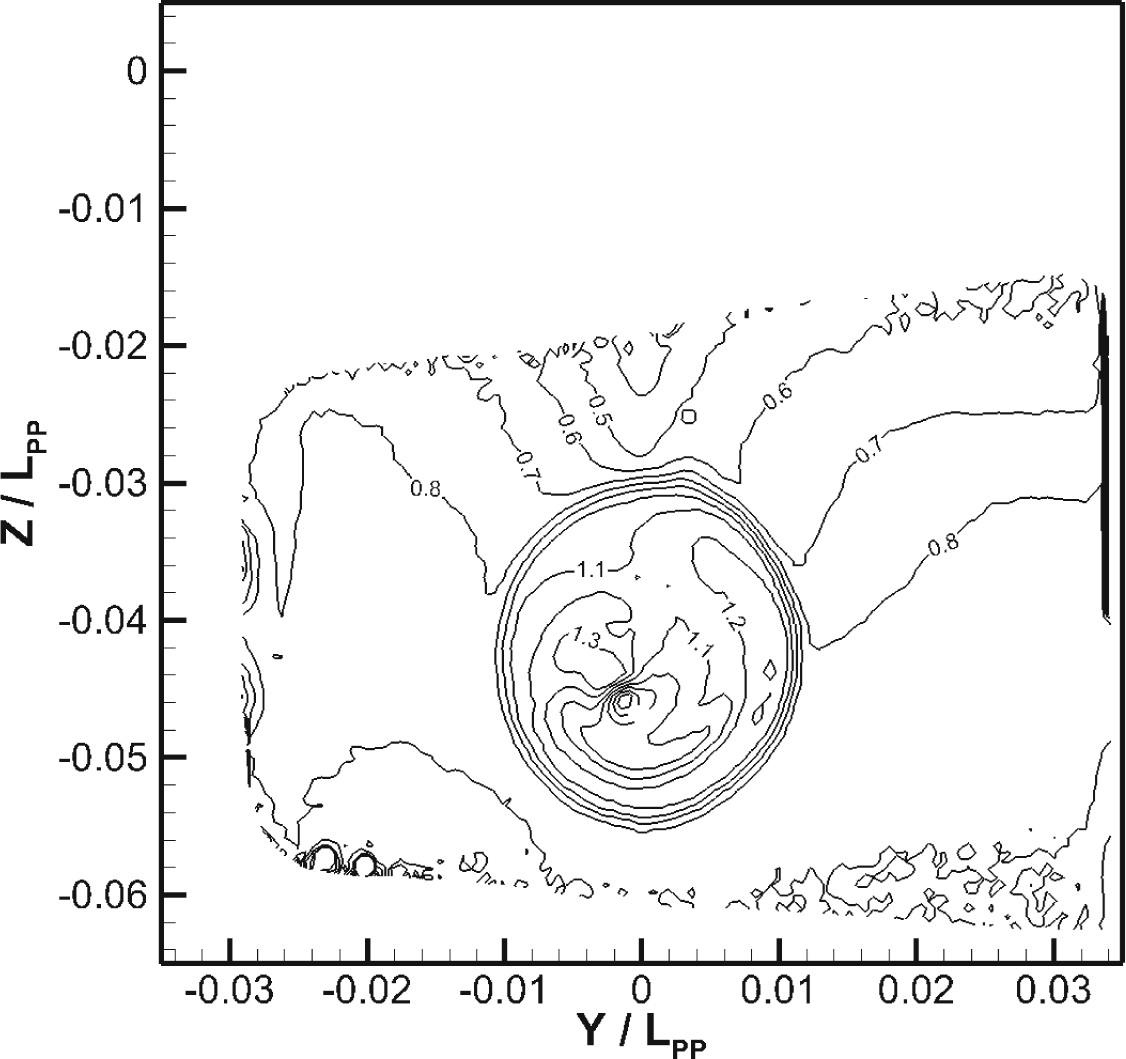}
        \captionsetup{width=0.9\linewidth}
        \caption{EFD}
    \end{subfigure}
    \begin{subfigure}[t]{0.32\linewidth}
        \includegraphics[width=1.0\linewidth]{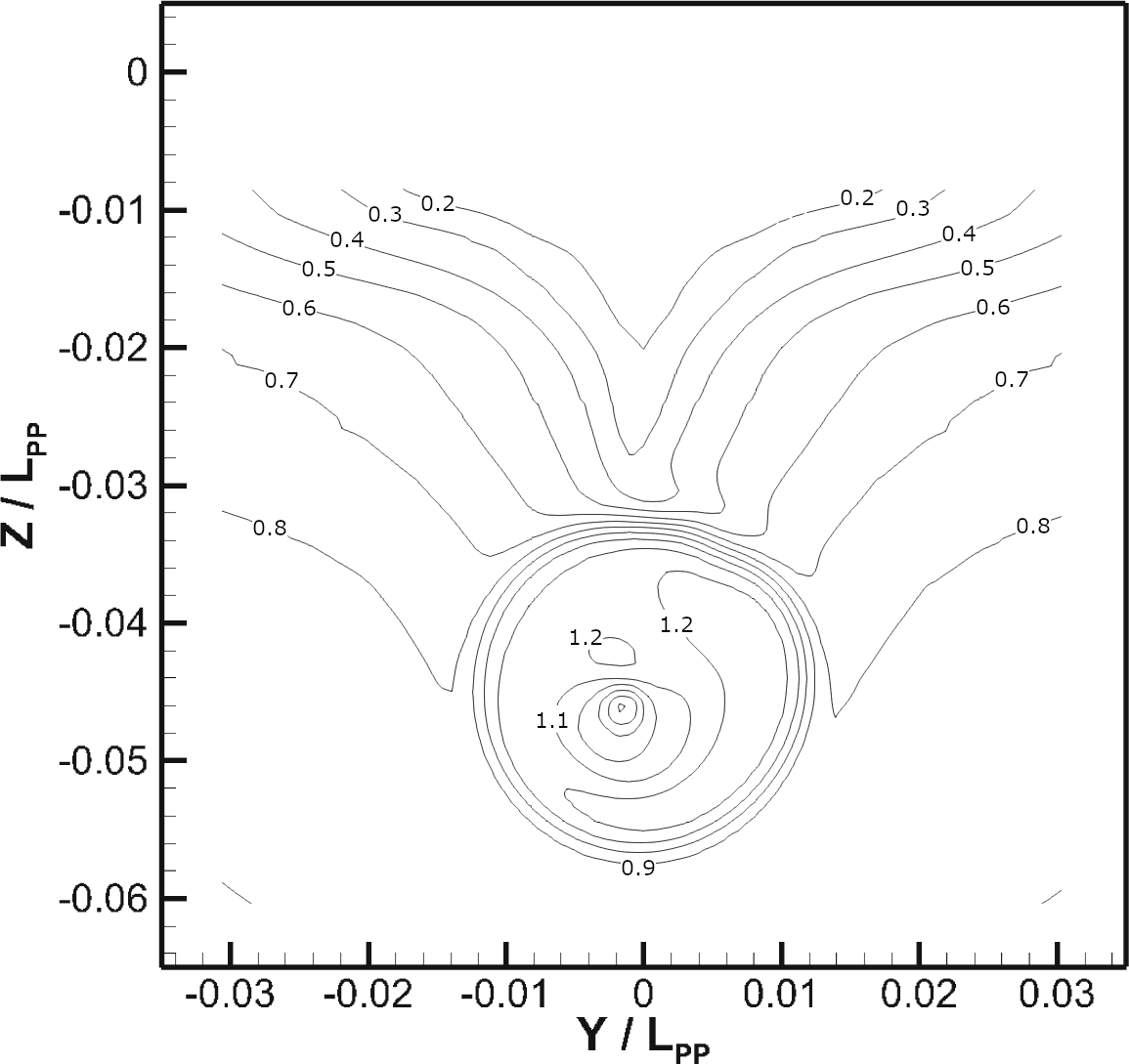}
        \captionsetup{width=0.9\linewidth}
        \caption{S1 ($n_{\mathrm{MRF}}/n = 0.0$)}
    \end{subfigure}
    \begin{subfigure}[t]{0.16\linewidth}
        \hfill
    \end{subfigure}
    \begin{subfigure}[t]{0.32\linewidth}
        \includegraphics[width=1.0\linewidth]{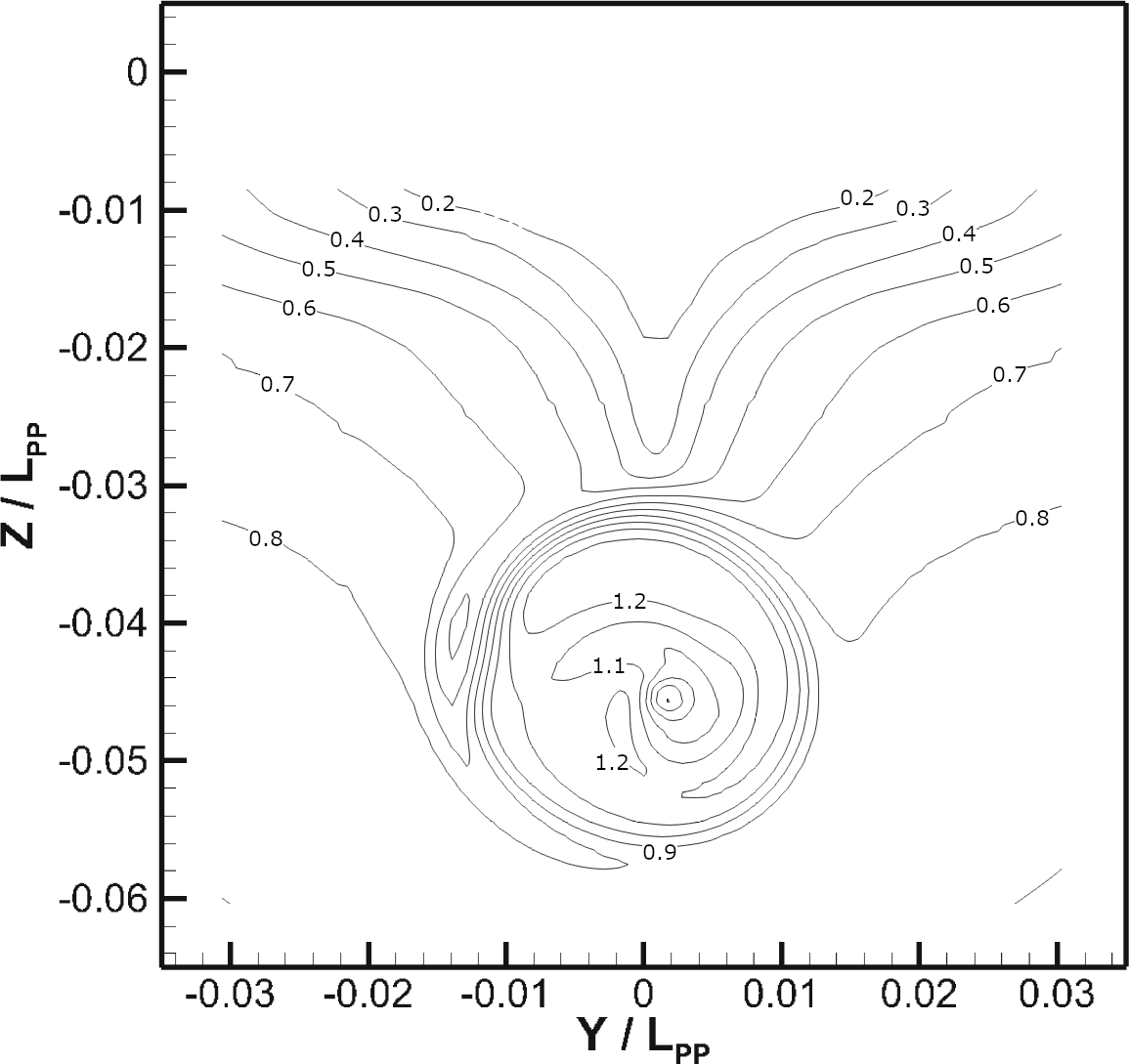}
        \captionsetup{width=0.9\linewidth}
        \caption{S3 (MRF, $n_{\mathrm{MRF}}/n = 0.5$)}
    \end{subfigure}
    \begin{subfigure}[t]{0.32\linewidth}
        \includegraphics[width=1.0\linewidth]{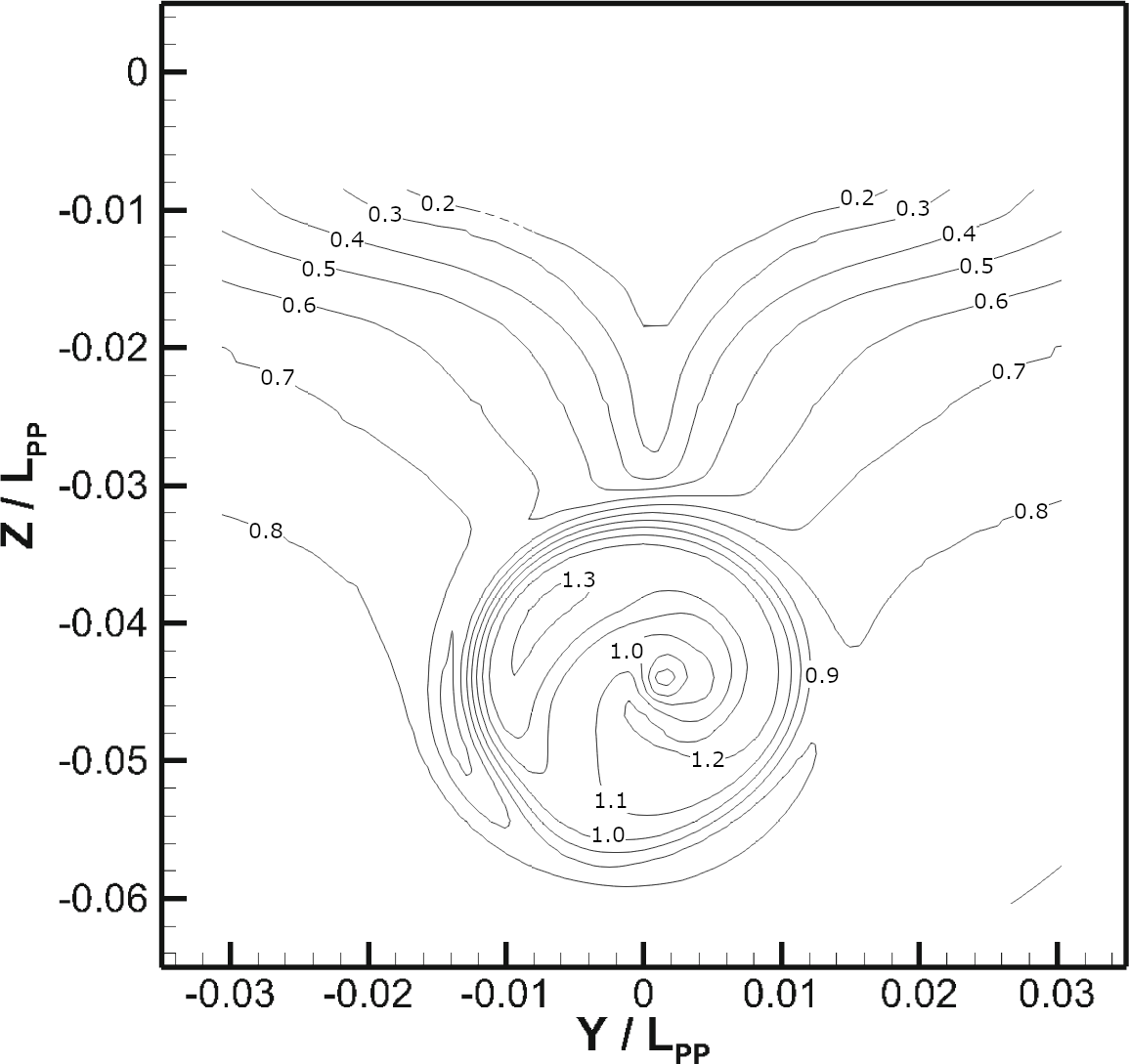}
        \captionsetup{width=0.9\linewidth}
        \caption{S4 (MRF, $n_{\mathrm{MRF}}/n = 0.75$)}
    \end{subfigure}
    \begin{subfigure}[t]{0.32\linewidth}
        \includegraphics[width=1.0\linewidth]{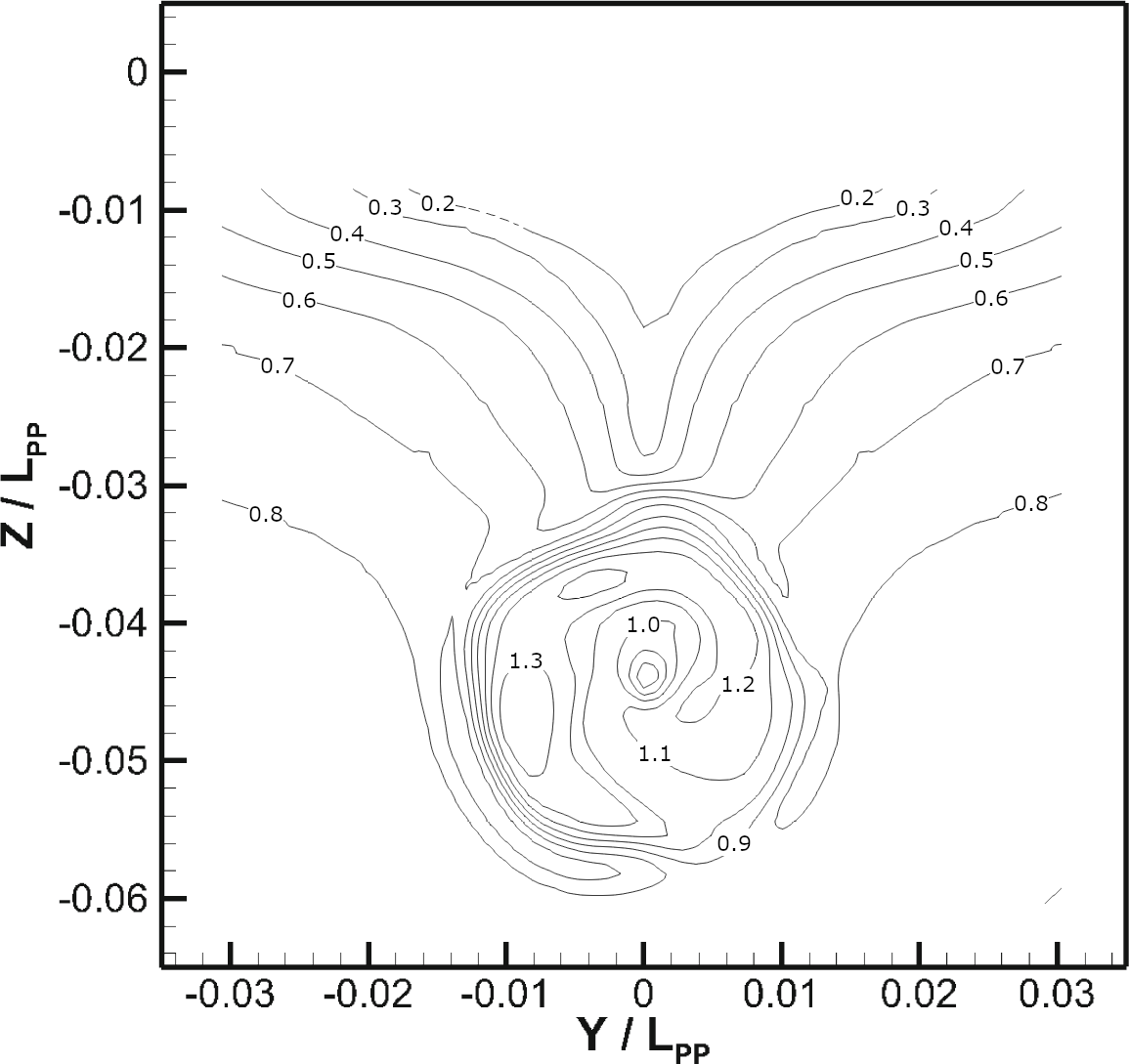}
        \captionsetup{width=0.9\linewidth}
        \caption{S5 (MRF, $n_{\mathrm{MRF}}/n = 1.0$)}
    \end{subfigure}
    \begin{subfigure}[t]{0.32\linewidth}
        \includegraphics[width=1.0\linewidth]{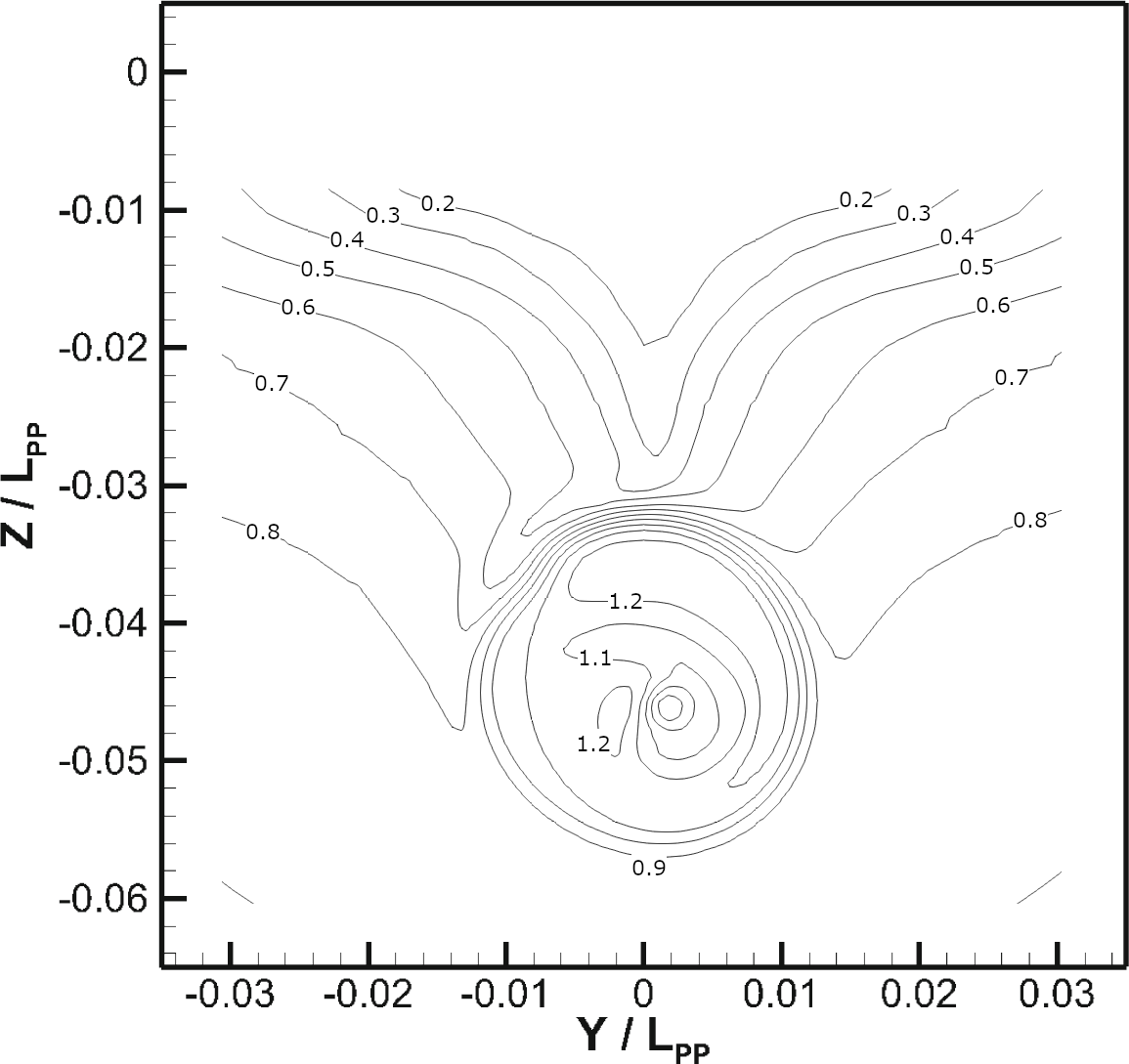}
        \captionsetup{width=0.9\linewidth}
        \caption{S7 (mMRF, $n_{\mathrm{MRF}}/n = 0.5$)}
    \end{subfigure}
    \begin{subfigure}[t]{0.32\linewidth}
        \includegraphics[width=1.0\linewidth]{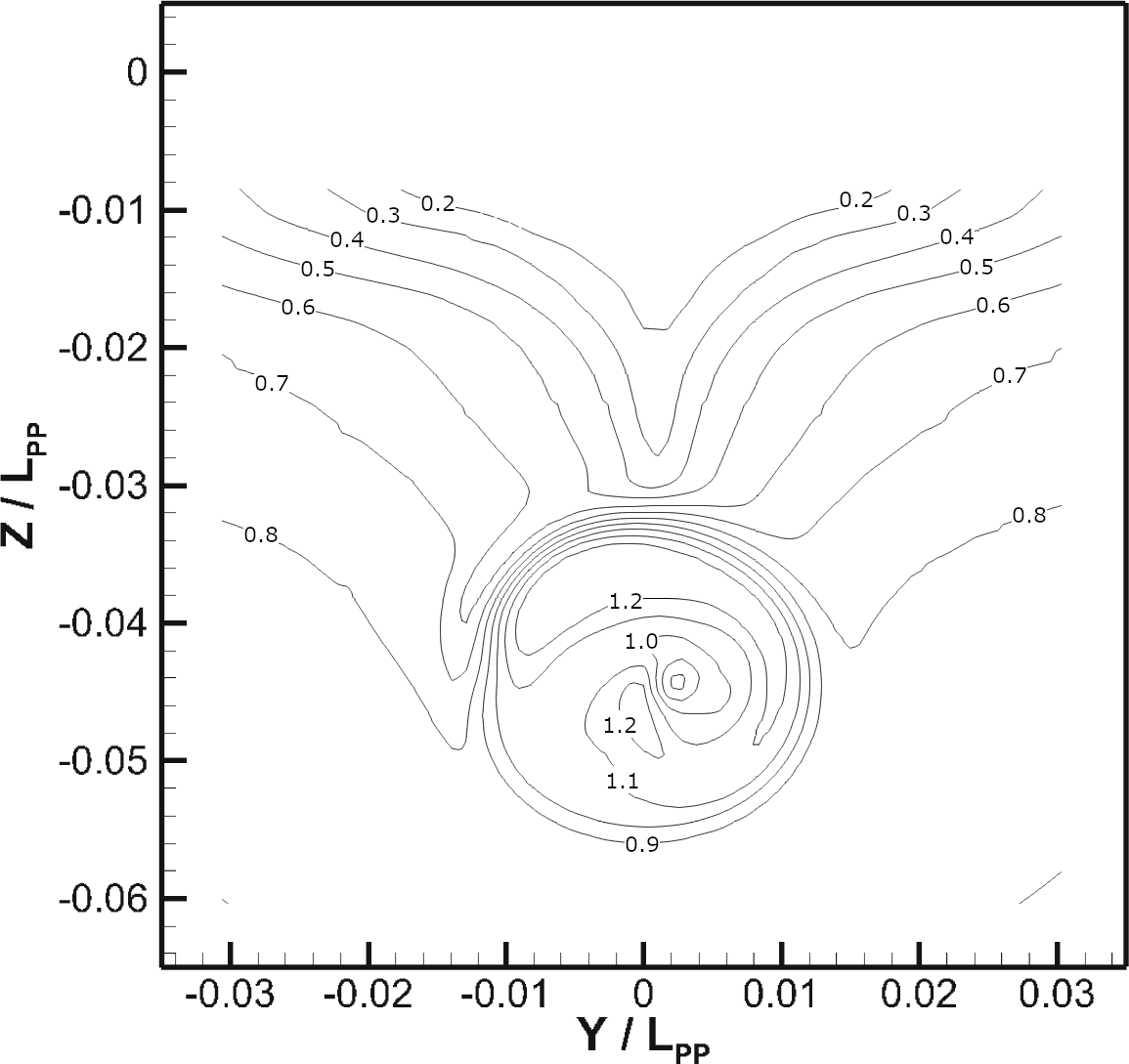}
        \captionsetup{width=0.9\linewidth}
        \caption{S8 (mMRF, $n_{\mathrm{MRF}}/n = 0.75$)}
    \end{subfigure}
    \begin{subfigure}[t]{0.32\linewidth}
        \includegraphics[width=1.0\linewidth]{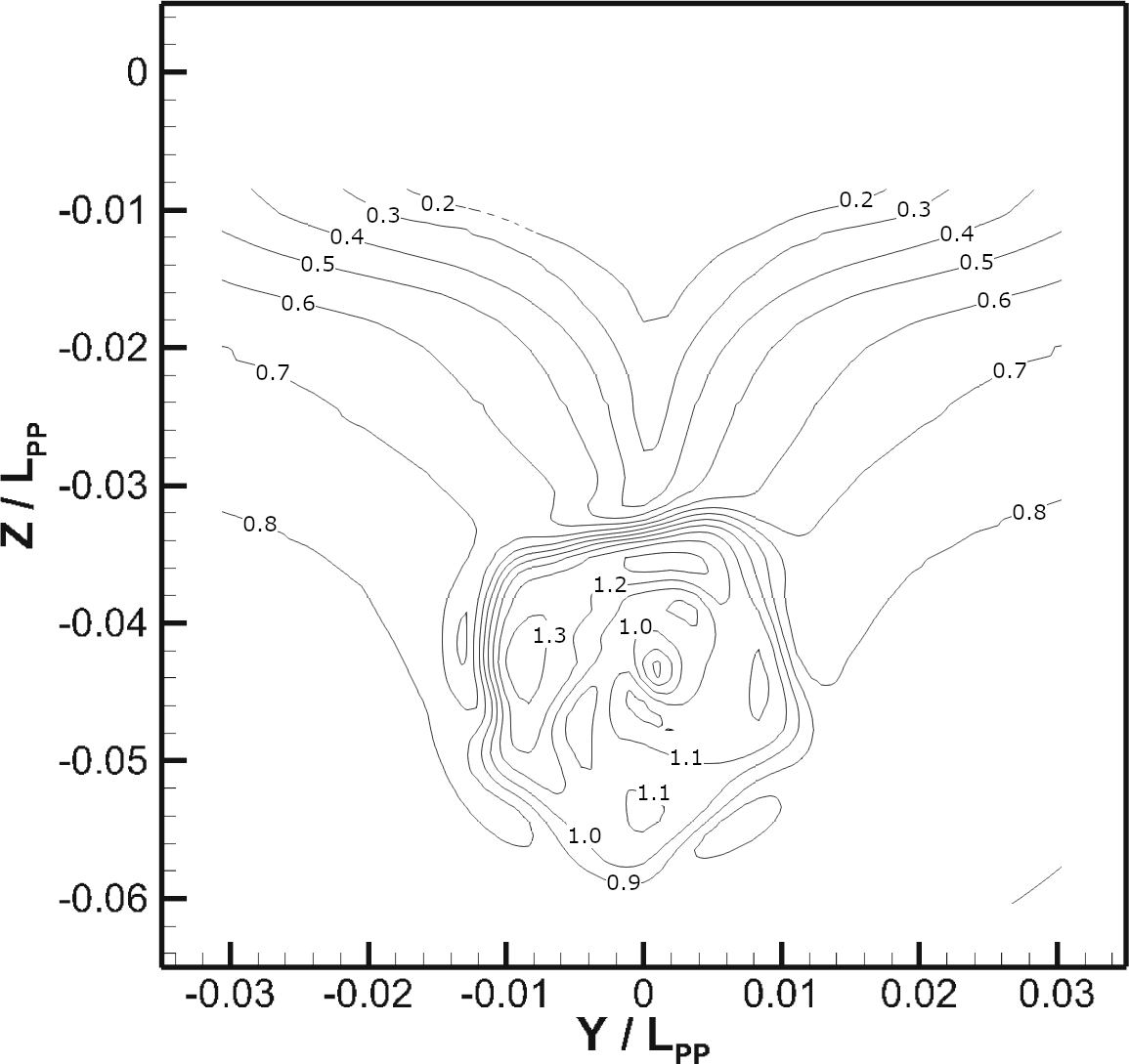}
        \captionsetup{width=0.9\linewidth}
        \caption{S9 (mMRF, $n_{\mathrm{MRF}}/n = 1.0$)}
    \end{subfigure}
    \caption{Axial velocity contours in the downstream plane ($x/L_{\mathrm{PP}}=0$) for the EFD data, the SI solution ($n_{\mathrm{MRF}}/n=0$), and the MRF and mMRF formulations at $n_{\mathrm{MRF}}/n=0.5$, $0.75$ and $1.0$.}
    \label{fig:contour_x0}
\end{figure}
\begin{figure}[htbp]
    \centering
    \begin{subfigure}[t]{0.16\linewidth}
        \hfill
    \end{subfigure}
    \begin{subfigure}[t]{0.32\linewidth}
        \includegraphics[width=1.0\linewidth]{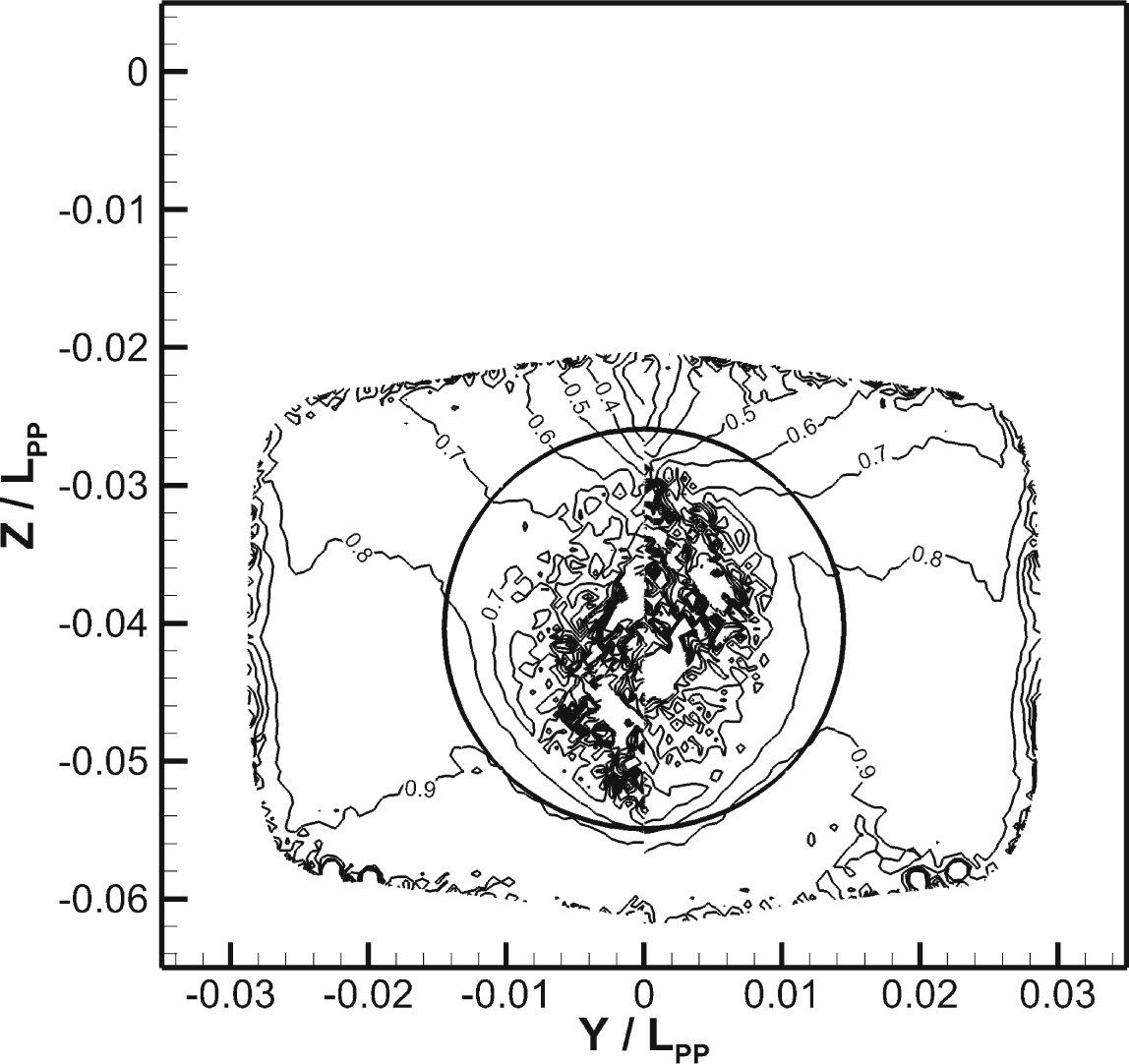}
        \captionsetup{width=0.9\linewidth}
        \caption{EFD}
    \end{subfigure}
    \begin{subfigure}[t]{0.32\linewidth}
        \includegraphics[width=1.0\linewidth]{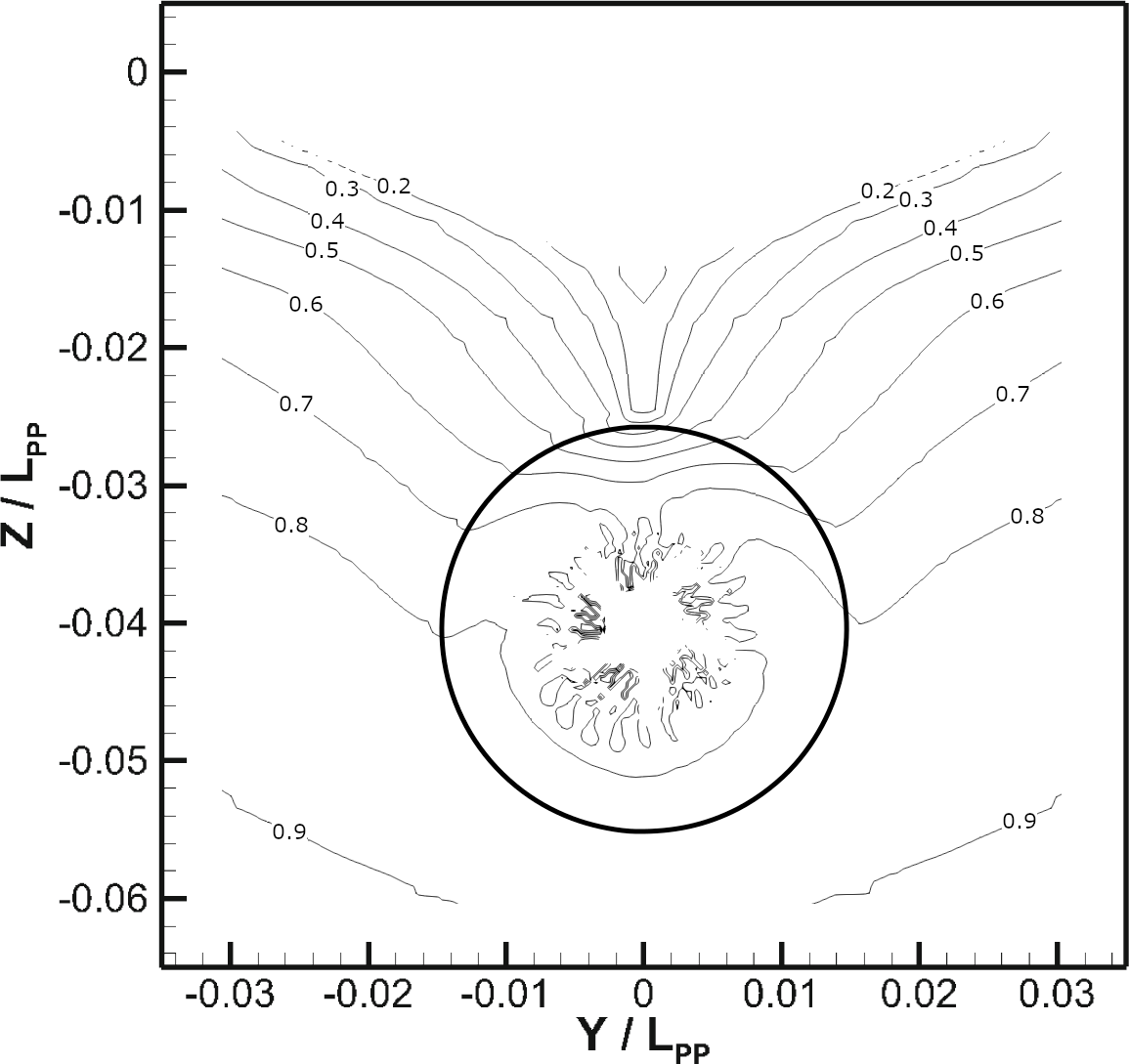}
        \captionsetup{width=0.9\linewidth}
        \caption{S1 ($n_{\mathrm{MRF}}/n = 0.0$)}
    \end{subfigure}
    \begin{subfigure}[t]{0.16\linewidth}
        \hfill
    \end{subfigure}
    \begin{subfigure}[t]{0.32\linewidth}
        \includegraphics[width=1.0\linewidth]{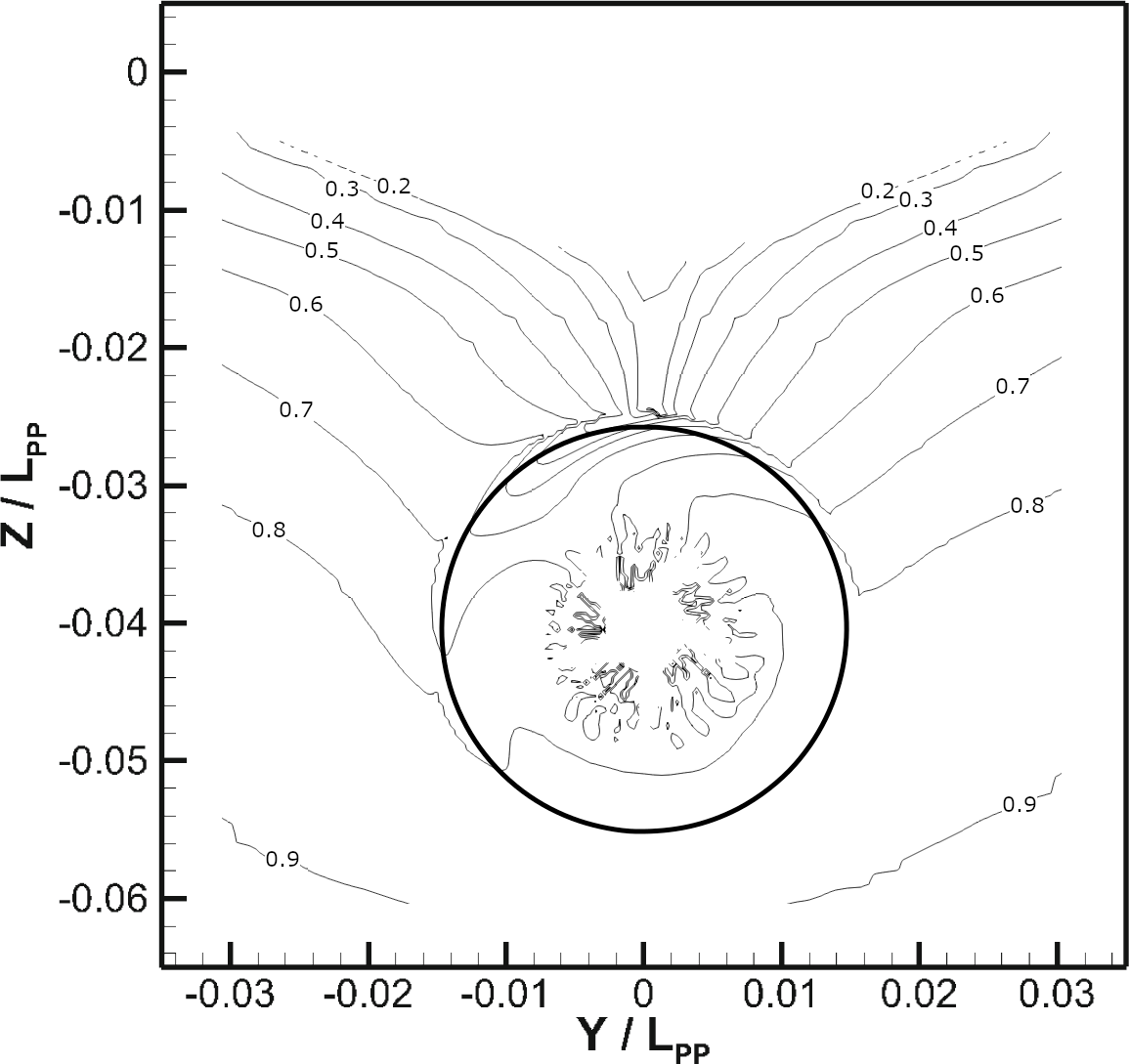}
        \captionsetup{width=0.9\linewidth}
        \caption{S3 (MRF, $n_{\mathrm{MRF}}/n = 0.5$)}
    \end{subfigure}
    \begin{subfigure}[t]{0.32\linewidth}
        \includegraphics[width=1.0\linewidth]{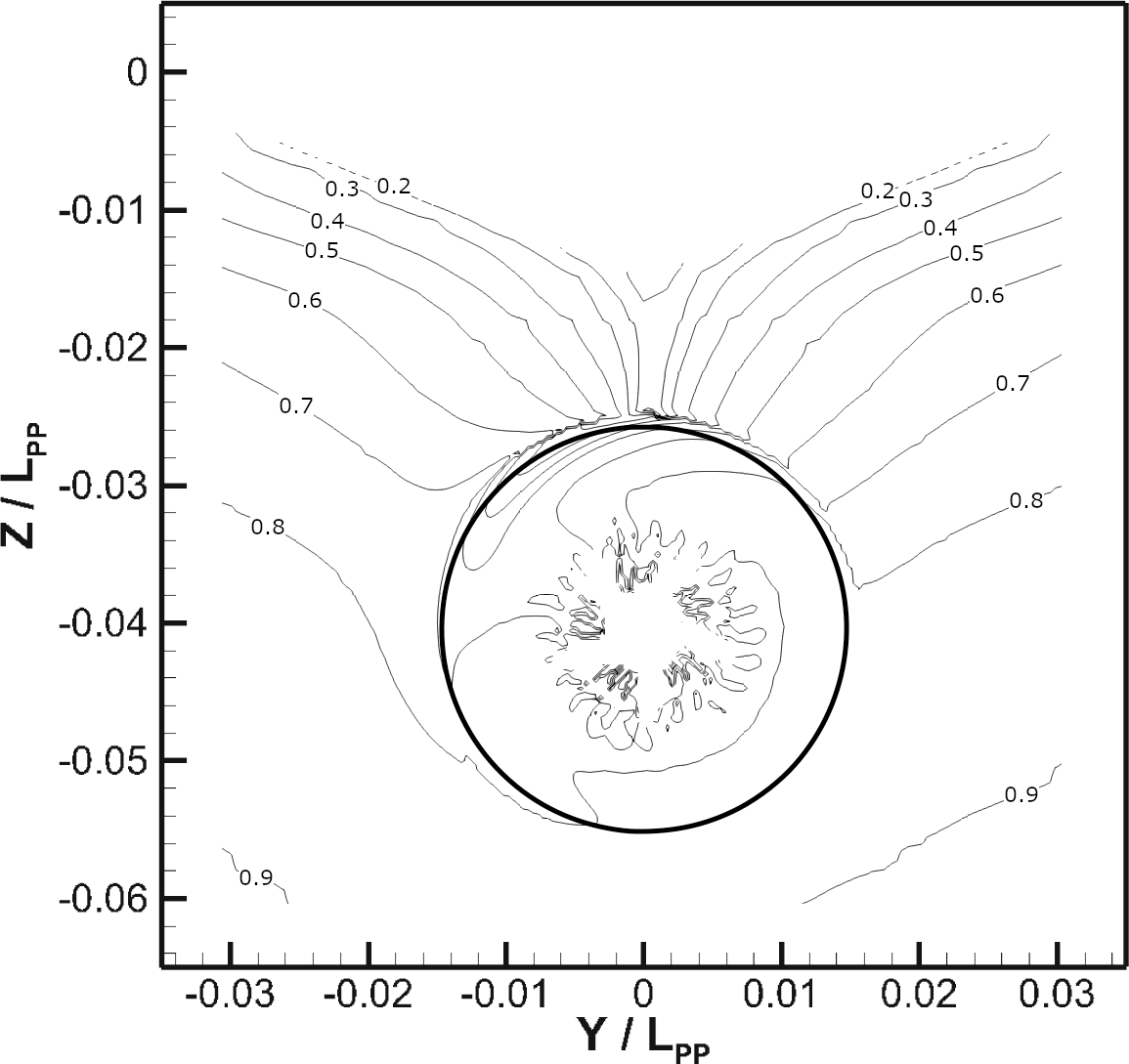}
        \captionsetup{width=0.9\linewidth}
        \caption{S4 (MRF, $n_{\mathrm{MRF}}/n = 0.75$)}
    \end{subfigure}
    \begin{subfigure}[t]{0.32\linewidth}
        \includegraphics[width=1.0\linewidth]{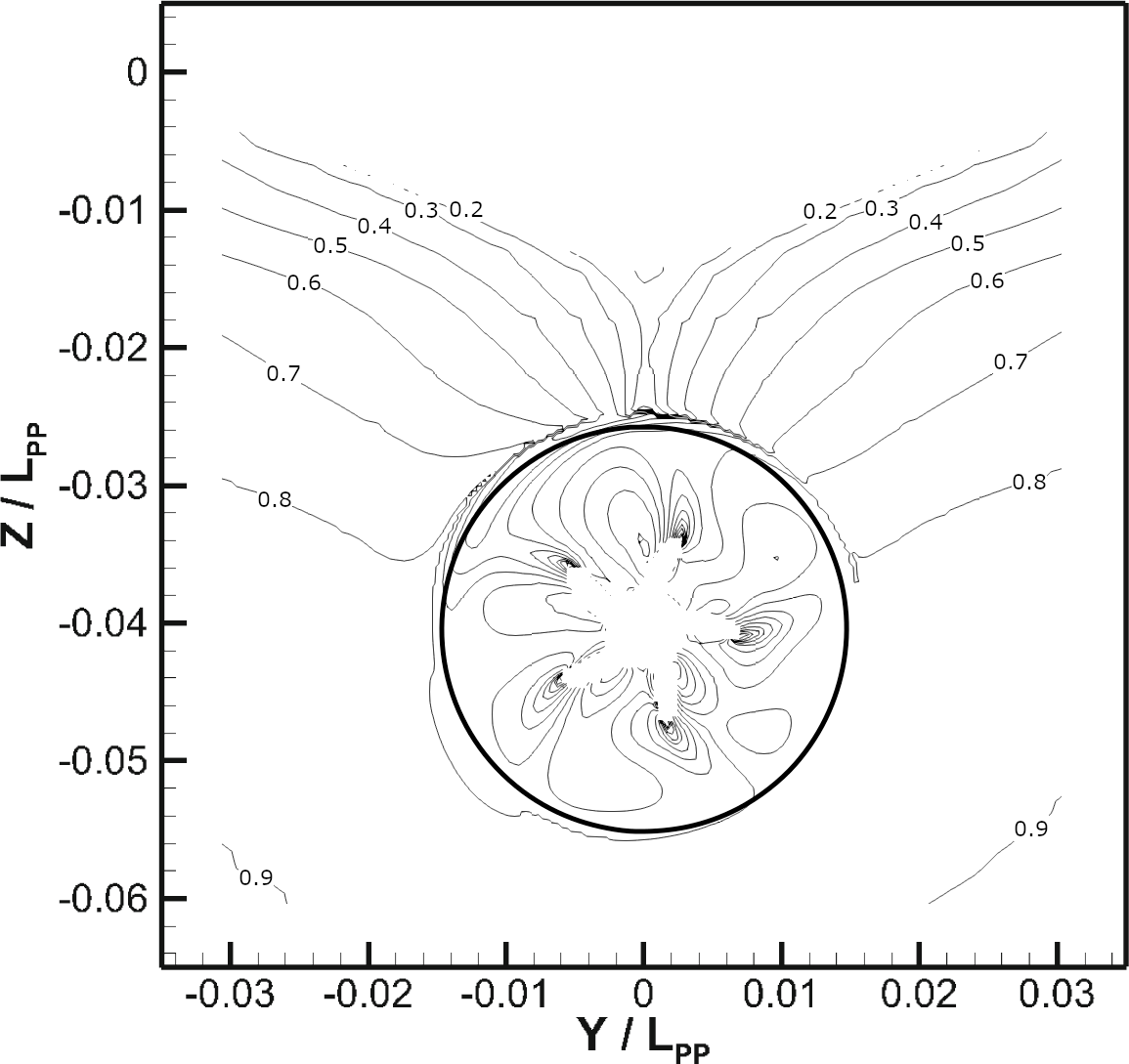}
        \captionsetup{width=0.9\linewidth}
        \caption{S5 (MRF, $n_{\mathrm{MRF}}/n = 1.0$)}
    \end{subfigure}
    \begin{subfigure}[t]{0.32\linewidth}
        \includegraphics[width=1.0\linewidth]{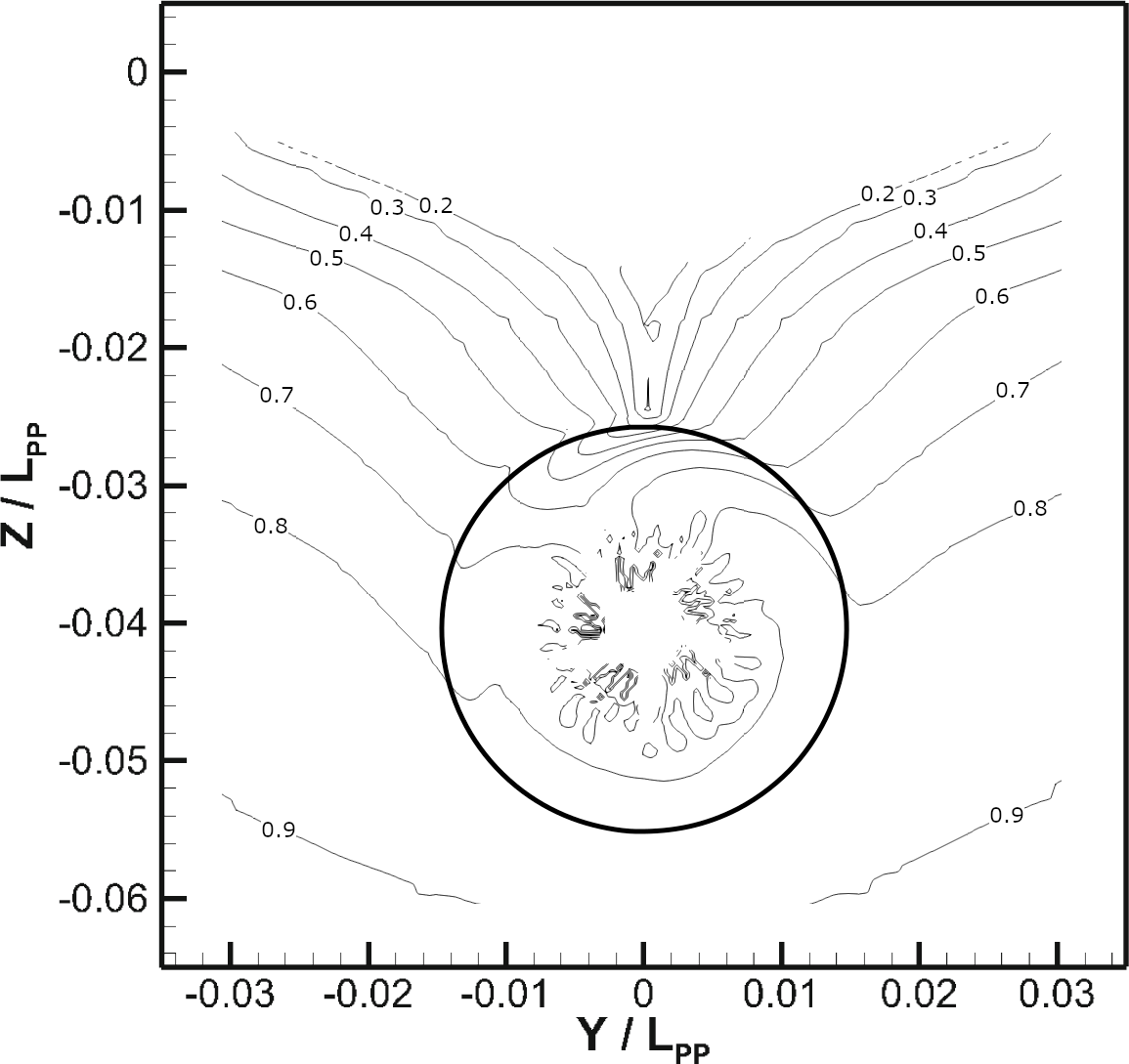}
        \captionsetup{width=0.9\linewidth}
        \caption{S7 (mMRF, $n_{\mathrm{MRF}}/n = 0.5$)}
    \end{subfigure}
    \begin{subfigure}[t]{0.32\linewidth}
        \includegraphics[width=1.0\linewidth]{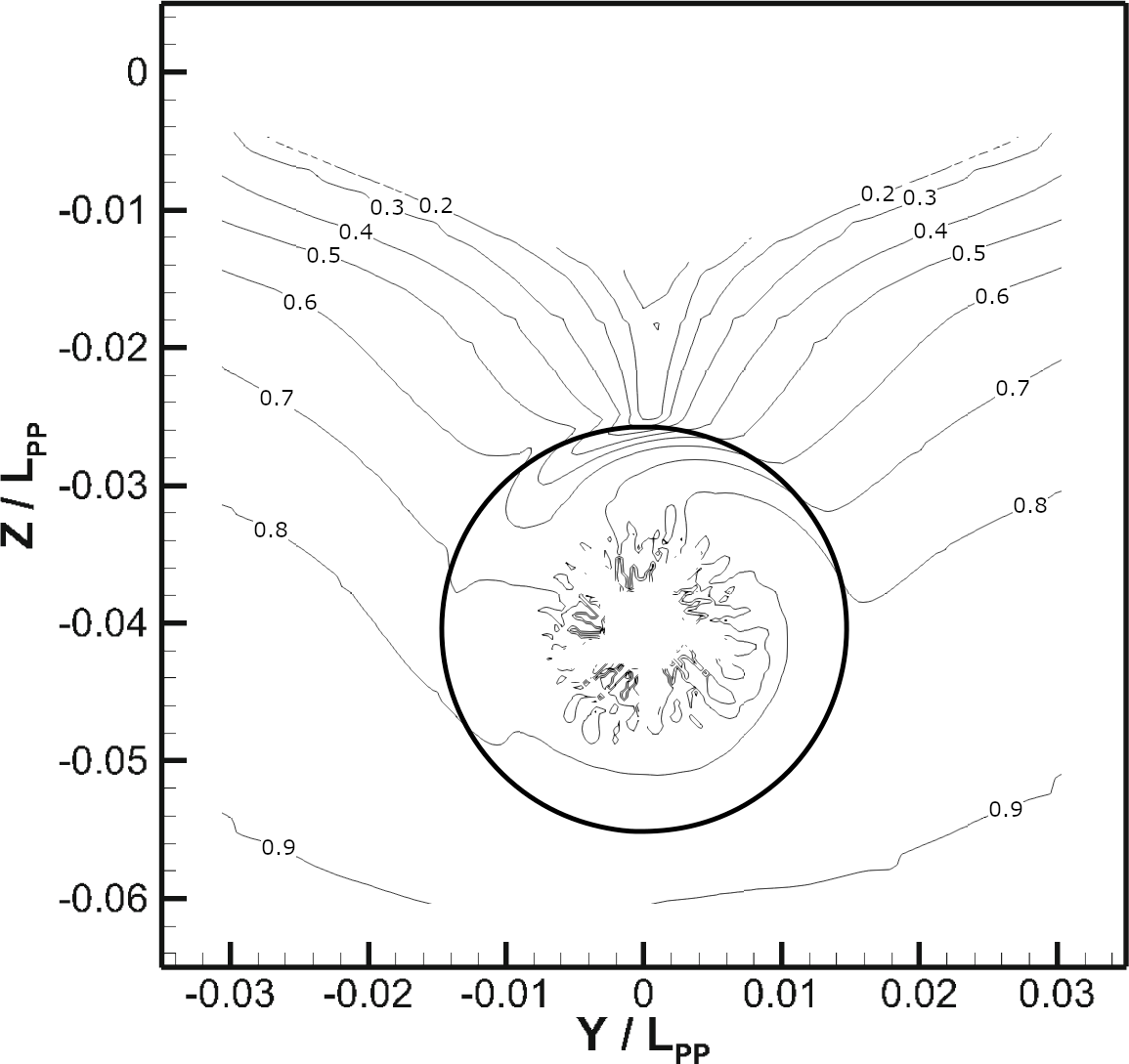}
        \captionsetup{width=0.9\linewidth}
        \caption{S8 (mMRF, $n_{\mathrm{MRF}}/n = 0.75$)}
    \end{subfigure}
    \begin{subfigure}[t]{0.32\linewidth}
        \includegraphics[width=1.0\linewidth]{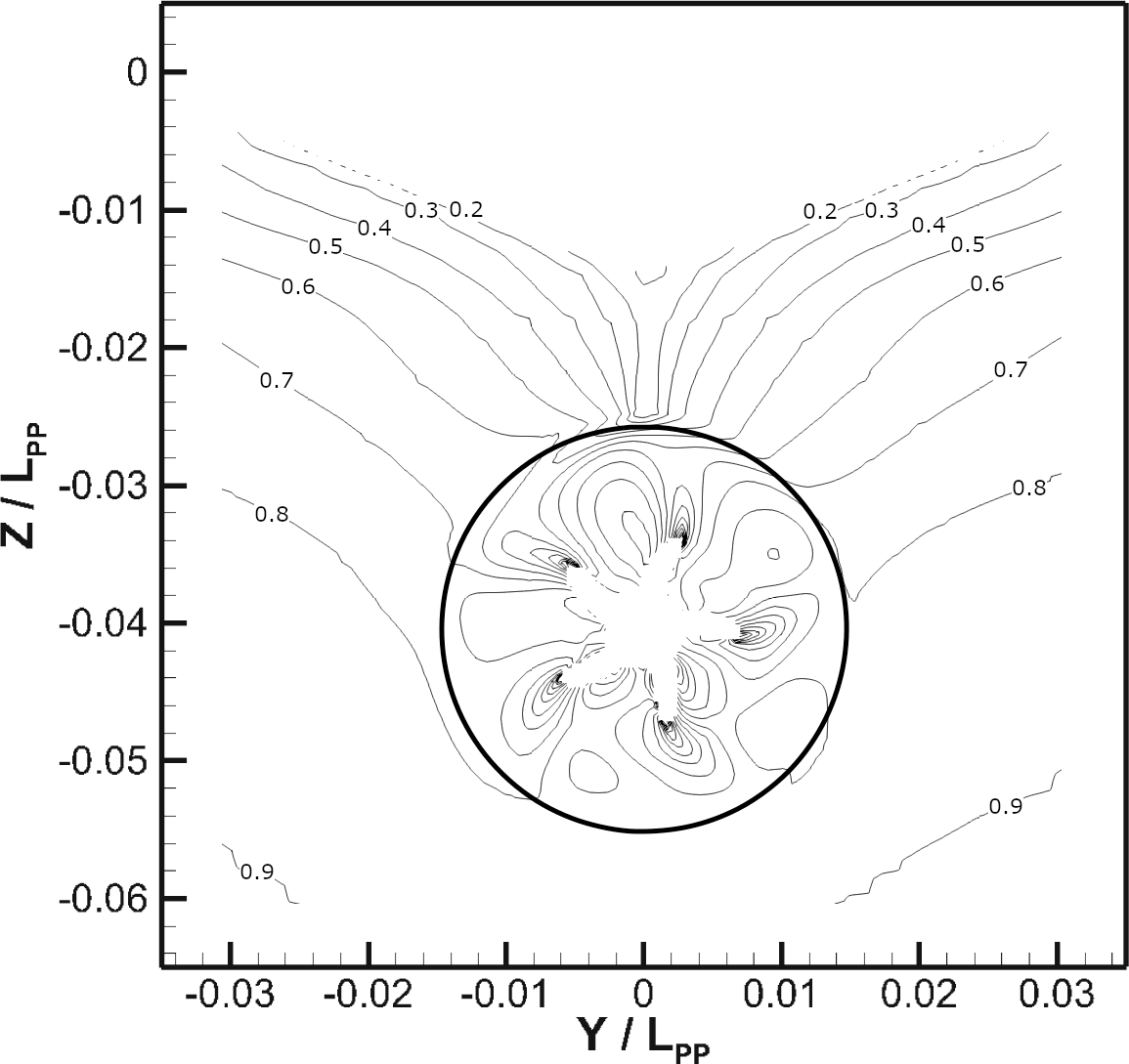}
        \captionsetup{width=0.9\linewidth}
        \caption{S9 (mMRF, $n_{\mathrm{MRF}}/n = 1.0$)}
    \end{subfigure}
    \caption{Axial velocity contours in the upstream plane ($x/L_{\mathrm{PP}}=0.0157$) for the EFD data, the SI solution ($n_{\mathrm{MRF}}/n=0$), and the MRF and mMRF formulations at $n_{\mathrm{MRF}}/n=0.5$, $0.75$ and $1.0$.}
    \label{fig:contour_x1}
\end{figure}
\begin{figure}[htbp]
    \centering
    \begin{subfigure}[t]{0.16\linewidth}
        \hfill
    \end{subfigure}
    \begin{subfigure}[t]{0.32\linewidth}
        \includegraphics[width=1.0\linewidth]{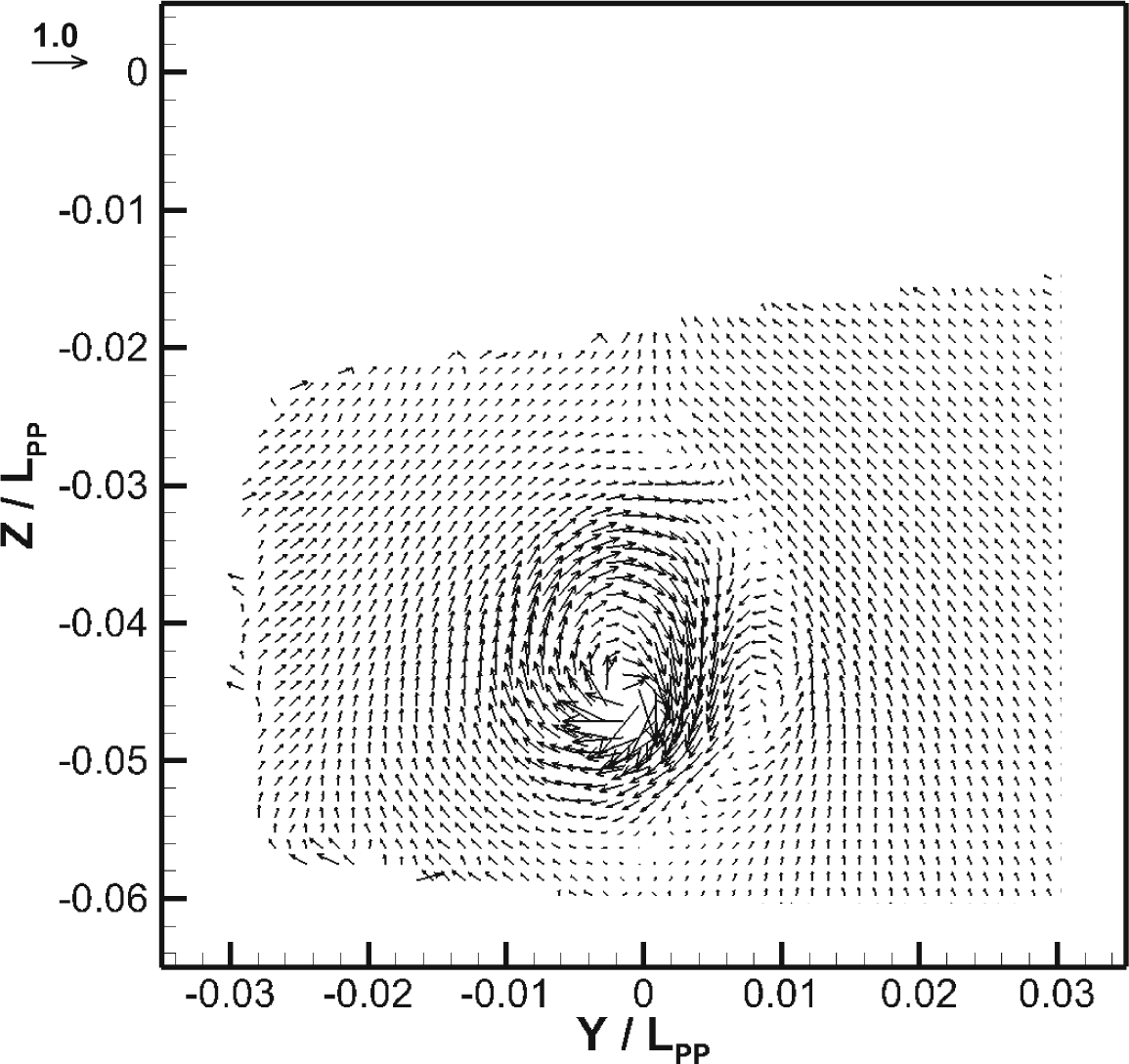}
        \captionsetup{width=0.9\linewidth}
        \caption{EFD}
    \end{subfigure}
    \begin{subfigure}[t]{0.32\linewidth}
        \includegraphics[width=1.0\linewidth]{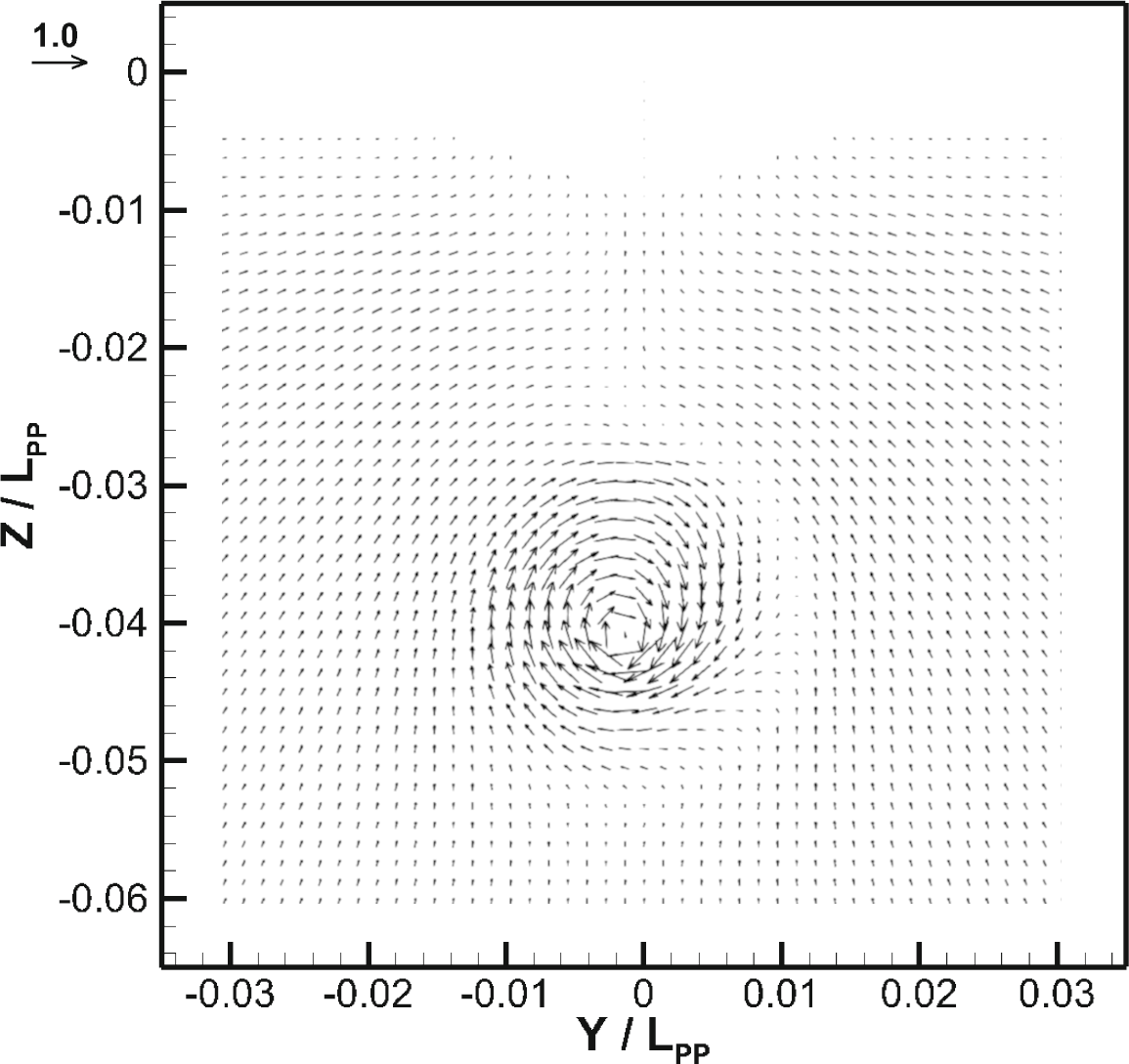}
        \captionsetup{width=0.9\linewidth}
        \caption{S1 ($n_{\mathrm{MRF}}/n = 0$)}
    \end{subfigure}
    \begin{subfigure}[t]{0.16\linewidth}
        \hfill
    \end{subfigure}
    \begin{subfigure}[t]{0.32\linewidth}
        \includegraphics[width=1.0\linewidth]{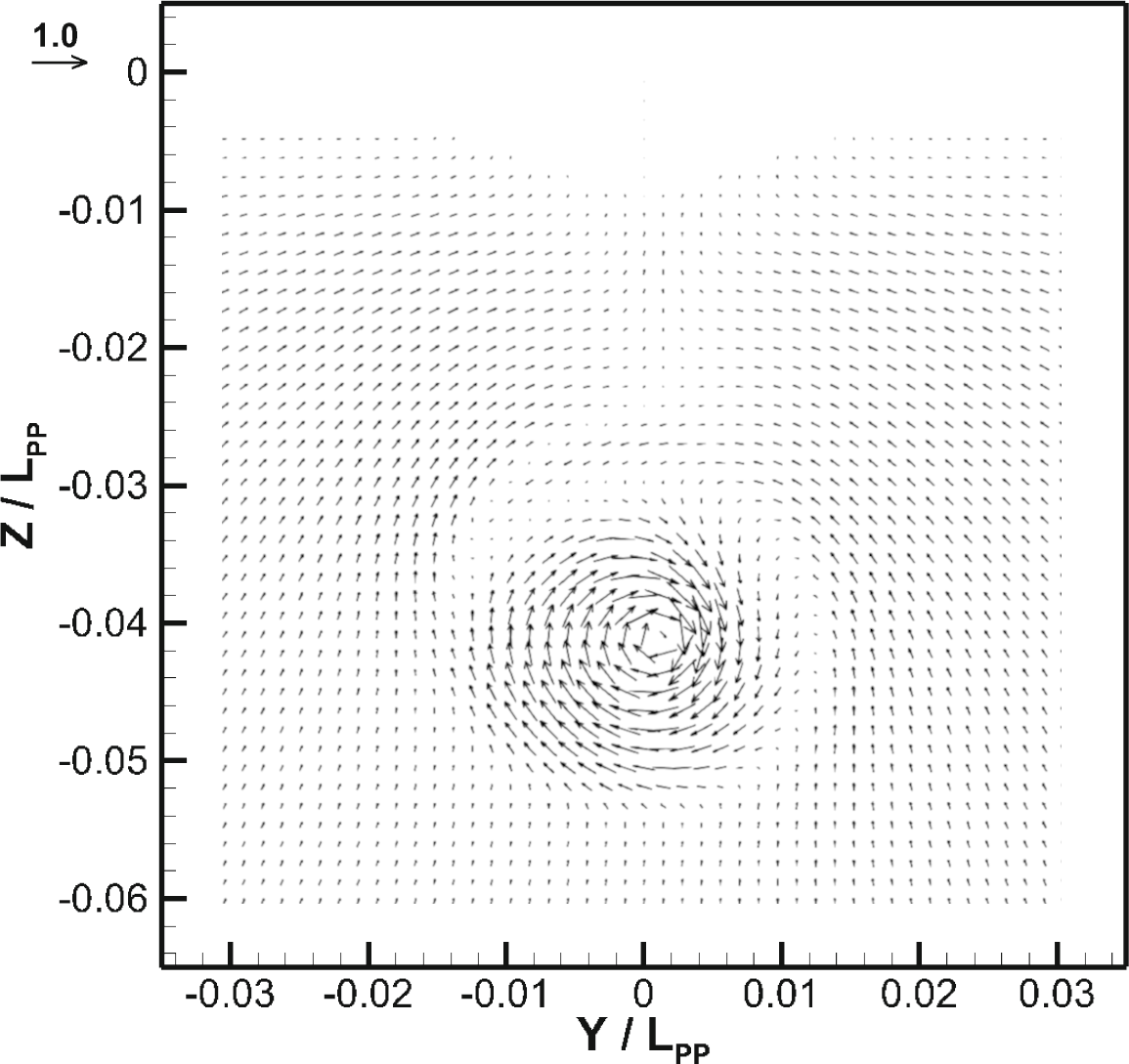}
        \captionsetup{width=0.9\linewidth}
        \caption{S3 (MRF, $n_{\mathrm{MRF}}/n = 0.5$)}
    \end{subfigure}
    \begin{subfigure}[t]{0.32\linewidth}
        \includegraphics[width=1.0\linewidth]{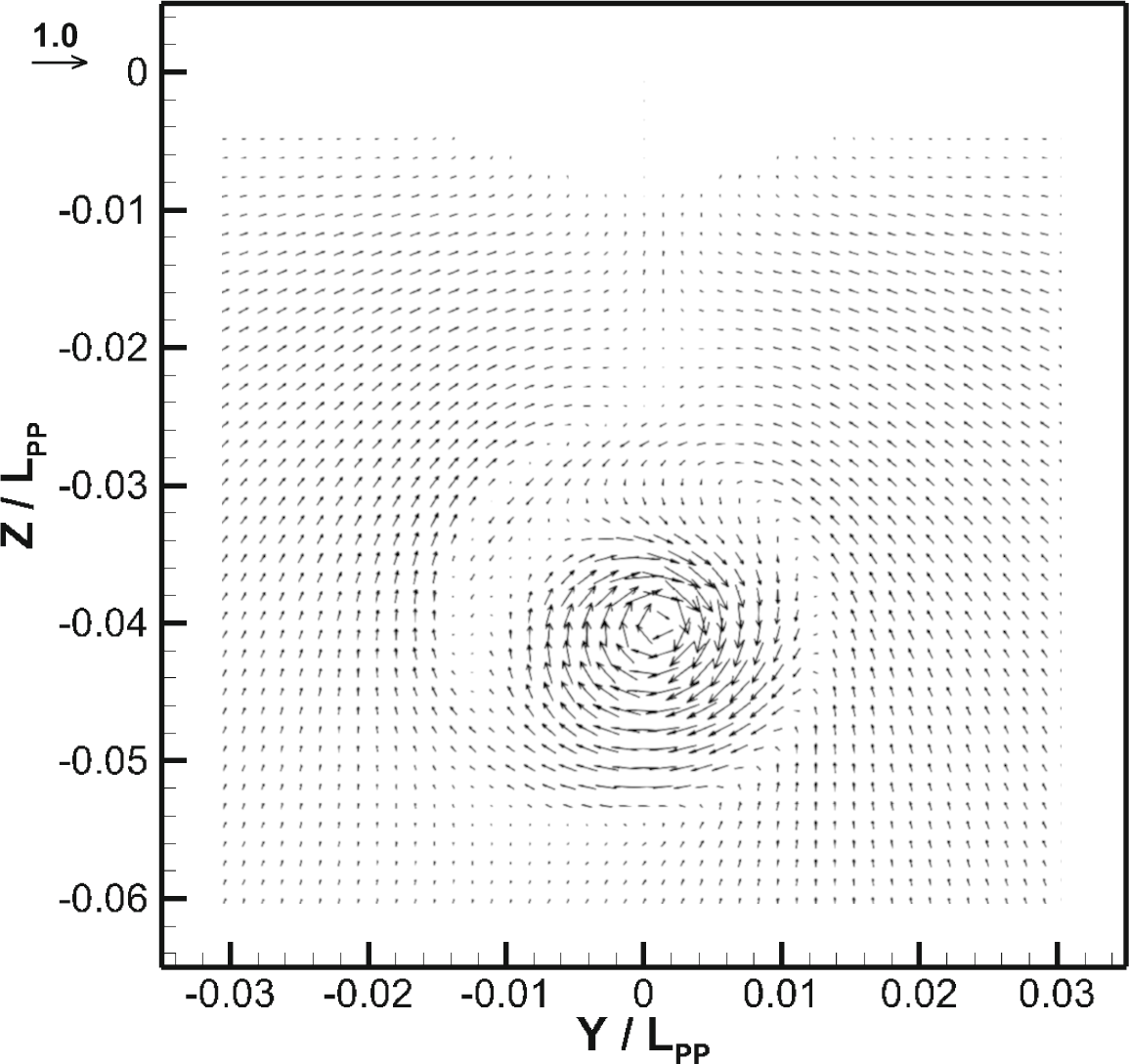}
        \captionsetup{width=0.9\linewidth}
        \caption{S4 (MRF, $n_{\mathrm{MRF}}/n = 0.75$)}
    \end{subfigure}
    \begin{subfigure}[t]{0.32\linewidth}
        \includegraphics[width=1.0\linewidth]{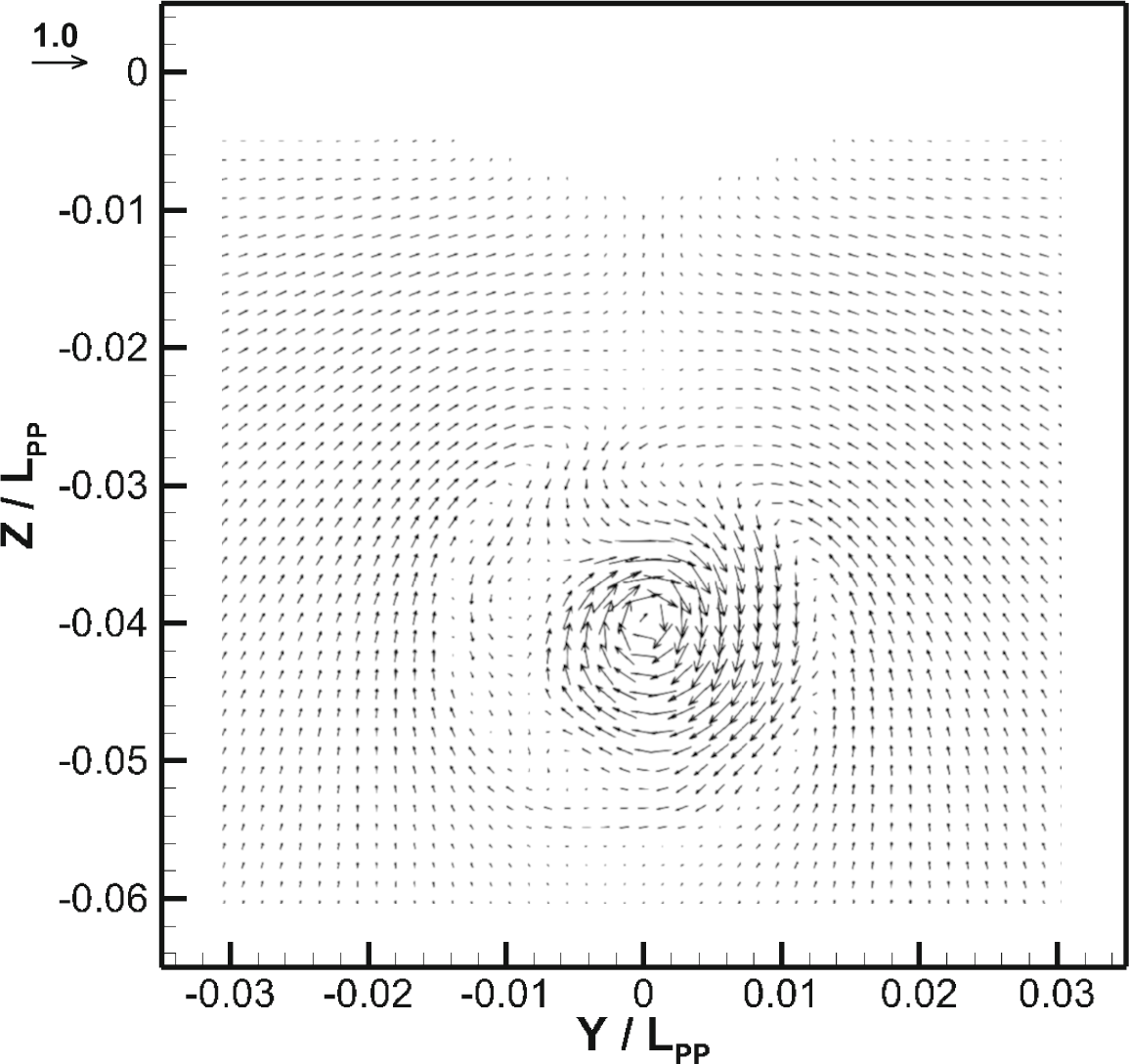}
        \captionsetup{width=0.9\linewidth}
        \caption{S5 (MRF, $n_{\mathrm{MRF}}/n = 1.0$)}
    \end{subfigure}
    \begin{subfigure}[t]{0.32\linewidth}
        \includegraphics[width=1.0\linewidth]{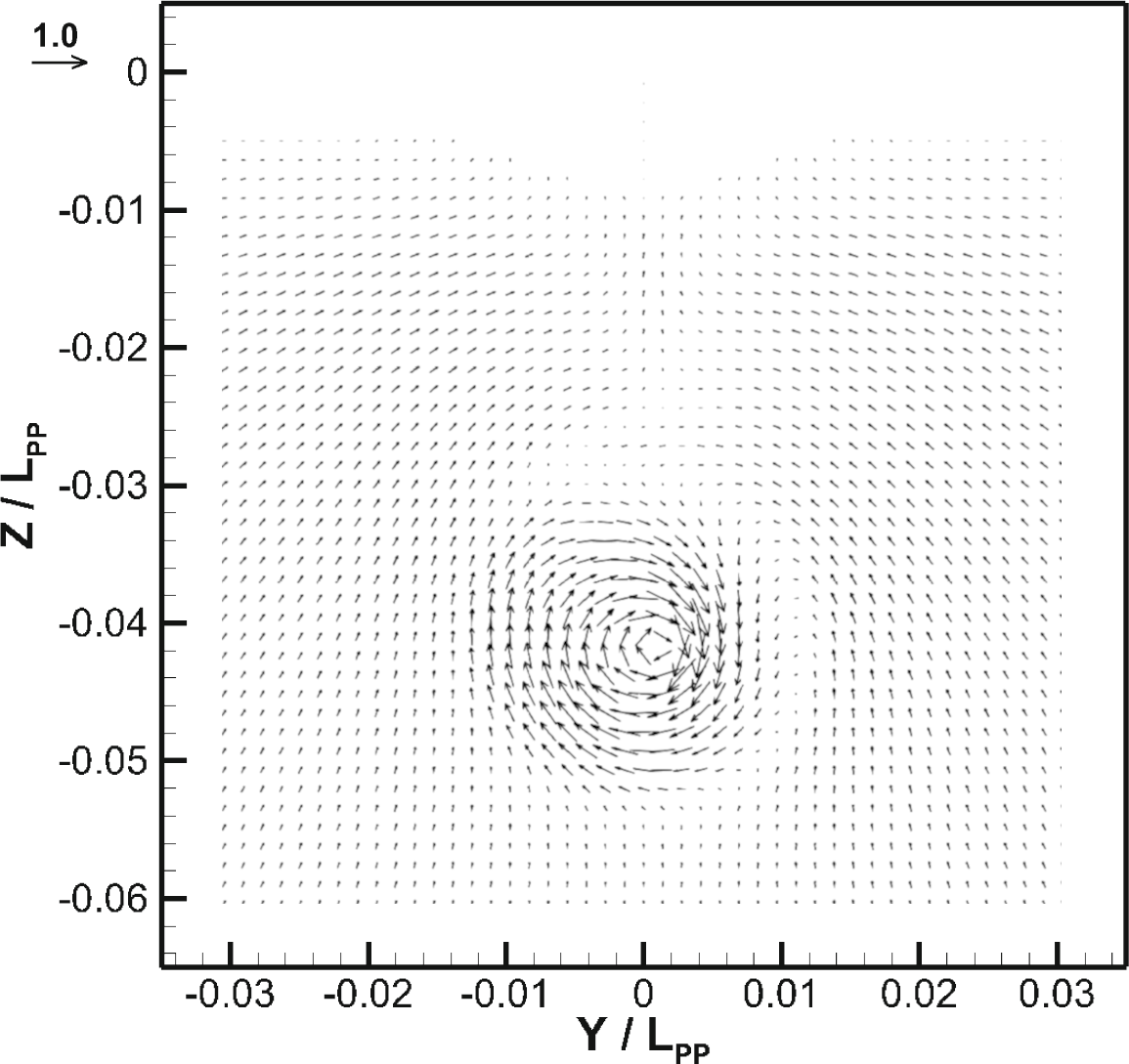}
        \captionsetup{width=0.9\linewidth}
        \caption{S7 (mMRF, $n_{\mathrm{MRF}}/n = 0.5$)}
    \end{subfigure}
    \begin{subfigure}[t]{0.32\linewidth}
        \includegraphics[width=1.0\linewidth]{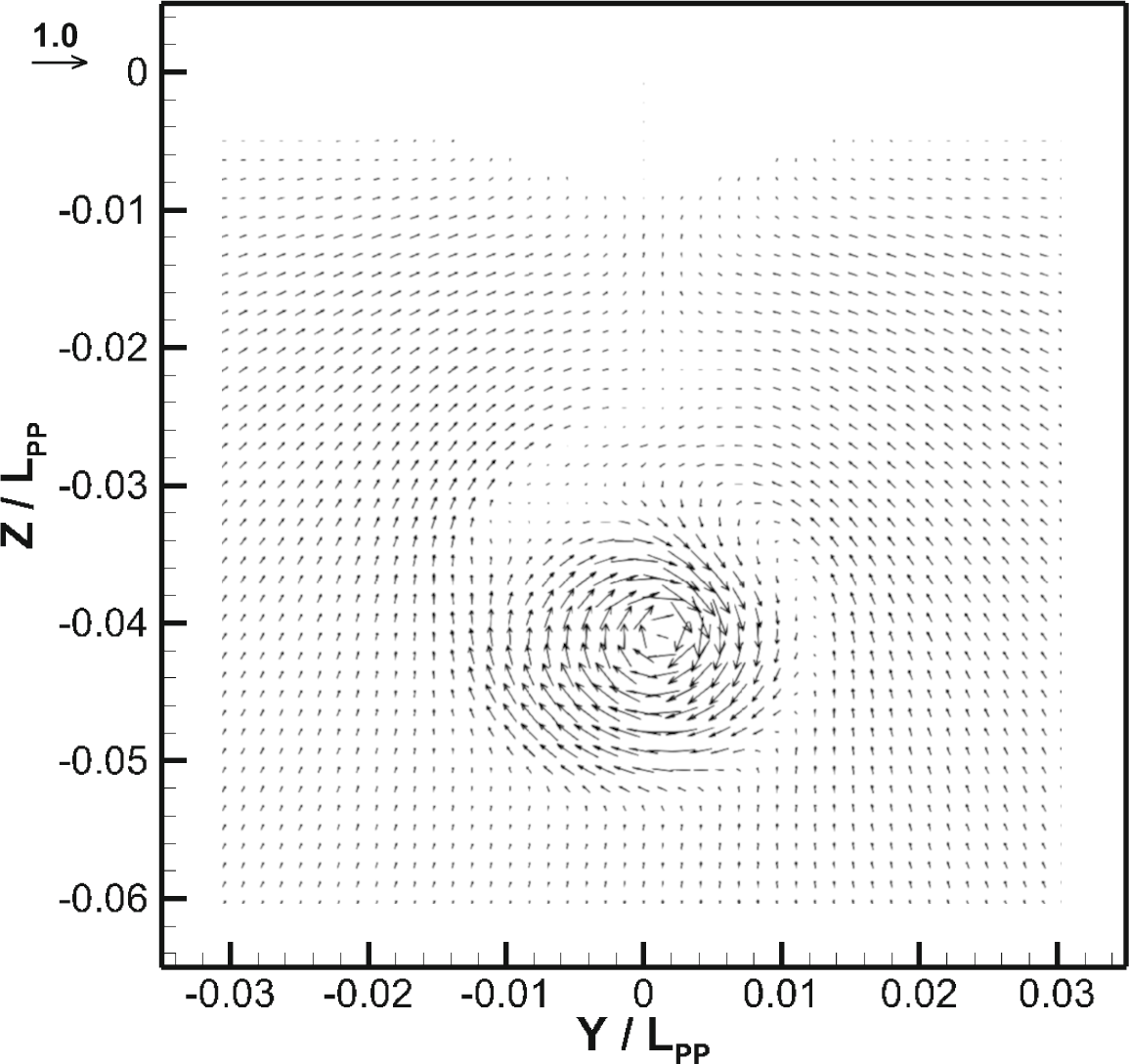}
        \captionsetup{width=0.9\linewidth}
        \caption{S8 (mMRF, $n_{\mathrm{MRF}}/n = 0.75$)}
    \end{subfigure}
    \begin{subfigure}[t]{0.32\linewidth}
        \includegraphics[width=1.0\linewidth]{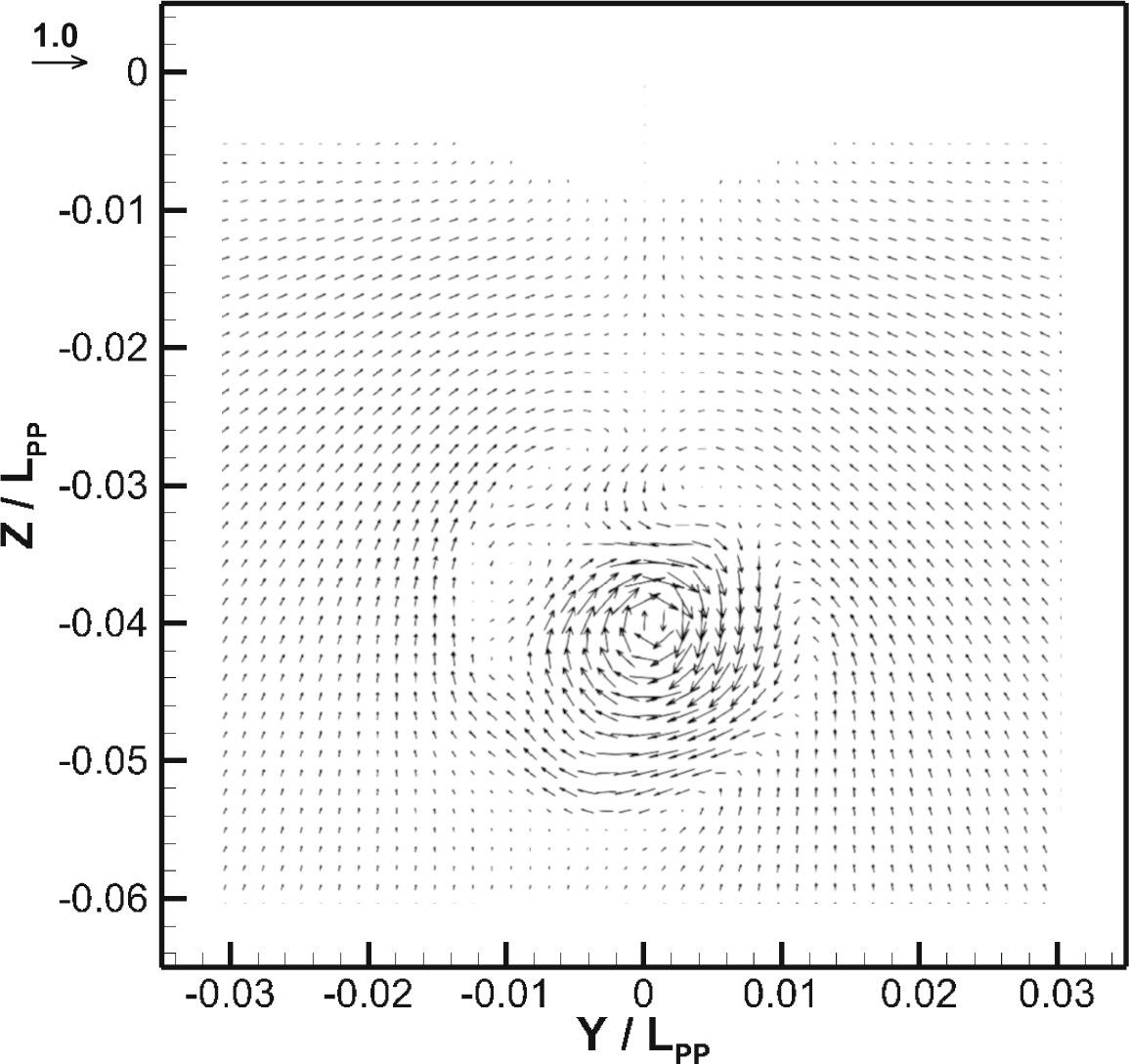}
        \captionsetup{width=0.9\linewidth}
        \caption{S9 (mMRF, $n_{\mathrm{MRF}}/n = 1.0$)}
    \end{subfigure}
    \caption{Cross-plane velocity vectors in the downstream plane ($x/L_{\mathrm{PP}}=0$) for the EFD data, the SI solution, and the MRF and mMRF formulations at $n_{\mathrm{MRF}}/n=0.5$, $0.75$ and $1.0$.}
    \label{fig:cross_x0}
\end{figure}
\begin{figure}[htbp]
    \centering
    \begin{subfigure}[t]{0.16\linewidth}
        \hfill
    \end{subfigure}
    \begin{subfigure}[t]{0.32\linewidth}
        \includegraphics[width=1.0\linewidth]{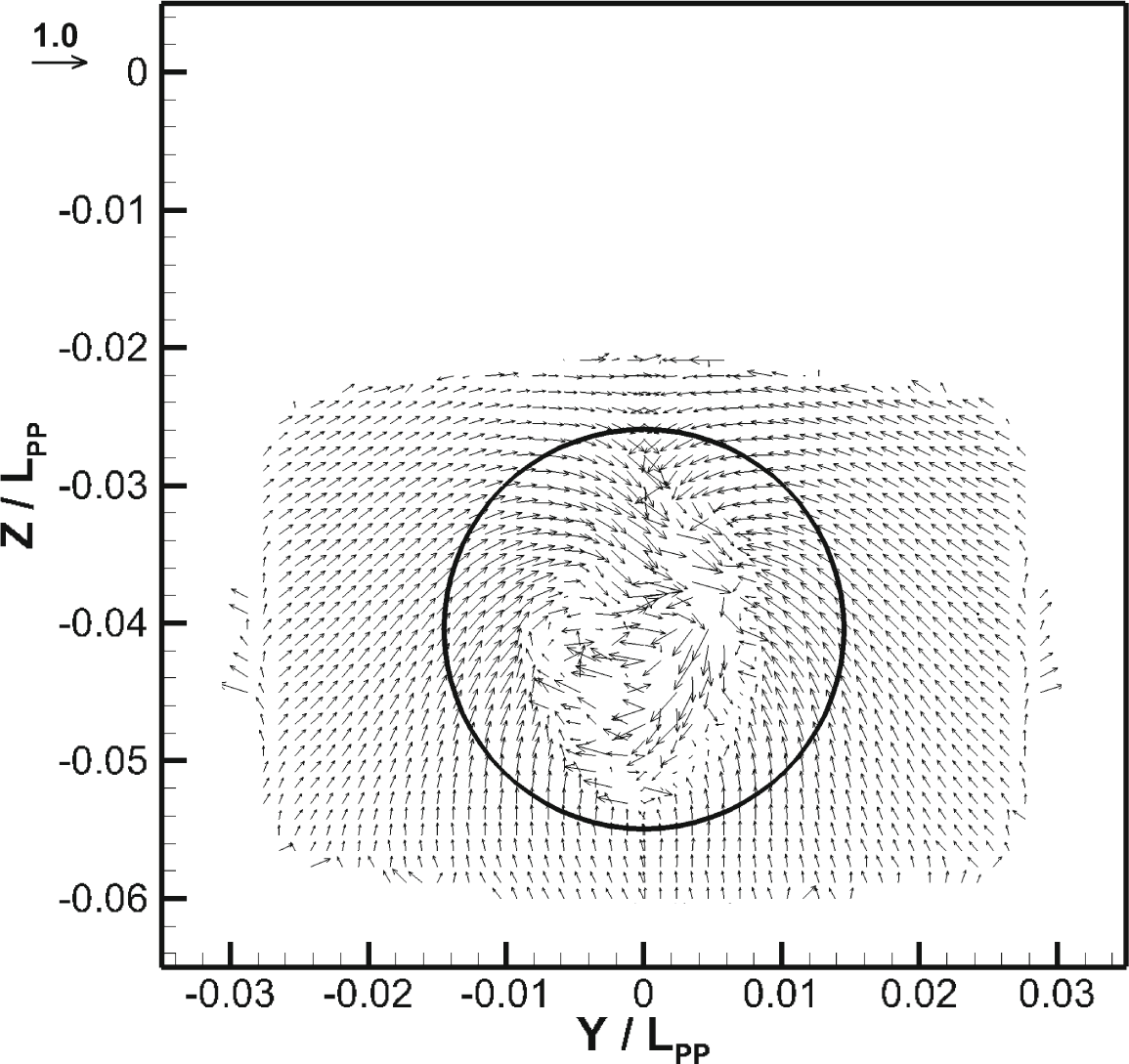}
        \captionsetup{width=0.9\linewidth}
        \caption{EFD}
    \end{subfigure}
    \begin{subfigure}[t]{0.32\linewidth}
        \includegraphics[width=1.0\linewidth]{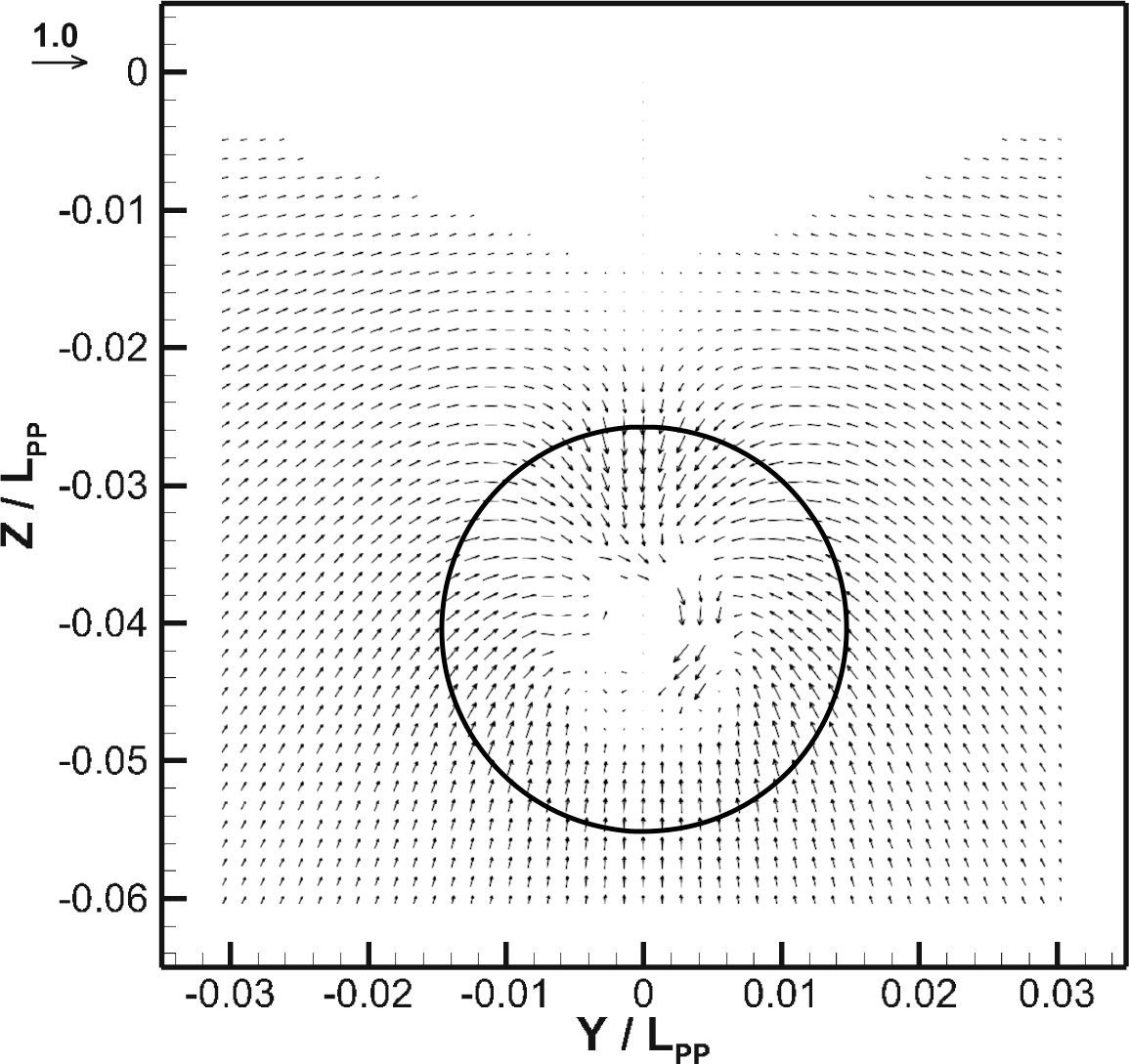}
        \captionsetup{width=0.9\linewidth}
        \caption{S1 (MRF, $n_{\mathrm{MRF}}/n = 0.0$)}
    \end{subfigure}
    \begin{subfigure}[t]{0.16\linewidth}
        \hfill
    \end{subfigure}
    \begin{subfigure}[t]{0.32\linewidth}
        \includegraphics[width=1.0\linewidth]{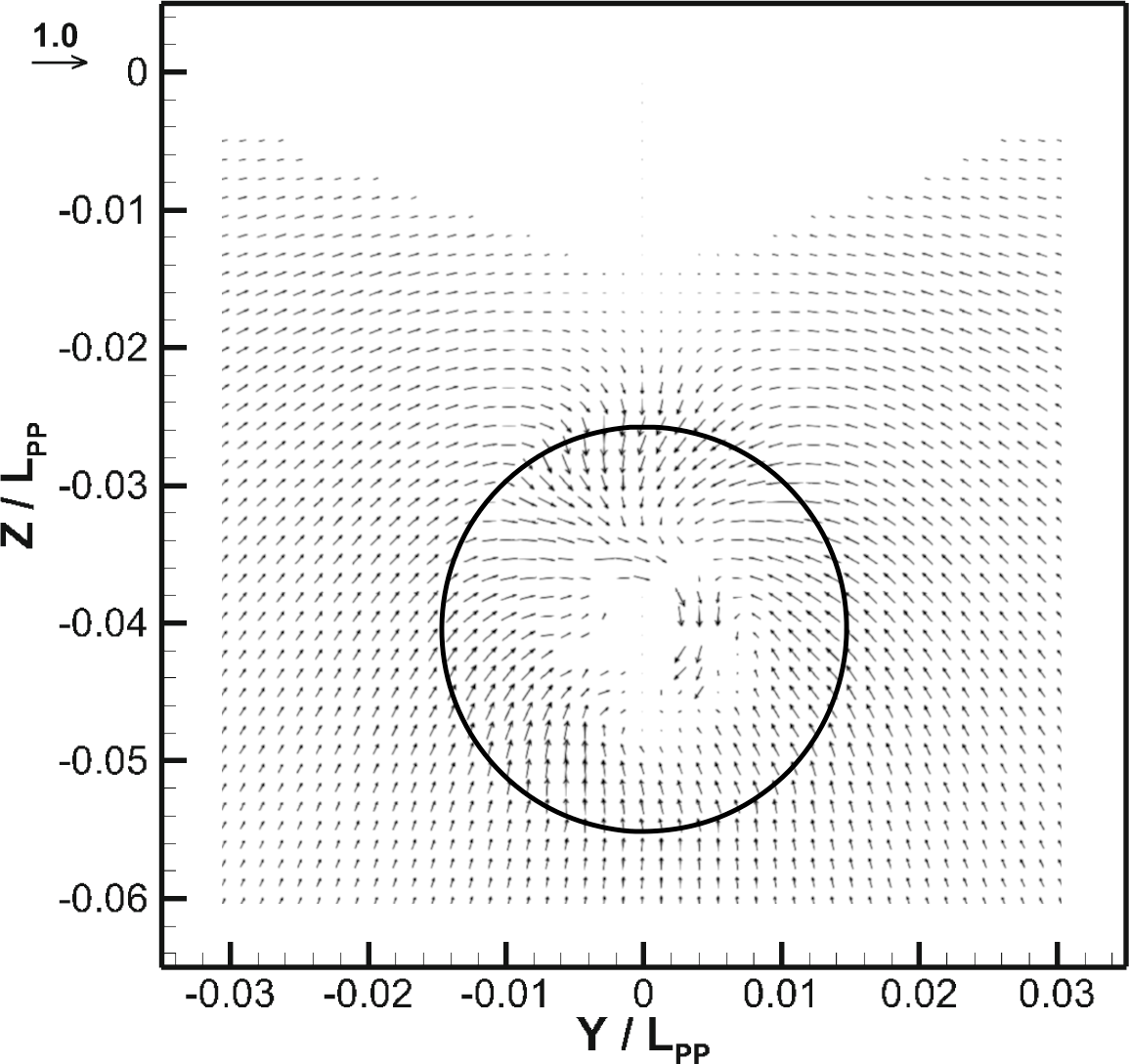}
        \captionsetup{width=0.9\linewidth}
        \caption{S3 (MRF, $n_{\mathrm{MRF}}/n = 0.5$)}
    \end{subfigure}
    \begin{subfigure}[t]{0.32\linewidth}
        \includegraphics[width=1.0\linewidth]{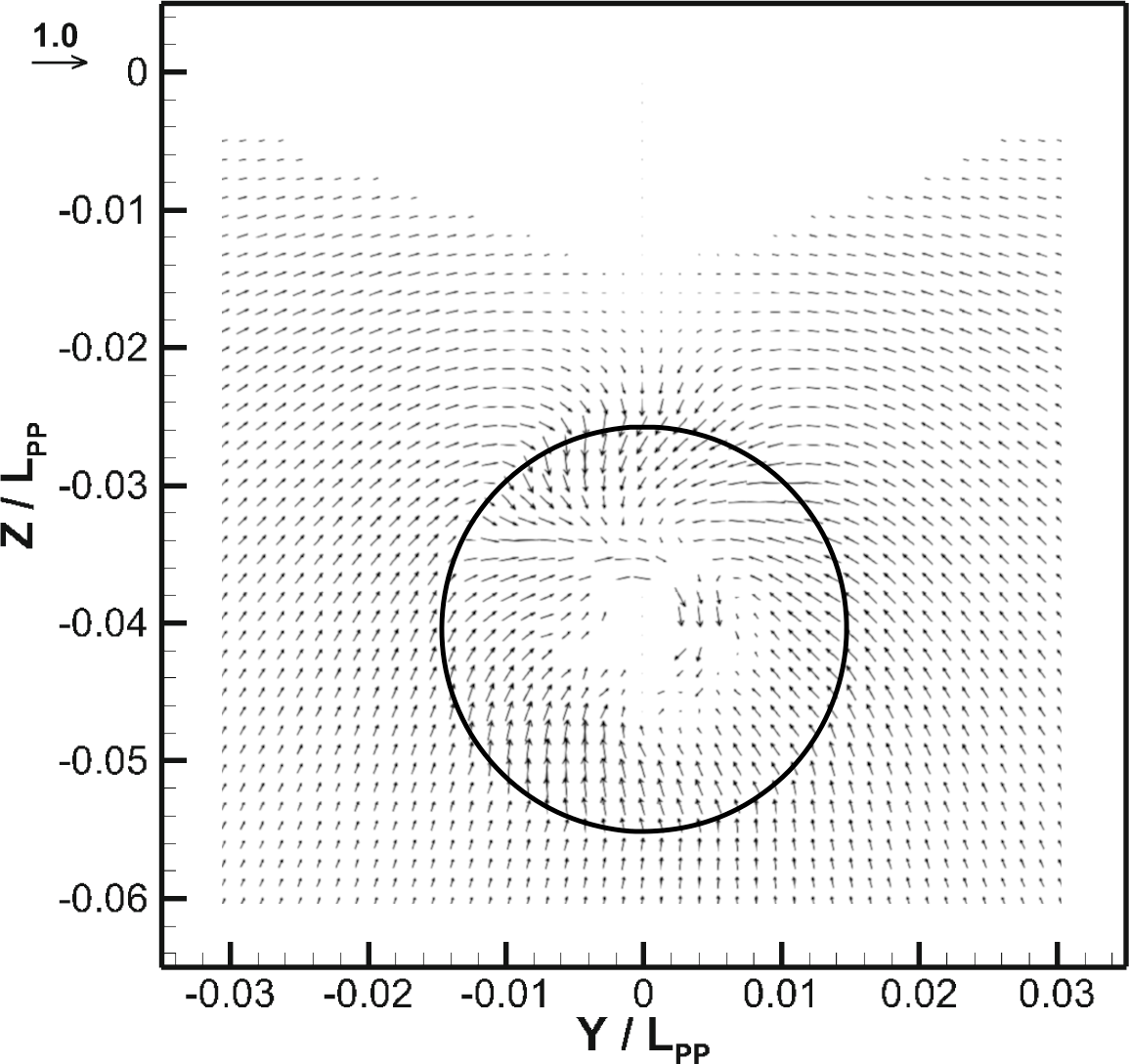}
        \captionsetup{width=0.9\linewidth}
        \caption{S4 (MRF, $n_{\mathrm{MRF}}/n = 0.75$)}
    \end{subfigure}
    \begin{subfigure}[t]{0.32\linewidth}
        \includegraphics[width=1.0\linewidth]{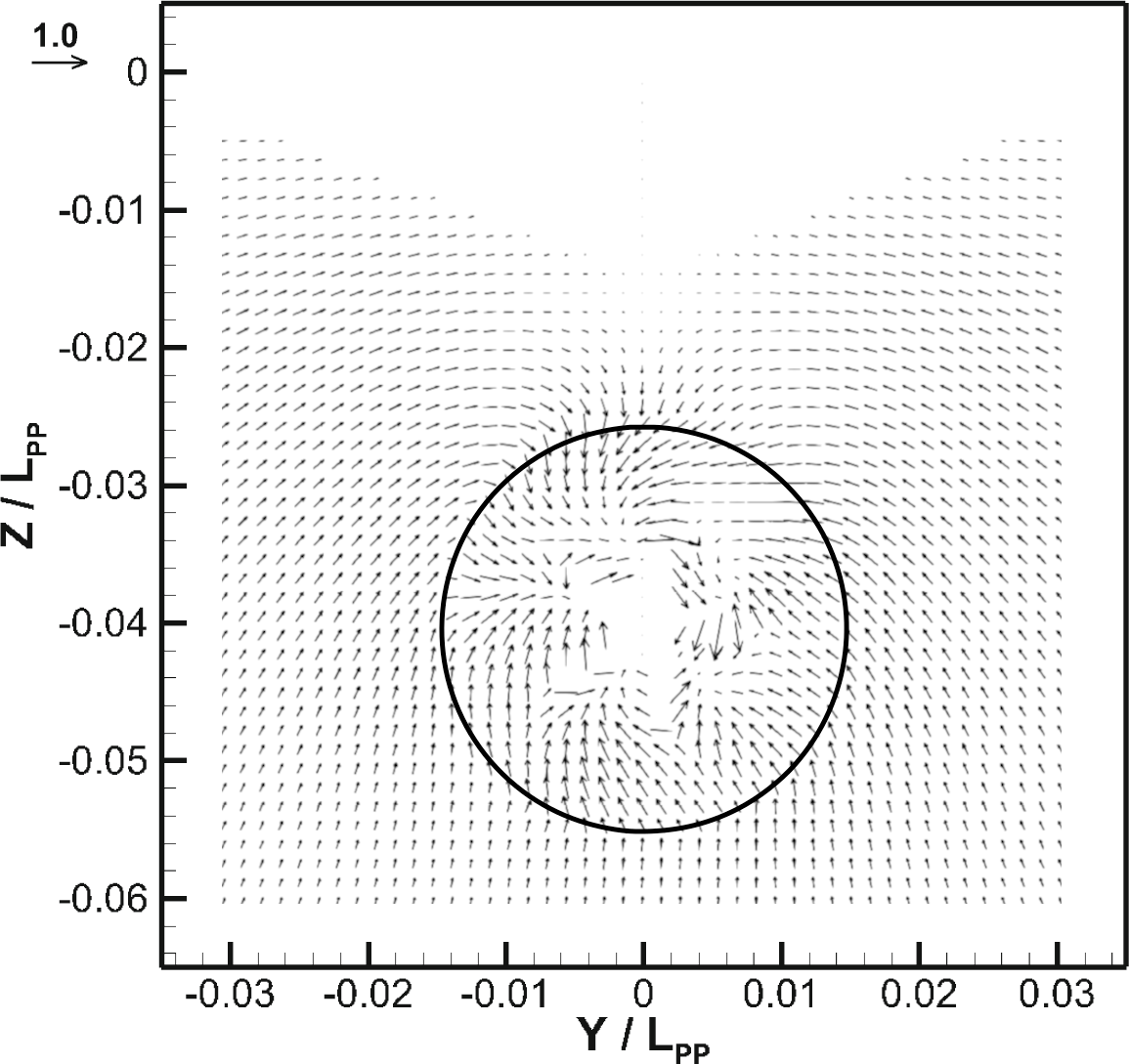}
        \captionsetup{width=0.9\linewidth}
        \caption{S5 (MRF, $n_{\mathrm{MRF}}/n = 1.0$)}
    \end{subfigure}
    \begin{subfigure}[t]{0.32\linewidth}
        \includegraphics[width=1.0\linewidth]{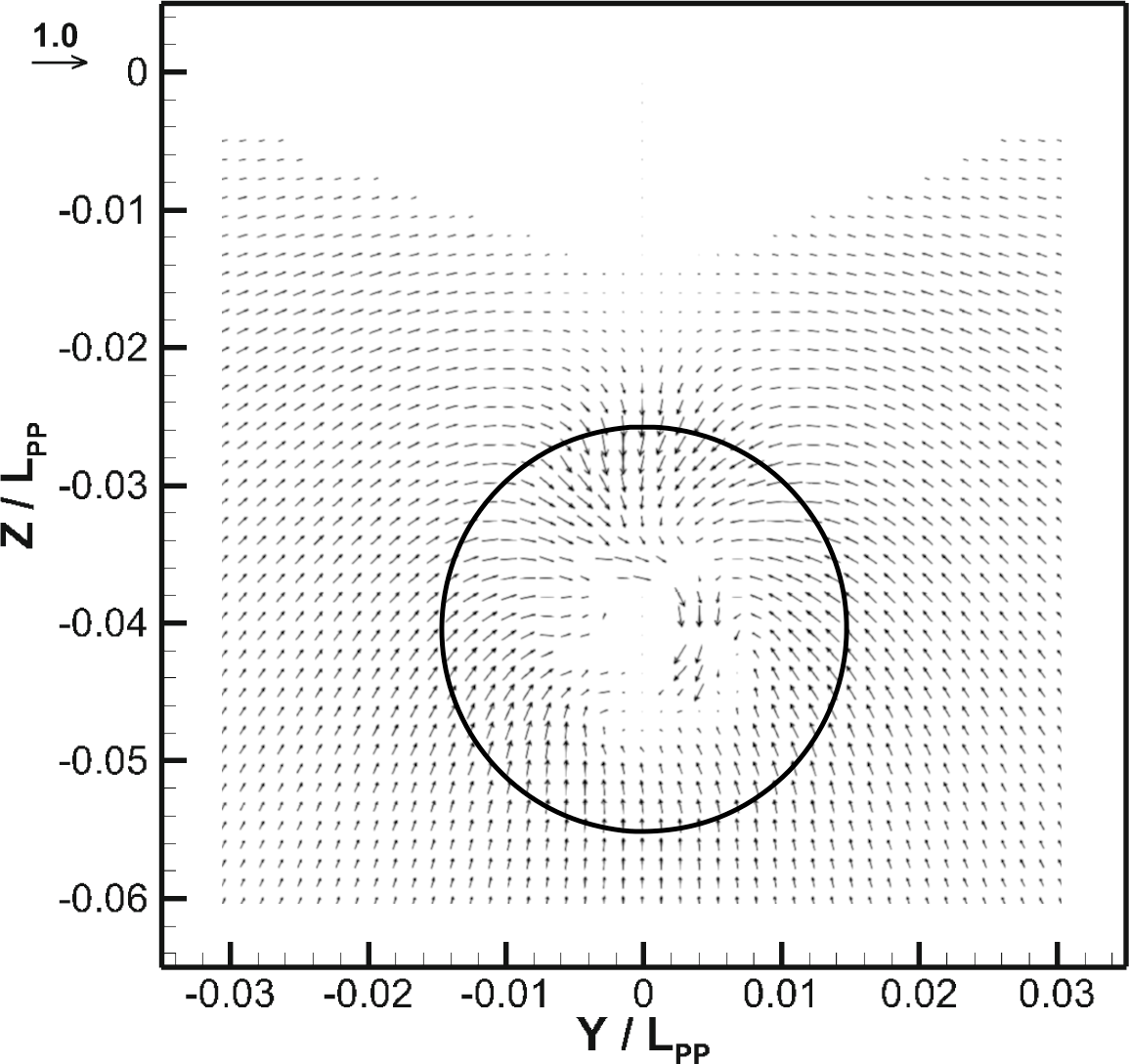}
        \captionsetup{width=0.9\linewidth}
        \caption{S7 (mMRF, $n_{\mathrm{MRF}}/n = 0.5$)}
    \end{subfigure}
    \begin{subfigure}[t]{0.32\linewidth}
        \includegraphics[width=1.0\linewidth]{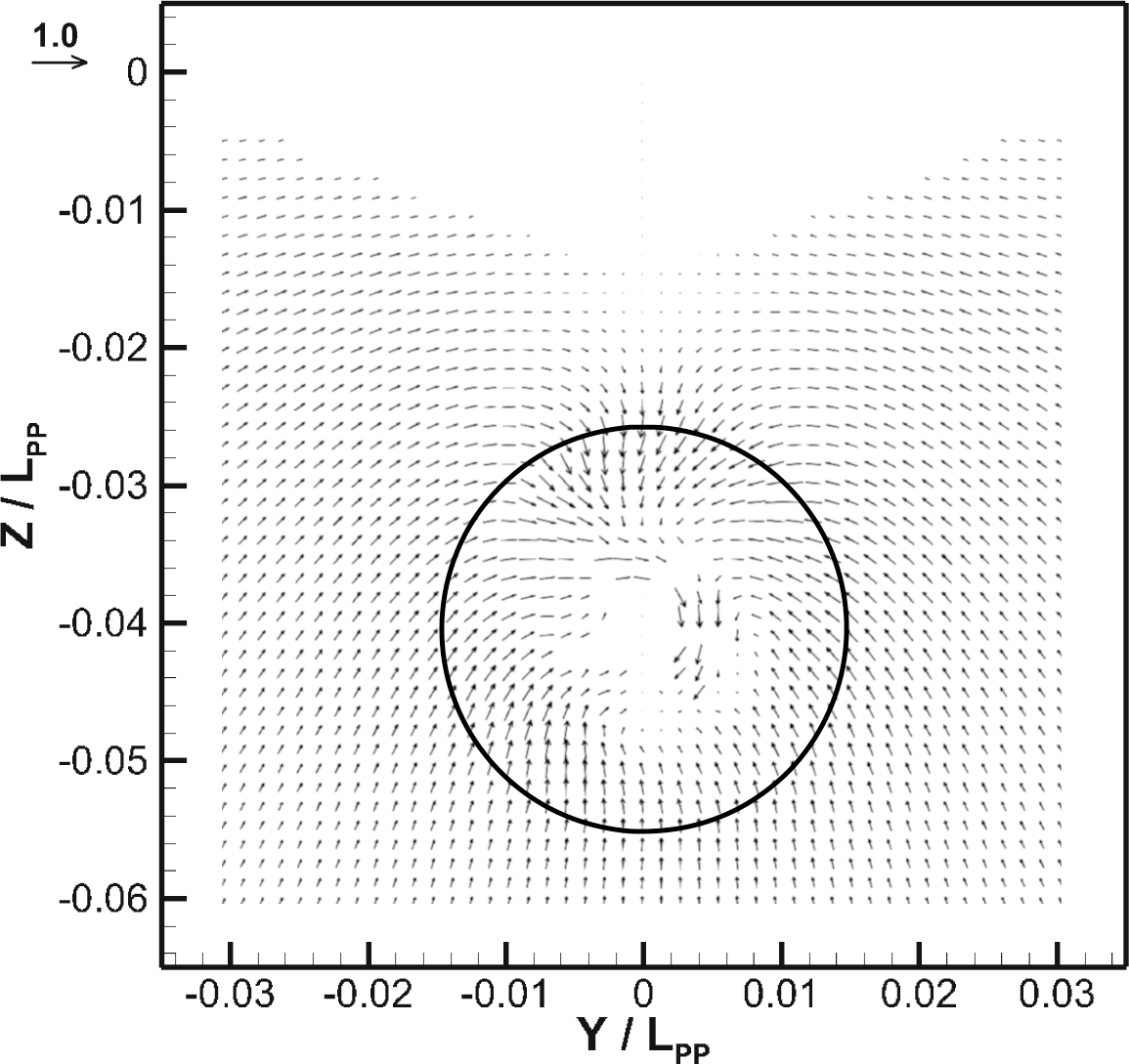}
        \captionsetup{width=0.9\linewidth}
        \caption{S8 (mMRF, $n_{\mathrm{MRF}}/n = 0.75$)}
    \end{subfigure}
    \begin{subfigure}[t]{0.32\linewidth}
        \includegraphics[width=1.0\linewidth]{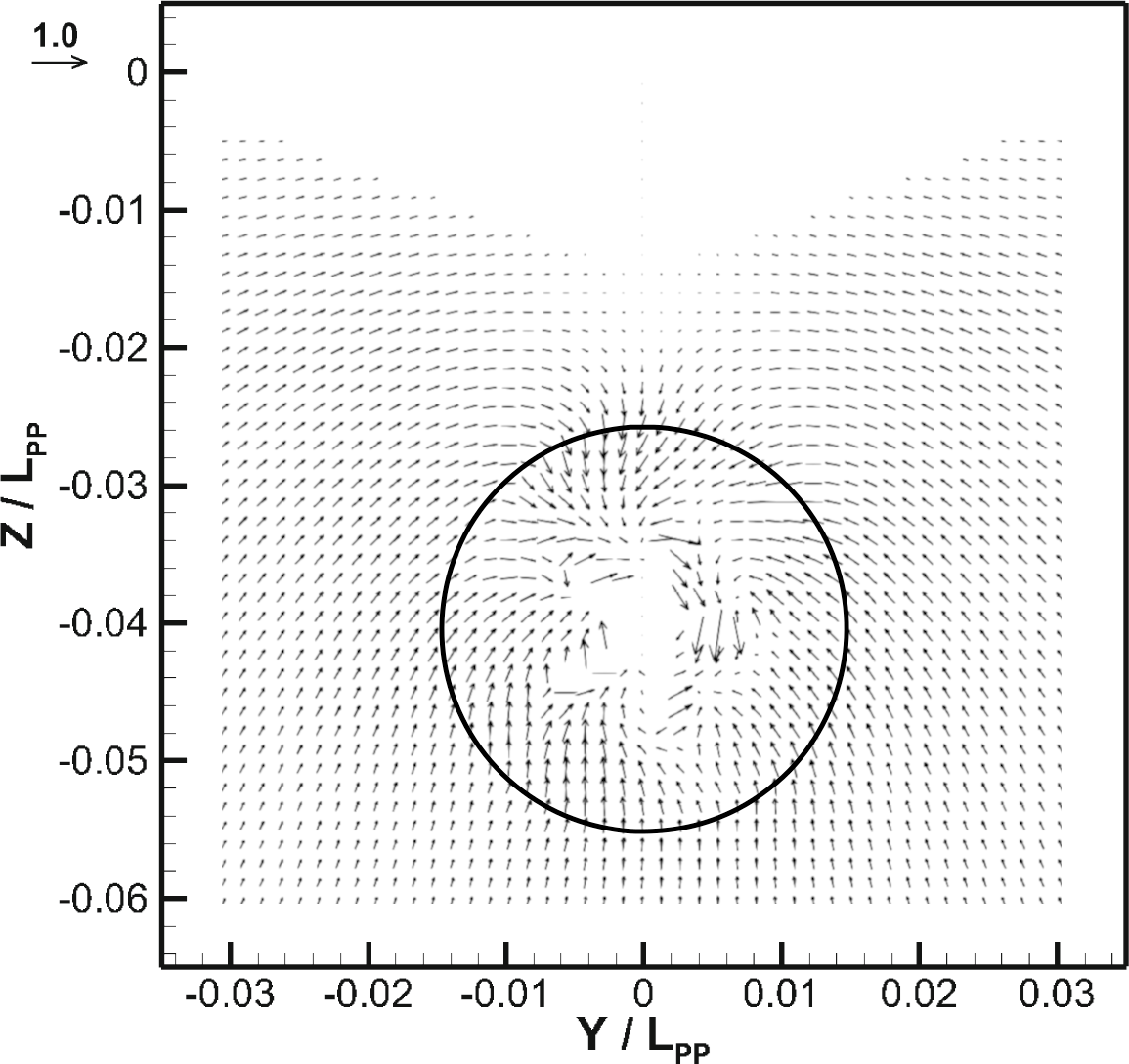}
        \captionsetup{width=0.9\linewidth}
        \caption{S9 (mMRF, $n_{\mathrm{MRF}}/n = 1.0$)}
    \end{subfigure}
    \caption{Cross-plane velocity vectors in the upstream plane ($x/L_{\mathrm{PP}}=0.0157$) for the EFD data, the SI solution, and the MRF and mMRF formulations at $n_{\mathrm{MRF}}/n=0.5$, $0.75$ and $1.0$.}
    \label{fig:cross_x1}
\end{figure}
For completeness, the axial velocity contours and the in-plane (cross-plane) velocity vectors at the two characteristic sections discussed in Section~\ref{sec:results} are collected here. Figure~\ref{fig:contour_x0} shows the axial velocity contours and Fig.~\ref{fig:cross_x0} the corresponding cross-plane vectors at the downstream plane $x/L_{\mathrm{PP}}=0$, while Fig.~\ref{fig:contour_x1} and Fig.~\ref{fig:cross_x1} show the same quantities at the upstream plane $x/L_{\mathrm{PP}}=0.0157$, each comparing EFD, the SI reference, and the MRF and mMRF formulations at the rotation ratios $n_{\mathrm{MRF}}/n=0.5$, $0.75$ and $1.0$. They support the trends reported for the disc-averaged deviation in Section~\ref{sec:results}: higher MRF ratios lead to larger deviations from the SI reference, which itself closely follows, though does not exactly reproduce, the EFD measurements, and the mMRF formulation consistently outperforms the classical MRF approach.
\FloatBarrier

\bibliography{
    refs/ref,
    refs/ref_ittc,
    refs/ref_imo,
    refs/ref_cfd
}

@book{hino-2020,
    title     = {Numerical Ship Hydrodynamics: An Assessment of the Tokyo 2015 Workshop},
    editor    = {Hino, T. and Stern, F. and Larsson, L. and Visonneau, M. and Hirata, N. and Kim, J.},
    doi       = {https://doi.org/10.1007/978-3-030-47572-7},
    publisher = {Springer Cham},
    year      = {2020},
    isbn      = {978-3-030-47572-7}
}

@article{durasevic-2022,
    title={Partially rotating grid method for self-propulsion calculations with a double body ship model},
    author={{\DJ}urasevi{\'c}, S. and  Gatin, I. and  Uroi{\'c}, T. and Jasak, H.}, 
    journal={Ocean Engineering},
    volume={266},
    pages={113105},
    year={2022},
    doi={10.1016/j.oceaneng.2022.113105},
    publisher={Elsevier}
}

@article{durasevic-2023,
    title = {Hydrodynamic performance of a full-scale ship with a Pre-Swirl Duct: A numerical study with partially rotating grid method},
    author={{\DJ}urasevi{\'c}, S. and  Gatin, I. and  Uroi{\'c}, T. and Jasak, H.}, 
    journal={Ocean Engineering},
    volume={283},
    pages = {115049},
    year={2023},
    doi={10.1016/j.oceaneng.2023.115049},
    publisher={Elsevier}
}

@incollection{luo-1994,
    author    = {Luo, J.Y. and Issa, R.I. and Gosman, A.D.},
    title     = {Prediction of Impeller Induced Flows in Mixing Vessels Using Multiple Frames of Reference},
    booktitle = {Institution of Chemical Engineers Symposium Series},
    volume    = {136},
    pages     = {549--556},
    year      = {1994},
    publisher = {Institution of Chemical Engineers}
}

@article{blades-2007,
    author = {Blades, Eric L. and Marcum, David L.},
    title = {A sliding interface method for unsteady unstructured flow simulations},
    journal = {International Journal for Numerical Methods in Fluids},
    volume = {53},
    number = {3},
    pages = {507-529},
    doi = {https://doi.org/10.1002/fld.1296},
    url = {https://onlinelibrary.wiley.com/doi/abs/10.1002/fld.1296},
    eprint = {https://onlinelibrary.wiley.com/doi/pdf/10.1002/fld.1296},
    year = {2007}
}

@inproceedings{wang-2019,
    author       = {Jianhua Wang and Decheng Wan},
    title        = {Numerical Simulations of Viscous Flows around {JBC} Ship Using Different Turbulence Models},
    booktitle    = {Proceedings of the 11th International Workshop on Ship and Marine Hydrodynamics},
    year         = {2019},
    doi          = {https://doi.org/10.15480/882.3364},
    address      = {Hamburg, Germany}
}

@techreport{huckins-1974,
    author       = {{Huckins III}, Earle K. and Turner, Richard E.},
    title        = {A General Form of the Co-Moving Tensorial Derivative},
    institution  = {NASA Langley Research Center},
    number       = {NASA TN D-7464},
    year         = {1974},
    url          = {https://ntrs.nasa.gov/citations/19740014139}
}

@inproceedings{benek-1983,
  author    = {Benek, J. A. and Steger, J. L. and Dougherty, F. C.},
  title     = {A Flexible Grid Embedding Technique with Application to the {Euler} Equations},
  booktitle = {Proceedings of the 6th AIAA Computational Fluid Dynamics Conference},
  year      = {1983},
  pages     = {373--382},
  address   = {Danvers, MA},
  month     = jul,
  note      = {AIAA Paper 83-1944},
  doi       = {10.2514/6.1983-1944}
}

@inproceedings{krasilnikov-2013,
    author       = {Vladimir I. Krasilnikov},
    title        = {Self-Propulsion {RANS} Computations with a Single-Screw Container Ship},
    booktitle    = {Proceedings of the Third International Symposium on Marine Propulsors (SMP'13)},
    editor       = {Jonathan Binns and Renee Brown and Neil Bose},
    organization = {Australian Maritime College, University of Tasmania},
    address      = {Launceston, Tasmania, Australia},
    year         = {2013},
    pages        = {430--438},
    url          = {https://www.marinepropulsors.com/proceedings/2013/9B.2.pdf}
}

@article{kawamura-1997,
    title        = {Numerical simulation of the flow about self-propelling tanker models},
    author       = {Kawamura, Takafumi and Miyata, Hideaki and Mashimo, Kohji},
    journal      = {Journal of Marine Science and Technology},
    volume       = {2},
    number       = {4},
    pages        = {245--256},
    year         = {1997},
    publisher    = {Springer},
    doi          = {10.1007/BF02491531},
}

@article{bakica-2019,
    title = {Accurate assessment of ship-propulsion characteristics using {CFD}},
    journal = {Ocean Engineering},
    volume = {175},
    pages = {149-162},
    year = {2019},
    issn = {0029-8018},
    doi = {https://doi.org/10.1016/j.oceaneng.2018.12.043},
    url = {https://www.sciencedirect.com/science/article/pii/S0029801818312782},
    author = {Andro Bakica and Inno Gatin and Vuko Vukčević and Hrvoje Jasak and Nikola Vladimir}
}

@article{taylor-1923,
    author    = {Taylor, G. I.},
    title     = {Stability of a Viscous Liquid Contained between Two Rotating Cylinders},
    journal   = {Philosophical Transactions of the Royal Society of London. Series A},
    volume    = {223},
    pages     = {289--343},
    year      = {1923},
    doi       = {10.1098/rsta.1923.0008}
}

@article{menter-2003,
    author  = {Menter, F. R. and Kuntz, M. and Langtry, R.},
    title   = {Ten years of industrial experience with the {SST} turbulence model},
    journal = {Turbulence, Heat and Mass Transfer},
    volume  = {4},
    pages   = {625--632},
    year    = {2003}
}

@article{demirdzic-1988,
  author    = {Demird{\v{z}}i{\'{c}}, I. and Peri{\'{c}}, M.},
  title     = {Space conservation law in finite volume calculations of fluid flow},
  journal   = {International Journal for Numerical Methods in Fluids},
  year      = {1988},
  volume    = {8},
  number    = {9},
  pages     = {1037--1050},
  doi       = {10.1002/fld.1650080906}
}

@article{hirt-1981,
    author  = {C. W. Hirt and B. D. Nichols},
    title   = {Volume of Fluid ({VOF}) Method for the Dynamics of Free Boundaries},
    journal = {Journal of Computational Physics},
    volume  = {39},
    number  = {1},
    pages   = {201--225},
    year    = {1981},
    doi     = {10.1016/0021-9991(81)90145-5}
}

@article{rung-2009,
  title={Challenges and perspectives for maritime {CFD} applications},
  author={Rung, T and W{\"o}ckner, K and Manzke, M and Brunswig, J and Ulrich, C and St{\"u}ck, A},
  journal={Jahrbuch der Schiffbautechnischen Gesellschaft},
  volume={103},
  pages={127--39},
  year={2009}
}

@misc{imo-2014,
    author  = {{International Maritime Organization}},
    shortauthor = {IMO},
    title   = {2014 Guidelines on the Method of Calculation of the Attained Energy Efficiency Design Index ({EEDI}) for New Ships},
    note    = {{IMO} Resolution MEPC.245(66)},
    year    = {2014},
    url     = {https://wwwcdn.imo.org/localresources/en/OurWork/Environment/Documents/245(66).pdf}
}

@misc{imo-2021,
    author  = {{International Maritime Organization}},
    shortauthor = {IMO},
    title   = {2021 Guidelines on the Method of Calculation of the Attained Energy Efficiency Existing Ship Index ({EEXI})},
    note    = {{IMO} Resolution MEPC.333(76)},
    year    = {2021},
    url     = {https://wwwcdn.imo.org/localresources/en/OurWork/Environment/Documents/Air%20pollution/MEPC.333(76).pdf}
}

@misc{ittc-perf-pred,
    author  = {{International Towing Tank Conference}},
    shortauthor = {ITTC},
    title   = {1978 {ITTC} Performance Prediction Method},
    note    = {{ITTC} Recommended Procedure 7.5-02-03-01.4},
    year    = {2021},
    url     = {https://www.ittc.info/media/9872/75-02-03-014.pdf}
}

@misc{ittc-uncertainty,
    author  = {{International Towing Tank Conference}},
    shortauthor = {ITTC},
    title   = {Uncertainty Analysis in {CFD} Verification and Validation, Methodology and Procedures},
    note    = {{ITTC} Recommended Procedure 7.5-03-01-01},
    year    = {2021},
    url     = {https://www.ittc.info/media/9765/75-03-01-01.pdf}
}

@misc{ittc-uncertainty-2,
    author  = {{International Towing Tank Conference}},
    title   = {Uncertainty Analysis, Example for Open Water Test},
    institution  = {International Towing Tank Conference},
    number       = {7.5-02-03-02.2},
    year         = {2014},
    note         = {{ITTC} Recommended Procedure 7.5-02-03-02.2, Revision 01}
}

\end{document}